\documentclass[]{interact}

\usepackage[numbers,sort&compress]{natbib}% Citation support using natbib.sty
\usepackage[cp1250]{inputenc}
\usepackage{enumerate}
\usepackage{amsfonts}
\usepackage{graphicx}    %for eps files
\usepackage{algorithmic} 
\usepackage{algorithm} 
\usepackage{amsmath,amsthm}
\usepackage{subfigure}
\usepackage{pdflscape}
\usepackage{ctable}
\usepackage[all,cmtip]{xy}
\usepackage{fancybox}
\usepackage{epstopdf}
\usepackage{array,arydshln}
\usepackage{lineno}
\usepackage[titletoc,title]{appendix}

\bibpunct[, ]{[}{]}{,}{n}{,}{,}% Citation support using natbib.sty
\theoremstyle{definition}% Theorem-like structures
\newtheorem{thm}{Theorem}[section]
\newtheorem{lemma}[thm]{Lemma}
\newtheorem{corollary}[thm]{Corollary}

\newtheorem{dfn}[thm]{Definition}

\newtheorem{pf}{Proof}

\newtheorem{rmk}{Remark}

\newcommand*{\vcenteredhbox}[1]{\begingroup
\setbox0=\hbox{#1}\parbox{\wd0}{\box0}\endgroup}

\newcommand{\argmax}{\operatornamewithlimits{argmax}}
\newcommand{\argmin}{\operatornamewithlimits{argmin}}
\newcommand{\avg}{\operatornamewithlimits{avg}}
\newcommand*{\uu}{\mathbf{u}}

\newcommand*{\VV}{\mathcal{V}}
\newcommand*{\EE}{\mathcal{E}}
\newcommand*{\FF}{\mathcal{F}}
\newcommand*{\NNN}{\mathcal{N}}
\newcommand*{\I}{\mathbb{I}}
\newcommand*{\III}{\mathcal{I}}

\newcommand*{\HAC}{C_{(\VV,\EE,\lambda)}}
\newcommand*{\FR}{\mathcal{F}_{\tilde{\IR}^+_0}}
\newcommand*{\RR}{\tilde{\mathbb{R}}^+_0}
\newcommand*{\PSI}{\tilde{\Psi}^2_{+\infty}} %nestables
\newcommand*{\NF}{\mathcal{N}^2_{\mathcal{F}}} 
\newcommand*{\FALL}{\FF_{\textrm{all}}} 
\newcommand*{\NALL}{\mathcal{N}_{\textrm{all}}}
\newcommand*{\FNEST}{\FF^2_{\textrm{nest}}}

\newcommand*{\eps}{\varepsilon}

\newcommand*{\IN}{\mathbb{N}}

\newcommand*{\IR}{\mathbb{R}}

\begin{document}

\articletype{ORIGINAL MANUSCRIPT}

\title{On structure, family and parameter estimation of hierarchical Archimedean copulas$^*$\footnote{$^*$This work has been submitted to Journal of Statistical Computation
and Simulation.}}

\author{
\name{Jan G\'{o}recki\textsuperscript{a,**}\thanks{$^{**}$ Corresponding author. Email: gorecki@opf.slu.cz}, Marius Hofert\textsuperscript{b} and Martin Hole\v{n}a\textsuperscript{c}}
\affil{
\textsuperscript{a}Department of Informatics, SBA in Karvin\'{a}, Silesian University in Opava, Univerzitn\'{i} n\'{a}m\v{e}st\'{i} 1934/3, 733 40 Karvin\'{a}, Czech Republic; \\
\textsuperscript{b}Department of Statistics and Actuarial Science,
University of Waterloo, 200 University Avenue West,
Waterloo, ON, Canada;\\
\textsuperscript{c}Institute of Computer Science, Academy of Sciences of the Czech Republic,
Pod vod\'{a}renskou v\v{e}\v{z}\'{i} 271/2, 182 07 Praha, Czech Republic
}
}

\maketitle

\begin{abstract}
Research on structure determination and parameter estimation of hierarchical Archimedean copulas (HACs) has so far mostly focused on 
the case in which all appearing Archimedean copulas belong to the same Archimedean family. The present work addresses this issue and proposes a new approach for estimating HACs that involve different Archimedean families. It is based on employing goodness-of-fit test statistics directly into HAC estimation. The approach is summarized in a simple algorithm, its theoretical justification is given and its applicability is illustrated by several experiments, which include estimation of HACs involving up to five different Archimedean families.
\end{abstract}


%Word count: 24~680

\section{Introduction} \label{sec:intro}

%Studying relationships among random quantities is an important task, which attracts researchers from many different fields. Having a dataset collected, the relationships among the observed variables can be studied by means of an appropriate measure of stochastic dependence. Under assumption of a multivariate continuous distribution of the variables, the famous Sklar's theorem \cite{Skl59} can be used to decompose the distribution in two components. While the first component describes the distributions of the univariate margins, the second component describes the copula of the distribution containing the whole information about the relationship among the variables. Thus, studying relationships among the random variables can be without loss of generality restricted to studying the copula.

%Despite the fact that copulas have most success in finance, e.g., \cite{Li01}, they are increasingly adopted also in other fields, where they are used due to their effective mathematical ability to capture even very complex dependence structures among variables. We can see applications of copulas in  water-resources and hydro-climatic analysis \cite{Gen07,Kao09,Kao08,Kuh07,Mai08}, gene analysis \cite{Las12,Yua08}, cluster analysis \cite{Cuv05,Koj10,Rey12} or in evolutionary algorithms, particularly in the estimation of distribution algorithms \cite{Gon12,Wan12}. For an illustrative example, we refer to \cite{Kao09}, where the task for anomaly detection in climate that incorporates complex spatio-temporal dependencies is solved using copulas.
 
Copulas, i.e., multivariate distribution functions with standard uniform univariate margins, establish a connection between general joint distribution functions (d.f.s) and their univariate margins through  Sklar's Theorem; see \cite{Skl59}. A popular class of copulas are Archimedean copulas (ACs)  \cite[p. 109]{Nel06} or their asymmetric generalization, \emph{hierarchical Archimedean copulas} (HACs); note that HACs are also called nested Archimedean copulas, see, e.g., \cite[p. 87]{Joe97}. For a motivation addressing the advantages of hierarchical over exchangeable Archimedean copulas in applications, see, e.g., \cite{Hofert13}. HACs are popular due to their flexibility but conveniently limited number of parameters. Fast sampling methods have already been proposed for them, see, e.g., \cite{Hof11,Hof2012stoch}, whereas efficient methods for their estimation are still a matter of research, see, e.g., \cite{Okh13,Gor14structure,goreckihofertholena2014a,goreckihofertholena2016approachjiis}, which concern both structure determination and parameter estimation of HACs. 
This research is largely restricted to the most commonly used HACs that are constructed by nesting of a single one-parametric family of Archimedean copulas. 
In what follows, we call a HAC consisting of ACs of the same family a \emph{homogeneous} HAC, otherwise, a \emph{heterogeneous} HAC. 
Although the above-mentioned papers briefly mention the possibility of heterogeneous HACs estimation, they are focused mainly on the case of homogeneous HACs and all reported experiments involve only homogeneous HACs. 
From this point of view, estimation of heterogeneous HACs is still an open task and our work is an attempt that  addresses this challenging case in detail. Note that the generalization of homogeneous HACs to heterogeneous HACs brings several advantages, such as the possibility of having bivariate margins from different Archimedean copula families rather than only having different  parameters but belonging to the same parametric Archimedean family. This is of interest when modeling pairwise tail dependence, see \cite{embrechts2016}, but also when Archimedean families are restricted in their range concerning concordance (e.g., the family of Ali-Mikhail-Haq only allows for Kendall's tau in $[0, \frac{1}{3})$).

In this work, we propose a new approach to estimating heterogeneous HACs, which is based on extending  a homogeneous HAC estimation approach by involving goodness-of-fit test statistics directly into the estimation. Whereas such an extension is rather simple and straightforward, assuring that a resulting estimate satisfies the \emph{sufficient nesting condition}, which guarantees that a proper copula results, is a substantially more complex problem compared to its counterpart for the homogeneous case. However, we provide an algorithm in  pseudo-code and  its theoretical justification represented by Theorem \ref{thm:alg_returns_HAC}, which shows that under weak conditions, the algorithm returns a function satisfying the sufficient nesting condition. Those weak conditions are additionally addressed by Theorems \ref{thm:L12} and \ref{thm:L123}, which explicitly present two general scenarios under which the algorithm guarantees to satisfy the sufficient nesting condition. 
We also show that the proposed extension to the heterogeneous case is not restricted to a specific homogenous HAC estimation approach and by application to another homogeneous estimator, we can easily construct an alternative heterogeneous HAC estimator, i.e., we actually introduce a whole new framework of HAC estimators.

Complementary to this theoretical part, the validity of the proposed approach is illustrated by experiments on simulated data which involve heterogeneous HAC models with up to five different parametric families of Archimedean generators. 
The results of these experiments confirm that the proposed approach is able to properly determine the structure and estimate the parameters of a HAC, as well as properly estimating different parametric families of Archimedean generators of a heterogeneous HAC.

Moreover, this work also contributes to the topic that is frequently called \emph{collapsing} of HAC structures, which, informally speaking, serves to turning binary HAC structures (binary trees), which often result from estimation processes, to non-binary ones, allowing to access all possible HAC structures.
Our contribution it to introduce a novel approach, which, by contrast to the collapsing approaches proposed, e.g., in \cite{Okh13,uyttendaele2016estimation}, does not force the user to specify any threshold \emph{before} a collapsing process begins and lets one to choose an appropriate collapsed HAC \emph{after} the collapsing process has been finished.
Complementary to this approach, an automated heuristic procedure for choosing an appropriate collapsed HAC is proposed, which is again not dependent on any pre-defined threshold.  Also, a new \emph{re-estimation} procedure for re-estimating the parameters of a collapsed HAC is introduced and compared to the approach proposed in \cite{uyttendaele2016estimation}. All these contributions are then intensively tested and the obtained results are reported in the experimental part of this work.

This work also addresses a close relationship between an existing HAC structure estimator and the sufficient nesting condition. We show that this structure estimator inherently assures the latter condition, and that such a property is unique among existing structure estimators. This finding is also theoretically justified by Theorem \ref{thm:alg_homo_returns_HAC}, which shows that by using the considered structure estimator, a proper homogeneous HAC can be obtained under relatively weak conditions.

As a by-product, two existing approaches for homogeneous HAC estimation are experimentally compared, showing not only their precision in structure and parameter estimation, but also their robustness against misspecification of the underlying families.

This work is structured as follows. Sections \ref{sec:ACs} and \ref{sec:HACs} recall the necessary theoretical concepts concerning ACs and HACs, respectively. An AC estimator based on the inversion of Kendall's tau is recalled in Section \ref{sec:param_estimation}, and an aggregated goodness-of-fit test statistic is introduced in Section \ref{sec:gof}.
Section \ref{sec:approach} represents the theoretical part of this work. In Section \ref{sec:collapsing}, the new approach to collapsing is addressed, in Section \ref{sec:structure_estimation},  the relationship between the existing structure estimator and the sufficient nesting condition is addressed, in Section \ref{sec:homo_HAC_estim}, an existing approach to homogeneous HAC estimation is recalled and 
the findings from Section \ref{sec:structure_estimation} are applied to this estimator, which results in Theorem \ref{thm:alg_homo_returns_HAC}. Finally, in Section \ref{sec:hetero_HAC_estim}, our new approach to heterogeneous HAC estimation is introduced. The proofs of all theorems and lemmas are presented in Appendix.  Section \ref{sec:exps} describes the design and the results of the performed experiments and Section \ref{sec:conclusion} concludes. Note that a part of the experimental results is included in an attachment.

\section{Archimedean Copulas} \label{sec:ACs}

To construct ACs in arbitrary dimensions, we need the notion of  an \emph{Archimedean generator} and of \emph{complete monotonicity}.

\begin{dfn}
\label{def:arch_gen}
An \emph{Archimedean generator} (shortly, \emph{generator}) is a continuous, non-increasing function $\psi : [0,+\infty] \rightarrow [0,1]$, which satisfies $\psi(0) = 1,~ \psi(+\infty) = \lim_{t \rightarrow +\infty}$ $\psi(t) = 0$ and which is strictly decreasing on $[0,\inf \{ t~|~\psi(t) = 0\}]$. We  denote the set of all generators by $\Psi$. If $\psi$ satisfies $(-1)^k \psi^{(k)}(t)$ $\geq 0,$ for all $k \in \mathbb{N}, ~t \in [0,+\infty)$, $\psi$ is called \emph{completely monotonic} (c.m.).  Note that $\psi^{(k)}$ denotes the $k$-th derivative of $\psi$. Finally, we denote the set of all c.m. generators by $\Psi_{+\infty}$.
\end{dfn}	

\begin{dfn}
\label{def:arch_cop}
Any $d$-dimensional  copula (simply $d$-copula) $C$ is an \emph{Archimedean copula} based on a generator $\psi \in \Psi$ (we denote it $d$-AC), if it admits the form
\begin{eqnarray}
\label{eq:arch_kopule}
C(\mathbf{u}) := C_{\psi}(\mathbf{u}) := \psi (\psi^-(u_1)+ ... + \psi^-(u_d)), \textbf{u} \in \mathbb{I}^d,
\end{eqnarray}
\noindent where $\psi^-:[0, 1] \rightarrow [0, +\infty]$ is defined by $\psi^-(s) = \inf \{t~|~\psi(t) = s\}, s \in \mathbb{I}$.
\end{dfn}

\noindent Note that if $\psi$ is strictly decreasing, $\psi^-$ is the ordinary inverse.

\begin{rmk} \label{rmk:exchangeable}
As can be seen from from their definition, ACs are invariant to permutation of their arguments, e.g., for any 2-AC $C$, $C(u_1, u_2) = C(u_2, u_1)$ for all $\uu \in \I^2$.
\end{rmk}

%As any function $C: ~\I^d \rightarrow \I$ that fulfills \eqref{eq:arch_kopule} is uniquely given by $\psi$, we will denote this function by $C_{\psi}$.
A condition sufficient for \eqref{eq:arch_kopule} to be a proper copula in arbitrary dimensions is stated in the following theorem.
\begin{thm} \cite{McNei09} If $\psi \in \Psi_{+\infty}$, then the function $C_{\psi}$ given by \eqref{eq:arch_kopule} is a copula for all $d \geq 2$.
\end{thm}

Note that in \cite{McNei09}, given $d > 2$, the authors present a condition that is both necessary and sufficient for $C_{\psi}$ to be a $d$-copula. In this work, we will however build HACs only from ACs based on completely monotonic generators. The reason for it, which is connected to a sufficient condition assuring that a HAC-like function is a proper copula, is addressed in detail in Section \ref{sec:HACs}.

%Assuming complete monotonicity, generators of ACs correspond to distribution functions on the positive real line via Laplace–Stieltjes transforms; see \cite{bernstein1929fonctions}. With the knowledge on how to sample those one-dimensional distributions, we are able to efficiently sample ACs in arbitrary dimensions, what is an essential property in simulations. 
Assuming complete monotonicity of generators implies that ACs are restricted to be models for positively dependent random vectors only, in other words, the pairwise rank correlations of such ACs are non-negative.
This follows from the fact that if a generator $\psi \in \Psi$ is c.m., then, according to \cite[p. 167]{widder1946}, $\psi$ is logarithmically convex. Thus $- \ln \psi$ is concave. Defining $\psi_{\Pi}(t) = \exp(-t)$ and considering that $C_{\psi_{\Pi}}$ is the independence copula, it follows from Theorem 2.3.1 in \cite{Hof10book} that $\psi_{\Pi}^{-1} \circ \psi$ is subadditive, which in turn implies that $C_{\psi}(\uu) \geq C_{\psi_{\Pi}}(\uu)$ for all $\uu \in \I^d$, 
and thus that pairwise rank correlations are non-negative; see Remark 2.3.2 in \cite{Hof10book}. %This fact closely relates to parameter ranges corresponding to the families of Archimedean generators we consider in this work.

In practice, one often works with families of generators (families of ACs). E.g., in \cite[p. 116]{Nel06}, the reader can find 22 c.m. families parametrized by a real parameter. 
%Even if \cite{Nel06} shows examples of generators with two-parameters, we restrict to by-far more popular one-parametric families of generators.
Generally, a one-parametric family of generators can be represented by a bivariate function $\Psi^{\Theta}, ~\Theta \subseteq \mathbb{R}$ such that for all $\theta \in \Theta$ the function $\Psi^{\Theta}(\cdot, \theta)$ is the generator from the considered family with the parameter $\theta$. This idea is formalized in the following definition.

\begin{dfn} \label{def:parametric_Arch_fam}
Let $\Theta \subseteq \IR$. If $\Psi^{\Theta} : [0, +\infty] \times \Theta \rightarrow \I$ is such that
\begin{eqnarray}
\label{eq:parametric_family}
\forall \theta \in \Theta,~\Psi^{\Theta}(\cdot, \theta) \in \Psi_{+\infty},
\end{eqnarray}
\noindent then $\Psi^{\Theta}$ is called \emph{parametric family} of Archimedean generators, and $\Theta$ is called its \emph{parameter range}. In order to formalize the relationship describing whether or not a given generator is from a given parametric family, we define a relationship (denoted $\bar{\in}$) between a generator $\psi \in \Psi_{+\infty}$ and a parametric family of Archimedean generators $\Psi^{\Theta}$ given by
\begin{eqnarray}
\label{eq:psi_belongs_to}
\psi~ \bar{\in}~ \Psi^{\Theta} \Leftrightarrow \exists \theta \in \Theta, \forall t \in [0, +\infty], ~\psi(t) = \Psi^{\Theta}(t, \theta).
\end{eqnarray}
\end{dfn}
\noindent Note that $\Psi^{\Theta}$ is a bivariate function, hence the term $\psi \in \Psi^{\Theta}$ cannot be used.

\begin{figure}[t]
	\centering
		\includegraphics[width=1\textwidth]{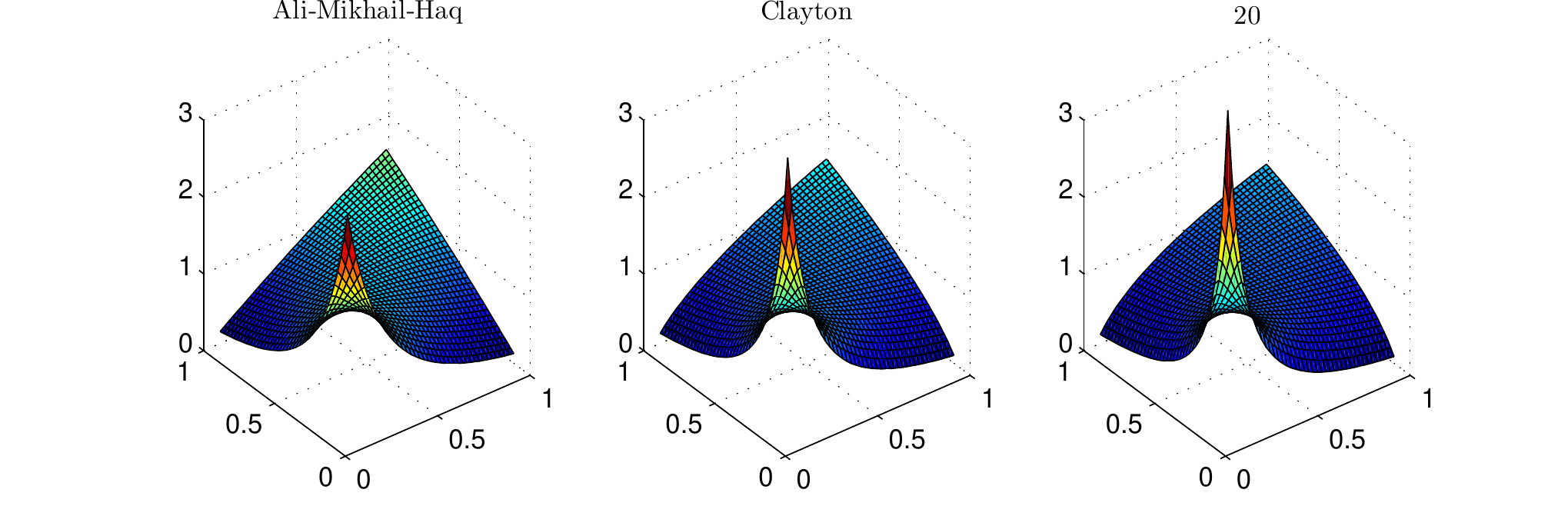}
		\caption{The density functions of 2-ACs belonging to the Ali-Mikhail-Haq family, the Clayton family and the family labeled 20. All depicted densities correspond to Kendall's tau = 0.25.}
	\label{fig:densities}
\end{figure}

\begin{figure}[th]
	\centering
		\includegraphics[width=1\textwidth]{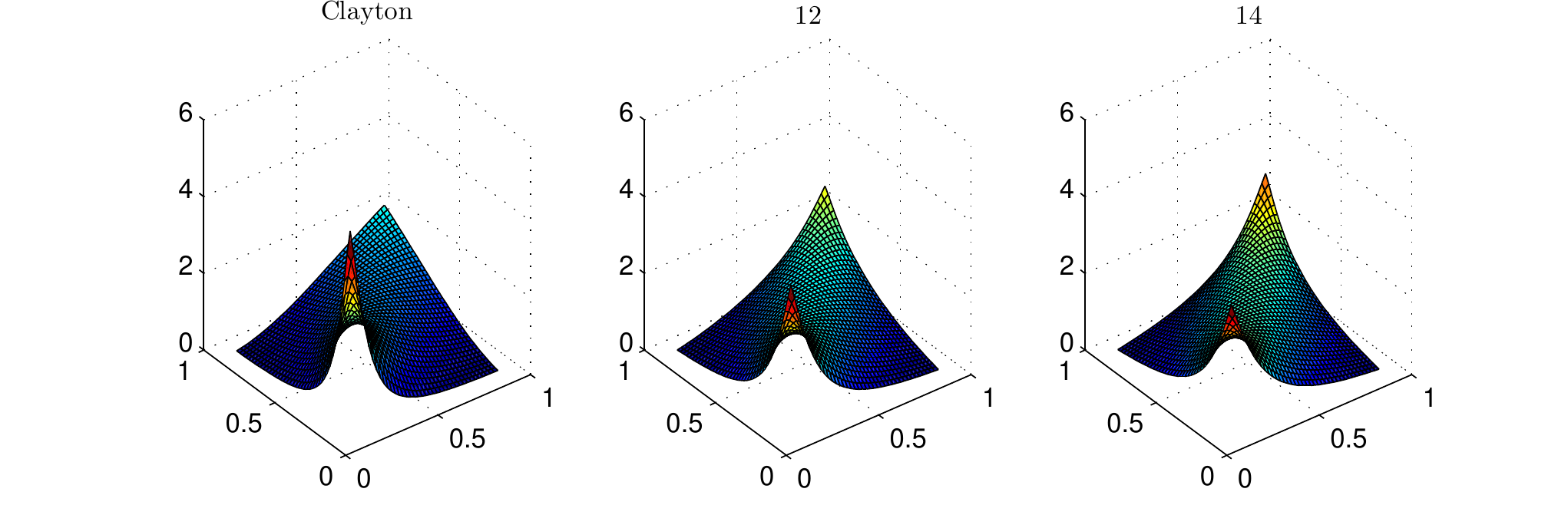}
		\includegraphics[width=0.75\textwidth]{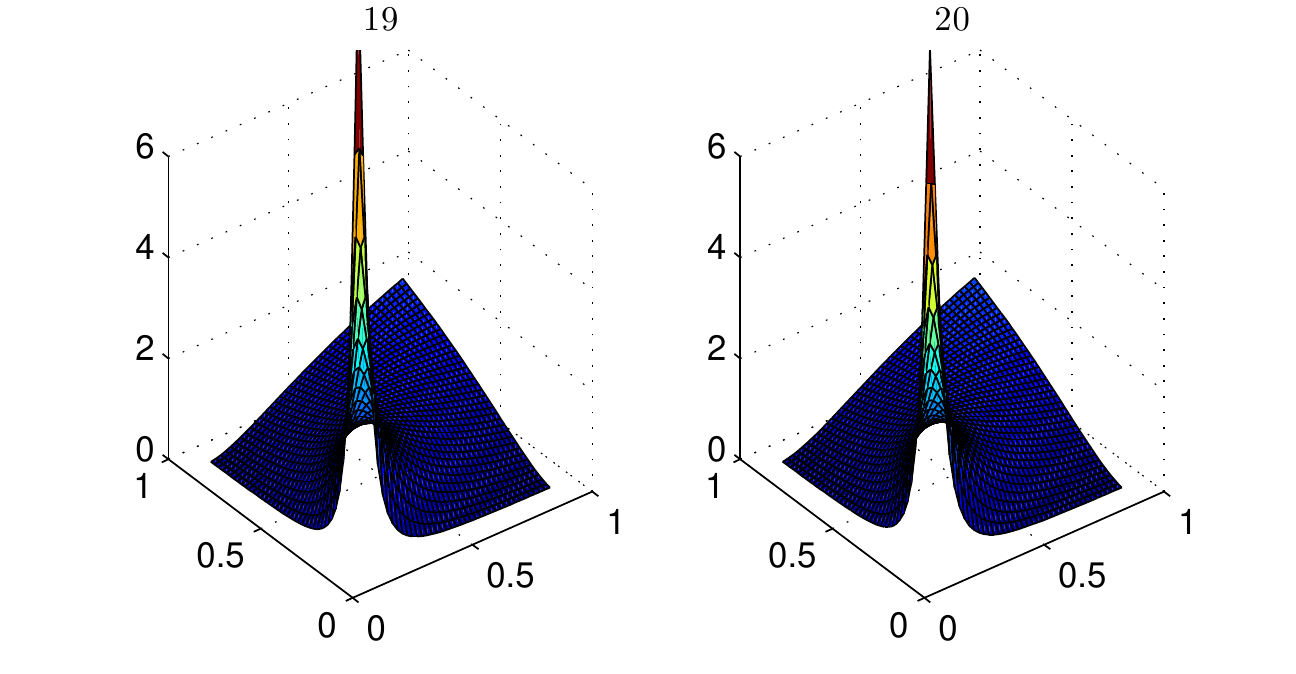}
		\caption{The density functions of 2-ACs belonging to the Clayton family and the families labeled 12, 14, 19 and 20, respectively. All depicted densities correspond to Kendall's tau = 0.5.}
	\label{fig:densities2}
\end{figure}

As we will work with several parametric families of Archimedean generators at the same time, we denote each of these families by $\Psi^{(a,\Theta_a)}$, where $a$ is a unique label (that will be called \emph{family label}) corresponding to the parametric family of Archimedean generators $\Psi^{\Theta_a}, ~\Theta_a \subseteq \IR$. \emph{Completely monotonic parametric families of Archimedean generators} (shortly, \emph{families}) we use in our work include Ali-Mikhail-Haq ($a =$ A) and Clayton ($a =$ C); see Table \ref{tab:geners}.  In connection with this fact, we denote $\FALL =$ \{A, C, 12, 14, 19, 20\}, where the numbers corresponds to the numbers of AC families presented in \cite[p. 116]{Nel06}.
The density functions of ACs $C_{\psi^{(a,\theta)}},~ a \in $ \{A, C, 20\} such that Kendall's tau equals 0.25 are depicted in Figure \ref{fig:densities}, and the density functions for $a \in $ \{C, 12, 14, 19 20\} such that Kendall's tau equals 0.5 are depicted in Figure \ref{fig:densities2}. These density functions illustrate that such ACs differ particularly in the tails. Note that dependence models which can capture tail dependence are of interest particularly in finance, see, e.g., \cite{Hof11CDO}.
%It is also noticeable that some of them are more similar than  others, e.g., 12 is similar to 14, or 19 is similar to 20, which is later discussed in Section \ref{sec:exps}. A relationship of the generators of the families 12, 14, 19 and 20 to the Clayton generator with the parameter $\theta=1$ is also discussed in \cite[p. 118]{Hof10book}.
\begin{rmk}
If $\psi~ \bar{\in}~ \Psi^{(a, \Theta_a)}$, then $\psi$ is uniquely given by the pair of the label $a$ and the parameter $\theta\in\Theta_a$, hence, we denote the generator $\psi$ by $\psi^{(a, \theta)}$.
\end{rmk}

As we work with c.m. generators only, the parameter ranges in Table \ref{tab:geners} are hence restricted to the values for which the corresponding ACs are models for a positively dependent random vector. This means that, e.g.,  for a Clayton generator $\psi^{(\textrm{C}, \theta)}$ with $\theta \in [-1, 0)$, although the function $C_{\psi^{(\textrm{C}, \theta)}}$ is a copula for $d=2$, we do not consider such a generator in this work, which is described later in Section \ref{sec:HACs}. 
Note that this restriction can often be (but not always) solved by appropriately transforming selected
input variables (through sign changes, for example) and using  the fact that, 
if $X$ and $Y$ are continuous random variables and $\tau_{X, Y}$ denotes the value of Kendall's tau for the random vector $(X, Y)$, then $\tau_{-X, Y} = \tau_{X, -Y} = -\tau_{X, Y}$. For more details on this approach allowing to model negatively dependent random vectors by HACs based on c.m. generators, see, e.g., the \emph{inverting procedure} described in Algorithm 4 in \cite{goreckihofertholena2016approachjiis}.

\begin{table}[t]
	\centering
			  \caption{Completely monotonic (c.m.) parametric families of Archimedean generators (taken over from \cite[p. 116]{Nel06}). The table contains the family labels ($a$), the corresponding parameter ranges $\Theta_a  \subset [0, +\infty)$, the explicit forms of $\psi^{(a,\theta)}~\bar{\in} ~\Psi^{(a,\Theta)}$, the sufficient nesting conditions, see Table 1 in \cite{Hof11}, where the sufficient nesting condition involves two generators $\psi^{(a,\theta_1)}, \psi^{(a,\theta_2)} ~\bar{\in} ~\Psi^{(a,\Theta_a)}$ and the set $\PSI$ is given by \eqref{eq:psi_nestable}, and the lower and upper tail-dependence coefficients $\Lambda_l (\theta) = \lim_{t \downarrow 0} C_{\psi^{(a,\theta)}}(t, t)/t$ and  $\Lambda_u(\theta) = \lim_{t \downarrow 0} (1-2t +C_{\psi^{(a,\theta)}}(t,t))/(1-t)$, respectively, where $C_{\psi^{(a,\theta)}}$ is a 2-AC, see Section 1.7.4 in \cite{Hof10book}.}
		\begin{tabular}{|lccccc|}
		\hline
			$a$ & $\Theta_a$ &$\psi^{(a,\theta)}(t)$ & $( \psi^{(a,\theta_1)}, \psi^{(a,\theta_2)}) \in \PSI$ & $\Lambda_l$ & $\Lambda_u$ \\
		\hline
			A & [0, 1) & $(1-\theta)/(\mathrm{e}^t - \theta)$ & $\theta_1 \leq \theta_2$ & 0 & 0\\
			C& $(0, +\infty)$ & $(1+t)^{-1/\theta}$ & $\theta_1 \leq \theta_2$ & $2^{-1/\theta}$ & 0\\
			%F & $(0, +\infty)$ & $-\log(1 -(1 - e^{-\theta})\exp(-t))/\theta$ & $\theta_1 \leq \theta_2$ \\
			%G  & $[1, +\infty)$ & $\exp(-t^{1/\theta})$ & $\theta_1 \leq \theta_2$\\
			%J & $[1, +\infty) $ & $1 - (1 - \exp(-t))^{1/\theta} $ & $\theta_1 \leq \theta_2$\\
			12 & $[1, +\infty)$ & $(1 + t^{1/\theta})^{-1}$ & $\theta_1 \leq \theta_2$ & $2^{-1/\theta}$ & $2 - 2^{-1/\theta}$\\
			14 & $[1, +\infty)$ & $(1 + t^{1/\theta})^{-\theta}$ & unknown & $1/2$ & $2- 2^{-1/\theta}$\\
			
			19 & $(0, +\infty)$ & $\theta / {\ln\left(t + \mathrm{e}^{\theta}\right)}$ & $\theta_1 \leq \theta_2$ & 1 & 0\\
			20 & $(0, +\infty)$ & $\ln ^{-1/\theta}(t + \mathrm{e})$ & $\theta_1 \leq \theta_2$  & 1 & 0\\
		\hline
		\end{tabular}
		  \label{tab:geners}
\end{table}

\section{Hierarchical Archimedean Copulas} \label{sec:HACs}
A $d$-copula $C$ is called \emph{hierarchical Archimedean} (HAC) if it is an AC with arguments possibly replaced by other HACs \cite{Hof11}; also, see, e.g., \cite{Hof2012stoch,Okh13,goreckihofertholena2016approachjiis}. 
In the last two articles, which address homogenous HAC estimation, the authors use, apart from the HAC definition, various auxiliary concepts addressing necessary properties of HAC structures, e.g., the HAC structures representation through \emph{sequences of reordered indices grouped through parentheses}.
In this work, we extend those articles to heterogeneous HACs.
This inevitably involves more notation and we consider introducing it directly in the HAC definition more transparent than introducing it as auxiliary concepts later.
Hence, we propose a definition of HACs, which is a slight modification of the one originally proposed in \cite{Holena13}, and we explicitly refer to the tree structure through concepts from graph theory. This is mainly motivated by the fact that
these concepts will be needed both in the construction of the algorithm for heterogeneous HACs estimation presented in Section \ref{sec:approach} and in its theoretical justification represented by Theorem \ref{thm:alg_returns_HAC}. 

\begin{figure}
	\centering
		\subfigure[]{%
            \label{fig:cop3a}
			        	\includegraphics[width=0.30\textwidth]{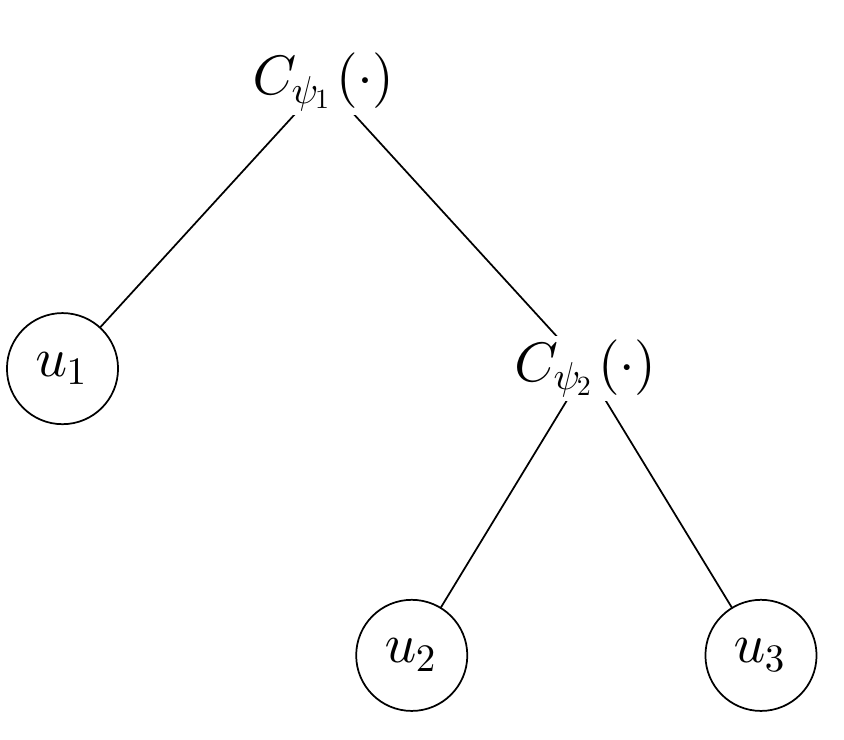}
						}
		\subfigure[]{%
            \label{fig:cop3b}
			        	\includegraphics[width=0.30\textwidth]{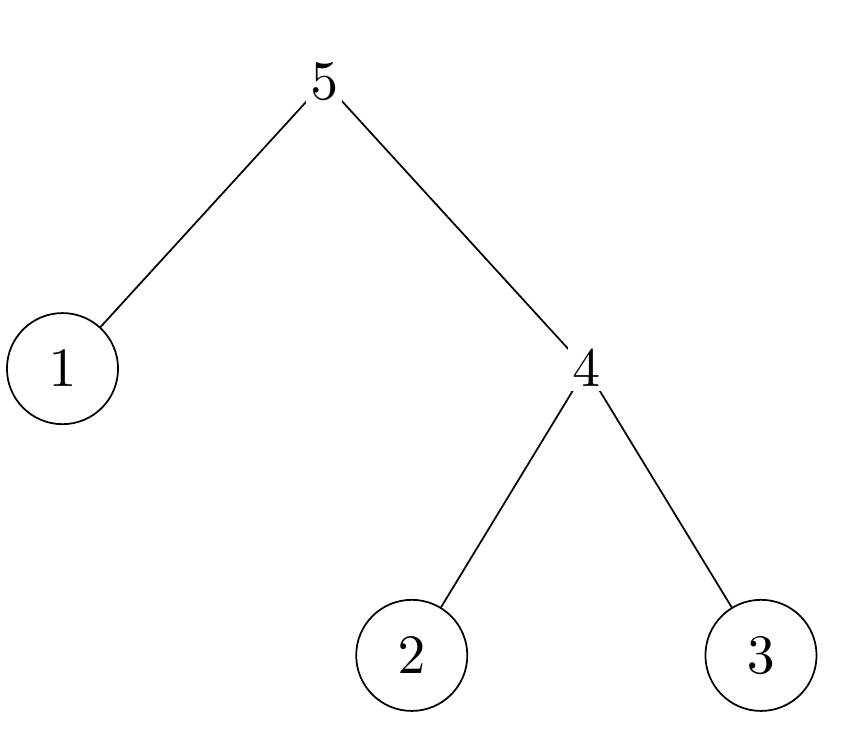}
						}
		\subfigure[]{%
            \label{fig:cop3c}
			        	\includegraphics[width=0.30\textwidth]{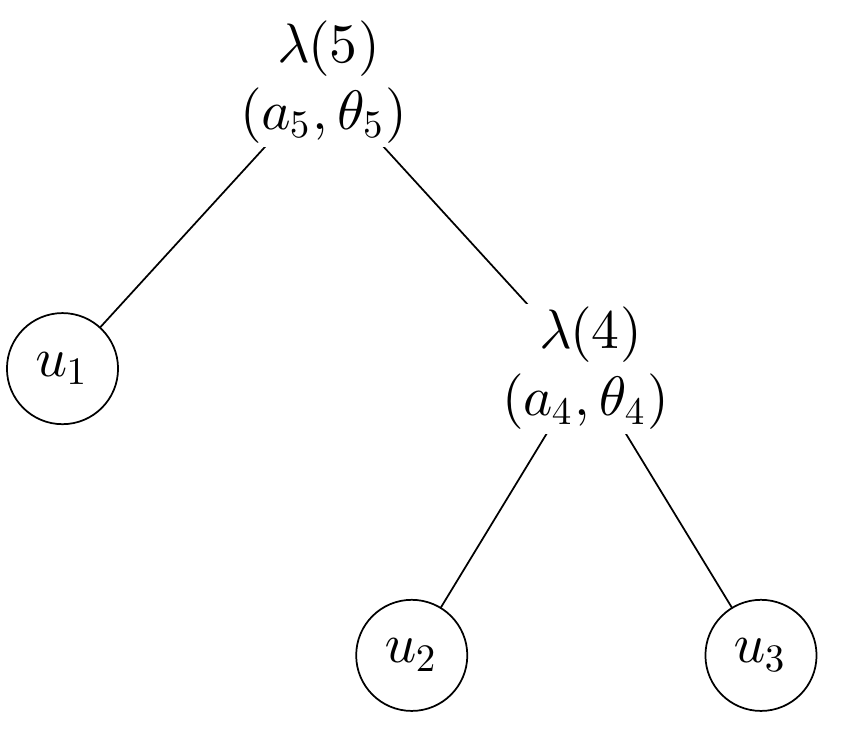}
						}
	\caption{(a) A tree-like representation of a 3-HAC given by $C_{\psi_1, \psi_2}(u_1, u_2, u_3) = C_{\psi_1}(u_1, C_{\psi_2}(u_2, u_3))$.	
	(b) A binary tree $(\VV,\EE), ~  \VV = \{1, ..., 5\},~  \EE = \{ \{1, 5\}, \{2, 4\}, \{3, 4\}, \{4, 5\} \}$ corresponding to the structure of the 3-FNAC $\HAC$. (c) Our representation of a $\{a_4, a_5\}$-heterogeneous 3-FNAC $\HAC$ with $\lambda(4) = \psi^{(a_4, \theta_4)}$ and $\lambda(5) = \psi^{(a_5, \theta_5)}$ and the structure $(\VV, \EE)$ corresponding to the left of this figure.}
	\label{fig:cop3}
\end{figure}

In order to get familiar with this alternative HAC definition, we first illustrate it in an example.
Consider a HAC $C_{\psi_1, \psi_2}(u_1, u_2, u_3) = C_{\psi_1}(u_1, C_{\psi_2}(u_2, u_3))$ 
for two generators $\psi_1,\psi_2 \in \Psi_{+\infty}$; its tree structure is depicted in Figure \ref{fig:cop3a}.
In the language of graph theory, an \emph{undirected tree} is a pair $(\VV,\EE)$, where $\VV$ is a set of nodes $\{1, ..., m\}, ~m \in \mathbb{N}$ and $\EE$ is a set of pairs of different nodes from $\VV$.
For the representation depicted in Figure \ref{fig:cop3a}, we can derive a tree with the same structure such as the one depicted in Figure \ref{fig:cop3b} just by assigning different numbers to all of its nodes. For this tree, we have $\VV = \{1, ..., 5\}$ and $\EE = \{ \{1, 5\}, \{2, 4\}, \{3, 4\}, \{4, 5\} \}$.
As we can observe, not all nodes correspond to the same objects. More precisely, the leaves $\{1, 2, 3\}$ correspond to the variables, whereas the non-leaf nodes $\{4, 5\}$ (we will call them \emph{forks}) correspond to the ACs (uniquely determined by the corresponding generators) nested in $C_{\psi_1, \psi_2}$. 
%Generally, given a $d$-HAC, will always have leaves ${1, ..., d}$ and forks $\{d+1, ..., d + k\}$, where $k$ is the number of generators in the $d$-HAC. 
As each fork corresponds to a generator, we represent this relationship using a labeling denoted $\lambda$, which maps the forks to corresponding generators. In our example, it would be $\lambda(4) = \psi_2$ and $\lambda(5) = \psi_1$. Using this notation, $C_{\psi_2}(u_2, u_3)$ turns into $C_{\lambda(4)}(u_2, u_3)$ and $C_{\psi_1}(u_1, C_{\psi_2}(u_2, u_3))$ into $C_{\lambda(5)}(u_1,  C_{\lambda(4)}(u_2, u_3))$.

Finally, having the structure and the involved generators encoded in $(\VV, \EE)$ and $\lambda$, respectively, we can for a given $\uu = (u_1, u_2, u_3) \in \I^3$, explicitly express the value $C_{\psi_1, \psi_2}(u_1, u_2, u_3)$, which we assign to the node 5 using the values assigned to its children. These values are defined by $x^{\uu}_i = u_i$ for $i \in \{1, ...,3\}$ for the variables, and, for the generators (or forks), going  from the bottom to the top of the structure, $x^{\uu}_4 = C_{\lambda(4)}(x^{\uu}_2, x^{\uu}_3)$  and $x^{\uu}_5 = C_{\lambda(5)}(x^{\uu}_1, x^{\uu}_4)$. Clearly, $x^{\uu}_{5} = C_{\psi_1, \psi_2}(u_1, u_2, u_3)$. Now consider that  $\{2, 3\}$ are the children of the node 4 and $\{1, 4\}$ are the children of the node 5. Using this notation, the (inner) copula values, $(\VV, \EE)$ and $\lambda$ are glued together through $x_4 = C_{\lambda(4)}(\x_{\{2,3\}})$ and $x_5 = C_{\lambda(5)}(\x_{\{1,4\}})$, where $\x_{\{i,j\}} = (x_i, x_j)$. For clarity, also note that, e.g., node 2 (or node 3) is not a child of the node 5.

\begin{dfn} \label{def:hier_arch_kopule} 
Let $d, ~k \in \mathbb{N},~ d \geq 2, ~k \in \{1, ..., d-1\}, ~(\VV,\EE)$ be a labeled tree with nodes $\VV = \{1, ..., d + k\}$, edges $\EE \subset \VV \times \VV$ and rooted in the node $d+k$. Let the nodes $ \{1, ..., d\}$ be the \emph{leaves} of $(\VV,\EE)$ and the nodes $ \{d+1, ..., d + k\}$, which will be called \emph{forks}, have at least two children each. In connection with $(\VV,\EE)$, the following notation will be used:
\begin{itemize}
	\item For $v \in \VV$, denote by $\wedge(v)$ the set of children of $v$; thus the cardinality of $\wedge(v)$ fulfills $\#\wedge(v) \geq 2$ for $v \in \{d+1, ..., d+k \}$, i.e., $v$ being a fork, and $\#\wedge(v) = 0$ for $v \in \{1, ..., d\}$, i.e., for $v$ being a leaf.
	\item For $S \subset \VV$ and $\x^{\uu} \in \I^{d+k}$, the simplified notation
	\begin{flalign}
	\x_S = (x_{v_1}, ..., x_{v_{\#S}}),~\textrm{where}~ S = \{v_1, ..., v_{\#S}\},
	\end{flalign}
	will be used with a further simplification $\x_v = \x_{\{v\}}$ for $v \in \VV$.
\end{itemize}
\noindent Finally, let $\lambda~:~\{d + 1, ..., d + k\} \rightarrow \Psi_{+\infty}$ be a labeling of forks with
c.m. Archimedean generators such that for each $\uu \in \I^d$, there exists $\x^{\uu} \in \I^{d+k}$ with the following two properties:
\begin{enumerate}[(i)]
	\item $\forall v \in \{d + 1, ..., d + k\},~ \x_v^{\uu} = C_{\lambda(v)}(\x^{\uu}_{\wedge(v)})$;
	\item $\forall v \in \{1, ..., d\},~ \x^{\uu}_v = u_v$.
\end{enumerate}
Then:
\begin{enumerate}[a)]
\item if the function $C_{(\VV,\EE,\lambda)}~:~ \I^d \rightarrow \I$, defined
 	\begin{flalign} 
	\forall {\uu} \in \I^d, ~ C_{(\VV,\EE,\lambda)}(\uu) = \x^{\uu}_{d+k},
	\label{eq:function_hac}
	\end{flalign}
	is a $d$-copula, it is called \emph{hierarchical Archimedean $d$-copula}  ($d$-HAC) with the (tree) structure $(\VV, \EE)$ and
the labeling $\lambda$;
	\item if $(\VV, \EE)$ is binary, then  $C_{(\VV,\EE,\lambda)}$ is called \emph{binary};
%	\item if $(\VV, \EE)$ is binary, there exists permutations $\pi_{\leq d}: \{1,...,d\} \rightarrow \{1,...,d\}$ and $\pi_{> d}: \{d+1,...,2d-1\} \rightarrow \{d+1,...,2d-1\}$ such that $\pi_{> d}(2d-1) = 2d-1$, $\wedge(v) = \{\pi_{\leq d}(2d-v), \pi_{> d}(v - 1)\}$ for all for $v \in \{d+2,..., 2d-1\}$	and $\wedge(d +1) = \{\pi_{\leq d}(d-1), \pi_{\leq d}(d)\}$, then $C_{(\VV,\EE,\lambda)}$ is called \emph{fully nested Archimedean copula} (FNAC), otherwise, $C_{(\VV,\EE,\lambda)}$ is called \emph{partially nested Archimedean copula} (PNAC). Note that the involved $\pi_{\leq d}$ and $\pi_{> d}$ just serve for saying that a HAC remains fully nested independently of a permutation of leaves or forks;
		
		\item given a finite set of family labels $\FF = \{a_1, ..., a_{\#\FF}\}$ such that $\Psi^{(a_i,\Theta_{a_i})}$ is a parametric family of Archimedean generators for all $i \in\{1, ..., \#\FF\}$, if $(\VV,\EE,\lambda)$  fulfill 	
	\begin{flalign}
	\forall v \in \{d+1,...,d+k\},~ \exists i_v \in \{1, ...,\#\FF\},~\lambda(v) ~ \bar{\in}~ \Psi^{(a_{i_v},\Theta_{i_v})},
	\end{flalign}
	then $C_{(\VV,\EE,\lambda)}$ is called $\FF-$\emph{parametric}; 
	
	\item  given that $C_{(\VV,\EE,\lambda)}$ is $\FF-$parametric, if $(\VV,\EE,\lambda)$  fulfill
	 	\begin{flalign}
	\exists i \in \{1, ...,\#\FF\},~\forall v \in \{d+1,...,d+k\},~\lambda(v) ~\bar{\in}~ \Psi^{(a_i,\Theta_{a_i})}, 
	\end{flalign}
	then $C_{(\VV,\EE,\lambda)}$ is called $a_i$-\emph{homogeneous}, otherwise, $C_{(\VV,\EE,\lambda)}$ is called $\FF$-\emph{heterogeneous}.
\end{enumerate}
\end{dfn}

\begin{rmk} \label{rem:binary_forks}
1) If $C_{(\VV,\EE,\lambda)}$ is a binary $d$-HAC then the number of forks in the structure $(\VV,\EE)$ (denoted by $k$ in Definition \ref{def:hier_arch_kopule}) is $d-1$. 
2) An $\mathcal{F}$-parametric $\HAC$ is a HAC with generators from the families in $\mathcal{F}$. A heterogeneous HAC involves generators from different families, whereas a homogenous HAC from a single family. 3) The vector variable $\x^{\uu}$ is used only for  clarity and notational convenience and is not used in the algorithms below.
\end{rmk}

\begin{rmk} \label{rem:convention}
In the rest of our work, especially in Section \ref{sec:dealing_nesting_condition}, given a set $\FF \subseteq \FALL$ and an $\FF$-parametric function $\HAC$,  we use the convention that $\lambda(i) = \psi^{(a_i, \theta_i)}$, where $a_i \in \FF,~ \theta_i \in \Theta_{a_i},~ i = d+1, ..., d +k$, unless stated otherwise.
\end{rmk}

If a function $\HAC$ is $\FF$-parametric, we use a graphical representation following from Remark \ref{rem:convention} that uniquely determines it. Given two family labels $\{a_4, a_5\} \subset\FALL$, an example of a $\{a_4, a_5\}$-heterogeneous trivariate function $\HAC$ is depicted in Figure \ref{fig:cop3c}. Clearly, its structure $(\VV, \EE)$ is the one depicted in Figure \ref{fig:cop3b}, and the convention stated in Remark \ref{rem:convention}, i.e., $\lambda(4) = \psi^{(a_4, \theta_4)}$ and $\lambda(5) = \psi^{(a_5, \theta_5)}$, is used. \\

As in the case of ACs, we can ask for necessary and sufficient conditions for the function given by \eqref{eq:function_hac}
to be a proper copula. As has already been mentioned before, such a condition is not known. 
However, several sufficient conditions are known. An early one has been proposed in \cite{McNeil08}, see Theorem 4.4  therein, which is proven for a subclass of HACs called \emph{fully nested Archimedean copulas}. Its generalization to all HACs, which has been proposed in \cite{holenascavnickybajer2015usingcopulas}, is recalled in the following theorem.

\begin{thm} \cite{holenascavnickybajer2015usingcopulas} 
\label{th:holena}
Let $(\VV,\EE)$ and $\lambda$ be the tree and the mapping from Definition \ref{def:hier_arch_kopule}, respectively. Define 
\begin{flalign} \label{eq:psi_nestable}
 \PSI = \{ (\psi_1, \psi_2)~ |  ~(\psi_1, \psi_2)  \in \Psi_{+\infty} \times \Psi_{+\infty} ~\& ~ (\psi_1^{-1} \circ \psi_2)' \textrm{ is c.m.}\}
\end{flalign}
and let $\lambda$ fulfills 
	 	\begin{flalign} \label{eq:holena_nesting_cond}
	\forall v \in \{d + 1, ..., d+k\},~\forall \tilde{v} \in \wedge(v) \cap \{d+1,..., d +k\},~ 
	  (\lambda(v), \lambda(\tilde{v})) \in \PSI,
	\end{flalign}
Then $C_{(\VV,\EE,\lambda)}$ is a copula.
\end{thm}

Another, weaker sufficient condition has appeared recently in \cite{rezapour2015construction}. In this condition, neither $\lambda(v)$ nor $\lambda(\tilde{v})$ nor $(\lambda(v)^{-1} \circ \lambda(\tilde{v}))' $ are required to be c.m., but it is sufficient that these three functions are $d$-\emph{monotone} or `less than'  $d$-\emph{monotone}, where $d < +\infty$, see \cite{rezapour2015construction} for details. Relaxing the requirement of the complete monotonicity allows to construct more HACs then under \eqref{eq:holena_nesting_cond}, e.g., \cite{rezapour2015construction} provides 151 new 5-HACs in Example 2.4. 
However, even if the article addresses parametric families of generators in that example, no simplification of the proposed condition in terms of the parameters of the generators similar to, e.g., the fourth column in Table \ref{tab:geners_comb}, is provided. 
Without it, $d$-monotonicity has to be directly checked for each pair of parent-child generators, which makes the estimation process at least challenging, especially in high dimensions and under consideration that the condition, by contrary to the complete monotonicity, depends on $d$. Also, no efficient sampling strategies, similar to the ones proposed, e.g., in \cite{Hof11,Hof2012stoch}, are known. Without these tools at hand, performing estimation experiments seems to be at least an intricate task.
Due to these reasons, we restrict to HACs satisfying the condition \eqref{eq:holena_nesting_cond}, which will be called the \emph{sufficient nesting condition} (\emph{s.n.c.}).

Observing the s.n.c.s in Table \ref{tab:geners}, we can see that, for an $a$-homogeneous function $C_{(\VV,\EE,\lambda)}$, \eqref{eq:holena_nesting_cond} simplifies to
	 	\begin{flalign} \label{eq:holena_nesting_cond_simple}
	\forall v \in \{d + 1, ..., d+k\},~\forall \tilde{v} \in \wedge(v) \cap \{d+1,..., d +k\}, 
~	  \theta_v \leq \theta_{\tilde{v}},
	\end{flalign}
for all $a \in\FALL \backslash \{14\}$.
Assuming an $\FF$-heterogeneous $\HAC$, where $\FF \subseteq \FALL$, such a simplification of \eqref{eq:holena_nesting_cond} is not possible as the s.n.c.s vary depending on the family combination $(a_v, a_{\tilde{v}})$ of a particular pair $(\lambda(v), \lambda(\tilde{v}))$, which can be seen from the last column in Table \ref{tab:geners_comb}. 
\begin{table}
	\centering
			  \caption{All family combinations of the completely monotone generators of \cite[pp. 116--119]{Nel06},  which result in proper
HACs according to the sufficient nesting condition given by \eqref{eq:holena_nesting_cond}. Taken over from \cite{Hof10book}, see Theorem 4.3.2 therein. The table contains the family labels (introduced in Section \ref{sec:ACs}) in a parent-child family combination $(a_1,a_2)$  with the corresponding parameter ranges  $\Theta_{a_1}$ and $\Theta_{a_2}$. The last column contains the sufficient nesting conditions in terms of the parameters of two generators $\psi^{(a_1,\theta_1)} ~\bar{\in} ~\Psi^{(a_1,\Theta_{a_1})}, \psi^{(a_2,\theta_2)} ~\bar{\in} ~\Psi^{(a_2,\Theta_{a_2})}$.
								}
		\begin{tabular}{|cccc|}
		\hline
			$(a_1,a_2)$ & $\Theta_{a_1}$ & $\Theta_{a_2}$ & $( \psi^{(a_1,\theta_1)}, \psi^{(a_2,\theta_2)}) \in \PSI$  \\
			\hline
			(A, C) & $[0, 1)$ & $(0, +\infty)$ & $\theta_2 \in [1, +\infty)$ \\
			(A, 19) & $[0, 1)$ & $(0, +\infty)$ & any $\theta_1, \theta_2$ \\
			(A, 20) & $[0, 1)$ & $(0, +\infty)$ & $\theta_2 \in [1, +\infty)$ \\
			(C, 12) & $(0, +\infty)$ & $[1, +\infty)$ & $\theta_1 \in (0, 1]$ \\
			(C, 14) & $(0, +\infty)$ & $[1, +\infty)$ & $\theta_1 \theta_2 \in (0, 1]$ \\
			(C, 19) & $(0, +\infty)$ & $(0, +\infty)$ & $\theta_1 \in (0,1]$ \\
			(C, 20) & $(0, +\infty)$ & $(0, +\infty)$ & $\theta_1 \leq \theta_2$ \\
		\hline
		\end{tabular}
\label{tab:geners_comb}
\end{table}
Table \ref{tab:geners_comb} lists
all family combinations of the completely monotone generators of \cite[pp. 116--119]{Nel06},  which result in proper
HACs according to the s.n.c., see \cite{Hof08} or Theorem 4.3.2 in \cite{Hof10book} for more details; see also Sections 4.2.3 and 4.3 in \cite{Hof10book}, which show how so-called \emph{general nesting transformations of generators} can be used for constructing such family combinations.
Hence, we will consider the following family combinations 
\begin{flalign} \label{eq:known_nestables}
\FNEST =   \{  & \textrm{(A, A), (C, C), (12, 12), (19, 19), (20, 20), }\nonumber\\
 &  \textrm{(A, C), (A, 19), (A, 20), (C, 12), (C, 14), (C, 19), (C, 20)\} }.
\end{flalign}

Finally, note that all generators compatible with a given generator can be constructed and characterized as in \cite{Hering2010}.

\section{An Archimedean copula estimator based on inversion of Kendall's tau} \label{sec:param_estimation}
Assume i.i.d. random vectors $\mathbf{X}_i = (X_{i1}, ..., X_{id}),~ i \in \{1, . . . , n\}$, distributed according to a joint distribution function
$H$ with continuous margins $F_j, ~j \in \{1, . . . , d\}$, and the copula $C$. Now consider that, as $\mathbf{U}_i = (U_{i1}, ..., U_{id}), ~ i = 1, ..., n$ is a random sample from $C$, where $U_{ij} = F_j(X_{ij}), ~ i \in \{1, . . . , n\}, j \in \{1, . . . , d\}$, one can base estimation of $C$ directly on $\mathbf{U}_i, ~ i = 1, ..., n$, if the margins $F_j, ~ j \in \{1, . . . , d\}$ are
known. In practice, the margins are typically unknown and must be estimated parametrically or
non-parametrically. In the following, we will work under unknown margins and thus we consider the \emph{pseudo-observations}
\begin{flalign}
U_{ij}= \frac{n}{n+1}\hat{F}_{n, j}(X_{ij}) = \frac{R_{ij}}{n+1}
\label{eq:pseudo_observations}
\end{flalign}
\noindent where $\hat{F}_{n, j}$ denotes the \emph{empirical distribution function} corresponding to the $j$th margin
and $R_{ij}$ denotes the \emph{rank} of $X_{ij}$ among $X_{1j}, ..., X_{nj}$. 

\begin{figure}%
	\centering
\includegraphics[width=0.49\textwidth]{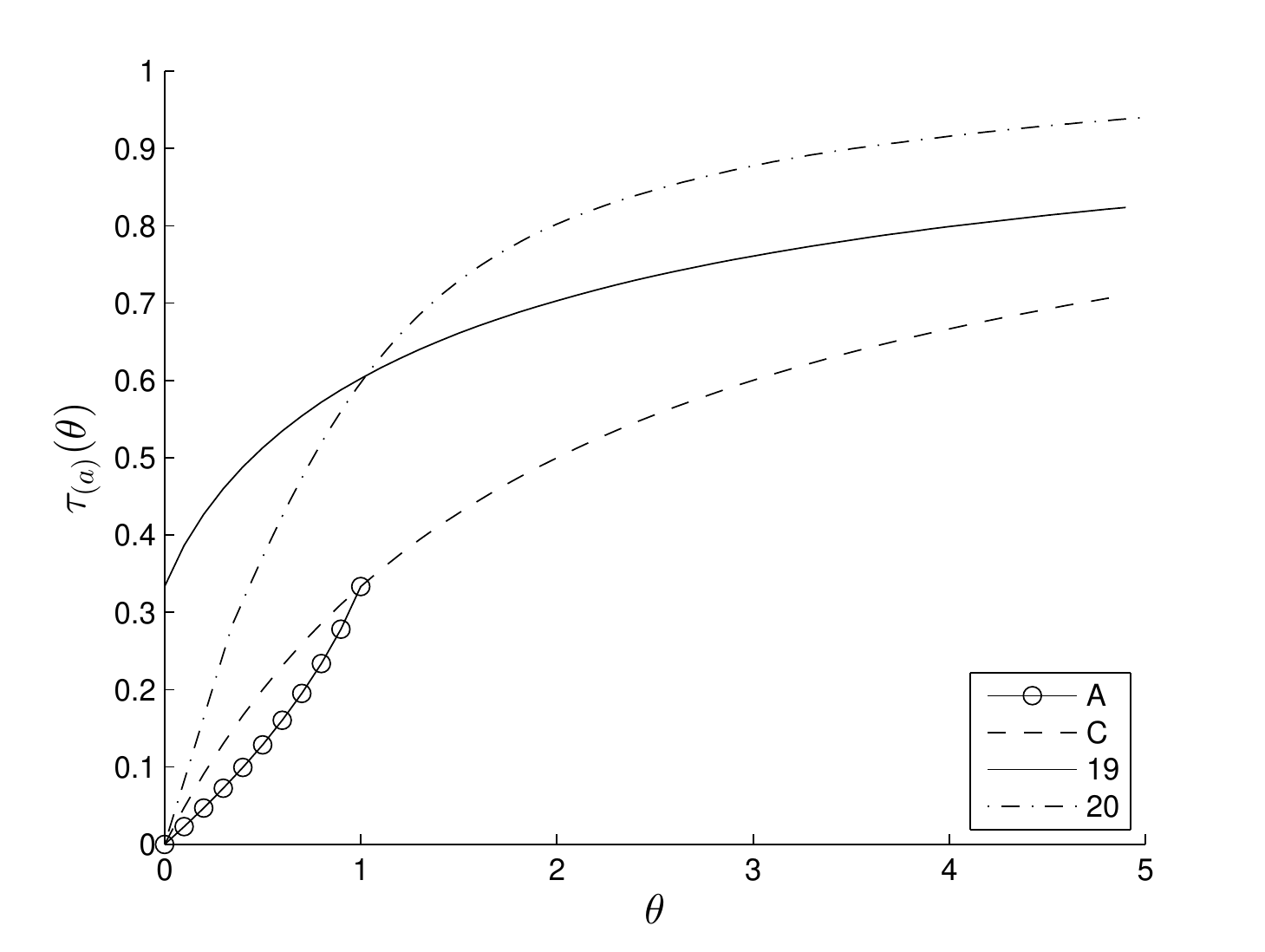}%
\includegraphics[width=0.49\textwidth]{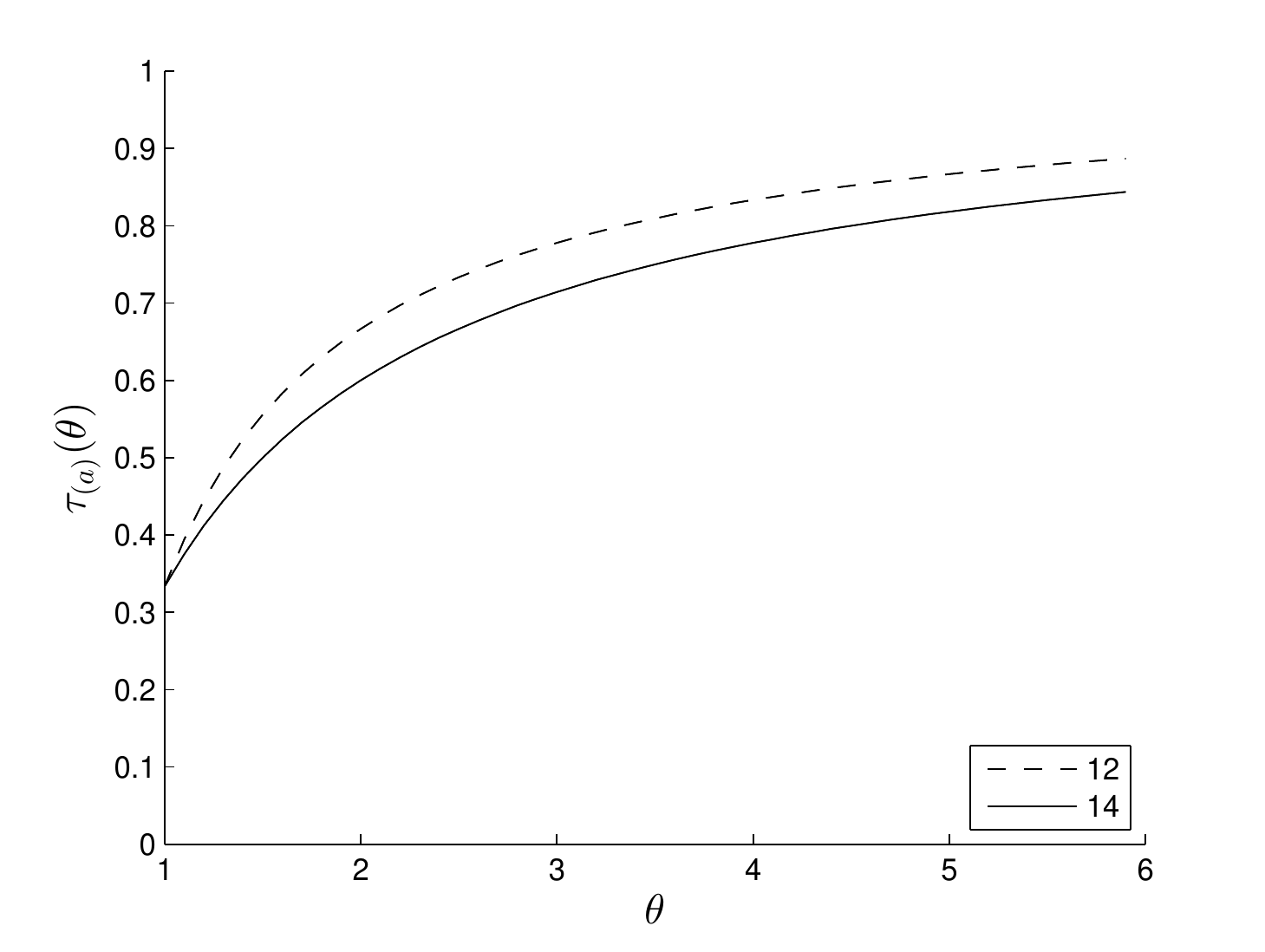}%
\caption{The graphs of $\tau_{(a)}(\theta)$ for $a \in$ \{A, C, 19, 20\} (left side) and $a \in$ \{12, 14\} (right side). 
}%
%generated by theta_tau_graphs_csda.mat
%
\label{fig:tau_theta}%
\end{figure}

In the following, we refer to
\begin{eqnarray} \label{eq:kendall_tau_for_copula}
\tau(C) = 4 \int_{\mathbb{I}^2}{C(u_1, u_2)dC(u_1, u_2)} - 1,
\label{eq:kendall}
\end{eqnarray}
as to the \emph{population version of Kendall's tau}. If $C_{\psi}$ is a 2-AC and $\psi$ is a twice continuously differentiable generator with $\psi(t) > 0$ for all $t \in [0 ,+\infty )$, Kendall's tau can be represented as \cite[p. 91]{Joe97}, \cite[p. 163]{Nel06}
\begin{eqnarray}
\tau(\psi) = \tau(C_{\psi}) = 1-4\int_0^{+\infty} t (\psi' (t) )^2dt = 1+4\int_0^1 \frac{\psi^-(t)}{(\psi^-)'(t)}  dt.
\label{eq:kendall_semi}
\end{eqnarray}
For a generator $\psi^{(a, \theta)}$ from a family $a$, 
 (\ref{eq:kendall_semi}) states a functional relationship between $\theta$ and $\tau$ (denoted by $\tau_{(a)}(\theta)$), which involves at most a one-dimensional integration, e.g., for $a \in$  \{19, 20\}, or even no integration, e.g., $\tau_{(a)}(\theta)$ is in a closed form for $a \in$ \{A, C, 12, 14\}, see Table \ref{tab:tau_theta} (the relationship between $\theta$ and $\tau$ for other generators can be found, e.g., in  \cite{Hof10book}). 
\begin{table}[t]
	\centering
	\caption{\cite[p. 65]{Hof10book} Kendall's tau for the parametric families of Archimedean generators in Table \ref{tab:geners}. The columns show: the family label $a$; the parameter range $\Theta_a$; $\tau_{(a)}(\theta)$; the range of  $\tau_{(a)}(\theta)$ for $\theta \in \Theta_a$.}
		\begin{tabular}{cccc}
			\hline
			$a$ & $\Theta_a$ & $\tau_{(a)}(\theta)$ & $\tau_{(a)}(\Theta_a)$\\
			\hline
			A & $[0,1)$ & $1 - 2(\theta + (1 - \theta)^2 \log(1 - \theta))/(3\theta^2)$ & $[0, \frac{1}{3})$\\
						C & $(0, +\infty)$ & $\theta/(\theta+2)$ & $(0, 1)$ \\
			12 & $[1, +\infty)$ & $1 - 2/(3\theta)$ & $[\frac{1}{3}, 1)$\\
			14 & $[1, +\infty)$ & $1 - 2/(1 + 2\theta)$ & $[\frac{1}{3}, 1)$\\
			19 & $(0, +\infty)$ & $1/3 + 2\theta(1-\theta \textrm{e}^\theta \int_{\theta}^{+\infty}\textrm{e}^{-t}/t~ dt)/3$ & $(\frac{1}{3}, 1)$\\
			20 & $(0, +\infty)$ & $1 - 4/\theta(1/(\theta+2) - \textrm{e}\int_0^1 t^{\theta+1}/\textrm{e}^{t^{-\theta}}~dt)$ & $(0, 1)$\\
			\hline
		\end{tabular}
		\label{tab:tau_theta}
\end{table}
Note that the forms of the functions $\tau_{(a)}(\theta)$ presented in Table \ref{tab:tau_theta} are taken from \cite[p. 65]{Hof10book} for all considered families except for the family 20, which is introduced here. The graphs of $\tau_{(a)}(\theta)$ for $a \in \FALL$ are depicted in Figure \ref{fig:tau_theta}. 

In connection to \eqref{eq:kendall_semi}, using Remark 2.3.2 in \cite{Hof10book} and its generalization given by 
Proposition 2 in \cite{goreckihofertholena2016approachjiis}, it follows that if the function $\HAC$ given by \eqref{eq:function_hac} satisfies \eqref{eq:holena_nesting_cond}, then 
\begin{flalign}
	(\forall v \in \{d + 1, ..., d+k\})(\forall \tilde{v} \in \wedge(v) \cap \{d+1,..., d +k\}) ~
	  \tau_v \leq \tau_{\tilde{v}},
\label{eq:nnc}
\end{flalign}
where $\tau_i = \tau(\lambda(i))$ for all $i \in \{d+1,..., d +k\}$. This condition is clearly not equivalent to the s.n.c., e.g., considering the function $\HAC$ in Figure \ref{fig:cop3} with $(a_5, \theta_5)$ = (A, 0.25)  and  $(a_4, \theta_4)$ = (C, 0.5), it satisfies \eqref{eq:nnc} as $\tau_5 \approx 0.06$ and $\tau_4 = 0.2$ but violates \eqref{eq:holena_nesting_cond}, see Table \ref{tab:geners_comb}. However, assuming a homogenous HAC from a family such that the s.n.c.~is in the form $\theta_1 \leq \theta_2$, see Table \ref{tab:geners}, \eqref{eq:nnc} combined with relatively weak assumptions turns to the s.n.c., as is shown in the following lemma.

\begin{lemma}
Let $(\VV, \EE)$ be a tree with $d$ leaves and $k$ forks,   
$(\tau_i)_{d+1}^{d+k} \in [-1, 1]^k$ and  
$a$ be a family with the parameter range $\Theta_a$ such that
1) $\tau_{(a)}(\theta) = \tau(\psi^{(a, \theta)})$ for all $\theta \in \Theta_a$,
2) $\tau_{(a)}$ is strictly increasing on $\Theta_a$, and 
3) $( \psi^{(a,\theta_1)}, \psi^{(a,\theta_2)}) \in \PSI$ is equivalent to $\theta_1 \leq \theta_2$ for all $\theta_1, \theta_2 \in \Theta_a$.
If 
\begin{flalign}
(\forall i \in \{d+1,..., d +k\}) ~\tau_i \in \tau_{(a)}(\Theta_a), 
\label{eq:in_Theta_a}
\end{flalign} 
then let $\HAC$ be an $a$-homogeneous function given by \eqref{eq:function_hac} with the labeling $\lambda$ defined $\lambda(i) = \psi^{(a, \tau_{(a)}^{-1}(\tau_i))}$ for all $i \in \{d+1,..., d +k\}$, and then it holds that \eqref{eq:nnc} implies \eqref{eq:holena_nesting_cond_simple}.
\label{lem:nnc_equals_snc}
\end{lemma}

The three assumption on $a$ in Lemma \ref{lem:nnc_equals_snc} holds for all $a \in \FALL \slash \{14\}$, see Table \ref{tab:geners}, but also for many other families, e.g., for the three popular families of Frank, Gumbel and Joe copulas, see Table 2.3 in \cite{Hof10book}. 
With this lemma at hand, one can aim, in homogenous HAC estimation, to assure \eqref{eq:nnc} instead of \eqref{eq:holena_nesting_cond_simple}, which is one of the key ideas we use in our approach to homogenous HAC estimation addressed in Section \ref{sec:homo_HAC_estim}. Also, assuming a heterogeneous HAC, note that satisfying \eqref{eq:nnc} does not impose, by contrary to \eqref{eq:holena_nesting_cond} (see Tables \ref{tab:geners} and \ref{tab:geners_comb}), any assumptions on the underlying families of the generators but just requires a particular ordering of the strength of the dependence of the generators in the structure.  Hence, even if there are not known any additional conditions (like in the homogeneous case) that would turn it into the s.n.c., one can find its particular use also in heterogeneous estimation, as we show in Section \ref{sec:hetero_HAC_estim}.

As noted in Section \ref{sec:intro}, some of the considered families cannot model dependencies for all $\tau \in (0, 1)$, e.g., $(0, 1) \not\subseteq \tau_{(a)}(\Theta_a)$ for all $a \in $ \{A, 12, 14, 19\}. This fact illustrated by Figure \ref{fig:tau_theta} explicitly addresses the fourth column in Table \ref{tab:tau_theta}, which shows the ranges of  $\tau_{(a)}(\theta)$ for $\theta \in \Theta_a, ~ a \in \FALL$. It also clarifies the selection of the families in Figures \ref{fig:densities} and \ref{fig:densities2}.
 
If it exists, the inversion of $\tau_{(a)}(\theta)$ establishes a method-of-moments-like estimator of the parameter $\theta$ given by $\hat{\theta}_n = \tau^{-1}_{(a)}(\tau^n)$, based on the \emph{sample version of Kendall's tau} 
\begin{eqnarray}
\tau^n = \frac{4}{n(n - 1)}\Biggl(\sum_{k=1, l=1}^{n}{\mathbf{1}_{\{(U_{k1} - U_{l1})(U_{k2} - U_{l2}) > 0\}}}\Biggr) - 1;
\label{eq:kendall_emp}
\end{eqnarray}
\noindent %where $U_{ij}, ~ i \in \{1, . . . , n\}, j \in \{1, 2\}$, are pseudo-observations of  $(U_1, U_2) \sim C$; 
see \cite{Gen93}. Note that we use the notation $\tau^n$ for notational convenience related to the sample version of Kendall correlation matrix defined in Section \ref{sec:collapsing}.
If $\tau_{(a)}(\hat{\theta}_n) = \tau^n$ has no solution, this estimation method does not lead to an estimator. Unless there is an explicit form for $\tau^{-1}_{(a)}$, $\hat{\theta}_n$ is computed by numerical root finding \cite{Hofert13}.

In \cite{Hofert13}, a comparison of 10 estimators for ACs in precision of estimation of the parameter is provided, which includes Kendall's tau, Blomqvist's beta, minimum distance estimators, the maximum-likelihood estimator, a simulated maximum-likelihood estimator, and a maximum-likelihood estimator based on the copula diagonal. There, e.g., in Figure 4, one can observe that the Kendall's tau-based estimator recalled above (there denoted by $\tau_{\bar{\hat{\tau}}}$) together with the ML estimator perform well when compared with the remaining estimators. For this reason, these two estimators are considered in this work. Also, in \cite{uyttendaele2016estimation}, an HAC estimator based on Hoeffding's $D$ is investigated, however, this estimator was clearly outperformed by its analogous based on Kendall's tau in the ability to estimate the structure of the true copula. Finally, note that for
Spearman's rho, there is no explicit formula analogous to \eqref{eq:kendall_semi} known for ACs \cite[p. 62]{Hof10book}, which makes its application in our context at least challenging.

\section{An aggregated goodness-of-fit test statistic} \label{sec:gof}
Once we have the parameters estimated, we can ask how well our estimated model fits the data, that is, we can conduct a \emph{goodness-of-fit test} (GoF test).
In this work, we use three GoF tests based on statistics similar to Cram\'{e}r-von Mises statistics \cite{cramer1928composition}. These are the statistics from \cite{Gen09} denoted by $S_n,~ S_n^{(K)}$ or $S_n^{(C)}$. In this work, we denote the (empirical copula based) statistic $S_n$ by $S_n^{(E)}$ and the (Rosenblatt's transformation based) statistic $S_n^{(C)}$ by $S_n^{(R)}$. 
Given pseudo-observations \eqref{eq:pseudo_observations} and a $d$-copula estimate $C_{\theta_n}$ with parameter(s) $\theta_n$, 
they are given
\begin{eqnarray}
S_n^{(E)} = \sum\limits^n_{i=1}(C_n(\mathbf{U}_i) - C_{\theta_n}(\mathbf{U}_i))^2,
\label{eq:empirical_gof}
\end{eqnarray}
where $C_n (\mathbf{u})= \frac{1}{n} \sum_{i=1}^n{\mathbf{1}_{\{U_{i1} \leq u_1, ..., U_{id} \leq u_d\}}}$  is the empirical copula and  $\mathbf{u} = (u_1, ..., u_d) \in \mathbb{I}^d$,
\begin{align}
S_n^{(K)} =~ & \frac{n}{3} + n \sum_{j=1}^{n-1} K^2_n \bigg(\frac{j}{n}\bigg	)\bigg\{ K_{\theta_n} \bigg(\frac{j+1}{n}\bigg) - K_{\theta_n}\bigg(\frac{j}{n}\bigg) \bigg\} \nonumber\\
& - n \sum_{j=1}^{n-1} K_n \bigg(\frac{j}{n}\bigg	)\bigg\{ K^2_{\theta_n} \bigg(\frac{j+1}{n}\bigg) - K^2_{\theta_n}\bigg(\frac{j}{n}\bigg) \bigg\},
\label{eq:cramer_emp}	
\end{align}
where $K_n(v) = \frac{1}{n}\sum_{i=1}^{n}\mathbf{1}_{\{V_i \leq v\}},~ v \in \mathbb{I}$, $V_1 = C_n(\mathbf{U}_1), ..., V_n = C_n(\mathbf{U}_n)$ and $K_{\theta_n}$ denotes the distribution function of $C_{\theta_n}(U_1, ..., U_d)$, and
\begin{eqnarray}
S_n^{(R)} = \sum_{i=1}^n \{ D_n(\mathbf{E}_i) - C_{\Pi}(\mathbf{E}_i) \}^2, 
\label{eq:cram_ros}
\end{eqnarray}
where $D_n(\mathbf{u}) = \frac{1}{n}\sum_{i=1}^{n}\mathbf{1}_{\{\mathbf{E}_i \leq \mathbf{u}\}}, ~\mathbf{u} \in \mathbb{I}^d$, $C_{\Pi}(\mathbf{u}) = u_1 u_2 ... u_d$ is the $d$-variate independence copula and $\mathbf{E}_1 = \mathcal{R}_{\theta}(\mathbf{U}_1), ..., \mathbf{E}_n = \mathcal{R}_{\theta}(\mathbf{U}_n)$, where $\mathcal{R}_{\theta}$ is the Rosenblatt's transform based $C_{\theta_n}$.

A large value of such statistics leads to the rejection of $H_0: C \in \mathcal{C}_0$, where $\mathcal{C}_0 = \{C_{\theta} : \theta \in \mathcal{O} \}$ and $\mathcal{O}$ is an open subset of $\mathbb{R}^k,~ k \geq 1$. Thus for measuring the fitting quality of copula models, we can, informally, evaluate copula models with lower value of such statistics as `better'. All these three test statistics performed well in a large scale simulation study conducted at \cite{Gen09} in the bivariate case. We thus choose them as candidates for our purpose of goodness-of-fit evaluation.

We now introduce an \emph{g-aggregated statistic}. Note that a slightly different version of this statistic has originally appeared in \cite{goreckihofertholena2016approachjiis}. We later use such generalized statistics in Section \ref{sec:hetero_HAC_estim}. The one that follows is used directly in the HAC estimation process, where it is applied at each stage for evaluating which generator from some considered families fits the data best. In contrast, the one proposed in \cite{goreckihofertholena2016approachjiis} is used for evaluating the fit of a whole HAC estimate after its estimation is done, i.e., the evaluations computed for \emph{all} generators involved in that HAC are gatherer and aggregated using a $[0, +\infty)$-aggregation function, which is defined as follows.

\begin{dfn} \label{def:aggreg_fnc}
Let either $r = [0, a]$, where $a \in \mathbb{R}$ and $a > 0$, or $r = [0, +\infty)$.
Any function $g : r^k \rightarrow r, ~k \in \mathbb{N}$, satisfying 1)  $g(x, ..., x) = x$ for all $x \in r$ and 2) $g$ is exchangeable (i.e., $g(x_{p_1}, ..., x_{p_k}) = g(x_1, ..., x_k)$ for all $x_1, ..., x_k \in \mathbb{I}$ and all permutations $(p_1, ..., p_k)$ of $(1, ..., k)$) is called an \emph{$r$-aggregation} function.
\end{dfn}

\noindent Examples of, e.g., $[0, +\infty)$-aggregation functions are the functions maximum, minimum or average restricted to $[0, +\infty)^k$. Note that this type of aggregation functions is used in Algorithm \ref{alg:hetero_HAC_estim} for the input $g$.

\begin{dfn}\label{def:agg_s_n}
Let $\mathbf{u}_{IJ} = \{(U_{\bullet i}, U_{\bullet j}) | (i, j) \in I \times J \}$, where $U_{\bullet k} = (U_{1k}, ..., U_{nk})^{\top}$ (a column vector) for all $k \in \{1, ..., d\}$. Also, let $g$ be an $[0, +\infty)$-aggregation function and $S_n((U_{\bullet i}, U_{\bullet j}), C_{\psi})$ be the statistic corresponding to a GoF test, e.g., $S_n^{(E)},~ S_n^{(K)}$ or $S_n^{(R)}$, for a 2-AC $C_{\psi}$ and a pair $(U_{\bullet i}, U_{\bullet j}), ~i \neq j$. A \emph{g-aggregated statistics} $S_n^g$ is defined
%\begin{flalign*}
 %S_n^g\big(\mathbf{u}_{IJ}, \psi \big) =  ~ S_n((U_{\bullet i}, U_{\bullet j}), \psi) ,~  \textrm{ if}~ I = \{i\}, ~ J = \{j\}, 
%\end{flalign*}
%\noindent or
%\begin{flalign}
	 %~S_n^g\big(\mathbf{u}_{IJ}, \psi \big) = ~&	 
%g\bigl( ~S_n((U_{\bullet i_1}, U_{\bullet j_1}), \psi ),  \nonumber\\ 
  %&S_n((U_{\bullet i_1}, U_{\bullet j_2}), \psi ),  ..., 
   %~S_n((U_{\bullet i_l}, U_{\bullet j_q}), \psi )\bigr)  \textrm{,  otherwise,}
	%\label{eq:agg_sn}
%\end{flalign}
\begin{flalign}
	 S_n^g\big(\mathbf{u}_{IJ}, \psi \big) = 
	 \left\{ \begin{array}{ll}
S_n((U_{\bullet i}, U_{\bullet j}), C_{\psi}) & \textrm{if~} I = \{i\}, J = \{j\}, \\
\\
g\bigl(  S_n((U_{\bullet i_1}, U_{\bullet j_1}), C_{\psi} ),\\  
~~~S_n((U_{\bullet i_1}, U_{\bullet j_2}), C_{\psi} ), ..., \\
~~~S_n((U_{\bullet i_m}, U_{\bullet j_q}), C_{\psi} )\bigr), & \textrm{otherwise},\\ 
\end{array} \right.
	\label{eq:agg_sn}
\end{flalign}

\noindent where $I = \{i_1, ..., i_m\} , J = \{j_1, ..., j_q\}$ are non-empty disjoint subsets of $\{1, ..., d\}$.\\ 
\end{dfn}

Since the choice of $S_n^g$ substantially affects results of Algorithm \ref{alg:hetero_HAC_estim} and in turn the results of all experiments reported in Section \ref{sec:exps}, we now discuss the idea behind this concept in more detail.
Given a binary $d$-HAC $\HAC$, let $(U_1, ..., U_d) \sim C$. According to Proposition 3 in \cite{goreckihofertholena2016approachjiis}, it follows that, given leaves $i, j \in \{1, ..., d\}$, the bivariate margin $(U_i, U_j)$ is distributed according to the 2-AC $C_{\psi}$, where $\psi = \lambda(m)$ and $m$ is the \emph{youngest common ancestor} (see Definition 9 in \cite{goreckihofertholena2016approachjiis}) of the leaves $i$ and $j$. E.g., for the 3-HAC depicted in Figure \ref{fig:cop3}, since the node 5 is the youngest common ancestor of the pairs (1, 2) and (1, 3), it follows that $(U_1, U_2) \sim C_{\lambda(5)}$ and $(U_1, U_3) \sim C_{\lambda(5)}$. Similarly, as the node 4 is the youngest common ancestor of the pair (2, 3), $(U_2, U_3) \sim C_{\lambda(4)}$. Now, assume that we are estimating the generator $\lambda(5)$. 
Assuming a given family to consider for $\lambda(5)$, we can use some homogeneous HAC estimator, e.g., the one proposed in \cite{goreckihofertholena2016approachjiis} (and here recalled in Algorithm \ref{alg:HAC_estim}), which provides us with an estimate of its parameter. 
If we consider heterogeneous HACs and thus $\lambda(5)$ can be a member of one of more than one families, we can repeat the homogeneous HAC process for each of the possible families, which results in obtaining a corresponding number of generator estimates from these families. E.g., assume two possible families $a_1$ and $a_2$, which would result in two parameter estimates and in turn in two generator estimates, denoted by $\psi_{a_1}$, $\psi_{a_2}$. To decide which of these two estimates are more appropriate for modeling $\lambda(5)$, we can, e.g., evaluate how well a generator estimate fits the data. For this purpose, we suggest to use the aggregated statistic $S_n^g$ for this purpose. E.g., assume that $S_n = S_n^{(E)}$ and $g = \max$, i.e., $S_n^g = S_n^{(E), \max}$. 
As we know, both $(U_1, U_2)$ and $(U_1, U_3)$ are distributed according to $C_{\lambda(5)}$. Thus it is reasonable to evaluate an estimate of $\lambda(5)$ on the data that correspond to these two bivariate margins, i.e., on the two pairs of data columns $(U_{\bullet 1}, U_{\bullet 2})$ and $(U_{\bullet 1}, U_{\bullet 3})$ (note that we use the same idea also for the parameter estimation in Algorithms \ref{alg:HAC_estim} and \ref{alg:hetero_HAC_estim}). This can be done by setting $I = \{1\}$ and $J = \{2, 3\}$. Hence, for $\psi_{a_1}$, we evaluate $S_n^{(E), \max} \big(\mathbf{u}_{\{1\} \{2, 3\}}, \psi_{a_1} \big) = \max(S_n^{(E)}((U_{\bullet 1}, U_{\bullet 2}), \psi_{a_1}),  S_n^{(E)}(U_{\bullet 1}, U_{\bullet 3}), \psi_{a_1}))$, i.e., a rank-based Cram\'{e}r-von Mises statistic is computed for the two bivariate margins
and the maximum is returned. Similarly for $\psi_{a_2}$. As $S_n^{(E)}$ measures a certain type of a distance between the empirical copula corresponding to considered data and a 2-copula model (e.g., $C_\psi$), we choose the estimate with the lower value of the aggregated statistic.

\section{Estimation of hierarchical Archimedean copulas}
\label{sec:approach}
In this section, we propose a new approach to HACs estimation that allows to estimate both homogeneous and heterogeneous HACs. 
This approach is a generalization of the approach to homogeneous HACs estimation proposed in \cite{goreckihofertholena2016approachjiis}, which was experimentally compared with other state-of-the-art approaches to HACs estimation. This comparison shows that it outperforms the other approaches  in terms of the ability to determine the true structure, goodness-of-fit and run-time \cite{goreckihofertholena2016approachjiis}. In Section \ref{sec:homo_HAC_estim}, we first introduce its new version, where one superfluous step from the original version is removed, which can be done due to the findings summarized by Theorem \ref{thm:alg_homo_returns_HAC}. Then, we use it as a basis for the generalization to heterogeneous HACs proposed in Section \ref{sec:hetero_HAC_estim}.

Concerning only the structure estimation, a recent performance study comparing 11 available HAC structure estimators has been reported in \cite{uyttendaele2016estimation}. There, the estimator named \texttt{kt\_kagg} has shown the best performance in the ability to determine the true structure, as well as the lowest computation times.
This estimator  merges the estimation approach proposed in \cite{goreckihofertholena2016approachjiis} with the idea for collapsing tree structures proposed in \cite{Okh13}. Due to this, the estimation process of \texttt{kt\_kagg} is divided in two steps. In the first step, a binary tree is obtained using hierarchical agglomerative clustering. In the second step, it is collapsed if necessary, according to some strategy, see Algorithm 1 in \cite{uyttendaele2016estimation}. In this work, we also use such a two-step approach. First, we obtain a binary HAC using Algorithm \ref{alg:HAC_estim} or \ref{alg:hetero_HAC_estim}, and then, we use a collapsing strategy inspired in the strategy proposed in \cite{uyttendaele2016estimation}, which however differs in several of its aspects. This is discussed in Section \ref{sec:collapsing}.

As our definition of HACs explicitly employs a tree structure $(\VV,\EE)$, we can directly adopt some necessary concepts from graph theory and denote, for a node $v \in \VV$, by $\uparrow\!(v)$ its parent and by $\Uparrow\!(v)$ the set of all its ancestor forks. For a fork $v$, denote  by $\Downarrow\!(v)$ the set of all its descendant forks and by $\downarrow\!(v)$ the set of all its descendant leafs. For a leaf $v$, define $\downarrow\!(v) = \{v\}$. E.g., given the 3-HAC depicted in Figure \ref{fig:cop3}, $\uparrow\!(1) = \uparrow\!(4) = \Uparrow\!(1) = \Uparrow\!(4) = 5, ~ \uparrow\!(2) = \uparrow\!(3) = 4$, $\Uparrow\!(2) = \Uparrow\!(3) = \{4, 5\}$. Also, $\Downarrow\!(5) = 4, ~\Downarrow\!(4) = \emptyset$,  $\downarrow\!(5) = \{1, 2, 3\}$ and $\downarrow\!(4) = \{2,3\}$.

\subsection{Collapsing a binary structure to a non-binary one}
\label{sec:collapsing}

Briefly, all the collapsing strategies proposed in \cite{uyttendaele2016estimation} involve a process in which all parent-child pairs of forks that are close enough according to some parent-child distance $\delta$ are found, and each is collapsed into one fork. After a pair is collapsed into a fork, the parameter corresponding to this fork is re-estimated. Due to that, there might appear new parent-child pairs to collapse and thus the process repeats until there is nothing left to collapse. Concerning  $\delta$, the estimator \texttt{kt\_kagg} employs a distance given by $\delta(\tau_1, \tau_2) = |\tau_1 - \tau_2|$, where $\tau_1$ and $\tau_2$ are the Kendall's tau corresponding to the generators to collapse. The condition $\delta(\tau_1, \tau_2) < \tau_c$ for a threshold $\tau_c$  
 then determines whether these two nodes are close enough. However, no suggestions of how to set $\tau_c$ are provided in \cite{uyttendaele2016estimation}, which may make such an approach difficult to apply.

To overcome this drawback, we introduce a different approach inspired by the \emph{pruning} of decision trees proposed in \cite{CART1984}, in which we look for and collapse only one parent-child pair corresponding to the minimum according to $\delta$ (below, we refer to this minimum as to the \emph{minimal distance}), and repeat this process until there is nothing left to collapse, i.e., until a one-node structure is obtained. Note that if the families in a collapsing parent-child pair are different, we assign to the collapsed fork the family of the collapsing parent fork, which we do in an effort to satisfy the s.n.c. 
Once the collapsing process is finished, we obtain a set of structures with decreasing number of nodes, from which the user can choose, similarly to \cite{CART1984}, the most suitable structure according to his/her needs, e.g., according to goodness-of-fit or complexity of the model. This implies that the user is indeed not forced to specify any threshold before performing the collapsing step. 

However, in some cases, e.g., if one works with a lot of HAC estimates (see our Section \ref{sec:exps}), some automatized procedure that estimates the number of forks in an underlying non-binary HAC (which in turn determines which one to choose from the set of structures obtained in the collapsing step) would be desirable. For this purpose, we introduce a procedure based on a simple heuristic, which estimates the number of forks using the minimal distances generated during the collapsing process. 
Let $d \geq 3$ and $C_1$ be a binary $d$-HAC, i.e., $C_1$ has $d-1$ forks. Let $\delta_1 = 0$. Collapse $C_1$ to $C_2$ using the approach described in the previous paragraph, i.e., $C_2$ has $d-2$ forks, and store the minimal distance from from $C_1$ in $\delta_2$. 
Repeating this process until we get a $d$-AC $C_{d-1}$, we obtain the series $(\delta_1, ..., \delta_{d-1})$, which we use for the estimation of the number of forks in the true copula. To get this estimate, we simply find the lowest $\hat{i}$ from $\{1, ..., d-2\}$ such that $\delta_{\hat{i}+1} - \delta_{\hat{i}}  \geq \frac{\delta_{d-1}}{d-1}$. Note that such an $\hat{i}$ alway exists, which follows from the consideration that some of the steps $\delta_{\hat{i}+1} - \delta_{\hat{i}}, ~\hat{i} \in \{1,...,d-2\}$ have to be larger than or equal to the average step $\frac{\delta_{d-1}}{d-1}$.
Then $C_{\hat{i}}$ is considered to be the estimate of the true copula and thus $\hat{k} = d-\hat{i}$ the estimate of the number of forks in the true copula. 

It follows from the construction of the procedure that $\hat{k} \neq 1$, i.e., the procedure does not allow for collapsing an HAC into an AC. We thus suggest to always consider $C_{\hat{i}}$ and also $C_{d-1}$, where  $C_{d-1}$ might be preferred if the difference between the lowest and the highest $\tau$ of  the forks in $C_1$ is very `low', where `low' has to be determined by the user based on a particular data. Together with these two copulas, we also suggest to consider another one, namely $C_1$. This follows from our observation that if the true copula is binary, the suggested procedure frequently misses to estimate $\hat{k} = d-1$. Here, $C_1$ might be preferred if the lowest difference between the $\tau$ of a parent and a child in $C_1$ is very `large', where, again, `large' has to be determined by the user based on a particular data. One should thus always consider $C_{\hat{i}}$ and these two extremes. However, note that such a consideration can always be done \emph{after} a collapsing process has been finished, i.e., after some insights to the data have been gained by the user, on the contrary to the procedures suggested in \cite{Okh13,uyttendaele2016estimation}.

% these figures are obtained from nforksestimation.m
\begin{figure}[htb]
	\centering
		\subfigure[A \{C, 12, 14, 19, 20\}-heterogeneous non-binary 15-HAC model]{%
            \label{fig:nForks_model}
		\includegraphics[width=0.45\textwidth]{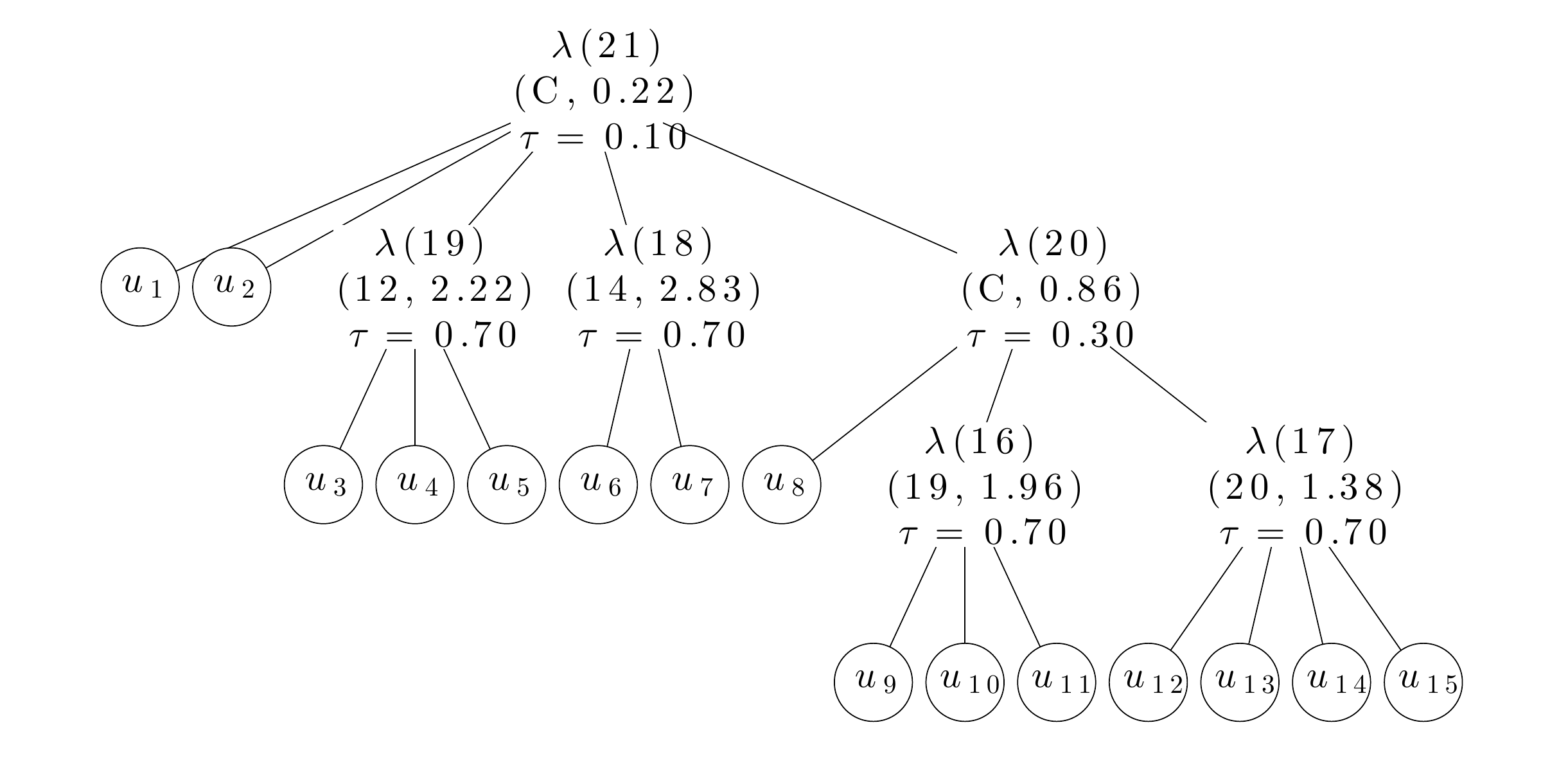}
						}
		\subfigure[A binary 15-HAC estimate]{%
            \label{fig:nForks_estimate}
		\includegraphics[width=0.45\textwidth]{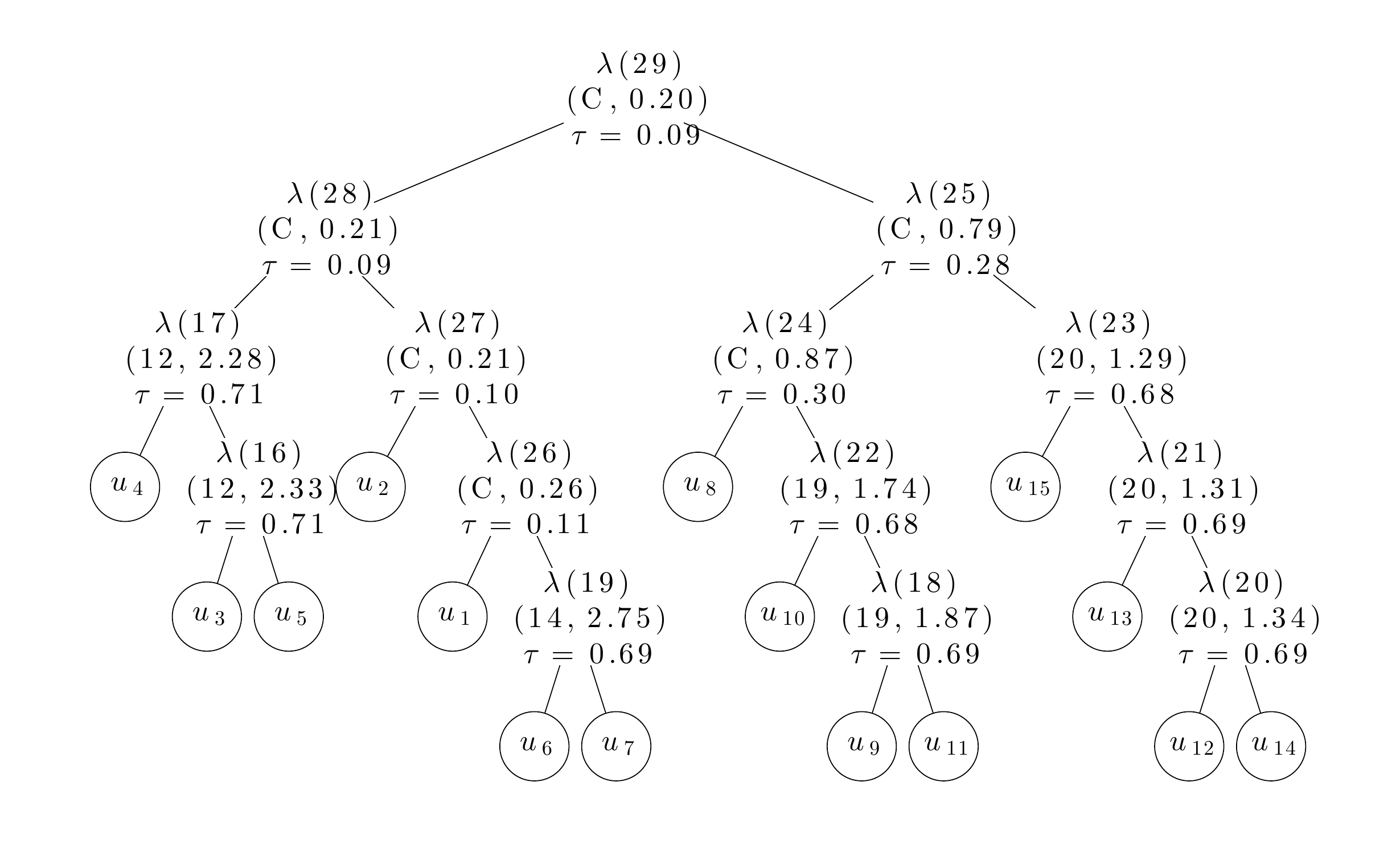}
						}
		\subfigure[The minimal distances obtained during the collapsing process.]{%
            \label{fig:nForks_deltas}
		 \includegraphics[width=0.4\textwidth]{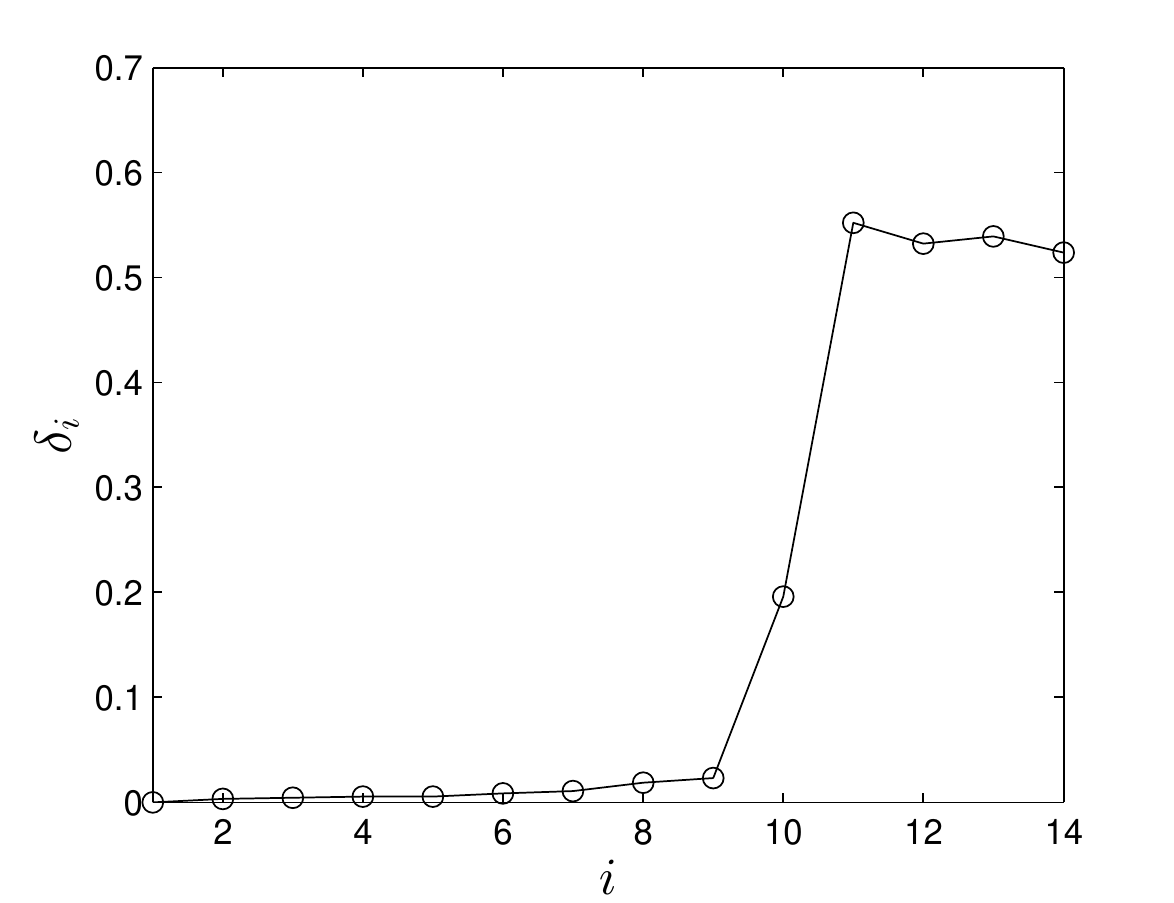}
								}		
		\subfigure[A collapsed 15-HAC estimate]{%
            \label{fig:nForks_collapsed}
		\includegraphics[width=0.45\textwidth]{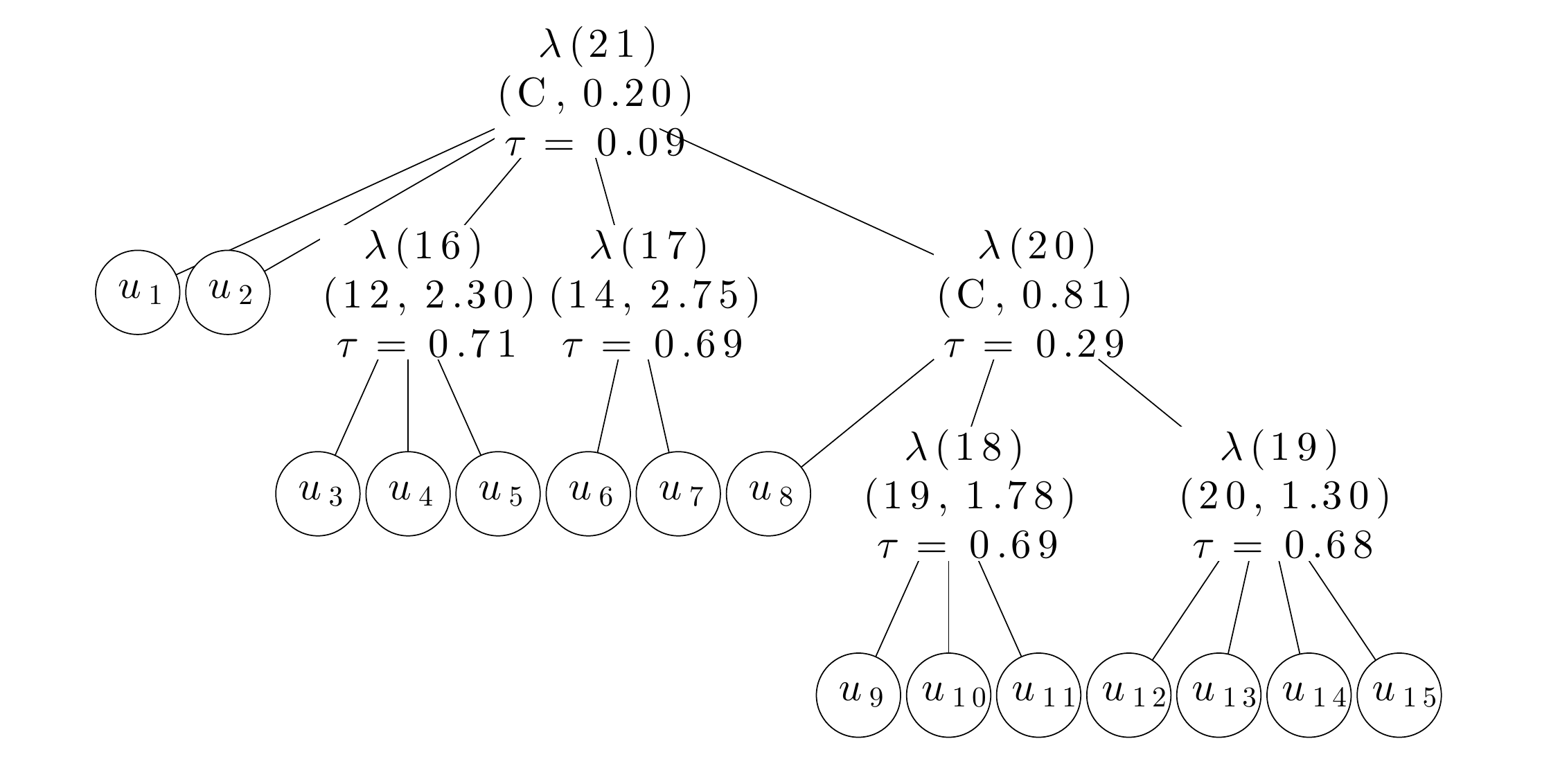}
						}
		\caption{Collapsing of a 15-HAC.}
		\label{fig:forks_estimation}
\end{figure}

\begin{figure}[b]
	\centering
		\includegraphics[width=1\textwidth]{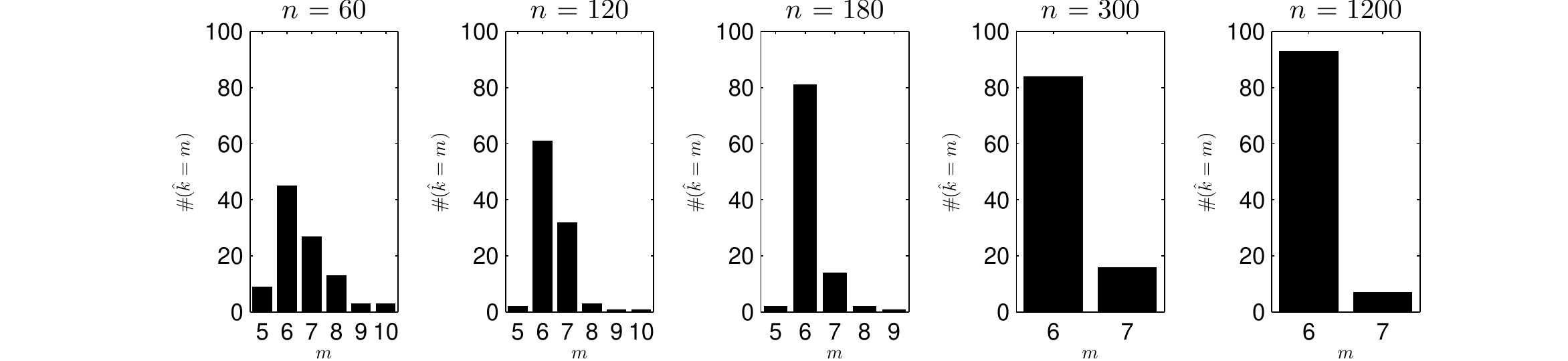}
		\caption{Distributions of $\hat{k} ~(= d - \hat{i})$ for 100 repetitions and for $n \in \{60, 120,  180, 300, 1200\}$.}
		\label{fig:forks_hist}
\end{figure}

To illustrate our procedure, we generated a sample of $n = 1200$ observations from the \{C, 12, 14, 19, 20\}-heterogeneous 15-HAC depicted in Figure \ref{fig:nForks_model}, and, using one of the binary HAC estimation approaches introduced below in Section \ref{sec:hetero_HAC_estim}, we obtained the binary 15-HAC depicted in Figure \ref{fig:nForks_estimate}.
Then, using the collapsing approach described above, we generated the series $(\delta_1, ..., \delta_{14})$  depicted in Figure \ref{fig:nForks_deltas}. Observe the large difference between $\delta_9$ and $\delta_{10}$. This difference means that the minimal distance is substantially larger when collapsing from $C_9$ to $C_{10}$  than when collapsing from $C_8$ to $C_9$. Less formally, when collapsing from $C_9$ to $C_{10}$, there might be collapsed forks that are far from each other in the $\delta$ distance and possibly should not be collapsed. Such a statement can be supported by the fact that $\hat{i} = 9$, and thus the estimate of the number of forks in the model is $\hat{k} = 6$. The HAC $C_9$ is depicted in Figure \ref{fig:nForks_collapsed}. For deeper insight, we repeat this estimation process 100 times for each $n \in \{60, 120, 180, 300, 1200\}$ (multiplications of 60, which are taken from the experiments described in Section \ref{sec:exps}) and show the results in the histograms depicted in Figure \ref{fig:forks_hist}.
% the histograms are obtained from nforksestimation.m (saved in nforksestimation.mat)
Observe that the frequency of $\hat{k}$ = 6 is getting closer to 100 as $n$ grows.
In Section \ref{sec:exps} we show, how successful such an approach is compared to the (unrealistic) assumption that the number of forks in the true copula is known.

As another improvement, we propose the re-estimation of the parameter of a collapsed fork.
The re-estimation process used in \texttt{kt\_kagg} is such that the Kendall's tau corresponding to the collapsed fork $\tau_{\tilde{v}}$ is set to $\tau_{\tilde{v}} = \min\{\tau_v, \tau_w\}$, where $\tau_v$ is the Kendall's tau corresponding to the collapsing parent fork and $\tau_w$ corresponds to the Kendall's tau of the collapsing child fork. 
If the s.n.c.~holds in the collapsing HAC, then $\tau_v \leq \tau_w$, which implies that $\tau_{\tilde{v}} =\tau_v$. On the one hand, such an approach (denoted by \emph{TauMin} in the following) guarantees the s.n.c.~to hold also for the collapsed HAC.
On the other hand, it does not take into account the value of the child parameter, which might unnecessarily lead to biasing the collapsed HAC estimate. Hence, in the following paragraph, we introduce a new re-estimation procedure (denoted \emph{KTauAvg}), which involves all $\tau^n_{ij}$ values related to the collapsed node and thus takes into account also the value of the child parameter. These two approaches are compared below in Section \ref{sec:exps}. Considerations about assuring the s.n.c. for the collapsed HAC while using our collapsing strategy described above are addressed in Section \ref{sec:homo_HAC_estim} (the paragraph containing \eqref{eq:tau_tilde_v_is_weighted_avg}) and Section \ref{sec:hetero_estim_others}.

Denote the \emph{sample version of Kendall correlation matrix} by $(\tau^n_{ij})$, where $\tau^n_{ij}$ denotes the sample version of Kendall's tau between $(U_{1i}, ..., U_{ni})$ and $(U_{1j}, ..., U_{nj})$ defined by \eqref{eq:kendall_emp}. Let $v$ be a fork and $\wedge(v) = \{w_1, ..., w_k\}, ~k \geq 2$ be the set of its children. Then the proposed (re-)estimator is given by
\begin{flalign}
\hat{\theta}_v = \tau_{(a)}^{-1}\Bigl(\avg\bigl((\tau^n_{\tilde{i}\tilde{j}})_{(\tilde{i},\tilde{j}) \in \downarrow(w_1) \times \downarrow(w_2)}, (\tau^n_{\tilde{i}\tilde{j}})_{(\tilde{i},\tilde{j}) \in \downarrow(w_1) \times \downarrow(w_3)}\bigr), ..., (\tau^n_{\tilde{i}\tilde{j}})_{(\tilde{i},\tilde{j}) \in \downarrow(w_{k-1}) \times \downarrow(w_k)}\Bigr),
	\label{eq:re-estimation}
\end{flalign}
where $a$ is the family of $\lambda(v)$ and $\avg$ is the average function. This estimator is just a generalization of the binary estimator based on Kendall's tau inversion proposed in \cite{goreckihofertholena2016approachjiis} to the non-binary case. Note that, given a node $v$ with two children, \eqref{eq:re-estimation} simplifies to the estimator used in Step 1 of Algorithm \ref{alg:HAC_estim}, see Section \ref{sec:homo_HAC_estim}.
Also note that if the employed values $\tau^n_{ij}$ have already been computed, which always happens in our approach as we use them to estimate the HAC structure, the computation time of this estimator is negligible. 
Let us illustrate the estimator using the 3-HAC depicted in Figure \ref{fig:cop3}. Given a family $a$, assume that $a_4 = a_5 = a$ and collapse the nodes 4 and 5. Denote the collapsed node by $v$ and consider that $\wedge(v) = \{1, 2, 3\}$. Then \eqref{eq:re-estimation} turns into $\hat{\theta}_v = \tau_{(a)}^{-1}(\avg(\tau^n_{12}, \tau^n_{13}, \tau^n_{23})).$

\subsection{HACs structure estimation}
\label{sec:structure_estimation}
Given an input matrix $(\tau^n_{ij})$, Algorithm \ref{alg:structure_estim} returns a binary structure estimate $(\hat{\VV}, \hat{\EE})$ and a sequence $(\hat{\tau}_k)_{d+1}^{2d-1} \in [-1, 1]^{d-1}$, where $\hat{\tau}_k$ is an estimate of $\tau(\lambda(k)), ~ k \in \{d+1, ..., 2d-1\}$.
This algorithm, which is based on the ideas from \cite{Gor13MME}, has been proposed in \cite{goreckihofertholena2016approachjiis} (see Algorithm 2 therein).
\begin{algorithm}[t]
\floatname{algorithm}{Algorithm}
\caption{HACs structure estimation}
\label{alg:structure_estim}

\begin{algorithmic}
\renewcommand{\algorithmicrequire}{\textbf{Input:}}
\renewcommand{\algorithmicensure}{\textbf{The estimation:}}
\REQUIRE 
\STATE 1) $(\tau^n_{ij})$ ... the sample version of a Kendall correlation matrix
%\STATE 2) $g$ ... an $\I$-aggregation function

\STATE ~
\ENSURE
\STATE 1. $\hat{\VV} := \{1, ..., 2d-1\}, ~\hat{\EE} := \emptyset$, $\mathcal{I} := \{1, ..., d\}$ 
\STATE ~~~ recall that $\downarrow\!(i) = \{i\}$ for $i \in \{1, ...,d\}$
\FOR{$k = 1, ..., d - 1$}
\STATE 2. find two nodes from $\mathcal{I}$ to join, i.e, 
\STATE ~~~ $(i, j) := \argmax\limits_{\tilde{i} < \tilde{j},~ \tilde{i} \in \mathcal{I}, ~\tilde{j} \in \mathcal{I}} \avg((\tau^n_{\tilde{\tilde{i}}\tilde{\tilde{j}}})_{(\tilde{\tilde{i}},\tilde{\tilde{j}}) \in \downarrow(\tilde{i}) \times \downarrow(\tilde{j})})$
\STATE 3. set the children of the fork $d+k$ to $\{i, j\}$, i.e., 
\STATE ~~~ $\hat{\EE} := \hat{\EE} \cup \{\{i, d+k\}, \{j, d+k\}\}$,
\STATE ~~~ which implies that $\wedge(d+k) = \{i, j\}$ and $\downarrow\!(d+k) = \downarrow\!(i) \cup \downarrow\!(j)$
\STATE 4. remove the nodes $i$ and $j$ from the clustering process (as they have been
\STATE ~~~ already joined) and add the fork $d+k$ to be considered for joining in 
\STATE ~~~ the following steps, i.e., 
\STATE ~~~ $\mathcal{I} := \mathcal{I} \cup \{d + k\} \backslash \{i, j\}$
\STATE 5. estimate the Kendall's tau corresponding to the fork $d+k$, i.e., 
\STATE ~~~ $\hat{\tau}_{d+k} := \avg((\tau^n_{\tilde{i}\tilde{j}})_{(\tilde{i},\tilde{j}) \in \downarrow(i) \times \downarrow(j)})$
\ENDFOR
\STATE ~
\renewcommand{\algorithmicensure}{\textbf{Output:}}
\ENSURE
\STATE $(\hat{\VV}, \hat{\EE}, (\hat{\tau}_k)_{d+1}^{2d-1})$

\end{algorithmic}
\end{algorithm}	
According to our new findings summarized below by Lemma \ref{lem:D_rs_is_monotonic}, it is actually an agglomerative hierarchical clustering (AHC) method \cite[p. 414]{Clar09}, which is a member of a broad class of AHC methods defined as follows. If $D_{rs}$ denotes a dissimilarity between objects or clusters $r$ and $s$ and $D_{r(s,t)}$ is the dissimilarity between $r$ and the combined cluster $(s, t)$, then
\begin{flalign}
D_{r(s,t)} = \alpha_s D_{rs} + \alpha_t D_{rt} + \beta D_{st} + \gamma |D_{rs}-D_{rt}|
\label{eq:D_rst}
\end{flalign}
defines the clustering method for suitable coefficients $(\alpha_s, \alpha_t, \beta, \gamma)$.  Due to notational convenience,  we will use the similarity $-D_{rs}$ instead of a dissimilarity $D_{rs}$ in the following.

From the point of view of Algorithm \ref{alg:structure_estim}, objects in AHC refer to leaves and clusters refer to sets of descendant leaves (i.e., to $\downarrow(\cdot)$). 
Given two clusters or leaves $\downarrow(r), \downarrow(s)$, the similarity (simply denoted by $-D_{rs}$) between $\downarrow(r)$ and $\downarrow(s)$ used in the algorithm is 
\begin{flalign}
-D_{rs} = \avg((\tau^n_{\tilde{i}\tilde{j}})_{(\tilde{i},\tilde{j}) \in \downarrow(r) \times \downarrow(s)}), 
\label{eq:D_rs}
\end{flalign}
see Step 2.

Now consider that a dendrogram obtained by an AHC method is a binary tree where the value $D_v$ defined $D_v = D_{rs}$ is assigned to each fork $v$ in the tree assuming that $\wedge(v) = \{r,s\}$.
\emph{Monotonicity} of dendrogram means that if $v_{q_1}, ..., v_{q_k}$ is a branch in a dendrogram such that $v_{q_1}$ is a leave and $v_{q_k}$ is the root, then $D_{v_{q_2}} \leq D_{v_{q_3}} \leq \ldots \leq D_{v_{q_k}}$ holds. 
In other words, if a dendrogram with $d$ leaves is monotonic, then
\begin{flalign}
(\forall v \in \{d + 1, ..., 2d-1\})(\forall \tilde{v} \in \wedge(v) \cap \{d+1,..., 2d-1\}) ~	  D_v \geq D_{\tilde{v}}.
\label{eq:D_v_cond}
\end{flalign}

Due to connection of \eqref{eq:D_rs} and Step 5, i.e., $-D_v = \hat{\tau}_v$ for all $v \in \{d+1, ..., 2d-1\}$, \eqref{eq:D_v_cond} turns to \eqref{eq:nnc} providing $k = d - 1$ and $\tau_v = \hat{\tau}_v$ for all $v \in \{d+1, ..., 2d-1\}$.
In other words, if the clustering method based on the dissimilarity $D_{rs} = -\avg((\tau^n_{\tilde{i}\tilde{j}})_{(\tilde{i},\tilde{j}) \in \downarrow(r) \times \downarrow(s)})$ produces a monotonic dendrogram, then the output $(\hat{\VV}, \hat{\EE}, (\hat{\tau}_k)_{d+1}^{2d-1})$ of Algorithm \ref{alg:structure_estim} satisfies \eqref{eq:nnc}. The next question is clear: Can we somehow assure that an AHC method produces only monotonic dendrograms? Fortunately, there exists a sufficient and necessary condition answering this question.

In \cite{batagelj1981ahc}, it was shown that a hierarchical clustering method given by equation \eqref{eq:D_rst} is monotonic, if and only if the following conditions hold:
\begin{enumerate}
	\item $\gamma \geq -\min(\alpha_t, \alpha_s)$;
	\item $\alpha_s + \alpha_t \geq 0$;
	\item $\alpha_s + \alpha_t + \beta \geq 1$.
\end{enumerate}

The following lemma shows that $D_{rs}$ given by \eqref{eq:D_rs} admits the representation by \eqref{eq:D_rst}. 

\begin{lemma}
Let $r, s$ and $t$ be nodes in a binary tree with $d$ leaves such that $\downarrow(r), ~ \downarrow(s)$ and $\downarrow(t)$ are disjoint subsets in $\{1, ..., d\}$ and there exists a fork $p$ satisfying $\wedge(p) = \{s, t\}$.
Let $D_{rs}$ be given by \eqref{eq:D_rs}. Then 
\begin{flalign}
D_{r(s,t)} = \alpha_s D_{rs} + \alpha_t D_{rt},
\label{eq:D_rs_is_monotonic}
\end{flalign}
where $\alpha_s >0, ~\alpha_t > 0$ and $\alpha_s + \alpha_t = 1$.
\label{lem:D_rs_is_monotonic}
\end{lemma}

Observing that coefficients $(\alpha_s, \alpha_t, \beta, \gamma)$  such that $\alpha_s >0, ~\alpha_t > 0, ~\alpha_s + \alpha_t = 1$ and $\beta = \gamma = 0$ satisfy the conditions on monotonicity implies the following corollary.

\begin{corollary}
Given an matrix $(\tau^n_{ij})$, 
the output $(\hat{\VV}, \hat{\EE}, (\hat{\tau}_k)_{d+1}^{2d-1})$ of Algorithm \ref{alg:structure_estim} satisfies \eqref{eq:nnc} provided $\tau_i = \hat{\tau}_i$ for all $i \in \{d+1, ..., 2d-1\}$.
\label{cor:alg_satisfies_nnc}
\end{corollary}

As follows from Lemma \ref{lem:nnc_equals_snc}, \eqref{eq:nnc} can be turned to \eqref{eq:holena_nesting_cond_simple} under relatively weak assumptions for a lot of families. This fact is essential in the construction of a new homogenous HAC estimator described in the following section.

\subsection{Homogeneous HACs estimation}
\label{sec:homo_HAC_estim}
This section proposes an improved version of the approach for homogeneous HACs estimation introduced in \cite{goreckihofertholena2016approachjiis}. The new approach summarized by Algorithm \ref{alg:HAC_estim} is a version of Algorithm 3 proposed in \cite{goreckihofertholena2016approachjiis} rewritten using our notation and, as will be explained below, one step has been removed. 

\begin{algorithm}[t]
\floatname{algorithm}{Algorithm}
\caption{The homogeneous HAC estimation}
\label{alg:HAC_estim}

\begin{algorithmic}
\renewcommand{\algorithmicrequire}{\textbf{Input:}}
\renewcommand{\algorithmicensure}{\textbf{The estimation:}}
\REQUIRE 
\STATE 1) $(\hat{\VV}, \hat{\EE}, (\hat{\tau}_k)_{d+1}^{2d-1})$ ... i.e., an output of Algorithm \ref{alg:structure_estim}
\STATE 2) $a$ ... an Archimedean family 

\STATE ~
\ENSURE
\FOR{$k = 1, ..., d - 1$}
\STATE 1. estimate the parameter corresponding to the fork $d+k$, i.e., 
\STATE ~~~ $\hat{\theta}_{d+k} := \tau^{-1}_{(a)}(\hat{\tau}_{d+k})$ ... see Remark \ref{rem:tau_trim}
%\STATE 6. (a redundant step, as follows from Theorem \ref{thm:alg_homo_returns_HAC}) if necessary, trim the parameter in order to the s.n.c. be satisfied, i.e., 
%\STATE $\bigl(\hat{\theta}_{d+k} = \min(\hat{\theta}_{d+k}, \hat{\theta}_{i}, \hat{\theta}_{j})\bigr)$
\STATE 2. store the estimated parameter and the input family $a$ in the labeling 
\STATE ~~~ $\hat{\lambda}$, i.e.,
\STATE ~~~ $\hat{\lambda}(d+k) := \psi^{(a, \hat{\theta}_{d+k})}$ 
\ENDFOR
\STATE ~
\renewcommand{\algorithmicensure}{\textbf{Output:}}
\ENSURE
\STATE $(\hat{\VV}, \hat{\EE}, \hat{\lambda})$

\end{algorithmic}
\end{algorithm}

As mentioned above, Algorithm 3 in \cite{goreckihofertholena2016approachjiis} includes one more step that serves to guarantee that the resulting estimate is a copula. This step, which is stated as
\begin{flalign}
\hat{\theta}_{d+k}= \min\{\hat{\theta}_{d+k}, \hat{\theta}_{i}, \hat{\theta}_{j}\}, 
\label{eq:superfluous_step}
\end{flalign}
where $ \hat{\theta}_{i}$ and $ \hat{\theta}_{j}$ denote the parameters of the children of the fork $d+k$, would be placed between Step 1 and Step 2 in our new version. Such a step manually forces \eqref{eq:holena_nesting_cond_simple} to be fulfilled, which however might lead to a bias in estimation. Assuming a homogenous HAC from a family with the s.n.c. in the form $\theta_1 \leq \theta_2$, see Table \ref{tab:geners}, such a step is not necessary under relatively weak assumptions as the following theorem reveals.

\begin{thm} \label{thm:alg_homo_returns_HAC}
Let $a$ be a family such the assumptions 1)-3) from Lemma \ref{lem:nnc_equals_snc} hold.
Also, let $(\hat{\VV}, \hat{\EE}, (\hat{\tau}_k)_{d+1}^{2d-1})$ be a triplet obtained by Algorithm \ref{alg:structure_estim}. If 
\begin{flalign}
\forall i \in \{d+1,..., 2d -1\}, ~\hat{\tau}_i \in \hat{\tau}_{(a)}(\Theta_a) 
\label{eq:hat_in_Theta_a}
\end{flalign} 
then the triplet $(\hat{\VV}, \hat{\EE}, \hat{\lambda})$ returned by Algorithm \ref{alg:HAC_estim} satisfies \eqref{eq:holena_nesting_cond_simple}, i.e, $C_{(\hat{\VV}, \hat{\EE}, \hat{\lambda})}$ is a copula.
\end{thm}

Note that in Steps 2 and 5 of Algorithm \ref{alg:structure_estim}, one can consider also other aggregation functions instead of $\avg$, e.g., the minimum or maximum.
However, according to the experiments reported in \cite{goreckihofertholena2016approachjiis}, setting the aggregation function $\avg$ provides better results than, e.g., for the minimum or maximum. We thus do not provide analogues to Theorem \ref{thm:alg_homo_returns_HAC} for these alternative aggregation functions.

Considering other homogenous estimators available in the literature, analogues to Theorem \ref{thm:alg_homo_returns_HAC} do not exist and thus \eqref{eq:nnc} either must be manually forced or it cannot be assured that the resulting estimate satisfies \eqref{eq:holena_nesting_cond_simple}. To illustrate this fact, we now consider the following two estimators: 
\begin{enumerate}
\item Step 5 of Algorithm \ref{alg:structure_estim} together with Step 1 of Algorithm \ref{alg:HAC_estim} imply that $\hat{\theta}_{d+k} := \tau^{-1}_{(a)}\bigl(\avg((\tau^n_{\tilde{i}\tilde{j}})_{(\tilde{i},\tilde{j}) \in \downarrow(i) \times \downarrow(j)})\bigr)$. Another Kendall's tau based estimator can be constructed just by changing the ordering of  the use of $\avg$ and $\tau_{(a)}^{-1}$, i.e., by letting $\hat{\theta}_{d+k} := \avg\bigl((\tau^{-1}_{(a)}(\tau^n_{\tilde{i}\tilde{j}}))_{(\tilde{i},\tilde{j}) \in \downarrow(i) \times \downarrow(j)}\bigr)$ .
Such an estimator can be found in \cite{Koj10a}. In the context of ACs, it is experimentally shown in \cite{Hofert13}, see Tables 5, 6 and 7 therein, that although this estimator (denoted there by $\tau_{\bar{\hat{\theta}}}$) has a bias that is similar to the bias of our estimator (denoted there by $\tau_{\bar{\hat{\tau}}}$), $\tau_{\bar{\hat{\theta}}}$ is computationally less efficient than $\tau_{\bar{\hat{\tau}}}$. 
The following observation that this estimator does not assure that the resulting estimates satisfy \eqref{eq:holena_nesting_cond_simple} can thus be considered as another evidence against choosing this estimator for HAC estimation;
\item The HAC estimator introduced in \cite{Okh13} with an improvement proposed in \cite{goreckihofertholena2014a}. In this estimator, we use the maximum likelihood (ML) estimation method for the estimation of $\theta_{d+k}$ at each step $k \in \{1, ..., d-1\}$. An implementation of this estimator in \textsf{R} can be found in \cite{OkhrinRistig2014HACinR} and one of its applications is described in \cite{Okhrin2015Conditional}. For a detailed implementation in a pseudo-code (already generalized to the heterogeneous case), see Appendix \ref{app:diagonal}.
\end{enumerate}

To quantify how many times an estimator returns an estimate that violates \eqref{eq:holena_nesting_cond_simple}, we have done a small experiment, which is described below. According to our observations, the closer the parameters of the parent-child pairs in a HAC model are, the more the considered estimators are prone to violate \eqref{eq:holena_nesting_cond_simple} for a data sampled according to this model. When choosing a HAC model that would be appropriate for this experiment, we went to an extreme and set all its parameters equal, which results in turning this HAC model to an AC.
Hence, we generated 1000 times a sample of 100 ($n$ = 100) observations according to the 10-AC from the Clayton family with the parameter $\theta = 1$ ($\tau_{(\textrm{C})}(1) = \frac{1}{3}$). The first considered estimator violated \eqref{eq:nnc} 7 times and the latter 958 times. Our estimator does not show this problem. 
Clearly, these results do not imply anything about overall performance of the considered estimators but rather serve for illustrating a desirable property of our estimator that is unique among the considered ones. 
Also note that other HAC estimators could be created, e.g., by replacing the ML AC estimator in the latter HAC estimator by the inversion of Kendall's tau or by doing just the opposite in Algorithm \ref{alg:HAC_estim}. However, consider that such HAC estimators would be dependent directly on the pseudo-observations (due to the ML estimation or the diagonal transformation), in contrast to Algorithm \ref{alg:HAC_estim}, which directly depends only on the Kendall correlation matrix. This would imply that they would not admit \eqref{eq:D_rst} and thus the s.n.c. would not be guaranteed.

\begin{rmk} \label{rem:tau_trim}
%Despite that MLE is generally preferred, we use the estimator based on the inversion of Kendall's tau. It is because, as we anyway need to compute the Kendall correlation matrix for the structure determination and $\tau_{(a)}(\theta)$ is mostly known in a semi-closed form, we can estimate the parameters directly from the Kendall correlation matrix, which is substantially faster than using MLE. Note that other estimators could be used instead these two, see, e.g., the AC estimators proposed in \cite{Hofert13}. Their use are, however, out of the scope of this work.

As $\tau_{(\textrm{A})}([0, 1)) = [0, \frac{1}{3})$, i.e., the family A is unable to model dependencies that correspond to $\tau \geq \frac{1}{3}$, we  set arbitrarily $\hat{\theta}_{d+k} = 1 - \epsilon$ in cases where $\hat{\tau}_{d+k}\geq \frac{1}{3}$, where $1 - \epsilon$ denotes the highest real allowed by the computing system lower than 1. E.g., in MATLAB, $1 - \epsilon$ would be 1 - 2.2204e-16. A similar approach is used for $a  \in \{12, 14, 19\}$, i.e., if $\hat{\tau}_{d+k} \leq \frac{1}{3}$, we set arbitrarily $\hat{\theta}_{d+k} = \epsilon$ for $a$ = 19, and, if $\hat{\tau}_{d+k} < \frac{1}{3}$,  we set arbitrarily $\hat{\theta}_{d+k} = 1$ for $a \in \{12, 14\}$. Also, as HACs satisfying \eqref{eq:holena_nesting_cond} are unable to model pairwise dependencies such that the corresponding $\tau$ is negative, if $\hat{\tau}_{d+k} < 0$, we set arbitrarily $\hat{\theta}_{d+k} = 0$ for $a$ = A, and $\hat{\theta}_{d+k} = \eps$ for $a \in$ \{C, 20\}.
\end{rmk}

Clearly, setting the parameter arbitrarily to some value causes a bias. We distinguish two \emph{attitudes} to cope with this problem: 
\begin{enumerate}
	\item The \emph{optimistic} attitude, in which we allow, if necessary, to set the parameter to some arbitrary value, accepting that the resulting estimate is biased. This attitude is motivated by the fact that a) the parameter estimation is relatively fast (it is a matter of milliseconds rather than days or weeks) and b) there are instruments available that can measure how well the resulting estimate fits the data, e.g., the GoF statistics recalled in Section \ref{sec:gof}. Thus, in the case that we have a collection of biased estimates from different families, we can choose the best fitting one using these instruments;
	\item The \emph{pessimistic} attitude, in which, given a family $a$, we stop the estimation process whenever $\hat{\tau}_{d+k} \notin \tau_{(a)}(\Theta_a)$, accepting that the estimation process may not result in any (H)AC at all.
\end{enumerate}
It is important to note that in some cases, even if we set the parameter estimate arbitrarily to some value and thus caused bias, it might happen that the resulting estimate fits the data better than another estimate corresponding to some other family that does not need to be trimmed to some interval. E.g., let $\hat{\tau}_{d+k} = 0.34$. In this case, for the family A, we arbitrarily set $\hat{\theta}_{d+k}$ to $1-\eps$ , i.e., $\tau_{(\textrm{A})}(\hat{\theta}_{d+k}) = \tau_{(\textrm{A})}(1-\eps) < \frac{1}{3}$. For the family C, such a trimming is not needed and the estimate $\tau^{-1}_{(\textrm{C})}(0.34)$ is provided. But, if the copula underlying the data is more like a copula from the family A, e.g., in tails, GoF might show lower value,  i.e., better fit, for the estimate corresponding to the family A despite it is biased. 
Hence, to avoid omitting such biased estimates that provide better fit, even if they appear rather rarely, we use the optimistic attitude. For more examples in the heterogeneous case, see Section \ref{sec:hetero_example}.

Concerning the collapsing procedure we proposed in Section \ref{sec:collapsing}, we will now show that 
if \eqref{eq:hat_in_Theta_a} holds for the input of Algorithm \ref{alg:HAC_estim}, the construction of this re-estimator inherently assures that the collapsed HAC satisfies the s.n.c. Assume an $a$-homogeneous (possibly non-binary) $d$-HAC with $k$ forks, $k \leq d -1$ satisfying \eqref{eq:holena_nesting_cond_simple}, and denote by $\tau_m$ the quantity $\tau(\lambda(m))$ for all $m \in \{d+1, ..., d+k\}$. 
Let $v$ be a fork and $\wedge(v) = \{w_1, ..., w_s\}, ~s \geq 2$ be the set of its children.
Now denote by $\tilde{v}$ the fork that is created by collapsing the forks $v$ and (without a loss of generality) $w_1$, i.e., $\tau_{w_1}$ is closer (in $|\tau_v - \tau_{\cdot}|$) to $\tau_v$ than $\tau_{w_2}, ..., \tau_{w_s}$. Assume that $\wedge(w_1) = \{w_{1,1}, ..., w_{1,l}\}, ~l \geq 2$, which implies $\downarrow\!(w_1) = \downarrow\!(w_{1,1}) \cup ... \cup \downarrow\!(w_{1,l})$ and $\wedge(\tilde{v}) = \{w_{1,1}, ..., w_{1,l}, w_2, ..., w_s\}$. Now, rewriting $\tau_{\tilde{v}}, \tau_v$ and $\tau_{w_1}$ using \eqref{eq:re-estimation}, it can be observed (similarly to the proof of Lemma \ref{lem:D_rs_is_monotonic}) that (simplifying $(\tau^n_{\tilde{i}\tilde{j}})_{(\tilde{i}\tilde{j}) \in \downarrow(w_i) \times \downarrow(w_j)}$ to $(\cdot)_{w_i \times w_j})$
\begin{flalign}
\tau_{\tilde{v}} =  \avg\bigl( & (\cdot)_{w_{1,1} \times w_{1,2}}, ..., (\cdot)_{w_{1,1} \times w_{1,l}}, (\cdot)_{w_{1,1} \times w_2}, ..., (\cdot)_{w_{1,1} \times w_s}, \nonumber\\
 & (\cdot)_{w_{1,2} \times w_{1,3}}, ..., (\cdot)_{w_{1,2} \times w_{1,l}}, (\cdot)_{w_{1,2} \times w_2}, ..., (\cdot)_{w_{1,2} \times w_s}, \nonumber\\
&..., \nonumber\\
& (\cdot)_{w_{1,l-1} \times w_{1,l}}, (\cdot)_{w_{1,l-1} \times w_2}, ..., (\cdot)_{w_{1,l-1} \times w_s}, \nonumber\\
& (\cdot)_{w_{1,l} \times w_2}, ..., (\cdot)_{w_{1,l} \times w_s}, \nonumber\\
& (\cdot)_{w_2 \times w_3}, ..., (\cdot)_{w_2 \times w_s}, ..., (\cdot)_{w_{k-1} \times w_s}\bigr)
\label{eq:tau_tilde_v_is_weighted_avg}
\end{flalign}
is just a weighted average of $\tau_v$ and $\tau_{w_1}$, which in turn implies that $\tau_v \leq \tau_{\tilde{v}} \leq \tau_{w_i}$ as $\tau_v \leq \tau_{w_i}$ due to the s.n.c. 
The left hand inequality $\tau_v \leq \tau_{\tilde{v}}$ implies that the s.n.c. for the parent of $\tilde{v}$ (which is the same as the parent of $v$) remains satisfied after collapsing.
As $w_1$ is chosen in the way that it is the closest one (in $\tau$ distance) to $v$, it holds that
$\tau_{w_1} \leq \tau_{w_j}$ for all $j \in \{2, ..., k\}$.  Hence, as $\tau_{\tilde{v}} \leq \tau_{w_1}$, it follows that $\tau_{\tilde{v}}\leq \tau_{w_j}$ for all $j \in \{2, ..., k\}$. Thus the s.n.c. remains satisfied also for the children of $\tilde{v}$. 
Note that under the optimistic attitude, if \eqref{eq:hat_in_Theta_a} does not hold for the input of Algorithm \ref{alg:HAC_estim},
we apply the trimming described in Remark \ref{rem:tau_trim}.

Note that even if the estimation of the parameters depends on the input family $a$, the structure determination is \emph{independent} on  $a$ because Algorithm \ref{alg:structure_estim} aims to assure \eqref{eq:nnc} instead of \eqref{eq:holena_nesting_cond_simple}, which does not involve the families of the underlying generators. As shows Theorem \ref{thm:alg_homo_returns_HAC}, such an approach leads to a proper copula for a lot of families. Moreover, as no assumptions on families are necessary, this structure determination approach can be directly used for heterogeneous HAC estimation, which is addressed in the following section.
Note that this independence on the underlying families can also be seen in the approach presented in \cite{Segers2014nonparametric} or in the supertree-based approach in \cite{uyttendaele2016estimation}, which, however, do not cover parameter estimation.

Also, due to this independence, one can consider to collapse an estimated binary structure \emph{before} (below denoted as \emph{pre-collapsing}) or \emph{after} (below denoted as \emph{post-collapsing}) the parameters are estimated. To clarify, Algorithm \ref{alg:HAC_estim} corresponds to the latter approach, i.e., a binary HAC (=its structure + its parameters) is first estimated, and then it is collapsed using the approach introduced in Section \ref{sec:collapsing}. Here, it is important to consider that the parameters are re-estimated, e.g., using \eqref{eq:re-estimation} if one chooses the KTauAvg re-estimation procedure.
In the pre-collapsing approach, one first collapses the structure obtained by Algorithm \ref{alg:structure_estim}, and then it is passed as an input to Algorithm \ref{alg:HAC_estim}. Here, for the KTauAvg re-estimation procedure, we use an analogoue of \eqref{eq:re-estimation} given by $\tau_{(a)}(\hat{\theta}_v)$, where $\hat{\theta}_v$ is defined by \eqref{eq:re-estimation}. 
Hence, given a data sample, two different collapsed HAC estimates can result (one using pre-collapsing and another using post-collapsing), particularly if one consider the attitudes. E.g, assuming the family A and the pessimistic attitude, it can happen that the Kendall's tau for the root in the binary structure obtained by Algorithm \ref{alg:structure_estim} is estimated to be -0.05. Using the post-collapsing approach, this implies that Algorithm \ref{alg:HAC_estim}, as $-0.05 \notin \tau_{(\textrm{A})}(\Theta_{\textrm{A}})$, \emph{necessarily} ends up without returning any HAC estimate. However, using the pre-collapsing approach, that binary structure is first collapsed, and it is possible that during the re-estimation process, the Kendall's tau estimate corresponding to the root has changed to 0.05, which a value from  $\tau_{(\textrm{A})}(\Theta_{\textrm{A}})$, and thus Algorithm \ref{alg:HAC_estim} \emph{does not necessarily} ends up without returning any HAC estimate. For this reason, we also consider these two approaches in the experiments described in Section \ref{sec:exps}.

\subsection{Heterogeneous HACs estimation}
\label{sec:hetero_HAC_estim}
This section describes, how a HAC that possibly involves generators from different \emph{completely monotonic parametric families of Archimedean generators} (simply, \emph{families}) can be estimated. 
As the structure estimator implemented by Algorithm \ref{alg:structure_estim} does not require any assumptions on the families of the underlying generators, we use it as a basis of our new heterogeneous estimator, which is summarized by Algorithm \ref{alg:hetero_HAC_estim} at page \pageref{alg:hetero_HAC_estim}.
Simply speaking, given a set of families $\FF$, the algorithm returns an $\FF$-heterogeneous HAC estimate
for observations $(U_{i1}, ..., U_{id}),~ i = 1, ..., n$. If $\#\FF = 1$, then the algorithm turns into a homogenous estimator.
In the rest of this section, we will describe in detail all parts of the algorithm.

Steps 1 and 2 of the algorithm imply that the resulting structure estimate $(\hat{\VV}, \hat{\EE})$ is computed by Algorithm \ref{alg:structure_estim}. Then, for the parameter estimation, we use the approach from Algorithm \ref{alg:HAC_estim} (see Step 1 there and Step 5 here) and extend it by involving the aggregated GoF statistics introduced in Definition \ref{def:agg_s_n}. This extension, which is rather straightforward, is described in Section \ref{sec:choosing_the_family}. In Section \ref{sec:dealing_nesting_condition}, we deal with the main problem arising from the extension to heterogeneous HACs, which is how to assure the s.n.c.~and in turn that a proper copula results. For this reason, we introduce new concepts that enable assuring the s.n.c.~under the optimistic approach, which is justified by Theorem \ref{thm:alg_returns_HAC}. Two major cases allowing for up to five different families in a single HAC are then discussed in Sections \ref{sec:L24} and \ref{sec:L123}. Finally, in Section \ref{sec:hetero_example}, we provide an example illustrating the whole heterogeneous estimation process.

\subsubsection{Choosing the appropriate family}
\label{sec:choosing_the_family}
Contrary to the homogeneous case, the family of each generator in a heterogeneous estimate can be chosen from a whole set of families. To choose the appropriate family, we evaluate all admissible ones using the GoF test statistic $S^g_n$ given as an input of Algorithm \ref{alg:hetero_HAC_estim} (inputs 5 and 6) and choose the best fitting one, see also Section \ref{sec:gof} for an example.
To implement this approach, we added the inner loop (for $l = 1, ..., \#\NNN$), where $\#\NNN$ is related to the s.n.c. and is addressed in detail in Section \ref{sec:dealing_nesting_condition}. In this inner loop, in Step 5, we compute the parameter estimates $\hat{\theta}_1, ..., \hat{\theta}_{\#\NNN}$ for all admissible families $a_1, ..., a_{\#\NNN}$ in the same way as we do for homogenous HACs.
Then, having estimated the generators $\psi^{(\hat{\theta}_1, a_1)}, ..., \psi^{(\hat{\theta}_{\#\NNN}, a_{\#\NNN})}$, we select the best fitting one and in turn the best fitting family according to the GoF evaluation of these generators, which is performed in Step 7. Note that Step 6 is closely related to the s.n.c. and is thus described in the following section.

\subsubsection{Dealing with the s.n.c.}
\label{sec:dealing_nesting_condition}

In homogenous estimation, one can use Theorem \ref{thm:alg_homo_returns_HAC} for a lot of families to deal with the s.n.c.,  i.e., having a triplet $(\hat{\VV}, \hat{\EE}, (\hat{\tau}_k)^{2d-1}_{d+1})$ obtained by Algorithm \ref{alg:structure_estim}, checking \eqref{eq:hat_in_Theta_a} to hold assures that a proper copula results, which can be done \emph{before} Algorithm \ref{alg:HAC_estim} has been performed.

In the heterogeneous estimation, no analogue to Theorem \ref{thm:alg_homo_returns_HAC} is known, i.e., there is no condition known that would assure that a proper copula results \emph{before} the estimation process has been started and one have to check the s.n.c. \emph{during} the estimation process.

In our approach, such a checking process is performed after Step 5. Under the pessimistic attitude (not included in the algorithm as it could be easily derived from the optimistic version described below), if $\hat{\theta}_l \notin r_l$, we remove the family $a_l$ from the set of families admissible for the actual generator. If there is no admissible generator left, the algorithm stops without any result.
Under the optimistic attitude, in Step 6, we assure the s.n.c.~by forcing that $\hat{\theta}_l$ lies in the interval $r_l$ using an auxiliary function defined by
%\begin{flalign}
%\textrm{trim}(\theta, r, \epsilon)\!=\! 
	 %\left\{ \begin{array}{ll}
%\max\{\min\{\theta, \beta\}, \alpha \},  & \textrm{if~} r = [\alpha, \beta], ~\alpha, \beta \in \mathbb{R}, \\
%\max\{\min\{\theta, \beta\}, \min\{\alpha + \epsilon, \beta\} \},  & \textrm{if~} r = (\alpha, \beta], ~\alpha, \beta \in \mathbb{R}, \\
%\max\{\min\{ \theta, \max\{\beta - \epsilon, \alpha\}, \alpha \},  & \textrm{if~} r = [\alpha, \beta), ~\alpha, \beta \in \mathbb{R}, \\
%\max\{\min\{\beta - \epsilon , \theta\}, \alpha + \epsilon \},  & \textrm{if~} r = (\alpha, \beta), ~\alpha, \beta \in \mathbb{R}, \beta - \alpha  > 2\epsilon,\\
%\{\alpha + \beta\}/2, & \textrm{if~} r = (\alpha, \beta), ~\alpha, \beta \in \mathbb{R},   \beta - \alpha \leq 2\epsilon,\\
%\end{array} \right.
	%\label{eq:trim_function}
%\end{flalign}
\begin{flalign}
\textrm{trim}(\theta, r)\!=\! \max\{\min\{\theta, \beta\}, \alpha \},  \textrm{where~} r = [\alpha, \beta], ~\alpha, \beta, \theta \in \mathbb{R}, 
	\label{eq:trim_function}
\end{flalign}
If $r$ is an opened or semi-opened interval, e.g., $r = (\alpha_0, \beta]$, we first turn it to the closed interval $[\alpha, \beta]$, where $\alpha$ is the smallest real allowed by the computing system higher than $\alpha_0$, and then we use the trim function given by \eqref{eq:trim_function}.
%In order to reduce biasing of the parameter estimate, we recommend to set $\epsilon$ to the lowest positive real allowed by the computing system. %, e.g., we use, in all experiments described in Section \ref{sec:exps}, $\epsilon$ set to  \texttt{eps} = 2.2204e-16, which corresponds to the lowest positive real for a double in MATLAB. 

Even under the optimistic attitude, assuring the s.n.c.~is not a trivial task, as can be anticipated from different forms of the s.n.c.s listed in Table \ref{tab:geners_comb} or can be seen in the example below Lemma \ref{lem:third_class}. However, in Theorem \ref{thm:alg_returns_HAC}, we propose specific conditions under which it is assured that a proper copula results before the estimation process has been started, i.e., a situation similar to Theorem \ref{thm:alg_homo_returns_HAC} but taking into consideration that the estimate might be biased. In addition to Theorem \ref{thm:alg_returns_HAC}, we will derive explicit forms of these specific conditions for two major cases, in which one is allowed to nest up to 4 or 5 families into a HAC, which are presented in Theorems \ref{thm:L123} and \ref{thm:L12}, respectively.

As follows from the previous paragraphs, the selection of the interval $r_l$ plays a crucial role in our approach to dealing with the s.n.c. Now, we will introduce several concepts that help to select $r_l$ that, on the one hand, assures the s.n.c. to hold, and, on the other hand, is as wide as possible in order to reduce biasing of the parameter estimates.

Regarding the nature of the parameter constraints following from the s.n.c. and the iterative approach of the estimation process, we introduce a parameter constraint representation which considerably simplifies dealing with different forms of the s.n.c. The idea is to represent the parameter range of a given family by a pair (family label, parameter range), e.g., for the six families from $\FALL$, this representation would be a set with six pairs $ \{ (\textrm{A}, [0, 1)), (\textrm{C}, (0,+\infty ))$, (12, $[1,+\infty )$), (14, $[1,+\infty )$),  $(19, (0, +\infty)),$ $ (20, (0, +\infty)) \}$, which is obviously derived from Table \ref{tab:geners}. The following definition introduces  a semigroup that contains such set representations of the parameter ranges. 

%An example explaining its use is provided below Remark \ref{rem:psi_interval_equivalence}.

\begin{dfn} \label{def:nesting_semigroup}
1) Let $\mathcal{F} \subseteq \FALL$, $\RR := \{[\alpha, \beta], (\alpha, \beta], [\alpha, \beta), (\alpha, \beta) ~|~$ $ \alpha \in [0, +\infty), \beta \in [0, +\infty],~ \alpha \leq \beta \}$,  and  
$\FR =  \bigl\{ \{(a_1, r_1), ..., (a_m, r_m)\} ~|~ \{a_1, ..., a_m\} \subseteq \mathcal{F} ~\&~ r_1, ..., r_m \in \RR ~\&~ m \in \IN \bigr\}$.
2) Let $\tilde{\cap}$ be the operation on $\FR$ defined for all $\mathcal{N}_1 \in \FR,~ \mathcal{N}_2 \in \FR$ by 
\begin{eqnarray}
\mathcal{N}_1 \tilde{\cap} \mathcal{N}_2 = \{ (a, r) | (\exists r_1 \in \RR, r_2 \in \RR)\nonumber\\ ( r_1 \cap r_2 = r ~\&~ (a, r_1) \in \mathcal{N}_1 ~\&~ (a, r_2) \in \mathcal{N}_2)   \}.
\label{eq:nesting_operator}
\end{eqnarray}
Then we call the ordered pair $(\FR, \tilde{\cap})$ the \emph{nesting semigroup}.
\end{dfn}

\begin{rmk} 
\label{rmk:semigroup}
$(\FR, \tilde{\cap})$ is a commutative semigroup, i.e., the operation $\tilde{\cap}$ is commutative, associative and the identity element is $\FR$. This also means that the ordering in which two or more elements of $\FR$ are intersected by $\tilde{\cap}$ is not important.
Also note that for all $\mathcal{N} \in \FR$ is $\mathcal{N} \tilde{\cap} \mathcal{N} = \mathcal{N}$. 
\end{rmk}

\noindent Hence, e.g., assuming $\mathcal{F} = \FALL$ and defining $\NALL = \{ (\textrm{A}, [0, 1)), (\textrm{C}, (0,+\infty ))$, (12, $[1,+\infty )$), (14, $[1,+\infty )$),  $(19, (0, +\infty)),$ $ (20, (0, +\infty)) \}$, it is clear that $\NALL  \in \FR$. Generally, an element from $\FR$ represents generators from different families of generators. However, note that an element from $\FR$ does not necessarily corresponds to a \emph{set} of generators, i.e., two or more representations can refer to the same generator. E.g., concerning $\NALL$, this follows from the fact that $\psi^{(\textrm{C}, 1)}(t) = \psi^{(12, 1)}(t) = \psi^{(14, 1)}(t)$ or $\psi^{(19, 1)}(t) = \psi^{(20, 1)}(t)$  for all $t \in [0, +\infty)$, see Table \ref{tab:geners}. But, removing the representation of these five generators out of $\NALL$, $\NALL$ becomes a representation of a set of generators, see Table \ref{tab:geners} again.

Now consider a generator $\psi^{(a, \theta)}$ and a set $\mathcal{N} \in \FR$. To decide, whether or not $\psi^{(a, \theta)}$ is one of the generators represented by the set $\mathcal{N}$, one needs only the pair $(a, \theta)$ (and does not need a functional form of $\psi^{(a, \theta)}$). Based on this observation, a relationship, similar to the relationship $\bar{\in}$ introduced in Definition \ref{def:parametric_Arch_fam}, is now introduced. 

\begin{dfn} \label{def:in_semigroup_element}
Let $\mathcal{F} \subseteq \FALL$,  $a \in \mathcal{F},~ \theta \in \mathbb{R}$ and $\mathcal{N} \in \FR$. Then define the relation $\tilde{\in}$  by 
\begin{flalign}
(a, \theta) \tilde{\in} \mathcal{N} \Leftrightarrow  \textrm (\exists r\in\RR)~ (a,r)\in\mathcal{N}~ \& ~\theta \in r.
\end{flalign}
\end{dfn}

\noindent By convention, $\neg((a, \theta) \tilde{\in} \mathcal{N})$ is simply denoted by $(a, \theta) \tilde{\notin} \mathcal{N}$.
It follows from the definition that,  e.g., (C, 0.5) $\tilde{\in} \NALL$ states that $\psi^{(\textrm{C}, 0.5)}$ \emph{is} one of the generators represented by $\NALL$. Similarly,  (C, -0.5) $\tilde{\notin} \NALL$ states that $\psi^{(\textrm{C}, -0.5)}$ \emph{is not} one of the generators represented by $\NALL$

\begin{figure}
	\centering
		\includegraphics[width=0.7\textwidth]{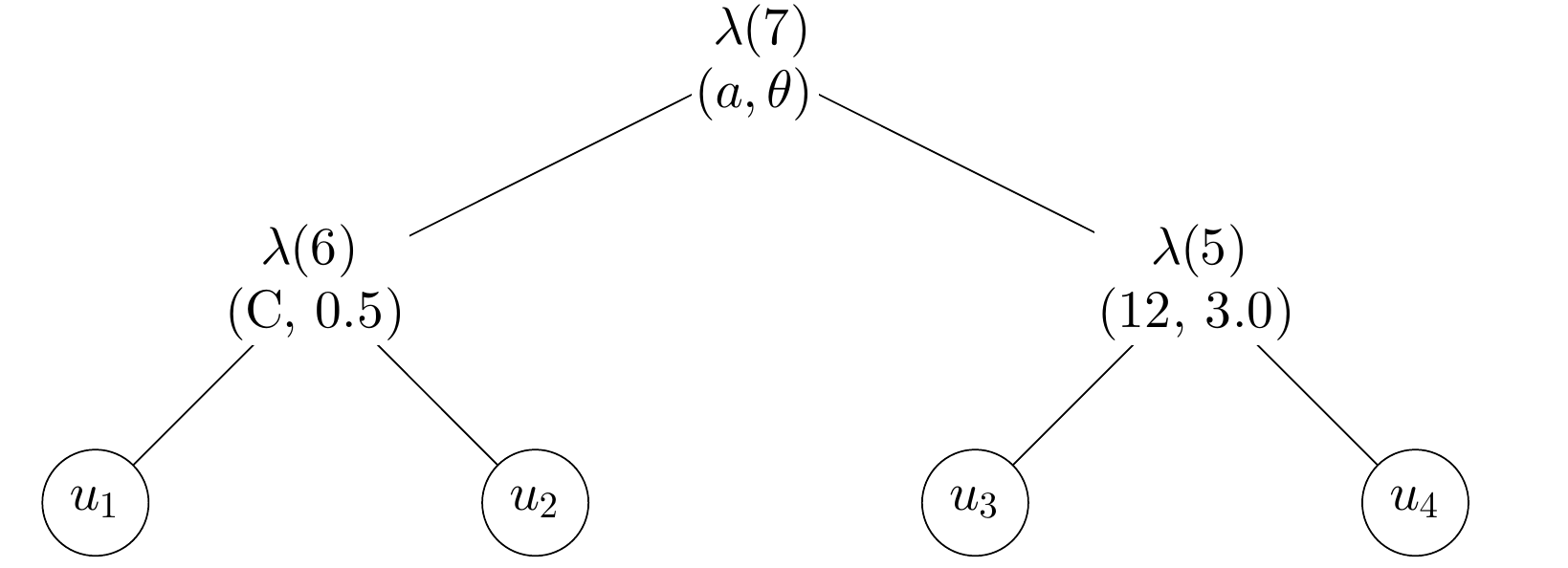}
	\caption{An \{C,12\}-heterogenenous 4-HAC with $\lambda(5) = \psi^{(\textrm{12}, 3.0)}$, $\lambda(6) = \psi^{(\textrm{C}, 0.5)}$ and $\lambda(7) = \psi^{(a, \theta)}$.}
	\label{fig:cop4example}
\end{figure}

Now, let us have a look at the operation $\tilde{\cap}$. Assume that we are estimating a \{C, 12\}-heterogeneous 4-HAC and we have already estimated its structure and two of its generators. E.g., let these two generators be $\psi^{(\textrm{C}, 0.5)}$ and $\psi^{(\textrm{12}, 3.0)}$, which are assigned in the estimation process to be the two children of a generator  $\psi^{(a, \theta)}$, which is still unknown, i.e., the values of the pair $(a, \theta)$ have not been estimated yet. Such a case corresponds to the one depicted in Figure \ref{fig:cop4example}.
%myHAC = HACopula({{'C', 0.5}, {{'C', 0.5}, 1,2} {{'12', 3}, 3,4} });
%plot(myHAC,'GraphType','dendrogram');
The question is, which family-parameter values $(a, \theta)$ are admissible in order to the function corresponding to Figure \ref{fig:cop4example} be a copula, i.e., the s.n.c to be satisfied for both parent-child pairs of generators. Considering   the pair ($\psi^{(a, \theta)}$, $\psi^{(\textrm{C}, 0.5)}$), 
according to Tables \ref{tab:geners} and \ref{tab:geners_comb}, if $(a, \theta) \tilde{\in} \{ (\textrm{C}, (0, 0.5])\}$, the s.n.c. to satisfied. Similarly, considering  the pair ($\psi^{(a, \theta)}$, $\psi^{(\textrm{12}, 3.0)}$), if $(a, \theta) \tilde{\in} \{ (\textrm{C}, (0, 1]),  (\textrm{12}, [1, 3.0])\}$, the s.n.c. is satisfied. According to Theorem \ref{th:holena}, if both conditions are satisfied, i.e., if $(a, \theta) \tilde{\in} \{  (\textrm{C}, (0, 0.5])\}$ as well as $(a, \theta) \tilde{\in} \{  (\textrm{C}, (0, 1]),  (\textrm{12}, [1, 3.0])\}$, then a copula results. To compose these conditions into one, we can use the $\tilde{\cap}$ operation. Observe that $\{ (\textrm{C}, (0, 0.5])\} \tilde{\cap}$ $\{ (\textrm{C}, (0, 1]),  (\textrm{12}, [1, 3.0])\} = \{ (\textrm{C}, (0, 0.5])\}$. Using Tables \ref{tab:geners} and \ref{tab:geners_comb} again, we can verify that for all $(a, \theta) \tilde{\in} \{ (\textrm{C}, (0, 0.5])\}$, the s.n.c. for both parent-child pairs is satisfied and thus a proper copula results. Such an approach can be generally used to assure that a proper copula results when a parent (generator) of one or more generators is estimated, which is later shown in connection with Theorem \ref{thm:alg_returns_HAC}.

Now, consider two special classes of elements from $\FR$.

\begin{dfn}
\label{def:set_of_generators}
Let $\mathcal{F} \subseteq \FALL$. An $\mathcal{N} \in \FR$ is called \emph{Archimedean}, if 
\begin{align}
\forall(a, \theta) \tilde{\in} \mathcal{N},~ \psi^{(a, \theta)} \in \Psi_{+\infty} 
\end{align}

\noindent Further, an $\mathcal{N} \in \FR$ is called $\mathcal{F}$-\emph{Archimedean}, if for all Archimedean $\mathcal{N}^+\in \FR$ holds that
\begin{align}
(a, \theta) \tilde{\in} \mathcal{N}^+ \Rightarrow  (a, \theta) \tilde{\in} \mathcal{N}.
\end{align}
\end{dfn}

In other words, an Archimedean $\mathcal{N}$ represents a set of c.m. generators from the families from $\mathcal{F}$. An $\mathcal{F}$-\emph{Archimedean} $\mathcal{N}$ represents the set of \emph{all} c.m. generators from the families from $\mathcal{F}$. E.g., $\NALL$ is $\FALL$-Archimedean.

Now, consider again the estimation process related to Figure \ref{fig:cop4example}. Having a child generator estimated, e.g., 
the generator $\psi^{(\textrm{12}, 3.0)}$, it would be desirable to have a simple tool that would provide us values of $(a, \theta)$ that are admissible in order to the s.n.c. be satisfied without the need of inspecting Tables \ref{tab:geners} and \ref{tab:geners_comb}. E.g., this tool would provide us the set $\{ (\textrm{C}, (0, 1]),  (\textrm{12}, [1, 3.0]))\}$ when concerning the child generator $\psi^{(\textrm{12}, 3.0)}$. Such a tool in a form of a mapping that relates to the pairs of families from $\FNEST$, see \eqref{eq:known_nestables}, is introduced in the following definition.

%Note that Figure \ref{fig:cop5nesting} on Page \pageref{fig:cop5nesting} shows a function $\HAC$ satisfying \eqref{eq:holena_nesting_cond} with generators from all families in $\FALL$ provided the parameters are set as described later in Section \ref{sec:L123}.

\begin{dfn} 
\label{def:n2}
Let $\mathcal{F} \subseteq \FALL$ and $\mathcal{N}_{\mathcal{F}} \in \FR$ be $\mathcal{F}$-Archimedean. 
Then the mapping $\mathcal{N}^2_{\mathcal{F}} : \mathcal{F} \times [0, +\infty) \mapsto \FR$ is defined for all $(a_2, \theta_2) \tilde{\in}  \mathcal{N}_{\mathcal{F}}$ by
\begin{align}
\mathcal{N}^2_{\mathcal{F}}(a_2, \theta_2) = \{(a_1, \theta_1) \tilde{\in} \mathcal{N}_{\mathcal{F}} ~|~  (a_1, a_2)  \in \FNEST ~\&~ (\psi^{(a_1, \theta_1)}, \psi^{(a_2, \theta_2)}) \in \tilde{\Psi}^2_{+\infty}  \}.
\end{align}
\end{dfn}

%Given a pair of parent-child generators, e.g., a pair $(\lambda(v), \lambda(\tilde{v}))$ considered in \eqref{eq:holena_nesting_cond}, the mapping $\NF$ represents the parameter constrains on the parameter of the parent generator following from the s.n.c. and the parameter of the child generator.
%E.g., if both these generators are from the Clayton family, i.e., $\FF = $ \{C\}, then the explicit expression of $\mathcal{N}^2_{\{\textrm{C}\}}$ is  $\mathcal{N}^2_{\{\textrm{C}\}}(\textrm{C}, \theta_j) = \{(\textrm{C}, (0, \theta_j])\}$, see Table \ref{tab:geners}. We can interpret the term $\{(\textrm{C}, (0, \theta_j])\}$ in the way that the parameter of the parent generator must belong to $(0, \theta_j]$ in order to the s.n.c. be satisfied for this pair of generators. The explicit expression of $\mathcal{N}^2_{\{a\}}$ can be derived similarly for all $a \in $ \{A, 12, 19, 20\}.

Hence, given a child generator representation $(a_2, \theta_2)$, $\NF(a_2, \theta_2)$ simply contains representations of all known c.m. generators that can be parents of $\psi^{(a_2, \theta_2)}$ such that the s.n.c.~is satisfied. Clearly, if there appear some other families of  c.m. generators that can be parents of $\psi^{(a_2, \theta_2)}$ such that the s.n.c.~is satisfied, a generalization of $\mathcal{N}^2_{\mathcal{F}}$ for these families can be simply done by adding these families to $\FNEST$.
Returning to the example related to Figure \ref{fig:cop4example}, one can easily see, using Tables \ref{tab:geners} and \ref{tab:geners_comb}, that, e.g., $\mathcal{N}^2_{\{\textrm{C}, 12\}}(12, 3.0) = \{ (\textrm{C}, (0, 1]),  (\textrm{12}, [1, 3.0]))\}$ or $\mathcal{N}^2_{\{\textrm{C}, 12\}}(\textrm{C}, 0.5) = \{ (\textrm{C}, (0, 0.5]) \}$.

\begin{algorithm}
\floatname{algorithm}{Algorithm}
\caption{The heterogenous HAC estimation}
\label{alg:hetero_HAC_estim}

\begin{algorithmic}
\renewcommand{\algorithmicrequire}{\textbf{Input:}}
\renewcommand{\algorithmicensure}{\textbf{Output:}}

\REQUIRE 
\STATE 1) $(U_{i1}, ..., U_{id}),~ i = 1, ..., n$ ... pseudo-observations 
\STATE 2) $\mathcal{F}$  ... a set of parametric Archimedean families
\STATE 3) $\mathcal{N}_0 = \{(a_1, r_1), ..., (a_{\#\mathcal{N}_0}, r_{\#\mathcal{N}_0})\} \in \mathcal{F}_{\mathbb{R}}$ ... defined in Definition \ref{def:nesting_semigroup}
\STATE 4) $\mathcal{N}^2_{\mathcal{F}}$  ... defined in Definition \ref{def:n2}
\STATE 5) $g$ ... an $[0, +\infty)$-aggregation function (used in the GoF evaluation)
\STATE 6) $S_n$ ... a GoF test statistic for a 2-AC, e.g. $S_n^{(E)}, S_n^{(K)}$ or $S_n^{(R)}$
%\STATE 7) $\epsilon > 0$ ... the maximal precision of the computing system

%\STATE ~
%\renewcommand{\algorithmicrequire}{\textbf{Idea:}}
%\REQUIRE 
%\STATE 1] determine the structure $(\hat{\VV}, \hat{\EE})$ from $(\tau^n_{ij})$
%\STATE 2] estimate the parameters from $(\tau^n_{ij})$ using the inverse of $\tau_{(a)}(\theta)$
%\STATE 3] estimate the best fitting families using $S_n^{g_2}$
%\STATE 4] assure, using $\NNN_0$ and $\NF$, that \eqref{eq:holena_nesting_cond} is satisfied and store the estimated families and parameters in $\hat{\lambda}$

\STATE ~
\renewcommand{\algorithmicensure}{\textbf{The estimation:}}
\ENSURE

\STATE 1. compute $(\tau^n_{ij})$ for  $(U_{i1}, ..., U_{id}),~ i = 1, ..., n$
\STATE 2. get $(\hat{\VV}, \hat{\EE}, (\hat{\tau}_k)_{d+1}^{2d-1})$ using Algorithm \ref{alg:structure_estim}
\STATE 3. set the admissible parents for each leave, i.e., $\mathcal{N}^1(k) := \mathcal{N}_0, ~k = 1, ...,d$

\FOR{$k = 1, ..., d - 1$}

\STATE 3. find two nodes to join, i.e., 
\STATE ~~~ $(i, j) := \wedge(d+k)$
\STATE 4. (*) intersect the sets of the admissible parents corresponding to 
\STATE ~~~ the nodes $i$ and $j$ and denote the resulting set $\mathcal{N}$
\STATE ~~~ by $\{(a_1, r_1), ..., (a_{\#\mathcal{N}}, r_{\#\mathcal{N}})\}$, i.e.,
\STATE ~~~ $\{(a_1, r_1), ..., (a_{\#\mathcal{N}}, r_{\#\mathcal{N}})\} = \mathcal{N} := \mathcal{N}^1(i) \tilde{\cap} \mathcal{N}^1(j)$ %... $\tilde{\cap}$ is defined in Definition \ref{def:nesting_semigroup}
\FOR{$l = 1, ..., \# \mathcal{N}$}
\STATE 5. estimate the parameter assuming the family $a_l$, i.e., 
\STATE ~~~ $\hat{\theta}_l :=  \tau^{-1}_{(a_l)}(\hat{\tau}_{d+k})$ ... see Remark \ref{rem:tau_trim}, where $\hat{\theta}_{d+k} = \hat{\theta}_l$ 
\STATE 6. assure the s.n.c. by trimming the parameter to the interval $r_l$, i.e., 
\STATE ~~~ $\hat{\theta}_l := \textrm{trim} (\hat{\theta}_l,r_l)$
\ENDFOR	
	\STATE 7. find the best fitting family $a_{l^*}$ according to $S_n^g$, i.e.,
	\STATE ~~~ $l^* := \argmin\limits_{l \in \{1, ..., \#\mathcal{N}\}} S_n^g(\mathbf{u}_{\downarrow(i) \downarrow(j)}, \psi^{(a_l, \hat{\theta}_l)})$ ... $S_n^{g}$ and $\mathbf{u}_{\{\cdot\}\{\cdot\}}$ are defined in Definition \ref{def:agg_s_n}
\STATE 8. store the best fitting family and the estimated parameter to $\hat{\lambda}$, i.e.,
\STATE ~~~ $\hat{\lambda}(d+k) := \psi^{(a_{l^*}, \hat{\theta}_{l^*})}$ 
\STATE 9. (**) compute the admissible parents for the fork $d+k$, i.e.,
\STATE ~~~ $\mathcal{N}^1(d+k) := \mathcal{N} \tilde{\cap} \mathcal{N}^2_{\mathcal{F}}(a_{l^*}, \hat{\theta}_{l^*})$ 
\ENDFOR
\STATE ~

\renewcommand{\algorithmicensure}{\textbf{Output:}}
\ENSURE
\STATE $(\hat{\VV}, \hat{\EE}, \hat{\lambda})$
\STATE ~
\STATE (*) if $\mathcal{N}$ computed in this step is empty, stop 
\STATE (**) if $\mathcal{N}^1(d+k)$ computed in this step is empty, stop

\end{algorithmic}
\end{algorithm}

Based on these ideas, i.e., using the nesting semigroup $(\FR, \tilde{\cap})$, the mapping $\mathcal{N}^2_{\mathcal{F}}$ and the function trim, we are able, for some appropriately selected $\mathcal{N}_0 \in \FR$ (its choice is discussed below), to assure that Algorithm \ref{alg:hetero_HAC_estim} \emph{always}
returns a triplet $(\hat{\VV}, \hat{\EE}, \hat{\lambda})$ satisfying \eqref{eq:holena_nesting_cond}, as states the main theorem of this work. 

\begin{thm} \label{thm:alg_returns_HAC}
Let $\mathcal{F} \subseteq \FALL$,  $\mathcal{N}_0 \in \FR$ be Archimedean, $\mathcal{N}^2_{\mathcal{F}}$ be given by Definition \ref{def:n2} and the condition
\begin{flalign}
\exists (a_1, \theta_1), a_1 \in \mathcal{F}, \theta_1 \in [0, +\infty) ~\textrm{such that}~
(a_1, \theta_1) ~ \tilde{\in} ~ \mathcal{N}_0 \tilde{\cap} \mathcal{N}^2_{\mathcal{F}}(a_2, \theta_2)  \tilde{\cap}   \mathcal{N}^2_{\mathcal{F}}(a_3, \theta_3)
\label{eq:non_empty_cond}
\end{flalign}
holds for all $(a_2, \theta_2) \tilde{\in} \mathcal{N}_0$ and $(a_3, \theta_3) \tilde{\in} \mathcal{N}_0$. Then, for given inputs 1), 5) and 6), Algorithm \ref{alg:hetero_HAC_estim} returns a triplet $(\hat{\VV}, \hat{\EE}, \hat{\lambda})$ satisfying \eqref{eq:holena_nesting_cond}, i.e., $C_{(\hat{\VV}, \hat{\EE}, \hat{\lambda})}$ is a copula.
\end{thm}

%tady pokracovat

In other words, Algorithm \ref{thm:alg_returns_HAC} returns a proper copula provided the condition \eqref{eq:non_empty_cond} holds. Note that this condition assures that, given two children (represented by $(a_2, \theta_2)$ and $(a_3, \theta_3)$) of a parent generator in an estimated HAC, there always exists an admissible pair $(a_1, \theta_1)$ representing the parent generator such that the s.n.c. holds with its child generators. Without satisfying this condition, it might happen that Algorithm \ref{thm:alg_returns_HAC} stops before the estimation process is finished without any resulting copula, which is discussed in detail in Section \ref{sec:L123}.

Theorem \ref{thm:alg_returns_HAC} opens two questions: 1) how to appropriately select $\mathcal{N}_0$ for a considered $\mathcal{F}$ in order to \eqref{eq:non_empty_cond} be satisfied? and 2) does there exist some simple, i.e., easy to implement, expression of the range of the mapping $\mathcal{N}^2_{\mathcal{F}}$? Both questions can be satisfactorily answered.

Consider the family combinations $\FNEST \backslash \{(a, a) | a \in  \FALL \backslash \{14\}\}$, i.e., the pairs in Table \ref{tab:geners_comb}. These pairs can be separated  according to the nature of the parameter constraints following from the corresponding s.n.c.~shown in the fourth column in Table \ref{tab:geners_comb}:% (note that we do not consider the parameter ranges for the corresponding generator families in the following):
\begin{enumerate}
	\item the $\mathcal{L}_1$ class -- no constraints on the parameters $\theta_1, \theta_2$ following from the corresponding s.n.c. $\mathcal{L}_1$ =\{(A, 19)\}. 
\item the $\mathcal{L}_2$ class  -- the s.n.c. results in the constraint on the parameter $\theta_1$ of the parent generator. $\mathcal{L}_2$ = \{(C, 12),  (C, 19)\}. 
\item the $\mathcal{L}_3$ class -- the s.n.c. results in the constraint on the parameter $\theta_2$ of the child generator. $\mathcal{L}_3$ = \{(A, C), (A, 20)\} .
\item the $\mathcal{L}_4$ class -- the s.n.c. results in the constraint both on the parameter $\theta_1$ of the parent generator and on the parameter $\theta_2$ of the child generator. $\mathcal{L}_4$ = \{(C, 14), (C, 20)\}.
\end{enumerate}

Using this classification, we answer the questions stated above for major two cases.

%The $\mathcal{N}^2_{\mathcal{F}}$ mapping can be expressed explicitly, what is a valuable property when implementing the algorithm. We show this for two important sets of different nestable Archimedean families. The approach for rest of the sets can be derived analogously.

\subsubsection{The $\mathcal{L}_{24}$ case}
\label{sec:L24}

We start with $\mathcal{F}_{24} = $ \{C, 12, 14, 19, 20\}, which contains as subsets all combinations from $\mathcal{L}_2$ and $\mathcal{L}_4$. 

\begin{lemma} \label{lem:n2f12}
Let $\mathcal{N}_{\mathcal{F}_{24}} \in \FR$ be $\mathcal{F}_{24}$-Archimedean. Then the mapping defined by
\begin{equation}
\mathcal{N}^2_{\mathcal{F}_{24}}(a, \theta) = \left\{ \begin{array}{ll}
\{ (\textrm{C}, (0, \theta] ) \} & 		\textrm{if} ~ a = \textrm{C}\\
\{ (\textrm{C}, (0, 1] ), (12, [1, \theta]\} &  \textrm{if} ~ a = 12\\
\{ (\textrm{C}, (0, \frac{1}{\theta}] )\} & \textrm{if} ~ a = 14\\
\{ (\textrm{C}, (0, 1] ), (19, (0, \theta]\} & \textrm{if} ~ a = 19\\
 \{ (\textrm{C}, (0, \theta] ), (20, (0, \theta])\} & \textrm{if}  ~ a = 20\\
\end{array} \right.
\end{equation}
for all $(a, \theta) \tilde{\in} \mathcal{N}_{\mathcal{F}_{24}}$ is the mapping given by Definition \ref{def:n2} for $\mathcal{F} = \mathcal{F}_{24}$.
\end{lemma}

Now we can discuss the appropriate $\mathcal{N}_0$. In the $\mathcal{L}_{24}$ case, we can allow that the appropriate $\mathcal{N}_0$, denote it $\mathcal{N}_{\mathcal{F}_{24}}$, is the broadest possible, i.e., $\mathcal{N}_{\mathcal{F}_{24}}$ is ${\mathcal{F}_{24}}$-Archimedean.  Explicitly, $\mathcal{N}_{\mathcal{F}_{24}} =  \{ (\textrm{C}, (0,+\infty )), (12, [1,+\infty )),$ $ (14, [1,+\infty )),$ $(19, (0, +\infty)),$ $(20, (0, +\infty)) \}$.
This is confirmed by the following theorem.

\begin{thm} \label{thm:L12}
Given the inputs $\mathcal{F} = \mathcal{F}_{24}, ~\mathcal{N}^2_{\mathcal{F}} = \mathcal{N}^2_{\mathcal{F}_{24}}, ~\mathcal{N}_0 = \mathcal{N}_{\mathcal{F}_{24}}$ and any inputs 1), 5) and 6), Algorithm \ref{alg:hetero_HAC_estim} returns the triplet $(\hat{\VV}, \hat{\EE}, \hat{\lambda})$ satisfying \eqref{eq:holena_nesting_cond}.
\end{thm}

\begin{rmk}
The corresponding version of Theorem \ref{thm:L12} for any $\mathcal{F} \subset \mathcal{F}_{24}$ containing a pair combination from $\mathcal{L}_2$ or $\mathcal{L}_4$ can be proved analogously. Note that such $\mathcal{F}$ always contains the family C.
\end{rmk}

\subsubsection{The $\mathcal{L}_{1234}$ case}
\label{sec:L123}

Answering the questions is more complicated, if we also involve some family combination from $\mathcal{L}_3$, i.e., if \{A, C\} $\subset \mathcal{F}$ or \{A, 20\} $\subset \mathcal{F}$. In this case, it is \emph{not} possible that the appropriate $\mathcal{N}_0$ is $\mathcal{F}$-Archimedean. This fact is explained by the following lemma, which generalizes to an arbitrary $d$-HAC the idea of nesting the family combinations \{A, C\} and \{A, 20\} in a 3-HAC proposed in Theorem 4.3.2 in \cite[p. 118]{Hof10book}. 

\begin{lemma} \label{lem:third_class} 
Let  $\mathcal{F}\subseteq \FALL$ be such that C $\in \FF$ or 20 $\in \FF$, $\HAC$ be an $\FF$-heterogeneous binary $d$-HAC satisfying \eqref{eq:holena_nesting_cond}. By convention, $\lambda(i) = \psi^{(a_i, \theta_i)}$ for all $i \in \{d+1, ..., 2d-1\}$. Also, let $(a_i, a_j) \in \FNEST$ (see \eqref{eq:known_nestables}) for all $i, j \in \{d+1, ..., 2d-1\}$ such that $i = \uparrow(j)$.
Then it holds that, if $a_i$ = C or $a_i$ = 20 for some $i \in \{d+1, ..., 2d-1\}$ and $\theta_i < 1$, then $a_j \neq $ A for all $j \in \{d+1, ..., 2d-1\}$.
\end{lemma}

In other words, Lemma \ref{lem:third_class} says that if we want to allow that the family A is involved in the output of Algorithm \ref{alg:hetero_HAC_estim}, we have to restrict the parameters of all generators from families C and 20 to be $\geq 1$. Otherwise, no copula function satisfying the s.n.c. can be constructed. Study the following example. Assume $\mathcal{F}$ = \{A, C\} and two child generators $\psi^{(\textrm{A}, 0.5)}$ and $\psi^{(\textrm{C}, 0.5)}$ in the same 4-HAC structure as in the one depicted in Figure \ref{fig:cop4example}. Observing that  $\mathcal{N}^2_{\{\textrm{A}, \textrm{C}\}}(\textrm{A}, 0.5) \tilde{\cap} \mathcal{N}^2_{\{\textrm{A}, \textrm{C}\}}(\textrm{C}, 0.5) = \emptyset$, it follows that there is no admissible choice of the family and the parameter of their parent in order to the s.n.c. be satisfied. Hence, for this choice of $\mathcal{F}$, if we want to assure that Algorithm \ref{alg:hetero_HAC_estim} returns a copula, we cannot allow (by setting of  the input $\mathcal{N}_0$) all generators from the $\mathcal{F}$-Archimedean set $\{($A$, [0, 1)), ($C$, (0, +\infty))\}$ and have to set the input $\mathcal{N}_0$ to some smaller subset. According to Lemma \ref{lem:third_class}, we suggest to choose  (as $\mathcal{N}_0$) the set 
$\{($A$, [0, 1)), ($C$, [\theta_{\textrm{C}}, +\infty))\}$, where $\theta_{\textrm{C}} = 1$. Obviously, setting $\theta_{\textrm{C}} < 1$, we cannot assure that a copula results with the same argument as above in this paragraph. By contrary, setting $\theta_{\textrm{C}} > 1$, we unnecessarily avoid some generators in the estimation process, which in turn could lead to an unnecessarily bias.

%Now consider another situation. Assume the 5-HAC depicted on the left in Figure \ref{fig:cop5nesting}. The generator $\lambda(8)$, apart from its role of the child in the 
%pair $(\psi^{(\textrm{A}, \theta_9)}, \psi^{(\textrm{C}, \theta_8)})$ corresponding to a family combination from $\mathcal{L}_3$, appears as the parent in the pair $(\psi^{(\textrm{C}, \theta_8)}, \psi^{(\textrm{20}, \theta_6)})$ corresponding to the family combination from $\mathcal{L}_2$. 
%Allowing $\theta_6$ to be $ < 1$, the s.n.c. enforce $\theta}_8$ to be $< 1$ also. But in this case, the s.n.c. for $(\psi^{(\textrm{A}, \theta_9)}, \psi^{(\textrm{C}, \theta_8)})$, i.e., that $\hat{\theta}_8$ is to be $\geq 1$, cannot be satisfied. It implies that to allow for a generator from the family A, we also must restrict all generators from family 20 to have their parameters $\geq 1$. 

%Matlab
%myHAC = HACopula({ {'A', 0.5}, { {'C', 1}, { {'12', 1}, 1, 2 }, { {'14', 1}, 3, 4 } }, { {'C', 1}, { {'19', 1}, 5, 6}, { {'20', 1}, 7, 8 } }  }); 

\begin{figure}
	\centering
		%\vcenteredhbox{\includegraphics[width=0.495\textwidth]{copula_5D_AC9D-eps-converted-to.pdf}}
		\vcenteredhbox{\includegraphics[width=0.7\textwidth]{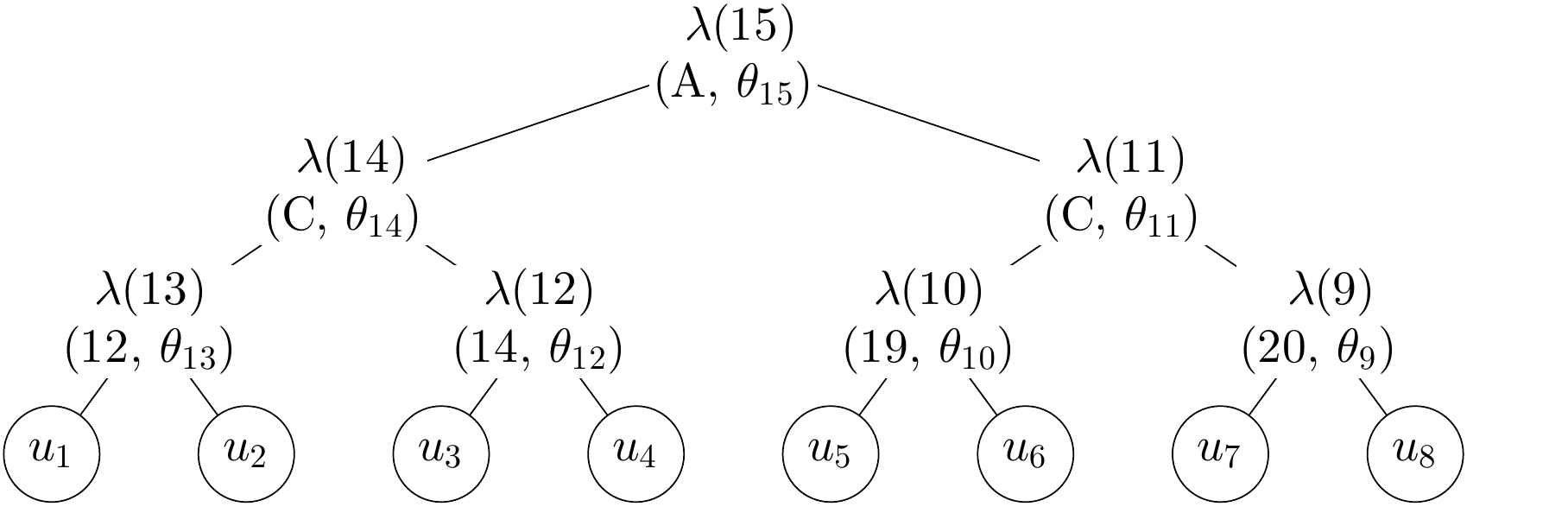}}
		%\vcenteredhbox{\includegraphics[width=1\textwidth]{3d_pdfs-eps-converted-to.pdf}}
		\caption{An $\FALL$-heterogeneous 8-HAC that involves family combinations both from $\mathcal{L}_2$ and $\mathcal{L}_3$.}
	\label{fig:cop5nesting}
\end{figure}

Our next consideration concerns the 8-HAC depicted in Figure \ref{fig:cop5nesting}. It is a $\FALL$-heterogeneous binary 8-HAC that involves family combinations both from $\mathcal{L}_2$ and  $\mathcal{L}_3$. Now consider the constraints for the parameters following from the s.n.c. As follows from Lemma \ref{lem:third_class}, $\theta_{11}$ and $\theta_{14}$ must be $\geq 1$ in order to allow for the family A in an output of Algorithm \ref{alg:hetero_HAC_estim}. Also, $\theta_{14}$ must be $\leq 1$ due to s.n.c. corresponding to the fact that one of its child is a generator from the family 12, see Table \ref{tab:geners_comb}. Thus $\theta_{14}$ must equal to 1. Next, $\theta_{12} \theta_{14} $ must be $\leq 1$ due to the s.n.c. for the combination (C, 14). As $\theta_{14} = 1$ and $\theta_{12} \geq 1$ (following the parameter range of the family 14), it implies that $\theta_{12} = 1$. Such a HAC with two parameters restricted to 1 is an extremely constrained model and is rarely feasible in practical applications. Considering other $d$-HACs, $d \geq 8$ that involve all families from $\FALL$, such HACs could be even more constrained, hence, we do not consider them in the following.

%5-HAC

We rather consider a smaller subset of $\FALL$, the set $\mathcal{F} = \mathcal{F}_{1234} := $ \{A, C, 19, 20\} that contains (as subsets) at least one family combination from each $\mathcal{L}_1$, ..., $\mathcal{L}_4$. 

\begin{lemma} \label{lem:n2f123}
Let $\tilde{\mathcal{N}}_{\mathcal{F}_{1234}} \in \FR$ be $\mathcal{F}_{1234}$-Archimedean. Then the mapping defined by
\begin{equation}
\mathcal{N}^2_{\mathcal{F}_{1234}}(a, \theta) = \left\{ \begin{array}{ll}
\{ (\textrm{A}, [0, \theta] ) \} & 		\textrm{if} ~ a = \textrm{A}\\
\{ (\textrm{C}, (0, \theta] ) \} & 		\textrm{if} ~ a = \textrm{C}, ~ \theta < 1 \\
\{ (\textrm{A}, [0, 1) ), (\textrm{C}, (0, \theta] ) \} & 		\textrm{if} ~ a = \textrm{C}, ~ \theta \geq 1 \\
\{ (\textrm{A}, [0, 1) ), (\textrm{C}, (0, 1] ), (19, (0, \theta]\} & \textrm{if} ~ a = 19\\
 \{ (\textrm{C}, (0, \theta] ), (20, (0, \theta])\} & \textrm{if}  ~ a = 20,~\theta < 1\\
 \{ (\textrm{A}, [0, 1) ), (\textrm{C}, (0, \theta] ), (20, (0, \theta])\} & \textrm{if}  ~ a = 20, ~ \theta \geq 1 \\
\end{array} \right.
\end{equation}
for all $(a, \theta) \tilde{\in} \tilde{\mathcal{N}}_{\mathcal{F}_{1234}}$ is the mapping from Definition \ref{def:n2} for $\mathcal{F} = \mathcal{F}_{1234}$.
\end{lemma}

Following the previous considerations, the broadest possible $\mathcal{N}_0$ is thus not $ \tilde{\mathcal{N}}_{\mathcal{F}_{1234}}$ used in Lemma \ref{lem:n2f123}, but $\mathcal{N}_{\mathcal{F}_{1234}} =  \{ (\textrm{A}, [0, 1)), (\textrm{C}, [\textbf{1},+\infty )), 
 (19,$ $(0, +\infty)), $ $(20, [\textbf{1},$ $+\infty)) \}$. The values in bold are the adjustments of the parameter ranges due to the consequences of Lemma \ref{lem:third_class}.

\begin{thm} \label{thm:L123}
Given the inputs $\mathcal{F} = \mathcal{F}_{1234},~ \mathcal{N}^2_{\mathcal{F}} = \mathcal{N}^2_{\mathcal{F}_{1234}}, ~\mathcal{N}_0 = \mathcal{N}_{\mathcal{F}_{1234}}$ and any inputs 1), 5) and 6), Algorithm \ref{alg:hetero_HAC_estim} returns the triplet $(\hat{\VV}, \hat{\EE}, \hat{\lambda})$ satisfying \eqref{eq:holena_nesting_cond}.
\end{thm}

\begin{rmk}
The corresponding version of Theorem \ref{thm:L123} for $\mathcal{F} = $ \{A, C, 19\} $ \subset \mathcal{F}_{1234}$ containing a pair combination from each $\mathcal{L}_1$, ..., $\mathcal{L}_4$ can be shown analogously. Also, the corresponding version of Theorem \ref{thm:L123} for any $\mathcal{F} \subset \mathcal{F}_{1234}$ containing a pair combination from each $\mathcal{L}_2$, ..., $\mathcal{L}_4$ can be shown analogously as the combination (A, 19) from $\mathcal{L}_1$ do not involve any constraints following from the s.n.c. that must be dealt with.
\end{rmk}

\subsubsection{A general strategy, assuring the s.n.c.~and pre-collapsing}
\label{sec:hetero_estim_others}

Finally, we propose a strategy for heterogeneous HACs estimation concerning the $\mathcal{L}_{24}$ and $\mathcal{L}_{1234}$ cases:
\begin{enumerate}
	\item If we want to allow a generator from the family C or 20 to have its parameter value lower than 1, execute the algorithm with $\mathcal{F} = \mathcal{F}_{24}$ and $\mathcal{N}_0 = \mathcal{N}_{\mathcal{F}_{24}}$, i.e., then the family A is not allowed in the resulting estimate;
	\item If we want to allow the family A in the resulting estimate, execute the algorithm with $\mathcal{F} = \mathcal{F}_{1234}$ and $\mathcal{N}_0 = \mathcal{N}_{\mathcal{F}_{1234}}$;
	\item If we aim to get the best possible fit of the resulting estimate no matter if the family A is involved or not, generate two HAC estimates using both of the previous cases and choose the one that better fits the data using, e.g., some of the GoF test statistics described in Section \ref{sec:gof}.
\end{enumerate}

Considering the problem of assuring the s.n.c. while collapsing a heterogeneous HACs using our collapsing strategy proposed in Section \ref{sec:collapsing}, we use an approach that is analogous to what we do in Step 6 of Algorithm \ref{alg:hetero_HAC_estim}. I.e., we compute $r_l$ (originally computed in Step 4) by 
\begin{flalign}
\{(a_1, r_1), ..., (a_{\#\mathcal{N}}, r_{\#\mathcal{N}})\} = \mathcal{N} := \mathcal{N}^1(i_1) \tilde{\cap} ... \tilde{\cap} \mathcal{N}^1(i_m), 
\label{eq:generalized_snc_assuring}
\end{flalign}
where $i_1, ..., i_m$ are the children of the collapsed node, which assures that the new collapsed fork satisfies the s.n.c with all of its children.
Also, for pre-collapsing (see Section \ref{sec:homo_HAC_estim}), Step 3 is substituted by $(i_1, ..., i_m) := \wedge(d+k)$, Step 4 is substituted by \eqref{eq:generalized_snc_assuring} and 
the term $\mathbf{u}_{\downarrow(i) \downarrow(j)}$ in Step 7 is substituted by $\mathbf{u}_{\downarrow(i_1) \downarrow(i_2)}, \mathbf{u}_{\downarrow(i_1) \downarrow(i_3)}, ..., \mathbf{u}_{\downarrow(i_{m-1}) \downarrow(i_m)}$, which is a construction analogous to \eqref{eq:re-estimation}.

\subsubsection{An example}
\label{sec:hetero_example}

Compared to the homogenous HAC estimation, the heterogeneous HAC estimation is significantly more complex partly due to the involvement of GoF testing but mainly due to the need to deal with the s.n.c. for different family combinations. Hence, to clarify the whole heterogeneous estimation process, we illustrate it by the following example.

\begin{figure}%
\centering
	\subfigure[All 2-dimensional projections of the example data (see Section \ref{sec:hetero_example}) with 1000 observations sampled according to the 5-HAC model depicted in Figure \ref{fig:hetero_5_HAC_model}]{
            \label{fig:hetero_5_HAC_data}
            \includegraphics[width=1\textwidth]{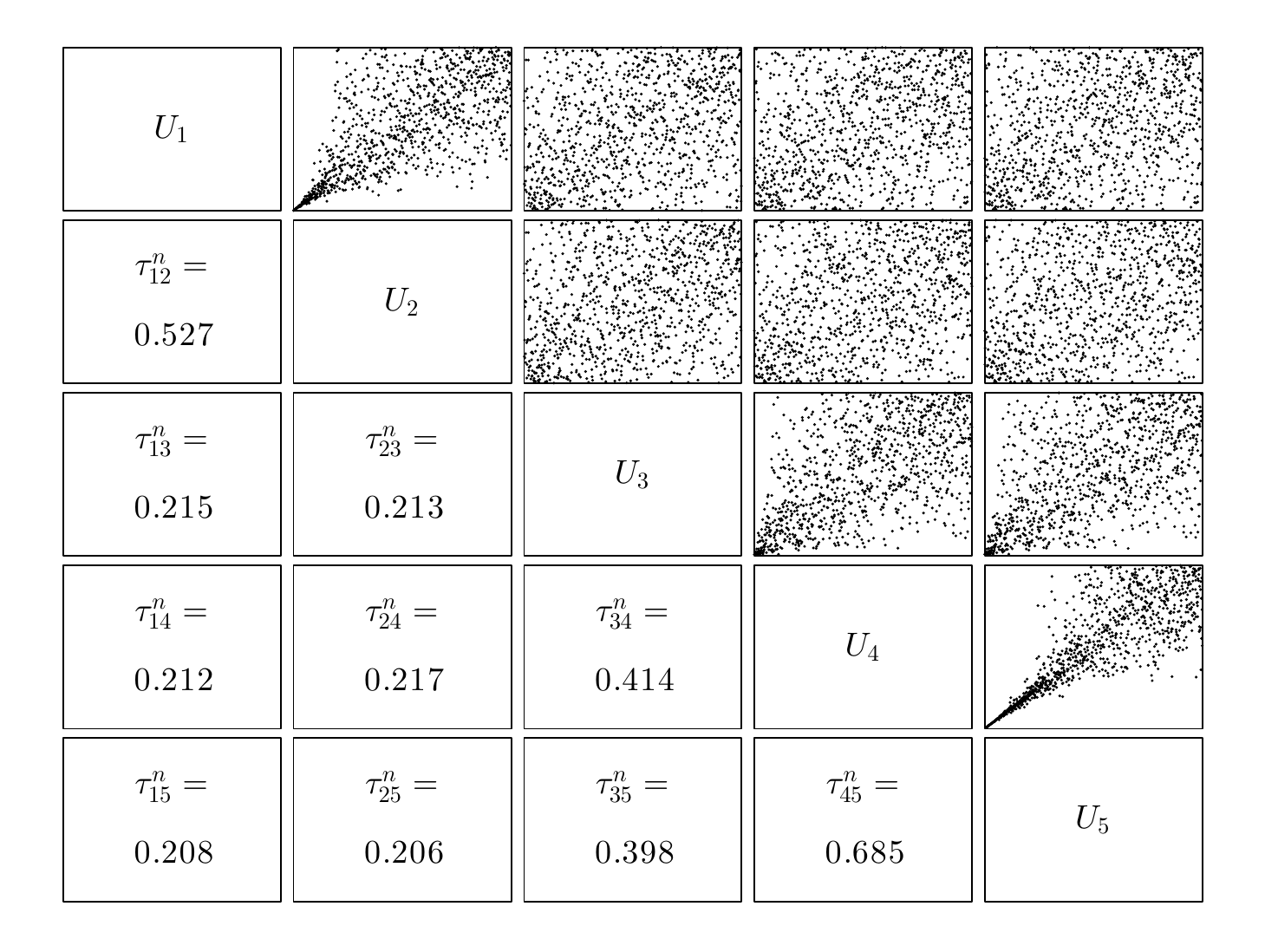}
						}
							\subfigure[The example 5-HAC model]{
            \label{fig:hetero_5_HAC_model}
            \includegraphics[width=0.45\textwidth]{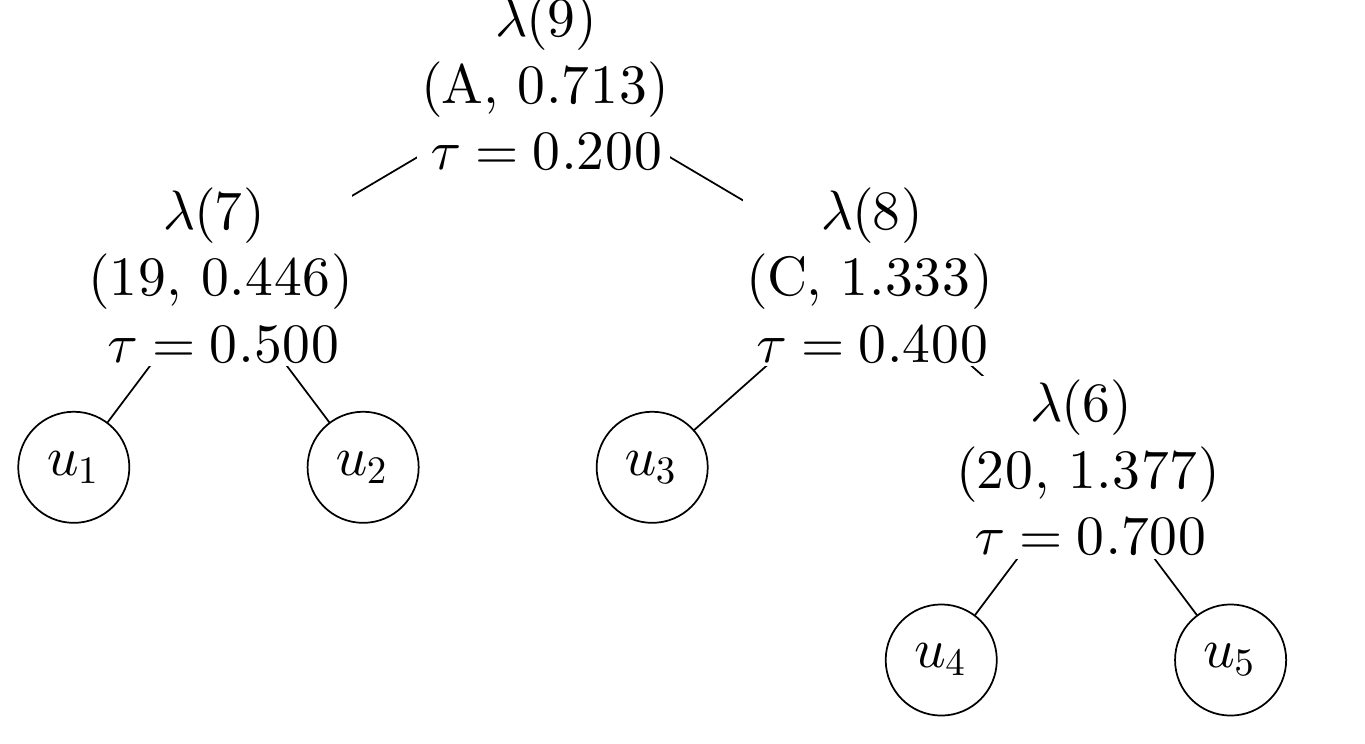}
						}
							\subfigure[The estimate obtained using Algorithm \ref{alg:hetero_HAC_estim} for the data depicted in Figure \ref{fig:hetero_5_HAC_data} provided $\mathcal{F}$ = \{A, C, 19, 20\} and under the optimistic attitude.]{
            \label{fig:hetero_5_HAC_estimate}
            \includegraphics[width=0.45\textwidth]{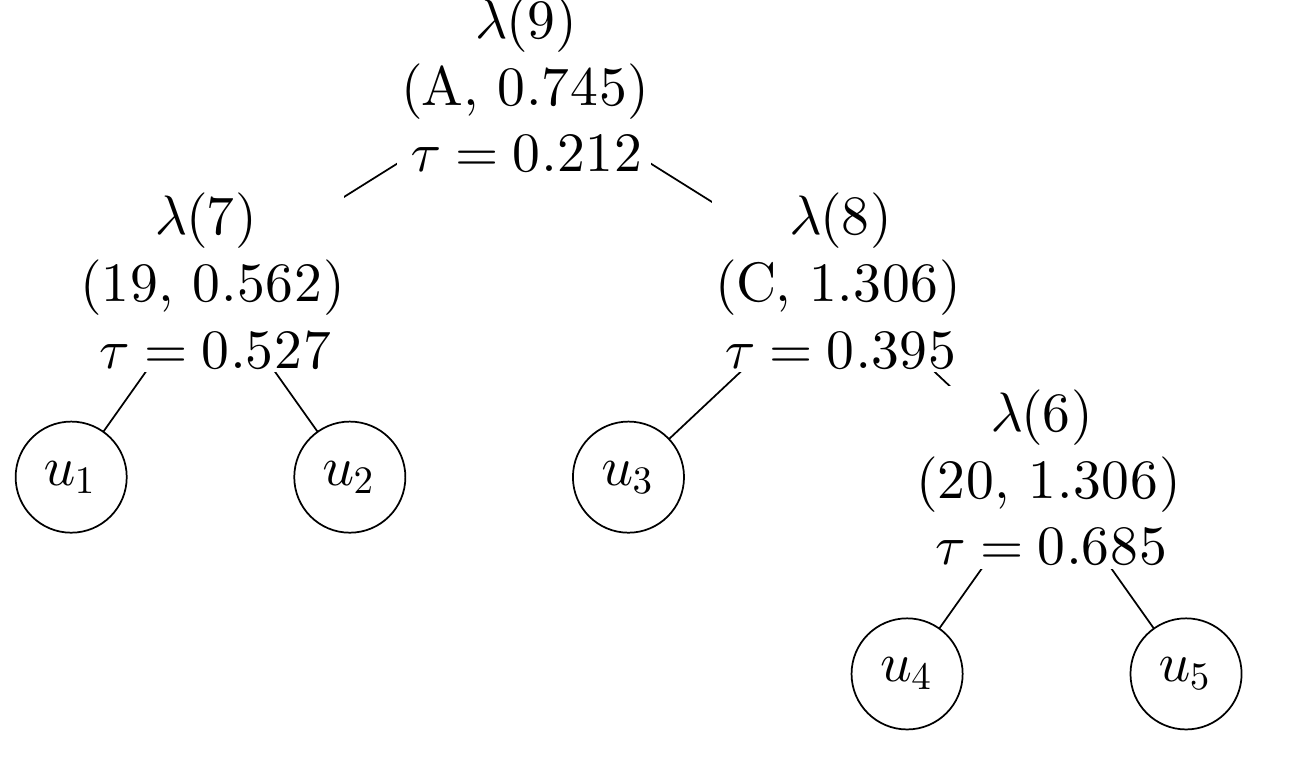}
						}

%\vcenteredhbox{\includegraphics[width=0.7\textwidth]{}}
%\vcenteredhbox{\includegraphics[width=0.49\textwidth]{}}%
%\vcenteredhbox{\includegraphics[width=0.49\textwidth]{}}%
\caption{The example data, their true 5-HAC model and its estimate described in Section \ref{sec:hetero_example}.}
\end{figure}

Let $d = 5$ and $\HAC$ be the  \{A, C, 19, 20\}-heterogeneous binary 5-HAC depicted in Figure \ref{fig:hetero_5_HAC_model}. We sample $n = 1000$ observations according to $\HAC$ using the approach described in \cite{Hof11}. These observations are depicted in Figure \ref{fig:hetero_5_HAC_data}. Now assume that $\HAC$ is unknown and we are interested in its estimate based on the data sample.

% generated by hetero_example.m
% data saved to hetero_example.mat

We set the inputs of Algorithm \ref{alg:hetero_HAC_estim} as follows: $\mathcal{F} = \mathcal{F}_{1234}$, $\mathcal{N}_0 = \mathcal{N}_{\mathcal{F}_{1234}}$ and $\mathcal{N}^2_{\mathcal{F}} = \mathcal{N}^2_{\mathcal{F}_{1234}}$. We also set $g$ to be the maximum function and $S_n$ to be $S_n^{(R)}$.
Recall that the algorithm is constructed under the optimistic attitude.

The Kendall's correlation matrix $(\tau^n_{ij})$ computed in Step 1 can be seen in Figure \ref{fig:hetero_5_HAC_data}. In Step 2, Algorithm \ref{alg:structure_estim} returns the triplet $(\hat{\VV}, \hat{\EE}, (\hat{\tau}_k)_{d+1}^{2d-1}) = (\{1, ..., 9\},$ $\{\{4,6\}, \{5,6\}, \{1,7\}, \{2,7\}, \{3,8\}, \{6,8\}, \{7,9\}, \{8,9\}\},$ $(0.685, 0.527, 0.406,$ $0.212))$. This result follows (recall that $\downarrow(m) = \{m\}$ for all $m \in \{1, ..., d\}$ by the definition) from the matrices depicted in Figures \ref{fig:hetero_5_HAC_data} and \ref{fig:tau_ij_matrix_k_is_23} corresponding to the loops $k = 1, 2, 3$ of Algorithm \ref{alg:structure_estim}, respectively, and the consideration that Algorithm \ref{alg:structure_estim} just joins two leaves or forks $\{i, j\}$ corresponding to the maximum in these matrices. Note that in the loop $k = 4$ of Algorithm \ref{alg:structure_estim}, $\III = \{7,8\}$ and thus $(i, j) = (7, 8)$ in Step 2 there and in turn $\avg((\tau^n_{\tilde{i}\tilde{j}})_{(\tilde{i},\tilde{j}) \in \downarrow(7) \times \downarrow(8)}) =$ $\avg((\tau^n_{\tilde{i}\tilde{j}})_{(\tilde{i},\tilde{j}) \in\{1, 2\} \times \{3, 4, 5\}}) = 0.212$.
Observe that $(\hat{\VV}, \hat{\EE})$ is the true structure $(\VV, \EE)$. 

\begin{figure}[th]
\begin{minipage}[b]{0.45\linewidth}
\centering
 \[ \arraycolsep=3.4pt\def\arraystretch{1.7}
\footnotesize
 \begin{array}{c|c|cccc}
& $j$ & 1 &  2 & 3 & 6\\
\hline
$i$ & \downarrow\!(\cdot)   & 1 & 2 & 3 & \{4, 5\} \\
\hline
1 & 1 & \cdot & \mathbf{0.527} & 0.215 & 0.210  \\
2  & 2 & \cdot & \cdot & 0.214 & 0.212 \\
3 & 3 & \cdot & \cdot & \cdot & 0.406 \\
 6 &\{4, 5\} & \cdot & \cdot & \cdot & \cdot 
\end{array} \]
\end{minipage}
\hspace{0.5cm}
\begin{minipage}[b]{0.45\linewidth}
\centering
 \[ \arraycolsep=3.4pt\def\arraystretch{1.7}
\footnotesize
 \begin{array}{c|c|ccc}
& j & 3 &  6 & 7\\
\hline
i & \downarrow\!(\cdot)   & 3 & \{4, 5\} & \{1, 2\} \\
\hline
3 & 3 & \cdot & \mathbf{0.406} & 0.214  \\
6  & \{4, 5\} & \cdot & \cdot & 0.211 \\
7 &\{1, 2\} & \cdot & \cdot & \cdot \\
\end{array} \]
\end{minipage}
\caption{The similarity $\avg((\tau^n_{\tilde{i}\tilde{j}})_{(\tilde{i},\tilde{j}) \in \downarrow(i) \times \downarrow(j)})$  computed in the loops $k = 2$ (at the left hand) and $k = 3$ (at the right hand) of Algorithm \ref{alg:structure_estim} for the data depicted in Figure \ref{fig:hetero_5_HAC_data}, where $i,j \in \III$ are shown in the first column and the first row, respectively, and $\downarrow\!(i), \downarrow\!(j)$ are shown in the second column and the second row, respectively. The maximum in a matrix is in bold.}
\label{fig:tau_ij_matrix_k_is_23}
\end{figure}

Continue with Algorithm \ref{alg:hetero_HAC_estim}. Step 3 says that each leave can be assigned to any parent generator from $\NNN_0$, i.e, generators (but not all of them) from the families A, C, 19 and 20.

Now, start the outer loop with $k = 1$. In Step 3, as $\wedge(6) = \{4,5\}$, it is $(i, j) = (4, 5)$. 
In Step 4, $\mathcal{N} := \mathcal{N}^1(4) \tilde{\cap} \mathcal{N}^1(5) = \mathcal{N}_0 \tilde{\cap} \mathcal{N}_0 = \mathcal{N}_0$. Thus, $\{(a_1, r_1), ..., (a_4, r_4)\} = 
\{ (\textrm{A}, [0, 1)), (\textrm{C}, [1,+\infty ))$, $(19, (0,$ $+\infty)), (20, [1, +\infty))\}$ assuming the lexicographical ordering of the families, i.e.,  $(a_1, .., a_4)$ = (A, C, 19, 20). As $\#\mathcal{N} = 4$, we compute in Step  5 the  estimates $\hat{\theta}_l , ~l = 1,...,4 $ for the families \{A, C, 19, 20\}. 
As $\hat{\theta}_l := \tau^{-1}_{(a_l)}(\hat{\tau}_6) = \tau^{-1}_{(a_l)}(0.685)$, we should
compute 
$\hat{\theta}_l$ for all $l = 1,...,4 $. 
But, as $\tau_{(\textrm{A})}(\Theta_{\textrm{A}}) = [0,  \frac{1}{3})$,  see Table \ref{tab:tau_theta}, and we assume the optimistic attitude, we artificially set the $\hat{\theta}_1 =  1 - \epsilon$, similarly to our approach to homogeneous HAC estimation,  see Section \ref{sec:homo_HAC_estim}. Note that $1 - \epsilon$ denotes the highest real allowed by the computing system (we used MATLAB for this example) lower than 1.
Hence, we get the parameter estimates $(a_1, ..., a_4)$ = (1 - $\epsilon$, 4.339,  1.761,  1.306).
Due to the fact that all the parameters belong to the corresponding parameter ranges stored in $\mathcal{N}$, the values
$\hat{\theta}_l, ~ l = 1, ...,4$ remain unchanged in Step 6.
In Step 7, we compute the statistics $S_n^{(R) \max}(\uu_{\{4\}\{5\}}, \psi^{(a_l, \hat{\theta}_l)}), ~l = 1, ..., 4$ and get the values (11.732,	3.002,	0.690,	0.355). Note that, assuming $(U_1, ..., U_5)\sim \HAC$, $\uu_{\{4\}\{5\}}$ are the observations of $(U_4, U_5)$.
Based on the GoF values, the generator $\psi^{(20, 1.306)}$ corresponding to the minimal value 0.355 is selected as the best fitting estimated generator. 
Thus, in Step 8, $\hat{\lambda}(6) :=\psi^{(20, 1.306)}$. In Step 9, as $\mathcal{N}^2_{\mathcal{F}}(20, 1.306) = \{(\textrm{A}, [0, 1)), (\textrm{C}, (0, 1.306], (20, (0, 1.306])\}$, the parameter constraints for the (still unknown) parent of the fork 6 are computed $\mathcal{N}^1(6) = \mathcal{N} \tilde{\cap} \mathcal{N}^2_{\mathcal{F}}(20,  1.306) = \{(\textrm{A}, [0, 1)),$ $(\textrm{C}, [1,  1.306], (20, [1,  1.306])\}$. The reader can compare this result with the s.n.c.s shown for the child generator from the family 20 in Tables \ref{tab:geners} and \ref{tab:geners_comb}. 

%the second loop 
Now consider the second loop ($k = 2$).
In Step 3, $(i, j) = (1, 2)$. 
In Step 4, again, $\mathcal{N} := \mathcal{N}^1(1) \tilde{\cap} \mathcal{N}^1(2) = \mathcal{N}_0 \tilde{\cap} \mathcal{N}_0 = \mathcal{N}_0$. Thus, $\{(a_1, r_1), ..., (a_4, r_4)\} = 
\{ (\textrm{A}, [0, 1)), (\textrm{C}, [1,+\infty ))$, $(19, (0,$ $+\infty)), (20, [1, +\infty))\}$.  In Step 5, as $\#\mathcal{N} = 4$, we compute the parameters estimates $(\hat{\theta}_1, ..., \hat{\theta}_4)$ = (1 - $\epsilon$, 2.225,  0.562,  0.788).
In Step 6, the trim function applies to $\hat{\theta}_4$ and changes it to 1. Recall that this adjustment follows from Lemma \ref{lem:third_class}, which says that if one wants to involve the family A in a HAC, it is not possible to have a generator from family 20 with parameter lower than 1.
In Step 7, we compute $S_n^{(R) \max}(\uu_{\{1\}\{2\}}, \psi^{(a_l, \hat{\theta}_l)}), ~l = 1, ..., 4$ resulting in the values (3.528, 0.656, 0.0527, 1.187) and thus $l^* = 3$. 
In Step 8, $\hat{\lambda}(7) :=\psi^{(19,  0.562)}$. In Step 9, $\mathcal{N}^1(7) = \mathcal{N} \tilde{\cap} \mathcal{N}^2_{\mathcal{F}}(19,  0.562) = \{(\textrm{A}, [0, 1)),$ $(\textrm{C}, [1,1], (19, (0,  0.562])\}$, see Tables \ref{tab:geners} and \ref{tab:geners_comb} to check this result.

%the third loop
In the third and fourth loop, the s.n.c.~comes into play more significantly. Consider the third loop ($k = 3$).
In Step 3, $(i, j) = (3, 6)$.
In Step 4, as $j = 6$ is a fork and not a leaf as in the previous loops, additional constraints following from the s.n.c.~come into play. Hence, 
 $\mathcal{N} := \mathcal{N}^1(3) \tilde{\cap} \mathcal{N}^1(6) =  
\{ (\textrm{A}, [0, 1)), (\textrm{C}, [1,+\infty ))$, $(19, (0, +\infty)),$ $ (20, [1, +\infty))\}$
$\tilde{\cap} \{(\textrm{A}, [0, 1)),$ $(\textrm{C}, [1,  1.306],$ $(20, [1,  1.306])\} = 
 \{(\textrm{A}, [0, 1)),$ $(\textrm{C}, [1,  1.306], (20, [1,  1.306])\}$. 
In Step 5, as $\#\mathcal{N} = 3$, we compute the estimates $(\hat{\theta}_1, ..., \hat{\theta}_3) = (1-\epsilon, 1.368,  0.534)$ assuming $(a_1, a_2, a_3) = $ (A, C, 20). Observe that two of the parameters are out of the corresponding parameter ranges, thus in Step 6, we set $\hat{\theta}_2$ to trim($1.368, [1,  1.306]) = 1.306$ and $\hat{\theta}_3$ to trim($0.534, [1,  1.306]) = 1$. Note that under the pessimistic attitude, the estimation process would stop here with no result, as all parameter estimates have been trimmed/biased.
In Step 7, we compute the statistic $S_n^{(R) \max}(\uu_{\{3\}\{4,5\}} \psi^{(a_l, \hat{\theta}_l)}), ~l = 1, 2, 3$  resulting in (0.287,	0.033,	6.331) and thus $l^* = 2$, i.e., $\hat{\lambda(8)}: = \psi^{(\textrm{C}, 1.306)}$ in Step 8. Recall that $S_n^{(R) \max}(\uu_{\{3\}\{4,5\}} \psi^{(a_l, \hat{\theta}_l)}) = \max\bigl(S_n^{(R)}((U_{\bullet 3}, U_{\bullet 4}), \psi^{(a_l, \hat{\theta}_l)}), S_n^{(R)}((U_{\bullet 3}, U_{\bullet 5}), \psi^{(a_l, \hat{\theta}_l)}) \bigr)$, i.e., two 2-AC GoF test statistics computed for the bivariate margins $(U_3,U_4)$ and $(U_3,U_5)$ are aggregated by the maximum function.
In Step 9, as $\mathcal{N}^2_{\mathcal{F}}(\textrm{C}, 1.306)) = \{(\textrm{A}, [0, 1)), (\textrm{C}, [0, 1.306]\}$, the parameter constraints for the parent of the fork 8 are computed as $\mathcal{N}^1(8) =  \{(\textrm{A}, [0, 1)),$ $(\textrm{C}, [1,  1.306], (20, [1,  1.306])\}
 \tilde{\cap}$ $ \{(\textrm{A}, [0, 1)), (\textrm{C}, [0, 1.306]\} = \{(\textrm{A}, [0, 1)), (\textrm{C}, [1, 1.306]\}$. The reader can compare this result with the s.n.c.s shown in Tables \ref{tab:geners} and \ref{tab:geners_comb}. 

%the last loop

Now consider the last loop ($k = 4$).
In Step 3, $(i, j) = (7, 8)$, i.e.,  both $i = 7$ and $j =8$ are forks.
Hence, in Step 4,  $\mathcal{N}^1(7) \tilde{\cap} \mathcal{N}^1(8)=$
$\{(\textrm{A}, [0, 1)),$ $(\textrm{C}, [1,1], (19, (0,  0.562])\}$
$\tilde{\cap} 
 \{(\textrm{A}, [0, 1)), (\textrm{C}, [1, 1.306]\}
= \{(\textrm{A}, [0, 1)), (\textrm{C},[1,1])\}$. 
In Step 5, we compute $(\hat{\theta}_1, \hat{\theta}_2) = (0.745,  0.538)$ assuming $(a_1, a_2) =$ (A, C) .
In Step 6, trimming applies to $\hat{\theta}_2$ and thus $\hat{\theta}_2$ = trim($0.538, [1, 1]) = 1$. Again, this change  relates to Lemma \ref{lem:third_class}.
In Step 7,  the statistic $S_n^{(R) \max}(\uu_{\{1, 2\}\{3, 4, 5\}} \psi^{(a_l, \hat{\theta}_l)}), ~l = 1, 2)$ results in (0.030,	0.674) and thus $l^* = 1$. Recall that $\uu_{\{1, 2\}\{3, 4, 5\}}$ are the observations corresponding to the bivariate margins $(U_1, U_3), (U_1, U_4),$ $(U_1, U_5),$ $(U_2, U_3), (U_2, U_4)$ and $(U_2, U_5)$.
In Step 8, $\hat{\lambda}(9)$ := $\psi^{(\textrm{A}, 0.745)}$. Step 9 in this loop does not influence the result.

Observe that we have obtained the same families for the same forks as in the model, and that the triplet $(\hat{\VV}, \hat{\EE}, \hat{\lambda})$ is satisfying \eqref{eq:holena_nesting_cond}, see also Figure \ref{fig:hetero_5_HAC_estimate}.

% generated by hetero_example.m
% data saved to hetero_example.mat
\begin{figure}[th]%
\centering
	\subfigure[The estimate obtained assuming $\mathcal{F}$ = \{C, 19, 20\} and the optimistic attitude.]{
            \label{fig:hetero_5_HAC_estimate_opt}
            \includegraphics[width=0.45\textwidth]{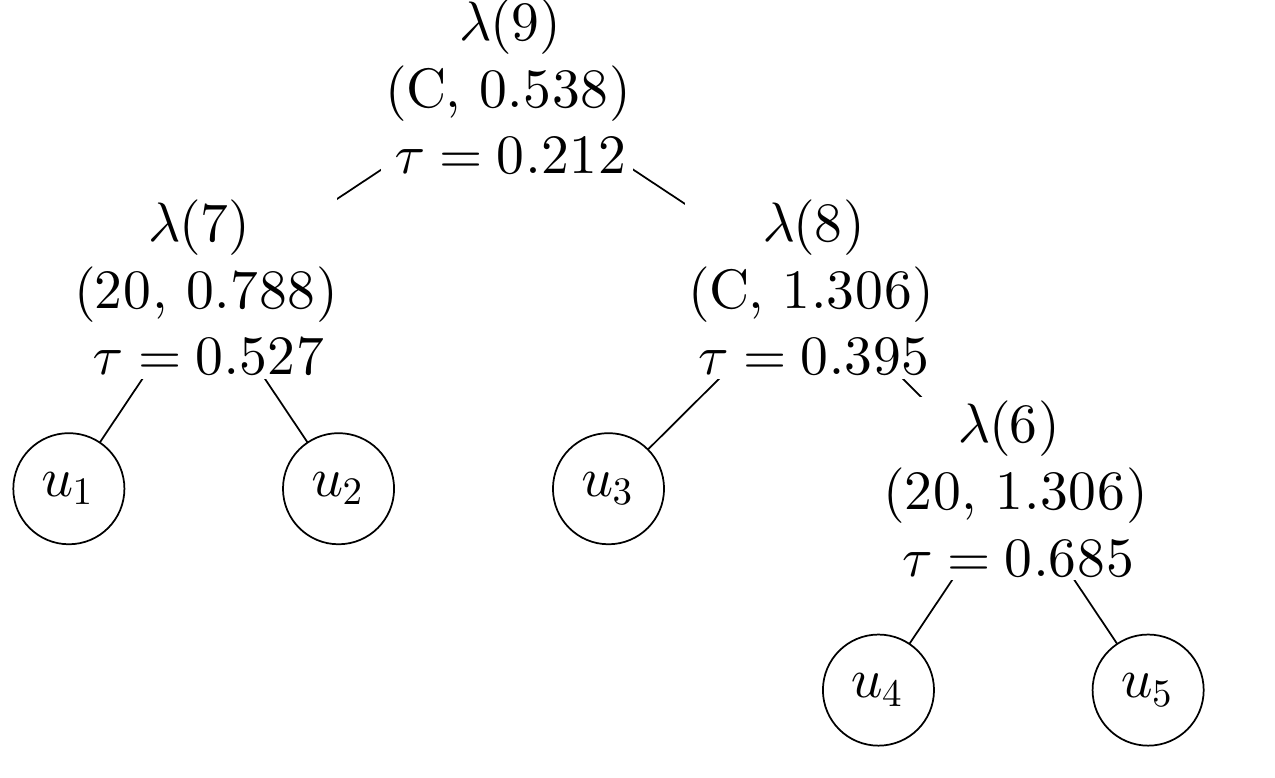}
						}
													\subfigure[The estimate obtained assuming $\mathcal{F}$ = \{C, 19, 20\} and the pessimistic attitude.]{
            \label{fig:hetero_5_HAC_estimate_pes}
            \includegraphics[width=0.45\textwidth]{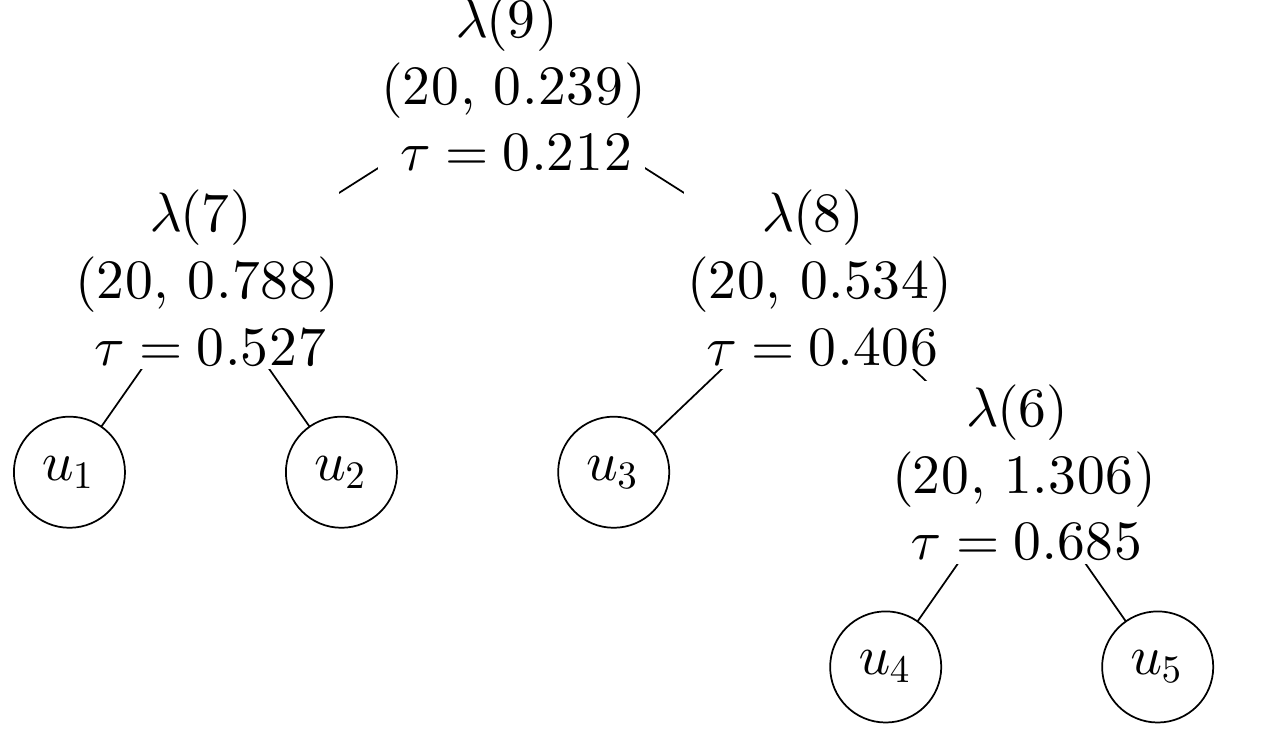}
						}
					
													\subfigure[The estimate obtained assuming $\mathcal{F}$ = \{A, C, 19\} and the optimistic attitude.]{
            \label{fig:hetero_5_HAC_estimate_opt_AC19}
            \includegraphics[width=0.45\textwidth]{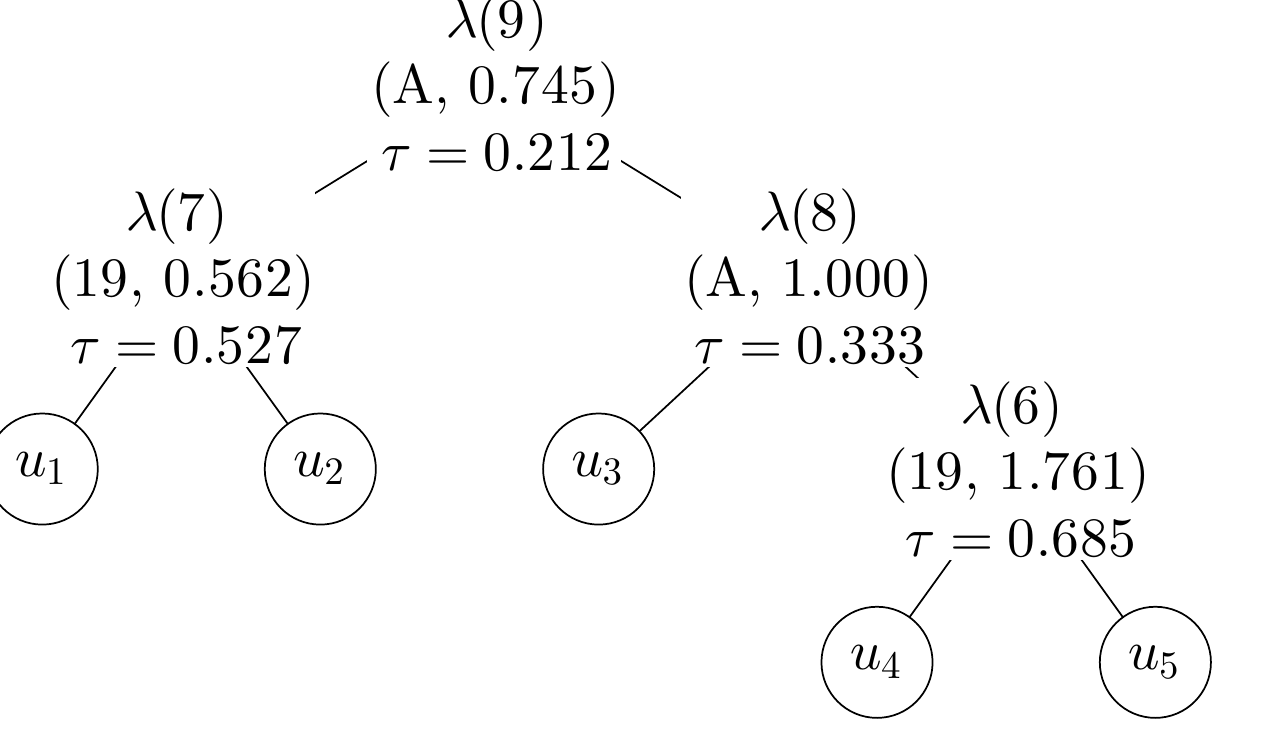}
						}
													\subfigure[The estimate obtained assuming $\mathcal{F}$ = \{A, C, 19\} and the pessimistic attitude.]{
            \label{fig:hetero_5_HAC_estimate_pes_AC19}
            \includegraphics[width=0.45\textwidth]{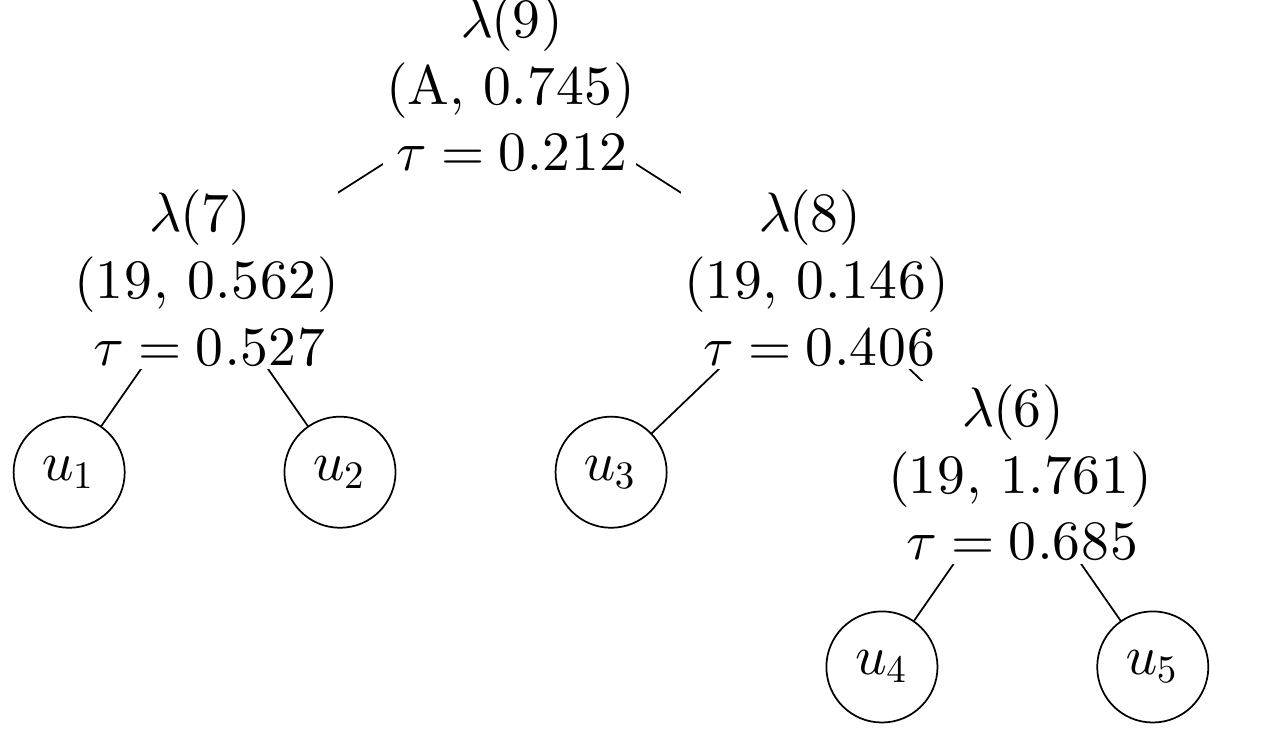}
						}

%\vcenteredhbox{\includegraphics[width=0.7\textwidth]{}}
%\vcenteredhbox{\includegraphics[width=0.49\textwidth]{}}%
%\vcenteredhbox{\includegraphics[width=0.49\textwidth]{}}%
\caption{Four other estimates obtained using Algorithm \ref{alg:hetero_HAC_estim} for the data depicted in Figure \ref{fig:hetero_5_HAC_data} for different settings of $\FF$ and attitudes.}
\label{fig:hetero_5_HAC_other_estimates}
\end{figure}

Finally, we evaluated the resulting estimate using $S^{(E)}_n$ and we obtained the value 0.032. As a comparison, we used Algorithm \ref{alg:hetero_HAC_estim} to obtain other 4 estimates assuming different settings of the set $\FF$ and attitudes. These estimates are depicted in Figures \ref{fig:hetero_5_HAC_estimate_opt}, \ref{fig:hetero_5_HAC_estimate_pes}, \ref{fig:hetero_5_HAC_estimate_opt_AC19} and \ref{fig:hetero_5_HAC_estimate_pes_AC19}, and the corresponding values of $S^{(E)}_n$ are  0.060,  0.110, 3.248 and 0.037, respectively.
As mentioned above, the estimation process for $\mathcal{F}$ = \{A, C, 19, 20\} under the pessimistic attitude terminated in the loop $k = 3$ without any result.
These results show that: 
\begin{enumerate}
	\item Allowing for more families enables for better flexibility and thus for better fit, which follows from the fact that even if the \{A, C, 19, 20\}-heterogeneous estimate is slightly biased (the generator $\hat{\lambda}(8)$ in the estimate depicted in \ref{fig:hetero_5_HAC_estimate}), it can fit the data better than the other estimates with a more restricted choice of families no matter if biased or unbiased;
	\item Comparing the values of $S^{(E)}_n$ for the two estimates in Figure \ref{fig:hetero_5_HAC_other_estimates} obtained under the optimistic attitude with the two estimates depicted in the same figure obtained under the pessimistic attitude, i.e., we observe that the optimistic one for $\mathcal{F}$ = \{C, 19, 20\} shows better fit that the pessimistic one but just the opposite for $\mathcal{F}$ = \{A, C, 19\},  one should always consider estimates obtained under both of the attitudes as none of the attitudes may assure the better fit.
\end{enumerate}

\section{Experiments} \label{sec:exps}
To show abilities of the proposed approach to heterogeneous HAC estimation represented by Algorithm \ref{alg:hetero_HAC_estim}, we perform an experimental study on data simulated from 12 different heterogeneous HACs with dimensions up to $d = 15$ and involving up to five parametric families of Archimedean generators in a single HAC. 
To the best of our knowledge, there does not exist any other implemented methods suitable for heterogeneous HACs estimation. to have at least one heterogeneous estimator to compare, we thus made use of the homogeneous estimator introduced in \cite{Okh13} with an improvement proposed in \cite{goreckihofertholena2014a}, and generalized it to the heterogeneous case applying the findings reported in Section \ref{sec:hetero_HAC_estim}. A detailed description of this alternative heterogeneous estimator can be found in Appendix \ref{app:diagonal}. 

In the rest of this section, several variants of these estimators, generated by different input settings and by using different collapsing approaches, are compared. This comparison concerns success in estimating the structure and the families of the true copula, precision of the estimated parameters and goodness-of-fit.
The comparison was implemented and performed in MATLAB. %and the source code can be obtained from the authors upon request.

\subsection{Design of the performed experiments}
The estimator represented by Algorithm \ref{alg:hetero_HAC_estim} will be denoted by \textbf{PT} as it is based on \textbf{p}airwise correlation coefficients and the inversion of Kendall's \textbf{t}au. Similarly, the estimator represented by Algorithm \ref{alg:diag_hetero_HAC_estim} will be denoted by \textbf{DM} as it is based on the \textbf{d}iagonal transformation and the \textbf{M}L estimation. 
Both these estimators (algorithms) depend, apart from other input settings addressed below, on $S_n$ (Input 6) and the PT estimator also depends on $g$ (Input 5). In the experiments, we use three settings of $S_n$ -- the GoF statistics $S_n^{(E)}, S_n^{(K)}$ and $S_n^{(R)}$ recalled in Section \ref{sec:gof} -- and two settings of $g$ -- the maximum and the average. Hence, these settings generate 9 \emph{basic} estimators  denoted \textbf{E PT avg}, \textbf{E PT max}, \textbf{E DM}, \textbf{K PT avg}, \textbf{K PT max}, \textbf{K DM}, \textbf{R PT avg}, \textbf{R PT max} and \textbf{R DM}, where each name obviously addresses the corresponding type of $S_n$, the underlying algorithm and, for the PT case, the setting of $g$ for a given estimator. 
We also consider the two attitudes -- the optimistic (denoted \textbf{opt}) and the pessimistic (denoted \textbf{pes}), which  doubles the number of these estimators. Finally, as these 18 estimators serve only for estimation of binary structured HACs, we use for each of these estimators the following collapsing strategies. 
These strategies are based on the re-estimation approaches KTauAvg and TauMin (denoted \textbf{Re-est=KTauAvg} and \textbf{Re-est=TauMin}, respectively); see Section \ref{sec:collapsing}. 
Using any of these two approaches, a set of HACs with decreasing number of generators is generated from a binary HAC estimate and it is up to the user, which one of these collapsed HACs will be used as the final estimate. To automatize this choice, we use the two following approaches. In the first one (denoted \textbf{\#Forks=known}), we assume that we know the number of generators (forks) in the true copula and simply choose out of the generated set the estimate with this number of forks. This approach is of course unrealistic, but useful for comparison of the KTauAvg and TauMin approaches independently on how one chooses a final collapsed HAC from the mentioned set of HACs. In the second approach  (denoted \textbf{\#Forks=unknown}), we assume the number of the generators to be unknown, i.e., a realistic assumption, and choose out of the generated set the estimate with the number of generators estimated by the heuristic procedure proposed in Section \ref{sec:collapsing}. Combining KTauAvg and TauMin with  \#Forks=known and \#Forks=unknown thus generates four collapsing strategies applicable for all 18 binary estimators. Also, as addressed in the last paragraph of Section \ref{sec:homo_HAC_estim}, for the PT-based estimators, a collapsing strategy can be applied \emph{before} (and hence denoted \textbf{Coll=pre}) the families and the parameters of a HAC is estimated, and the final non-binary HAC estimate is then obtained using Algorithm \ref{alg:hetero_HAC_estim}, where in Step 2, a non-binary structure can be assumed; see also Section \ref{sec:hetero_estim_others}. The approach where the HAC is collapsed \emph{after} its families and parameters are estimated is denoted \textbf{Coll=post}. Note that in Algorithm \ref{alg:diag_hetero_HAC_estim}, due to involvement of the diagonal transformation and the ML estimation, an assumption on the families have to be done before starting the algorithm, which implies that estimation of the structure cannot be done separately from estimation of the parameters and families. This in turn implies that no pre-collapsing is possible for the DM estimators and thus it is not considered in the experiments.

\begin{figure}[htb]
\centering
		\subfigure[The \{19, A, C\}-heterogeneous 5-HAC model]{%
            \label{fig:model_d5_AC19}
\includegraphics[width=0.45\textwidth]{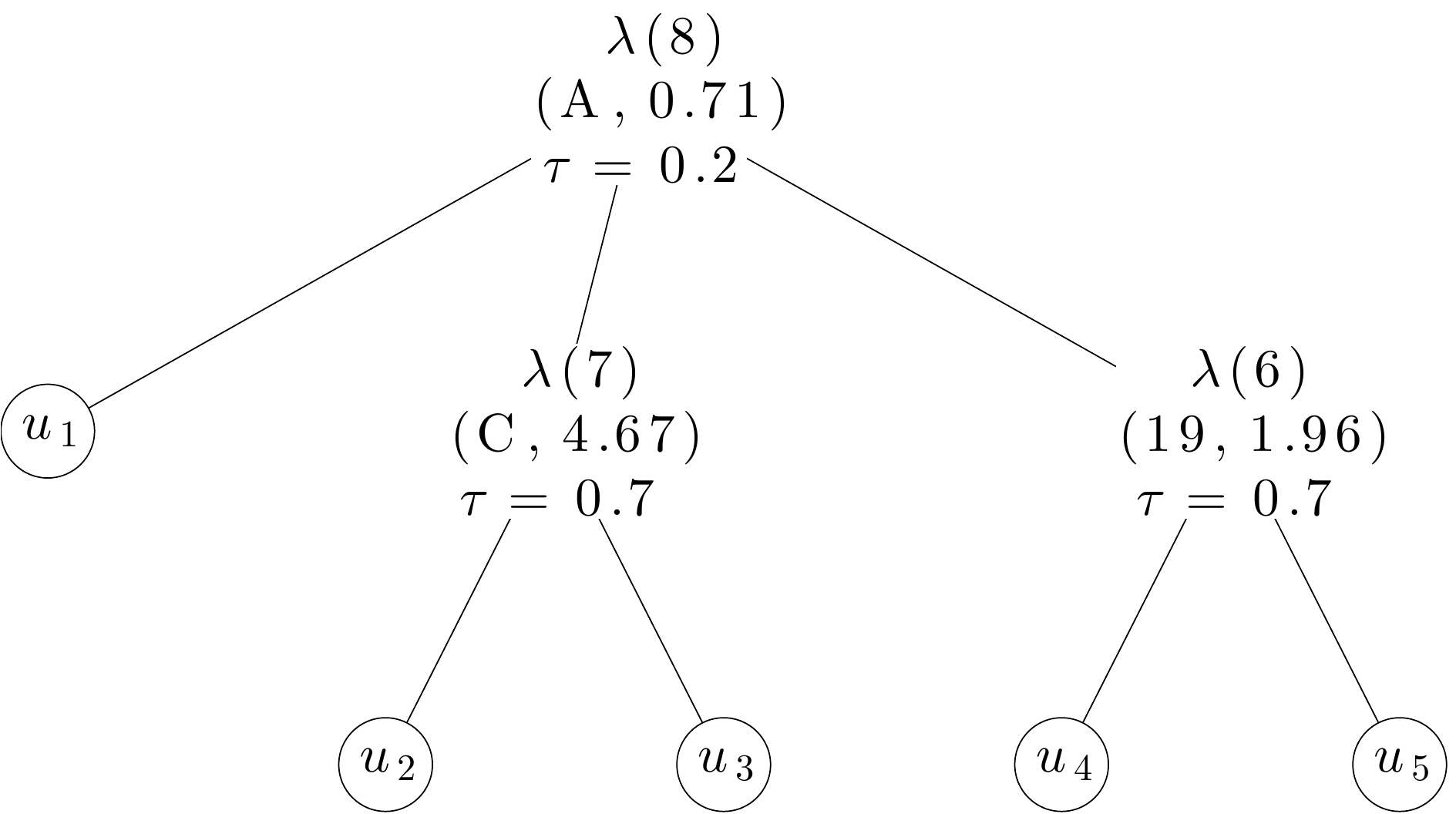}
						} 
		\subfigure[The \{19, A\}-heterogeneous 5-HAC model]{%
            \label{fig:model_d5_A19}
\includegraphics[width=0.45\textwidth]{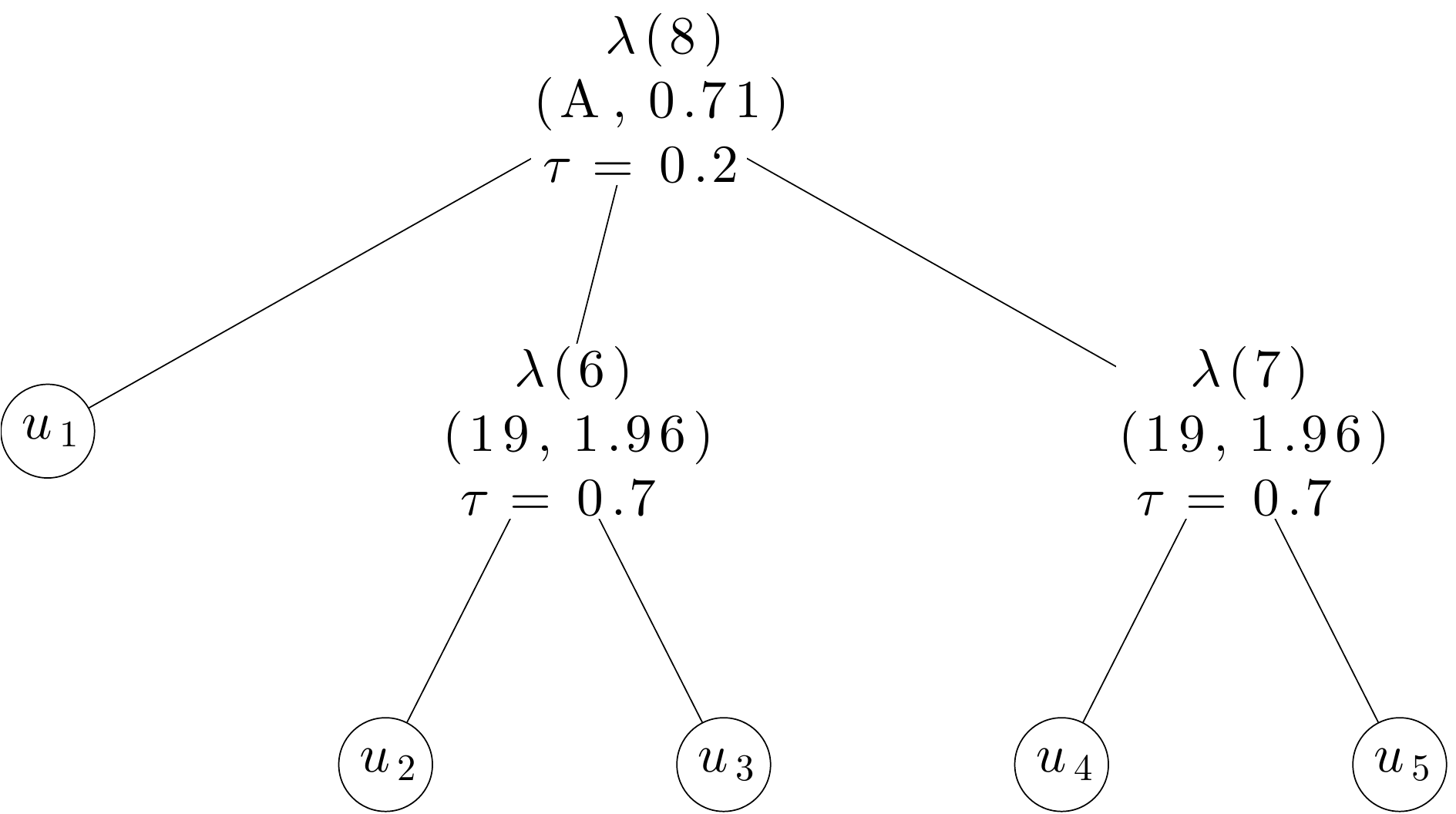}
						}
		\subfigure[The \{19, 20, A, C\}-heterogeneous 10-HAC model]{%
            \label{fig:model_d5_A19}
\includegraphics[width=0.45\textwidth]{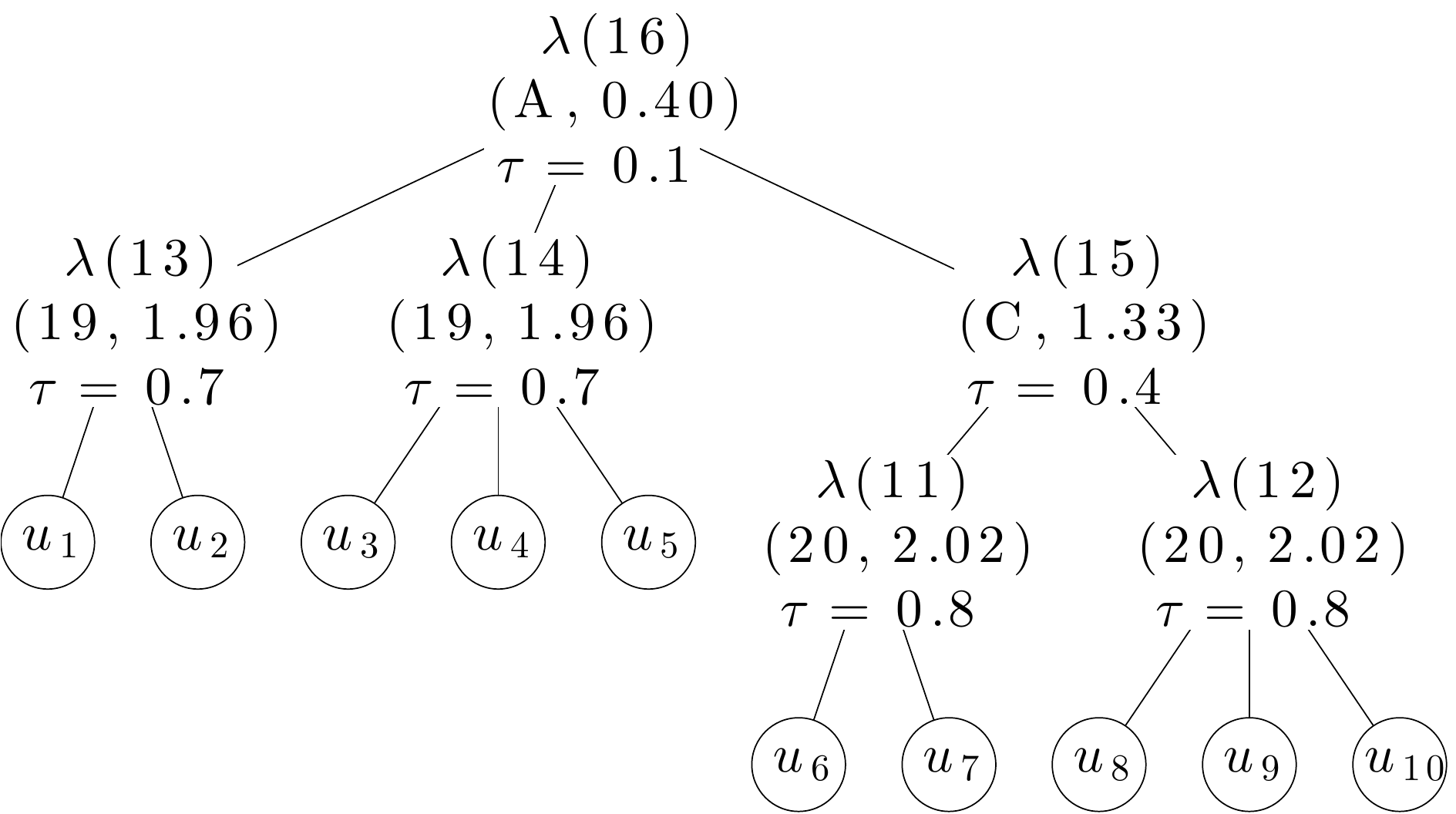}
						}
		\subfigure[The \{19, A, C\}-heterogeneous 10-HAC model]{%
            \label{fig:model_d5_A19}
\includegraphics[width=0.45\textwidth]{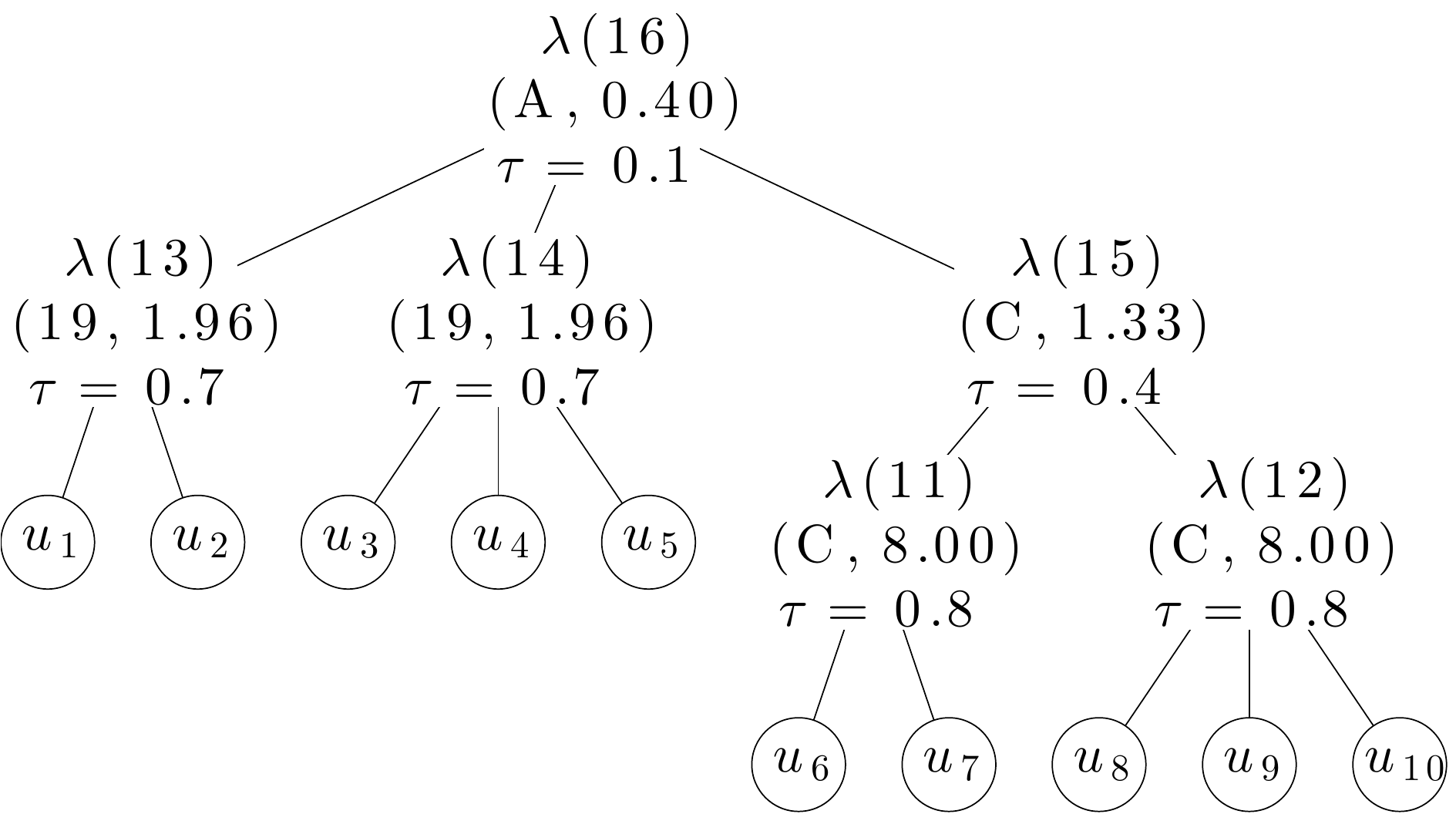}
						}
		\subfigure[The \{12, 14, 19, 20, C\}-heterogeneous 15-HAC model]{%
            \label{fig:model_d5_A19}
\includegraphics[width=0.45\textwidth]{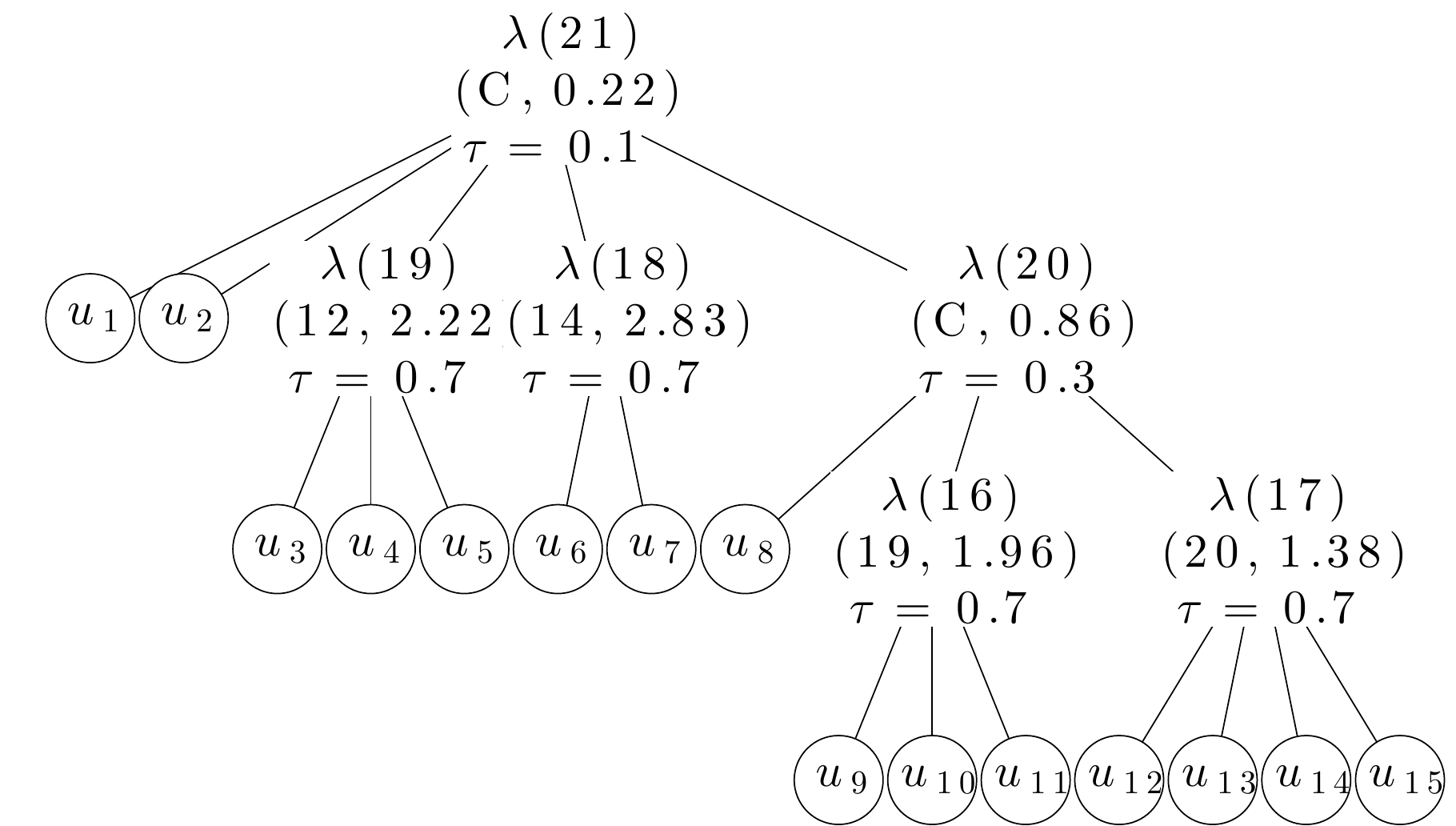}
						}
		\subfigure[The \{12, 20, C\}-heterogeneous 15-HAC model]{%
            \label{fig:model_d5_A19}
\includegraphics[width=0.45\textwidth]{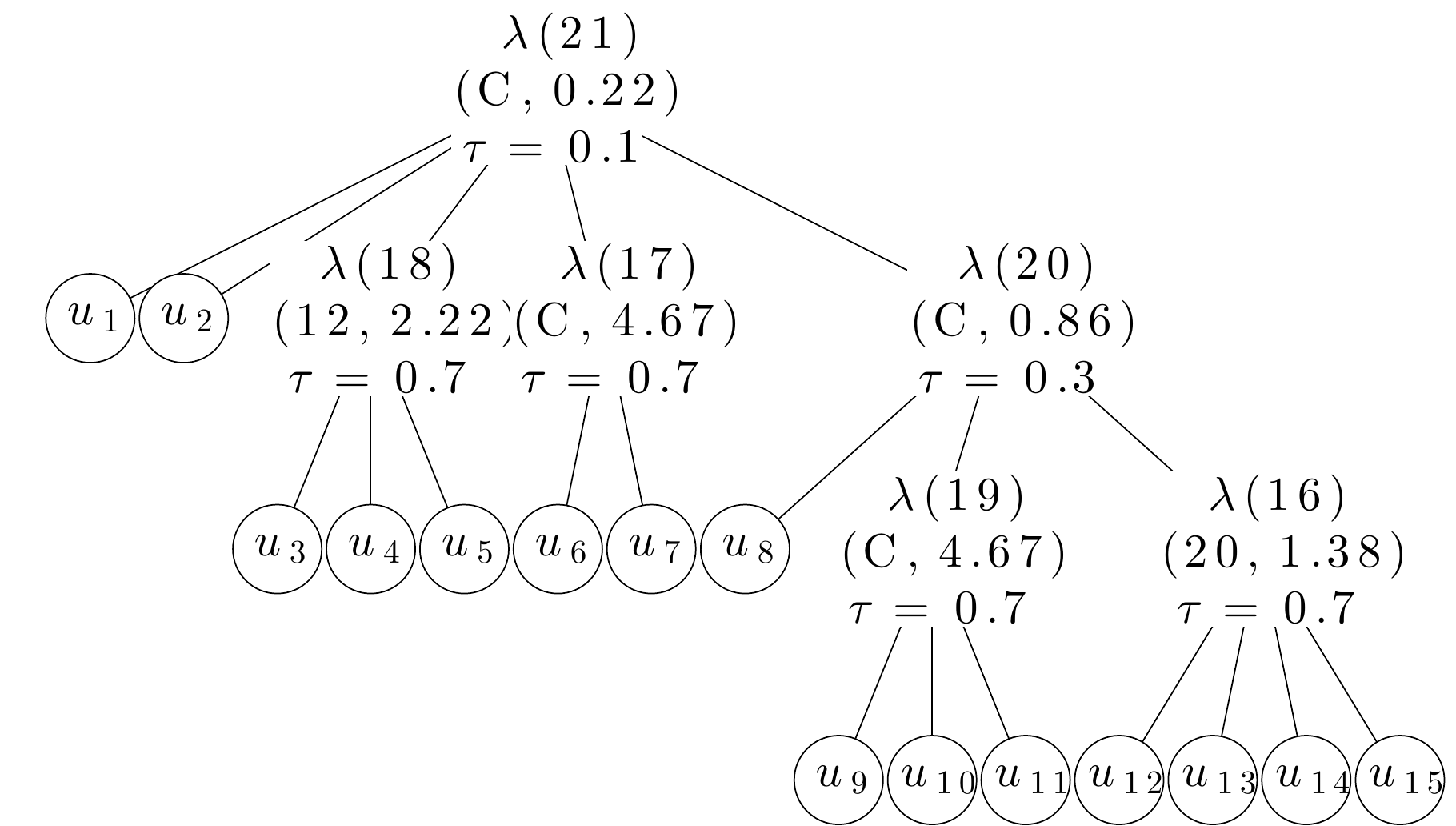}
						}
\caption{The $\tau$-spread=high models for $d \in \{5, 10, 15\}$.}
\label{fig:models_high}
\end{figure}

\begin{figure}[htb]
\centering
		\subfigure[The \{19, A, C\}-heterogeneous 5-HAC model]{%
            \label{fig:model_d5_AC19}
\includegraphics[width=0.45\textwidth]{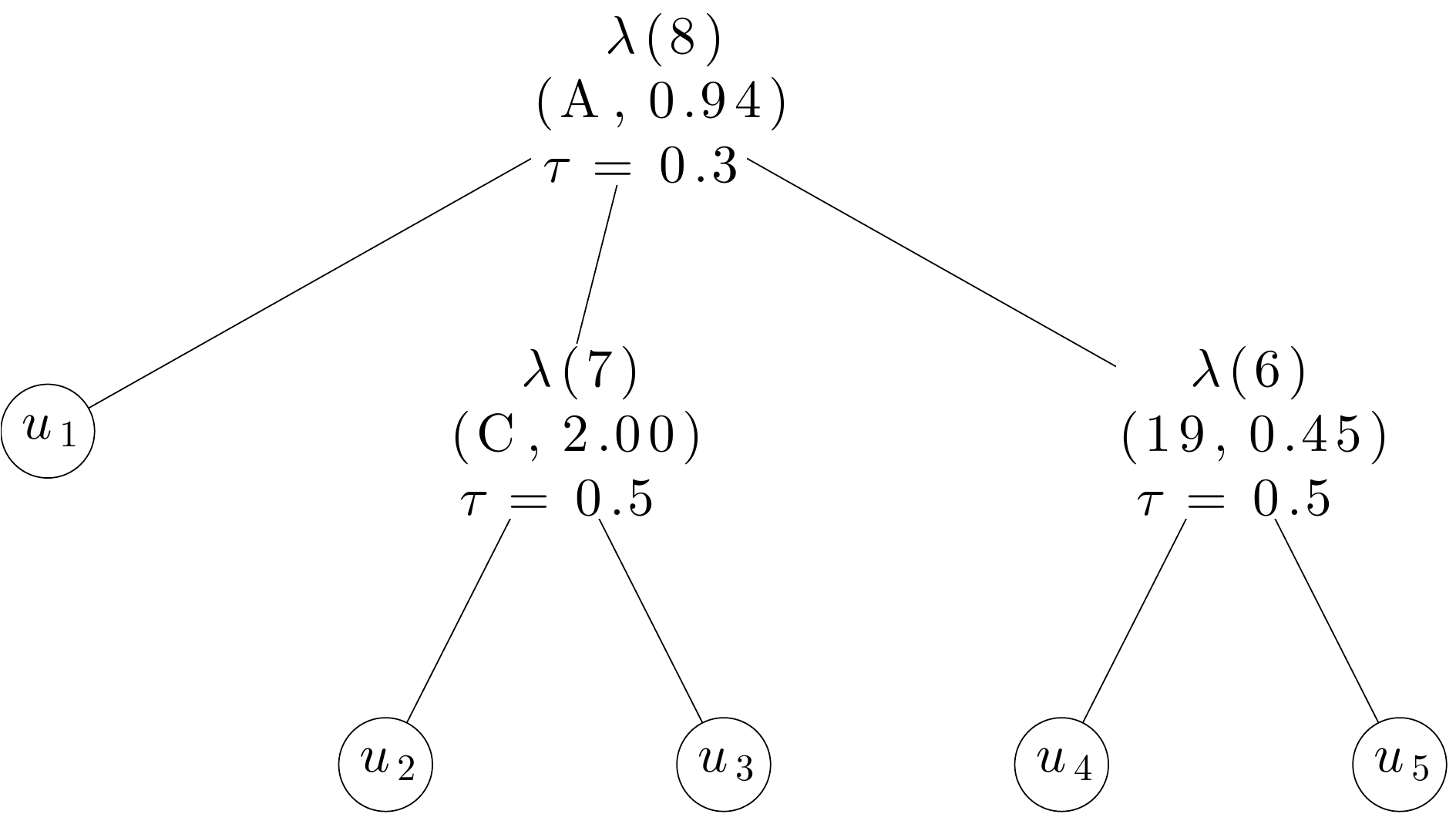}
						}
		\subfigure[The \{19, A\}-heterogeneous 5-HAC model]{%
            \label{fig:model_d5_A19}
\includegraphics[width=0.45\textwidth]{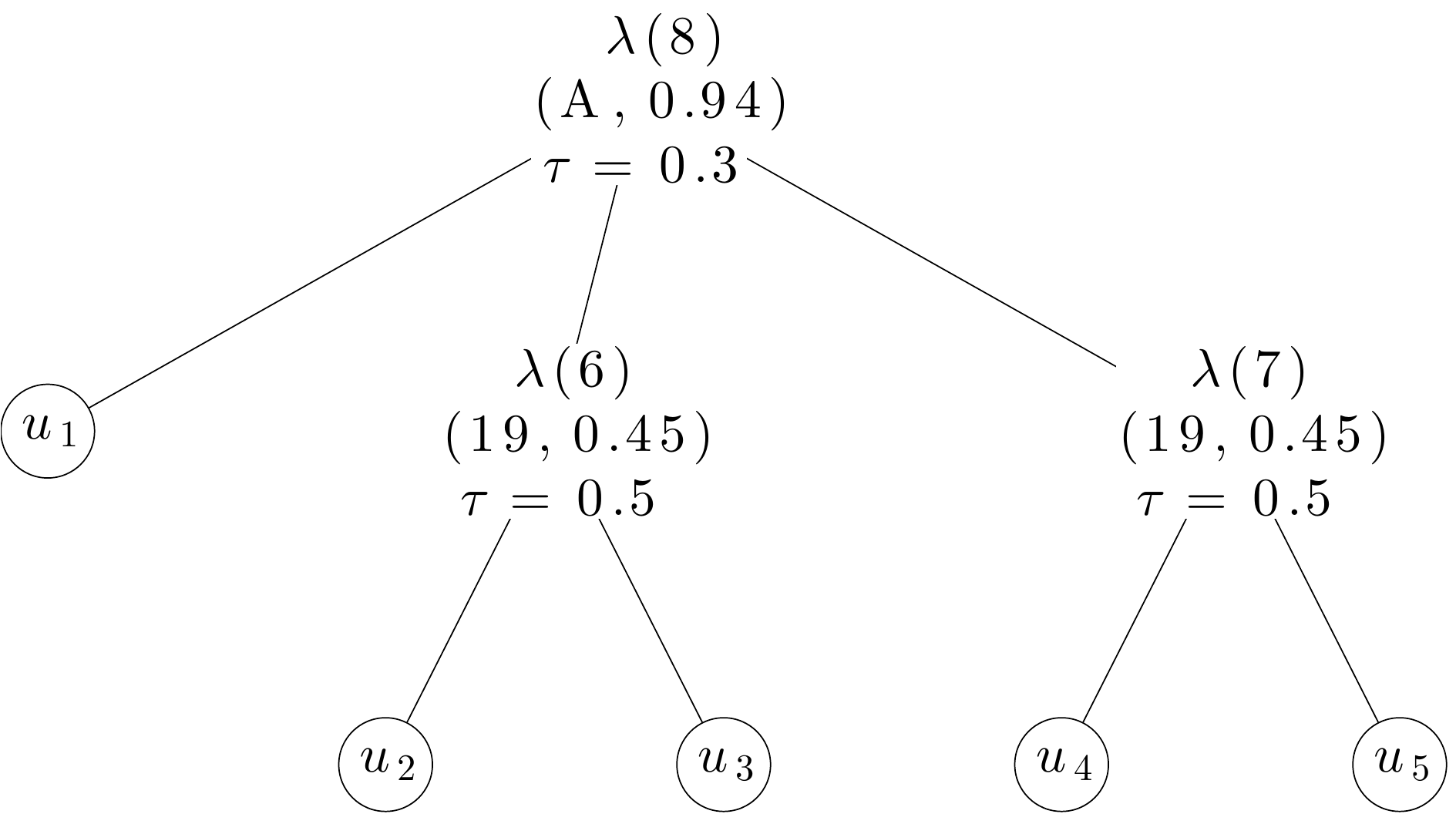}
						}
		\subfigure[The \{19, 20, A, C\}-heterogeneous 10-HAC model]{%
            \label{fig:model_d5_A19}
\includegraphics[width=0.45\textwidth]{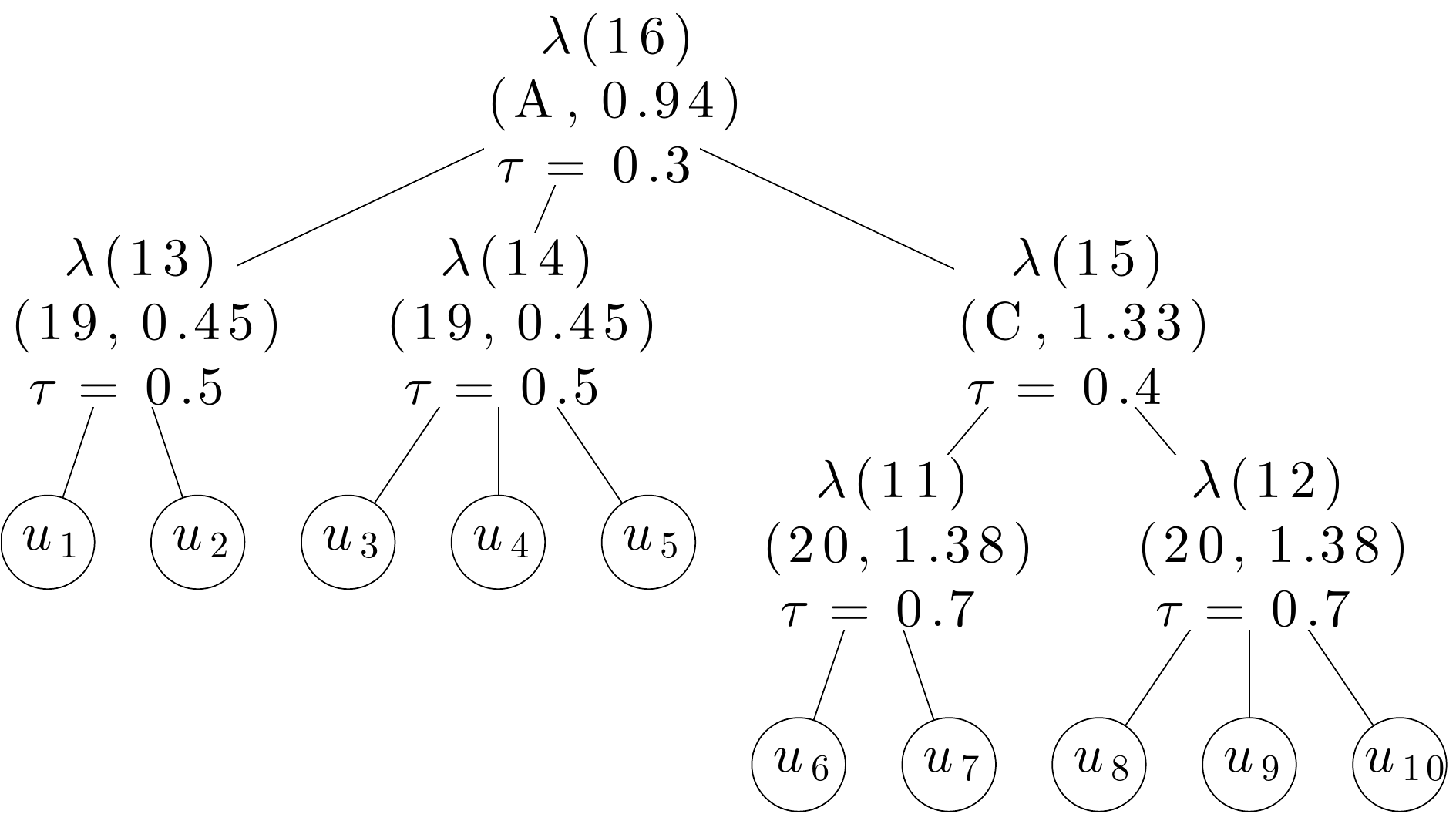}
						}
		\subfigure[The \{19, A, C\}-heterogeneous 10-HAC model]{%
            \label{fig:model_d5_A19}
\includegraphics[width=0.45\textwidth]{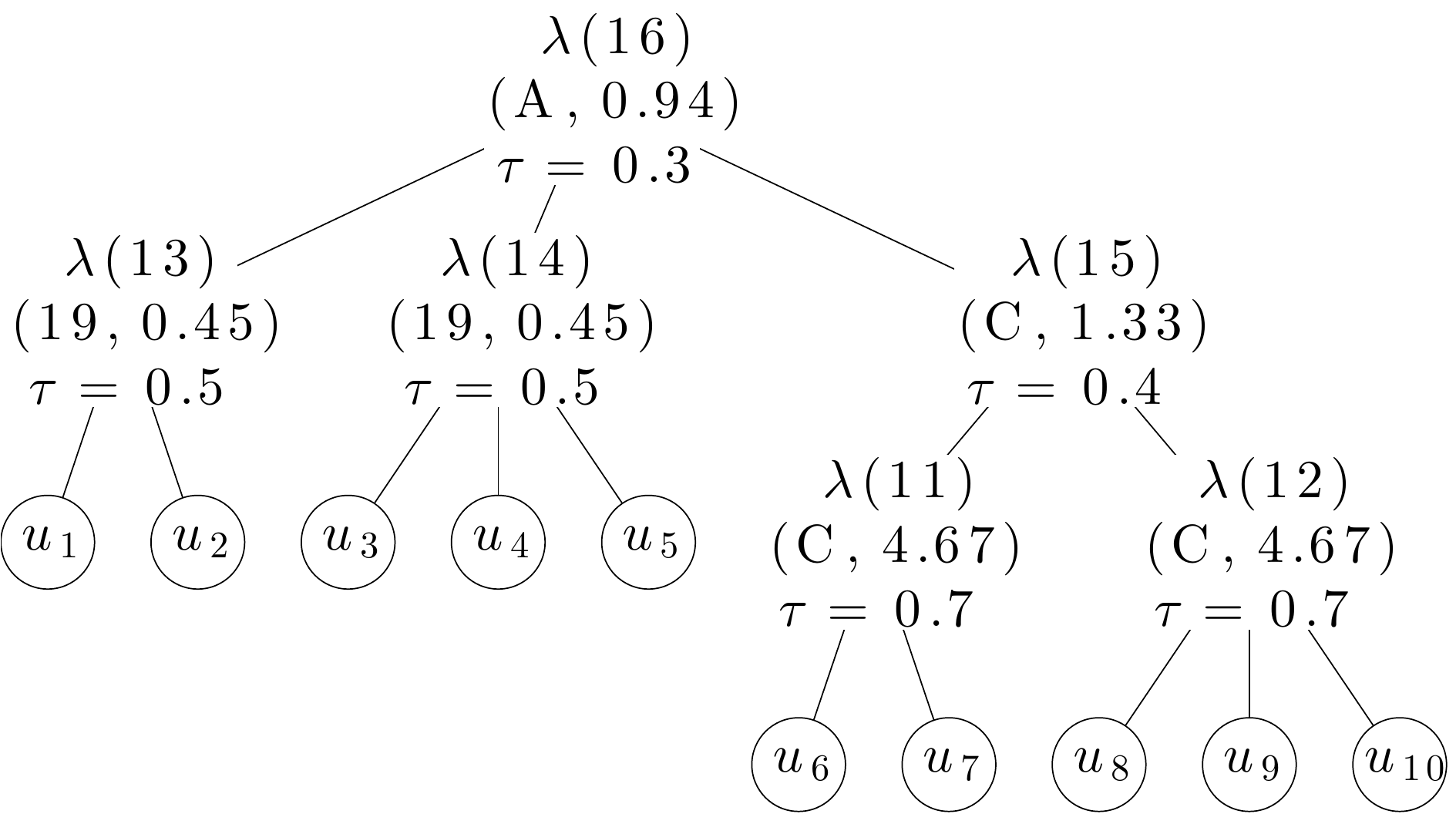}
						}
		\subfigure[The \{12, 14, 19, 20, C\}-heterogeneous 15-HAC model]{%
            \label{fig:model_d5_A19}
\includegraphics[width=0.45\textwidth]{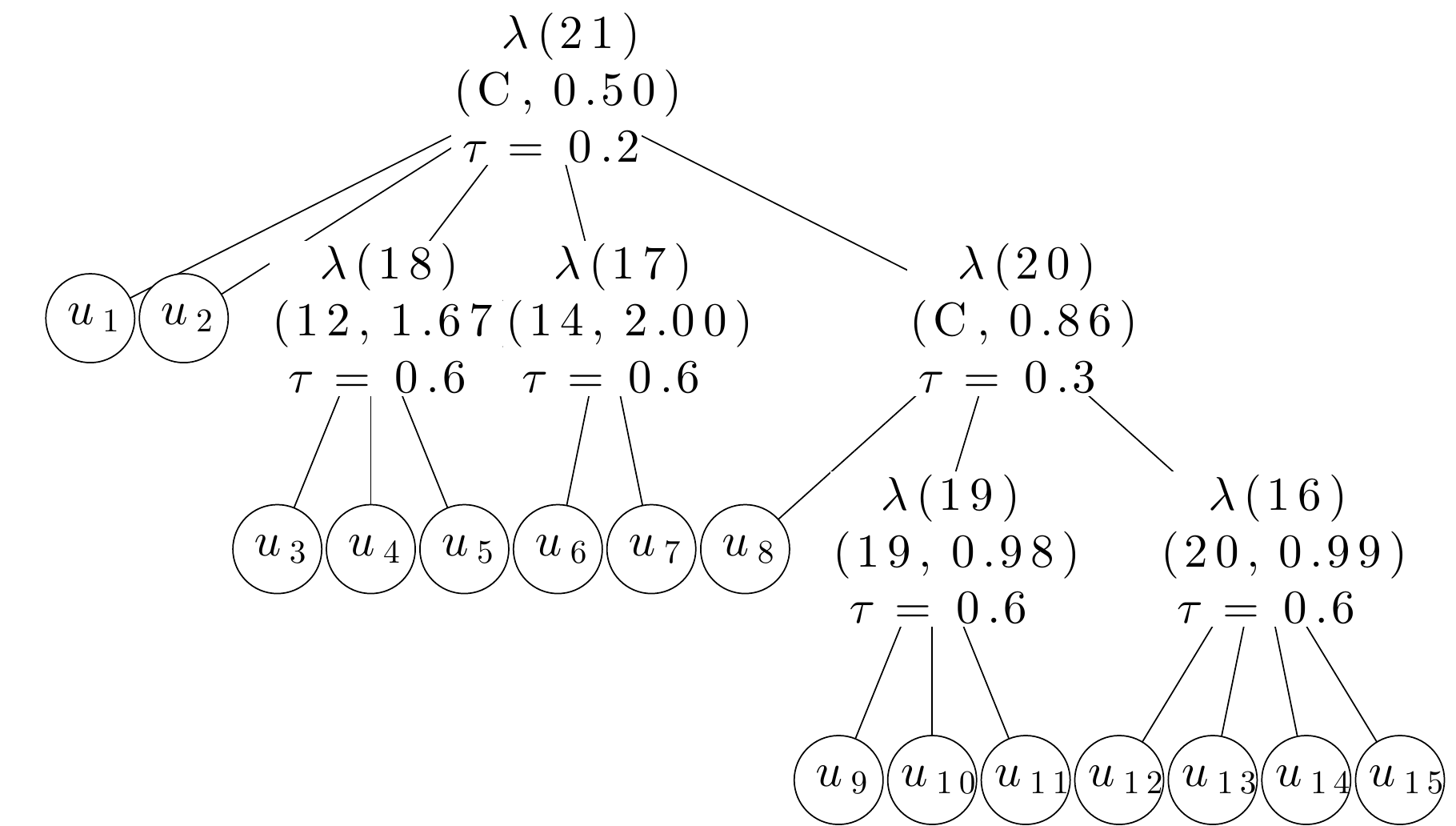}
						}
		\subfigure[The \{12, 20, C\}-heterogeneous 15-HAC model]{%
            \label{fig:model_d5_A19}
\includegraphics[width=0.45\textwidth]{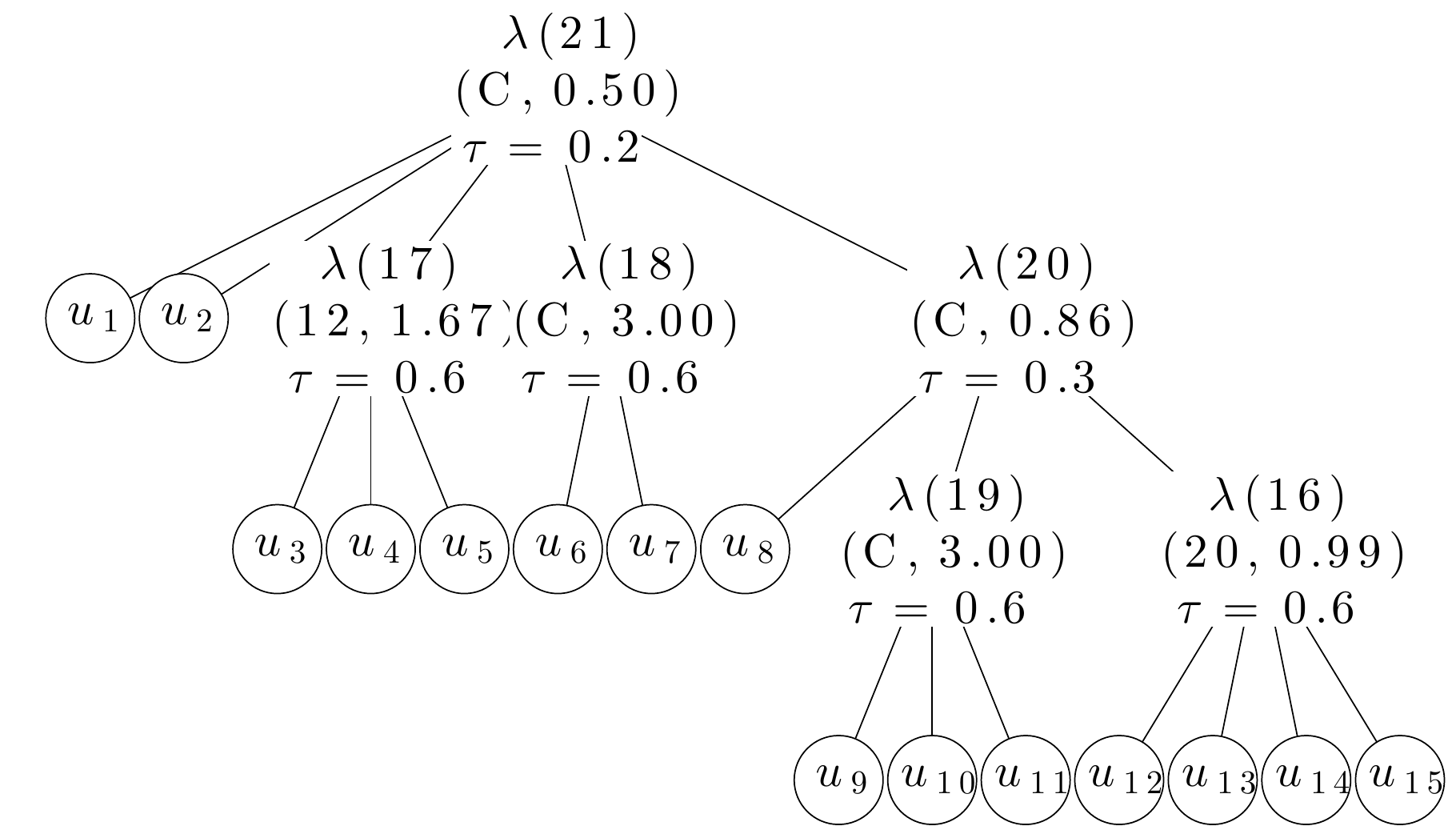}
						}
\caption{The $\tau$-spread=low models for $d \in \{5, 10, 15\}$.}
\label{fig:models_low}
\end{figure}

All in all, we consider $18 \times 4 = 72$ estimators for Coll=post plus $12 \times 4 = 48$ estimators for Coll=pre, i.e., 120 heterogeneous estimators, which we compare on data simulated from the 12 models depicted in Figures \ref{fig:models_high} and \ref{fig:models_low}. 

%The rest of the models is generated just by different parameter settings in the following way. 
These models are designed as follows.
As we want to study how the estimators behave when parameters in a model get closer (measured by the $\delta$, see Section \ref{sec:collapsing}) to each other, the parameters of the six models depicted in Figure \ref{fig:models_high} are set in a way that they differ more (we denote them by $\tau$-spread=high) than in the remaining six models (these are denoted by $\tau$-spread=low) depicted in Figure \ref{fig:models_low}. Observe that we mostly shift the parameters in the first and in the last levels of the structures keeping the parameters in the second level (for $d \in \{10, 15\}$) unchanged. Also note that the parameters were chosen in a way that the s.n.c.~is satisfied.

%Precisely, the $\tau$-spread=low models are set, for $d=5$, $\tau(\lambda(7)) = 0.3$ and $\tau(\lambda(6)) = 0.5$, 
%for $d=10$, $\tau(\lambda(16)) = 0.3, ~ \tau(\lambda(13)) = \tau(\lambda(14)) = 0.5, ~ \tau(\lambda(11)) = \tau(\lambda(12)) = 0.7$ and $\tau(\lambda(15))$ remains the same as for $\tau$-spread=high, 
%for $d=15$, $\tau(\lambda(21)) = 0.2, ~ \tau(\lambda(19)) = \tau(\lambda(18)) = \tau(\lambda(17)) = \tau(\lambda(16)) = 0.6$ and $\tau(\lambda(20))$ remains the same as for $\tau$-spread=high, 
%and for $d=20$, $\tau(\lambda(31)) = 0.2, ~ \tau(\lambda(23)) = \tau(\lambda(24)) = \tau(\lambda(25)) = \tau(\lambda(26)) = 0.6$, $\tau(\lambda(21)) = \tau(\lambda(22)) = 0.7$  and $\tau(\lambda(27)), ~ \tau(\lambda(28)),~\tau(\lambda(29))$ and $\tau(\lambda(30))$ remains the same as for $\tau$-spread=high. 

To allow for studying the behavior of the estimators depending on the number of involved families, each two models for a given $d$ and $\tau$-spread are chosen such that the number of different families in a model varies, precisely, three and two families for $d = 5$, four and three families for $d=10$ and five and three families for $d=15$. 
Note that some of these models involve all families from $\mathcal{F}_{1234}$ ($d = 10$) and also all families from $\mathcal{F}_{24}$ ($d = 15$).
Despite the fact that it is possible to nest even 6 different families in a HAC, we do not consider such models because of extreme restrictions of the parameters, as addressed in Section \ref{sec:hetero_HAC_estim}. From this point of view, the considered models thus contain the largest possible sets of families that are reasonable to nest into a single HAC. Also note that the leaves in all models are always ordered from 1 to $d$ (left to right) as their permutation does influence the results reported below.

Given a data sample and an estimator, we consider the following two scenarios for the input $\FF$ of the estimator:
\begin{enumerate}
	\item \textbf{$\FF$-known}: This scenario assumes that the set of families in the model underlying the data is known, and $\FF$ is thus set equal to this set. This scenario is unrealistic, however, it allows us to study the precision of the estimators if the underlying families are chosen correctly.
	
The inputs $\mathcal{N}_0$ and $\mathcal{N}^2_{\mathcal{F}}$ are set as proposed in  Section \ref{sec:hetero_HAC_estim},  e.g., see Lemmas \ref{lem:n2f12} and  \ref{lem:n2f123} for $\mathcal{F} = \mathcal{F}_{24}$ and for $\mathcal{F} = \mathcal{F}_{1234}$, respectively. 
For other sets of families, $\mathcal{N}_0$ and $\mathcal{N}^2_{\mathcal{F}}$ are narrowed from the ones corresponding to $\mathcal{F}_{1234}$ or $\mathcal{F}_{24}$ just by removing the families that are not in $\mathcal{F}$. Note that, if A $\in \FF$, $\mathcal{N}_0$ and $\mathcal{N}^2_{\mathcal{F}}$ are narrowed from the ones corresponding to $\mathcal{F}_{1234}$, otherwise, these inputs are narrowed from the ones corresponding to $\mathcal{F}_{24}$;
%Finally, the input $\epsilon$  is set to the lowest possible real for the primitive double in MATLAB, i.e., to \texttt{eps} = 2.2204e-16, in order to maximally reduce biasing of the parameter estimate in Step 6 of Algorithms \ref{alg:hetero_HAC_estim} and \ref{alg:diag_hetero_HAC_estim}. 
	\item \textbf{$\FF$-unknown}: $\FF$ is chosen to be a family from \{C, F, G\}, where F and G denote the Frank and Gumbel family of generators, respectively; see, e.g., \cite{Nel06,Hof11} for their definitions. This realistic scenario allows to study how robust the estimators are when misspecifying the underlying distribution (family). Note that we choose those three families as they cover a variety of different tail dependencies, more precisely, C allows for modeling lower tail dependency while being upper tail independent, F is both lower and upper tail independent and G represents the opposite of C. 

Hence, this scenario generates additional homogeneous estimators, which are constructed as follows. We consider the three families mentioned above, the two homogeneous version of the PT and ML estimators, and, similarly to the previous scenario, the two attitudes and the four collapsing strategies. Also we consider Coll=pre and Coll=post for the PT estimator. This gives $3\times 3$(=\{PT post, PT pre, ML\})$\times 2 \times 4 = 72$ estimators. Note that the homogeneous version of the PT estimator is not dependent on $g$; compare Algorithms \ref{alg:HAC_estim} and \ref{alg:hetero_HAC_estim}. Also note that the results for this scenario serve as benchmark results for the previous scenario.
\end{enumerate}

For each of the 12 considered models, we generate $N=100$ samples of sizes $n \in \{n_1, ..., n_{20}\}$ according to the sampling approach proposed in \cite{Hof11}. The value $n_i$ depends on $d$ and is chosen as $n_i = d*4*i, ~ i \in \{1, ..., 20\}$. This choice follows from our observation that the estimators' abilities, e.g, the ability to detect the structure of the true copula, strongly depend on the amount of available data. Also, this choice allows to see some progress (according to $n$) for most of the considered evaluations.
Thus we generate 100$\times$20 = 2~000 data samples for each model, i.e., 24~000 data samples in total. Applying the 120 heterogeneous estimators and the 72 homogeneous estimators for each of these samples, we finally obtained 4~608~000 HAC estimates, where each one is evaluated using the following criteria. Given an estimator, a model and $i \in \{1, ..., 20\}$, which correspond to $N = 100$ generated samples and estimates, we consider the following evaluation criteria:
\begin{itemize}
\item \emph{False structure ratio} (in \%): expresses how many times the estimator missed the true structure (= the structure of the model) out of all times  it has returned an estimate (note that under the pessimistic attitude, the estimator may not return any estimate, which is studied by the criterion called Rejection rate, which is described in next item);
	\item \emph{Rejection rate}: denotes the number of times the estimator has not returned (\emph{has rejected}) an estimate out of $N = 100$ under the pessimistic attitude;
	\item \emph{False families ratio} (in \%): expresses how many times the estimator missed the true families (= the families of the model) out of the times it has returned the true structure. To clarify, to miss the true families, it is enough that the estimated and the true family corresponding to a fork in the true structure differ at least once. Note that this ratio is not considered for the $\FF$-unknown scenario;
	\item \emph{Tau distance median}: denotes the median of the $N=100$ realizations of $\sum_{i=1}^d$ $\sum_{j=i+1}^d (\tau^n_{ij}-\hat{\tau}_{ij})^2$, where $(\tau^n_{ij})$ and $(\hat{\tau}_{ij})$ denote the Kendall correlation matrices corresponding to a data sample and the estimate based on this sample, respectively;
	\item \emph{GoF median}: denotes the median of the $S_n^{(E)}$ values, see \eqref{eq:empirical_gof}, computed for the $N=100$ pairs of the data samples and estimates. Note that for $d > 2$, particularly for the dimensions we consider in our experiments, using $S_n^{(K)}$ or $S_n^{(R)}$ directly instead of $S_n^{(E)}$ would be at least a challenging task. For more details on this problem, see Section 2.5 in \cite{goreckihofertholena2016approachjiis}. Also note the in Algorithms \ref{alg:hetero_HAC_estim} and \ref{alg:diag_hetero_HAC_estim}, we do not use the above-mentioned statistics directly, but we use their aggregated version; see Definition \ref{def:agg_s_n}.
	\end{itemize}

\subsection{Results of the performed experiments}
An evaluation of the 120 heterogeneous estimators for the four $d$-HAC models corresponding to a given $d \in \{5, 10, 15\}$ is always depicted in two figures,  e.g., see Figures \ref{fig:false_struc_ratio_opt} and \ref{fig:false_struc_ratio_pes} for the false structure ratio and $d=15$. One of these two figures corresponds to the optimistic estimators, the latter to the pessimistic ones. Given an attitude, 
the corresponding figure shows four sub-figures each corresponding to one of the four considered collapsing strategies. More precisely, the left-hand sub-figures correspond to \#Forks=known, whereas the right-hand to \#Forks=unknown, and the top sub-figures correspond to KTauAvg whereas the bottom sub-figures correspond to TauMin. 
Next, in each sub-figure, there are four plots, each of which corresponds to one of the four models for the given $d$. The corresponding models are identified by $\FF$ and $\tau$-spread. 
Finally, each plot shows the evaluation of the corresponding fifteen (nine for Coll=post and six for Coll=pre) estimators for all $n \in \{n_1, ..., n_{20}\}$. Note that showing all these results for all considered dimensions would make this work excessively large. We thus show only the results for $d=15$, and the results for $d \in \{5, 10\}$ are included in the attachment.

For the false structure ratio, tau distance median and GoF median, we also add a benchmark evaluation obtained for the $\FF$-unknown scenario. For simplicity, in each plot, we show only the evaluation of the estimator that has shown to be the best (the lowest) in average rank (along $n$) among the corresponding 9 homogeneous estimators according to a given criterion. This evaluation is denoted \textbf{homo min in avg}. Also, for the tau distance median and GoF median, we add a benchmark evaluation showing the corresponding result for the underlying copula model. This benchmark evaluation is denoted \textbf{true copula}. Note that for a given model, its graph is the same for both attitudes and all collapsing strategies.

\subsubsection{Structure and Rejections}
\label{sec:exps_struc}
\begin{figure}[t]
\centering
\includegraphics[width=0.49\textwidth]{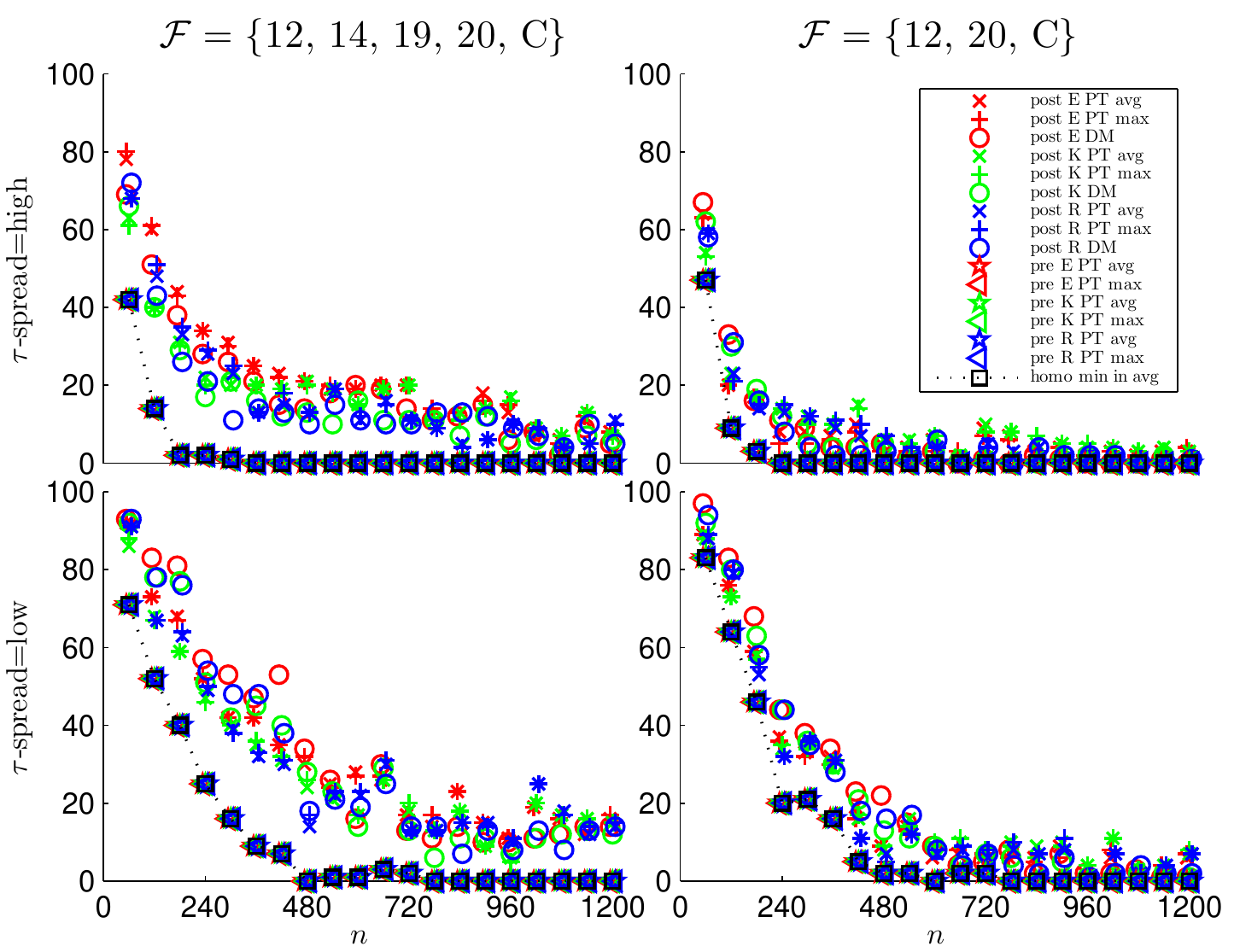}
\includegraphics[width=0.49\textwidth]{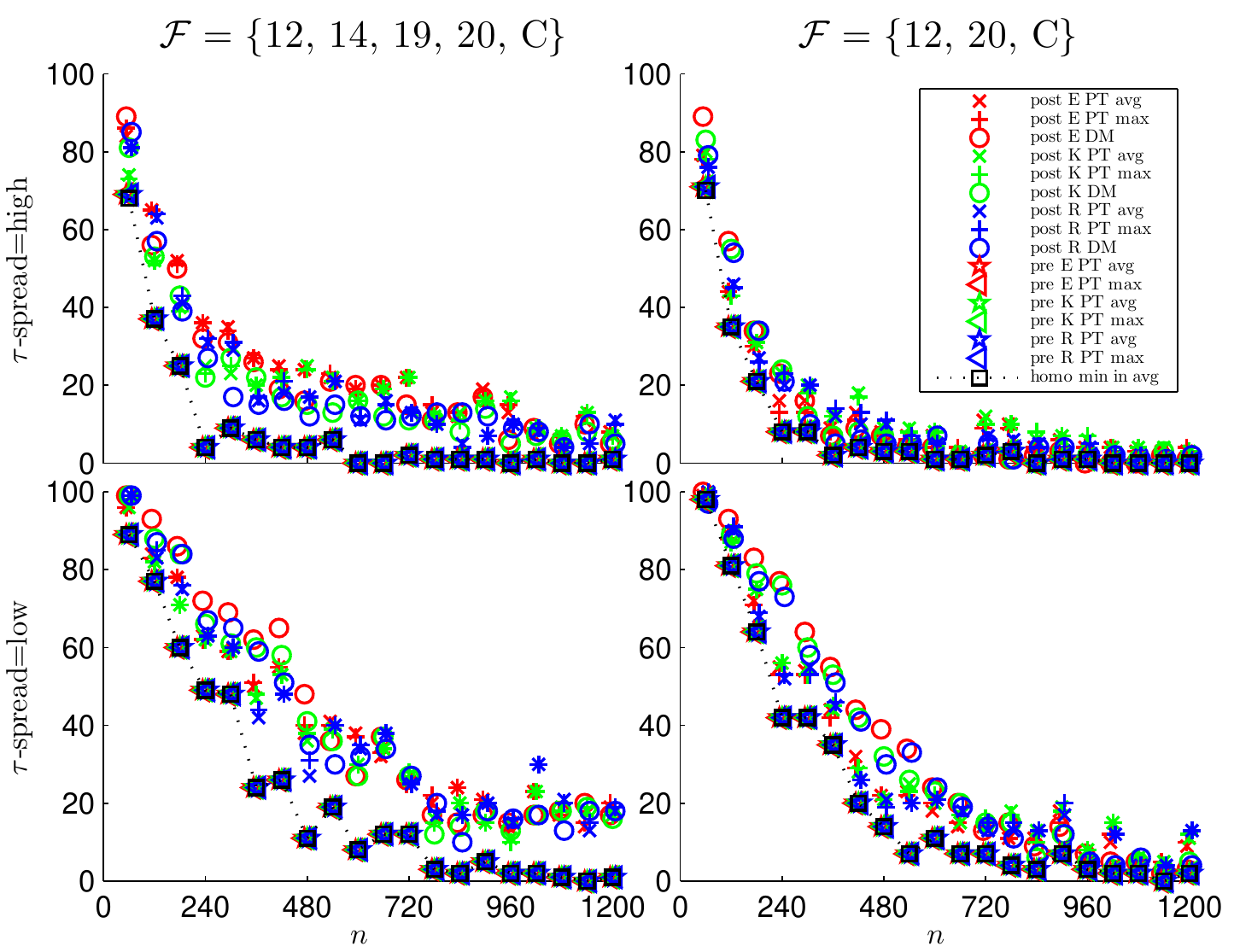}
\includegraphics[width=0.49\textwidth]{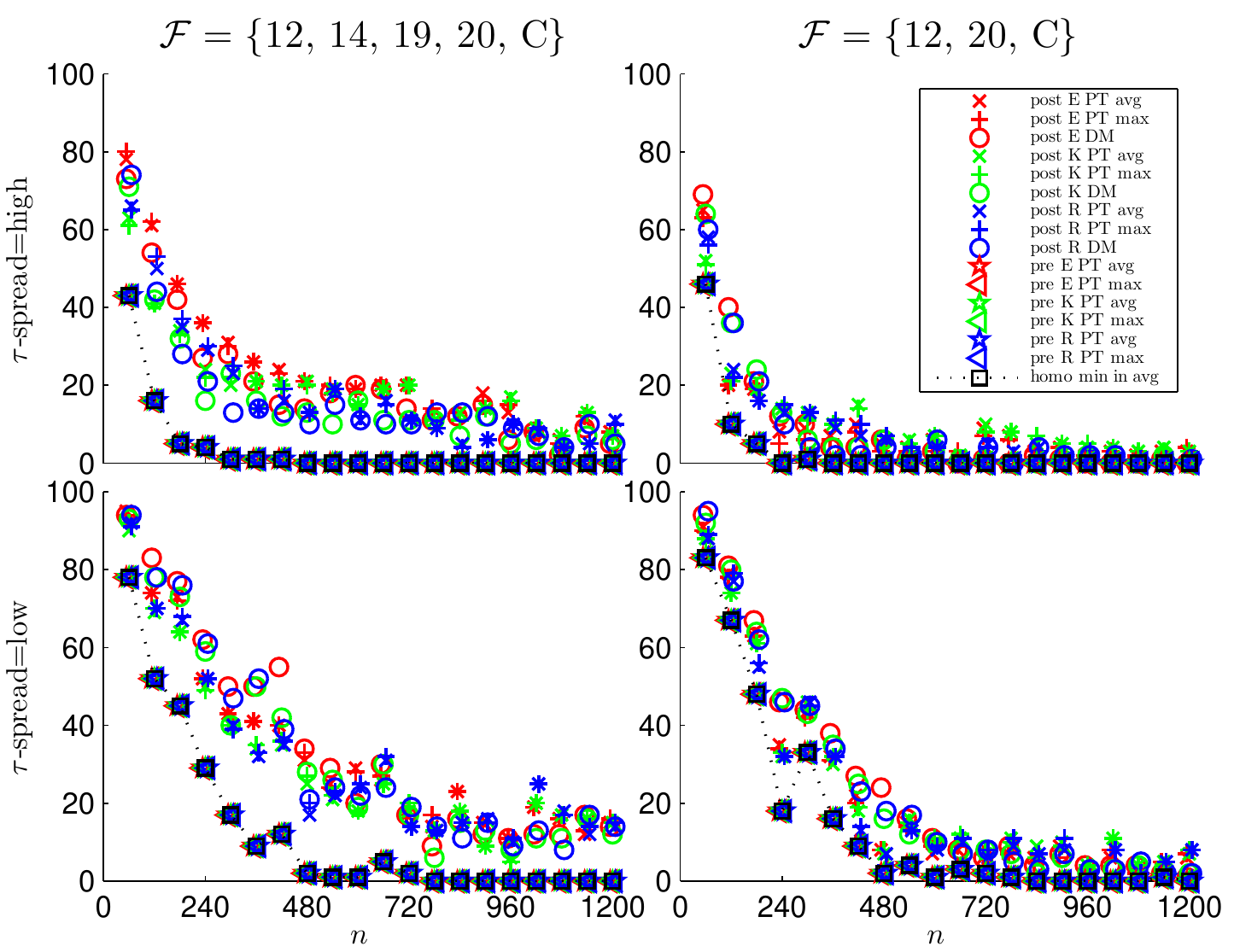}
\includegraphics[width=0.49\textwidth]{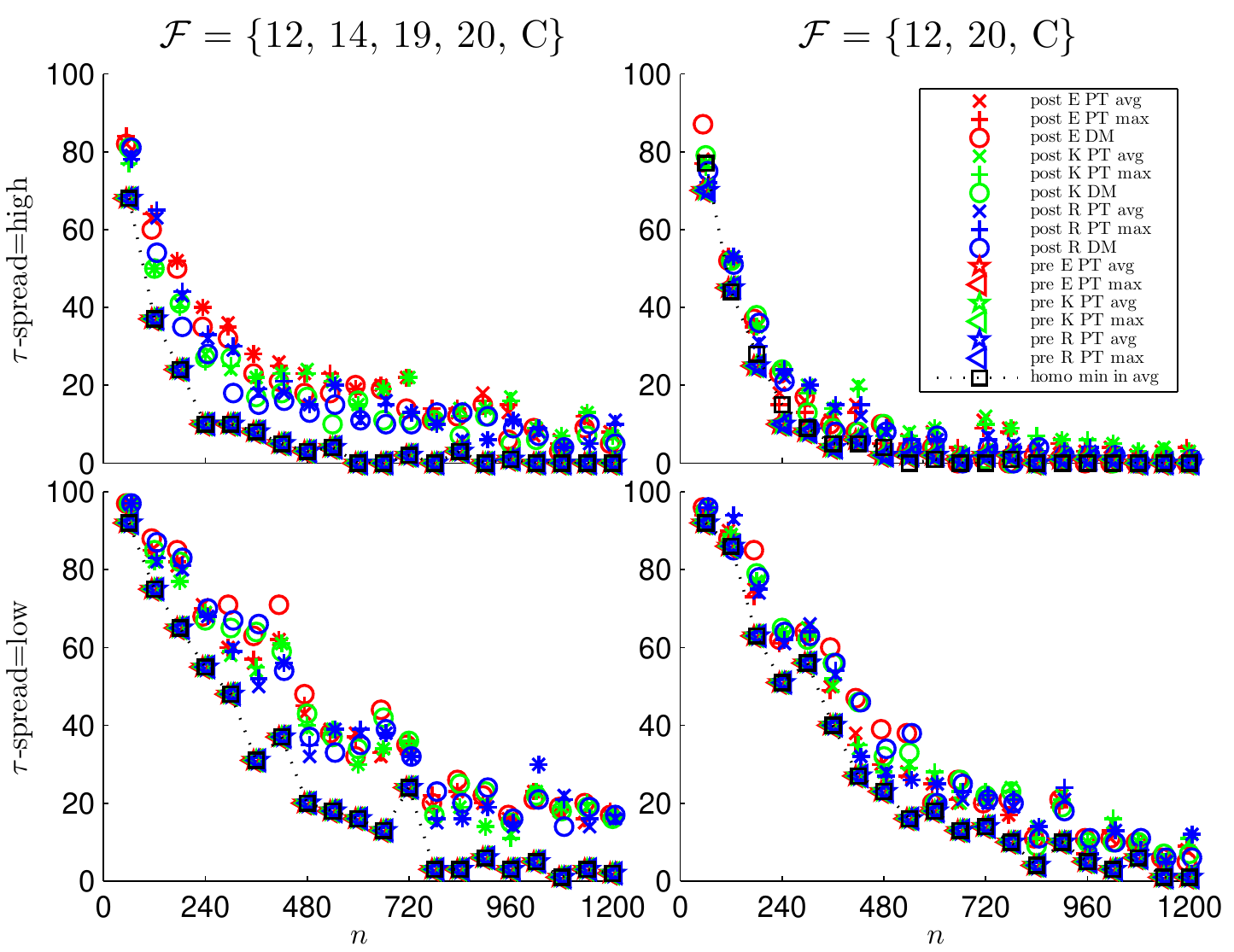}
\caption{False structure ratio (in \%) for the four 15-HAC models and the optimistic estimators. The left-hand and right-hand sub-figures correspond to \#Forks=known and to \#Forks=unknown, respectively, and the top and bottom sub-figures correspond to KTauAvg and to TauMin, respectively.}
\label{fig:false_struc_ratio_opt}
\centering
\end{figure}

\begin{figure}[t]
\centering
\includegraphics[width=0.49\textwidth]{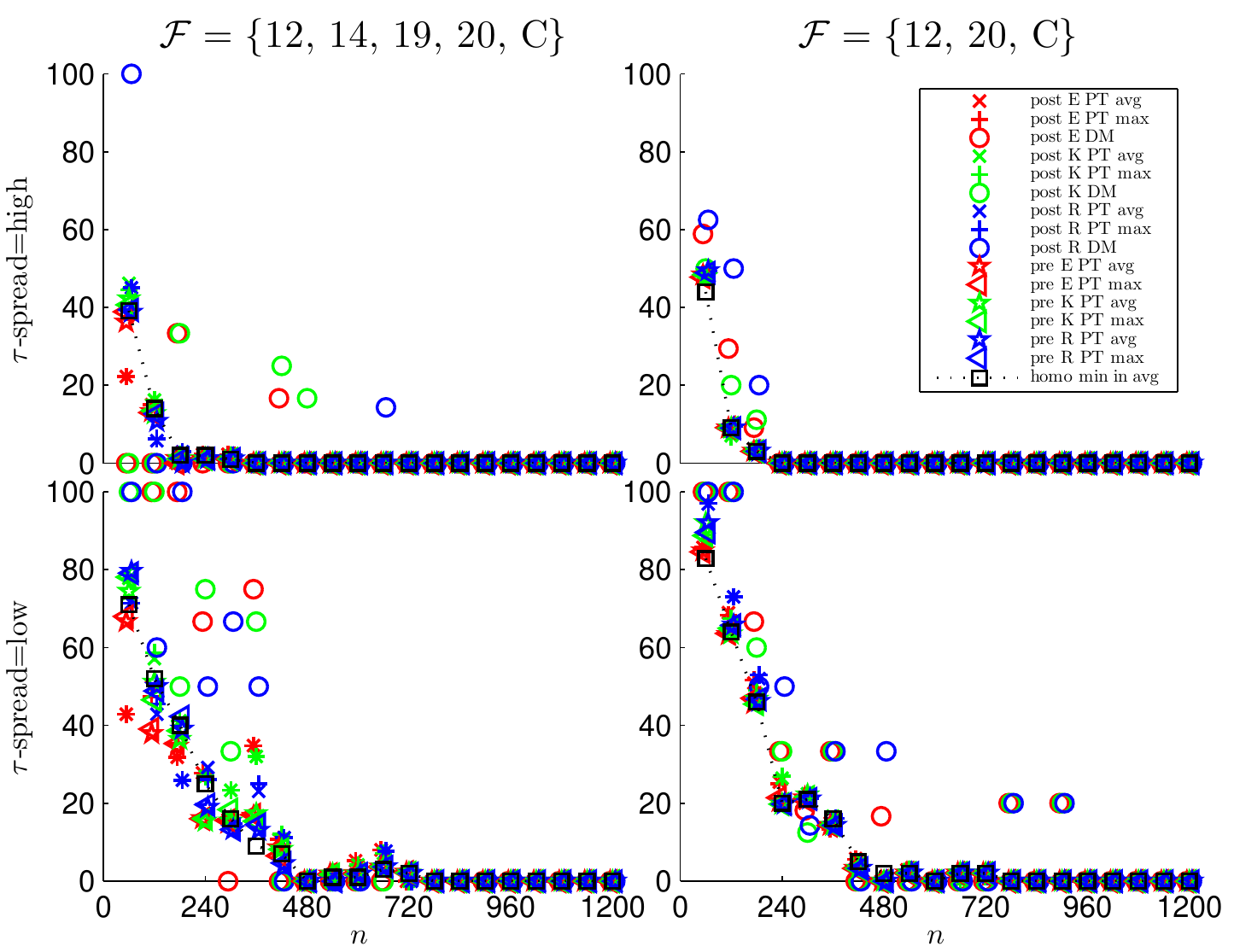}
\includegraphics[width=0.49\textwidth]{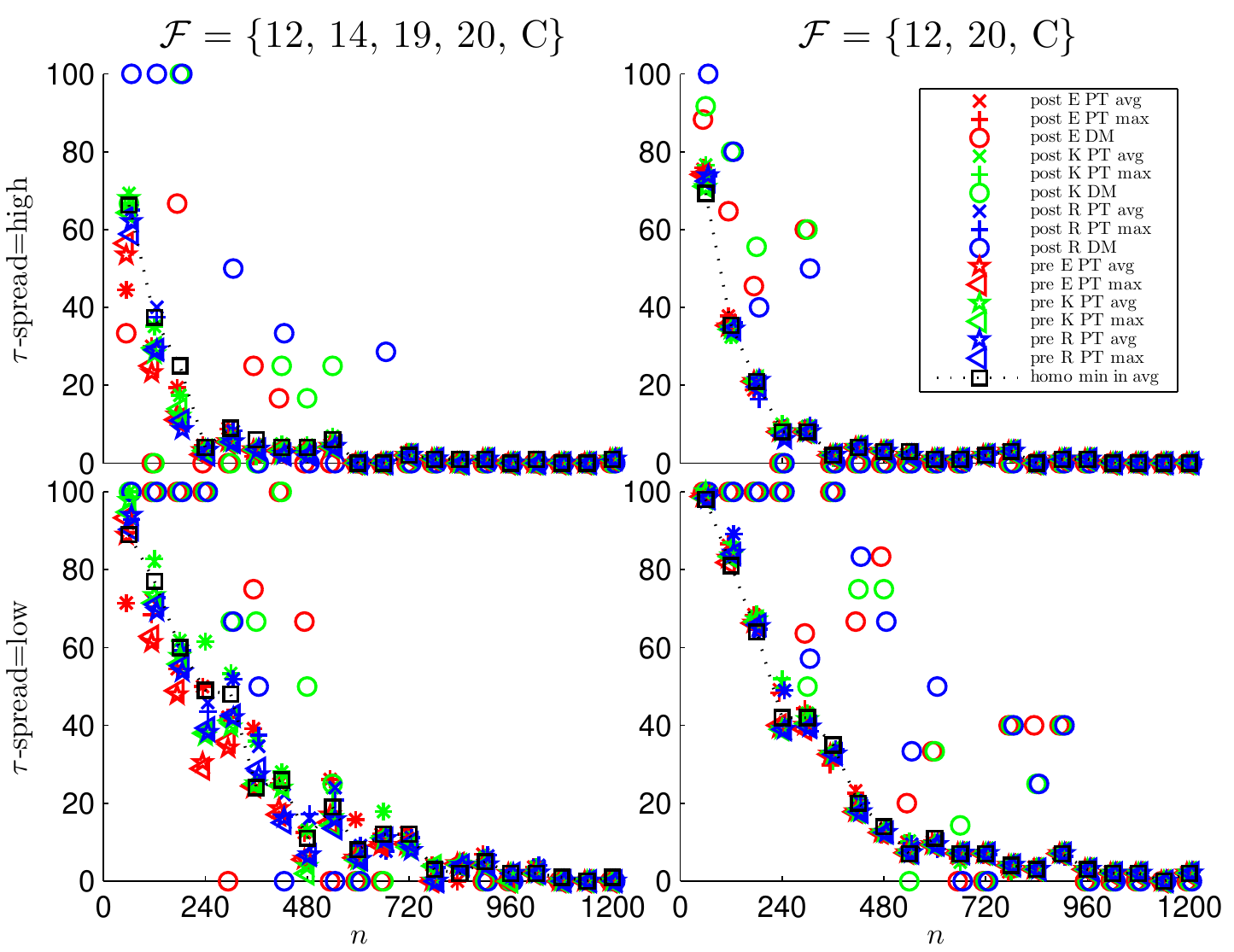}
\includegraphics[width=0.49\textwidth]{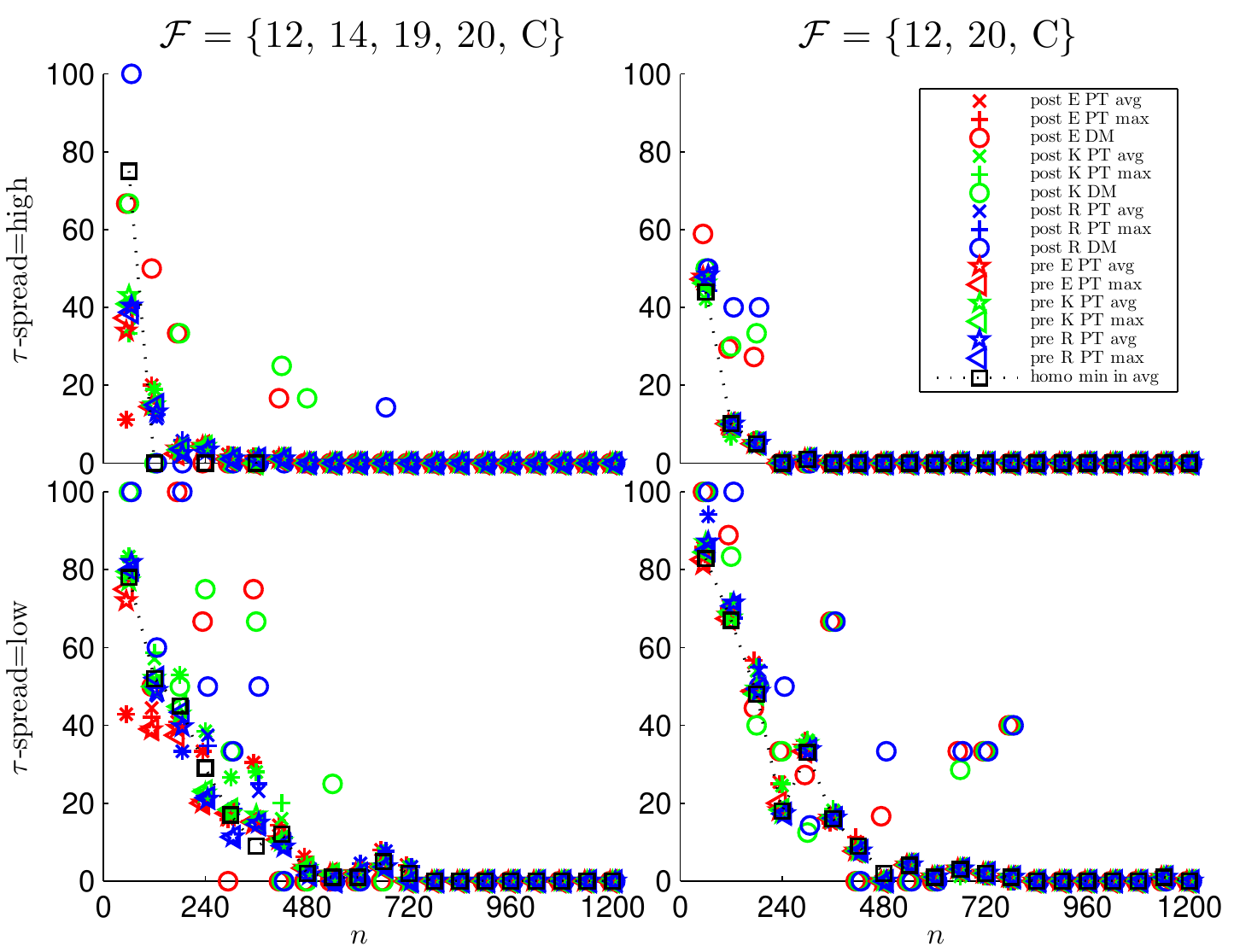}
\includegraphics[width=0.49\textwidth]{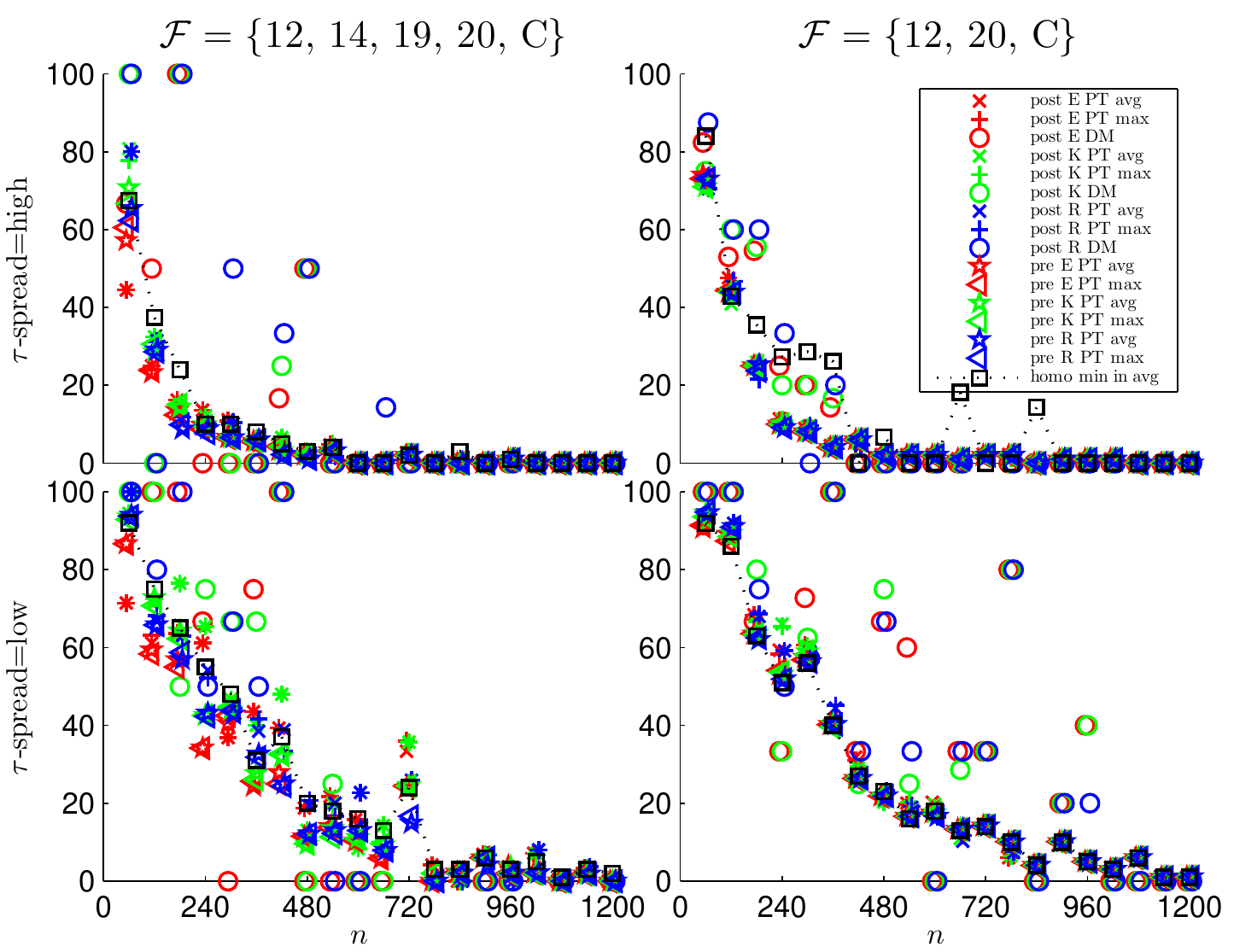}
\caption{False structure ratio (in \%) for the four 15-HAC models and the pessimistic estimators. The left-hand and right-hand sub-figures correspond to \#Forks=known and to \#Forks=unknown, respectively, and the top and bottom sub-figures correspond to KTauAvg and to TauMin, respectively.}
\label{fig:false_struc_ratio_pes}
\centering
\end{figure}

Success of the estimators in identifying the true structure is evaluated by the false structure ratio. This ratio is depicted in Figures \ref{fig:false_struc_ratio_opt} and \ref{fig:false_struc_ratio_pes} for $d = 15$. 

\begin{figure}
\centering
\includegraphics[width=0.49\textwidth]{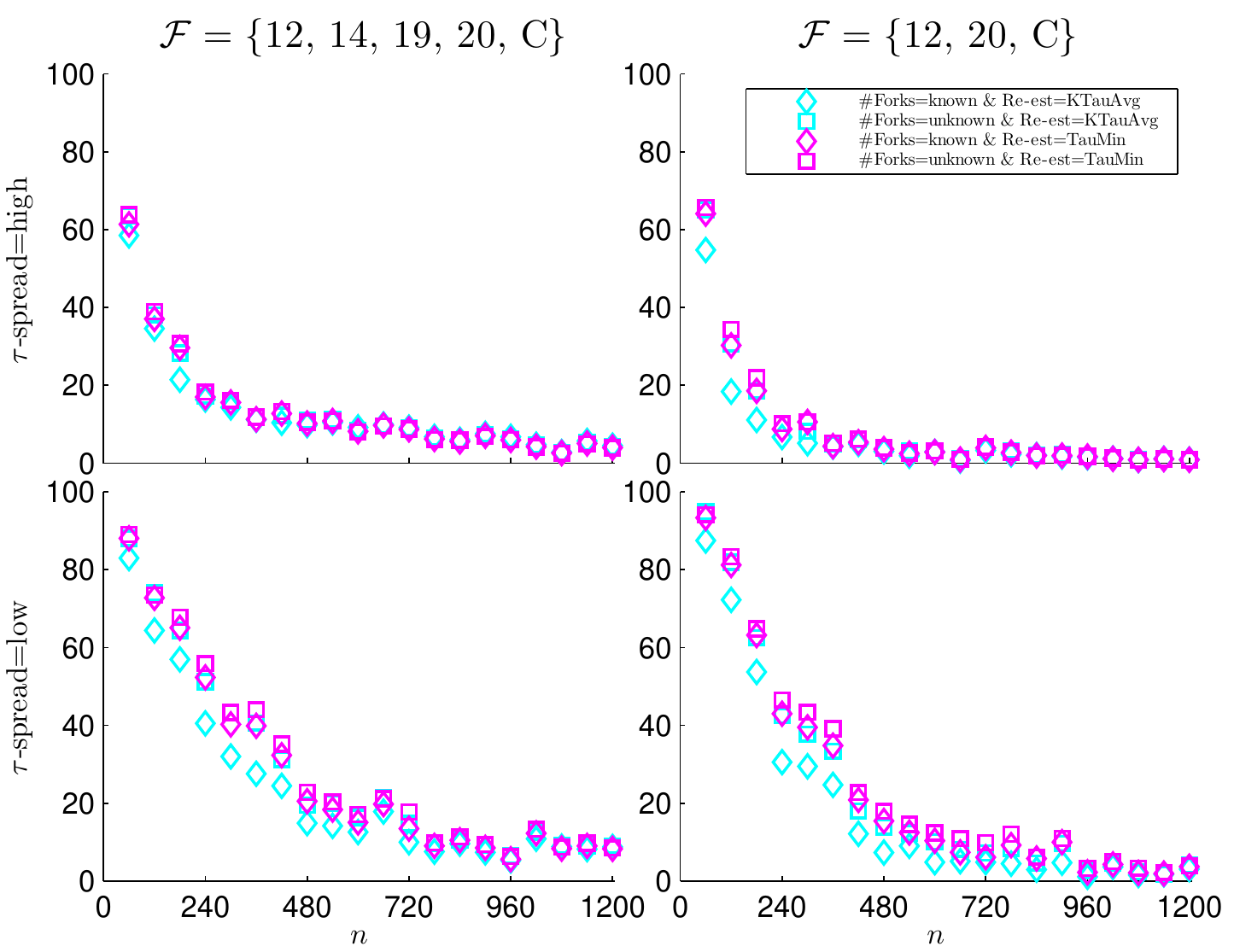}
\includegraphics[width=0.49\textwidth]{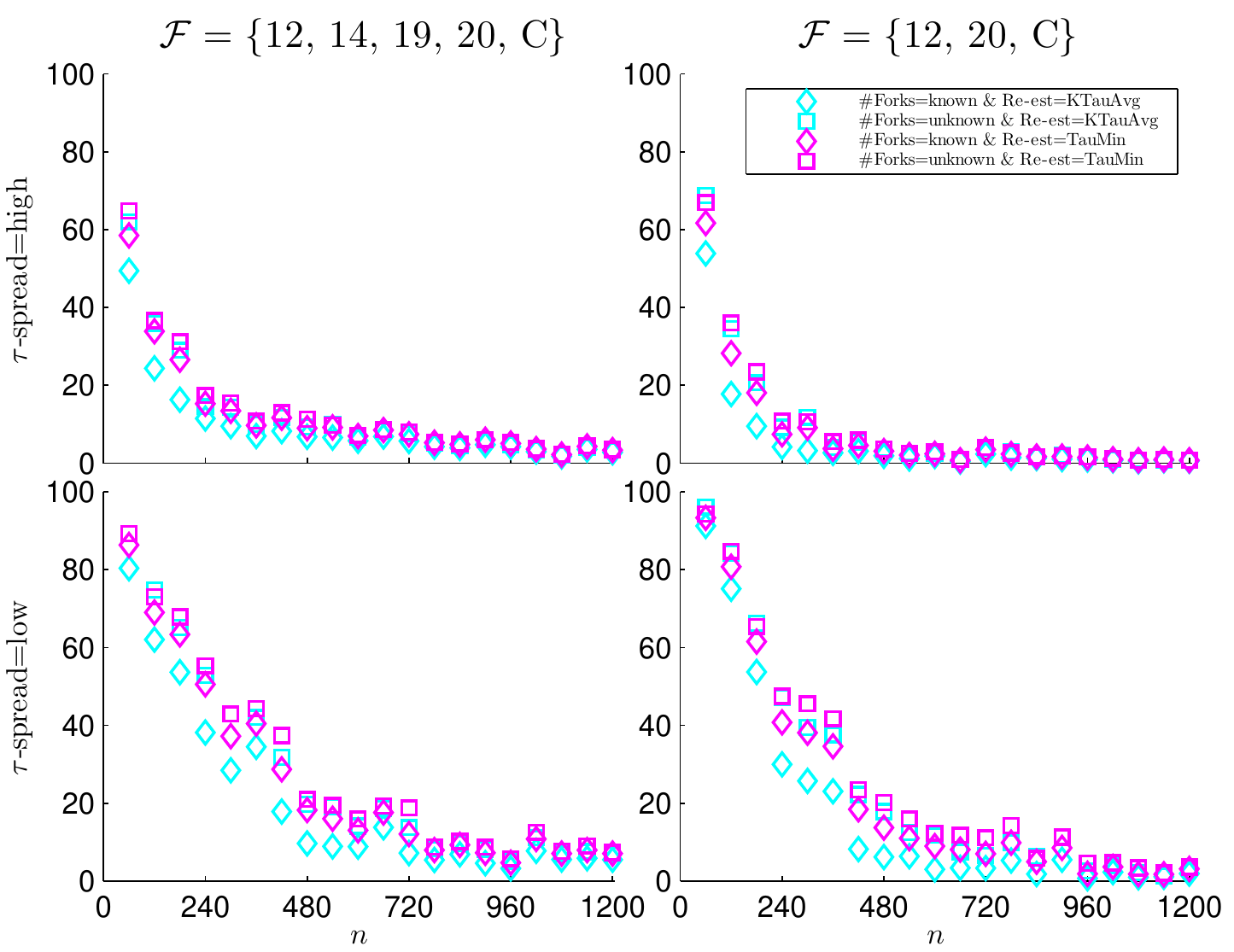}
\caption{Aggregated (by the average) false structure ratio computed for the evaluations depicted in Figure \ref{fig:false_struc_ratio_opt} - the optimistic attitude, here depicted at the left-hand - and Figure \ref{fig:false_struc_ratio_pes} - the pessimistic attitude,  here depicted at the right hand.}
\label{fig:nForks_false_structure_ratio}
\includegraphics[width=1\textwidth]{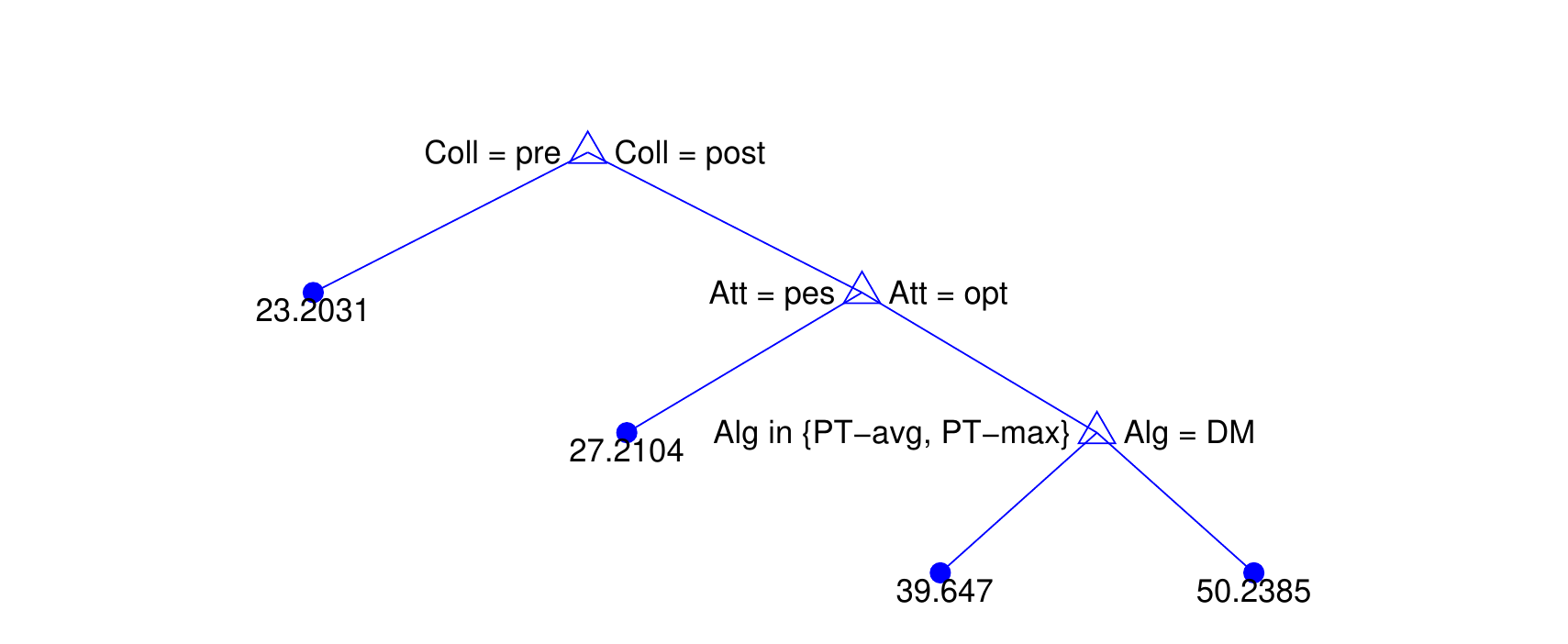}
\caption{The regression tree for the false structure ratio ranks based on the five features of the heterogeneous estimators.}
\label{fig:false_struc_ratio_tree}
\includegraphics[width=0.49\textwidth]{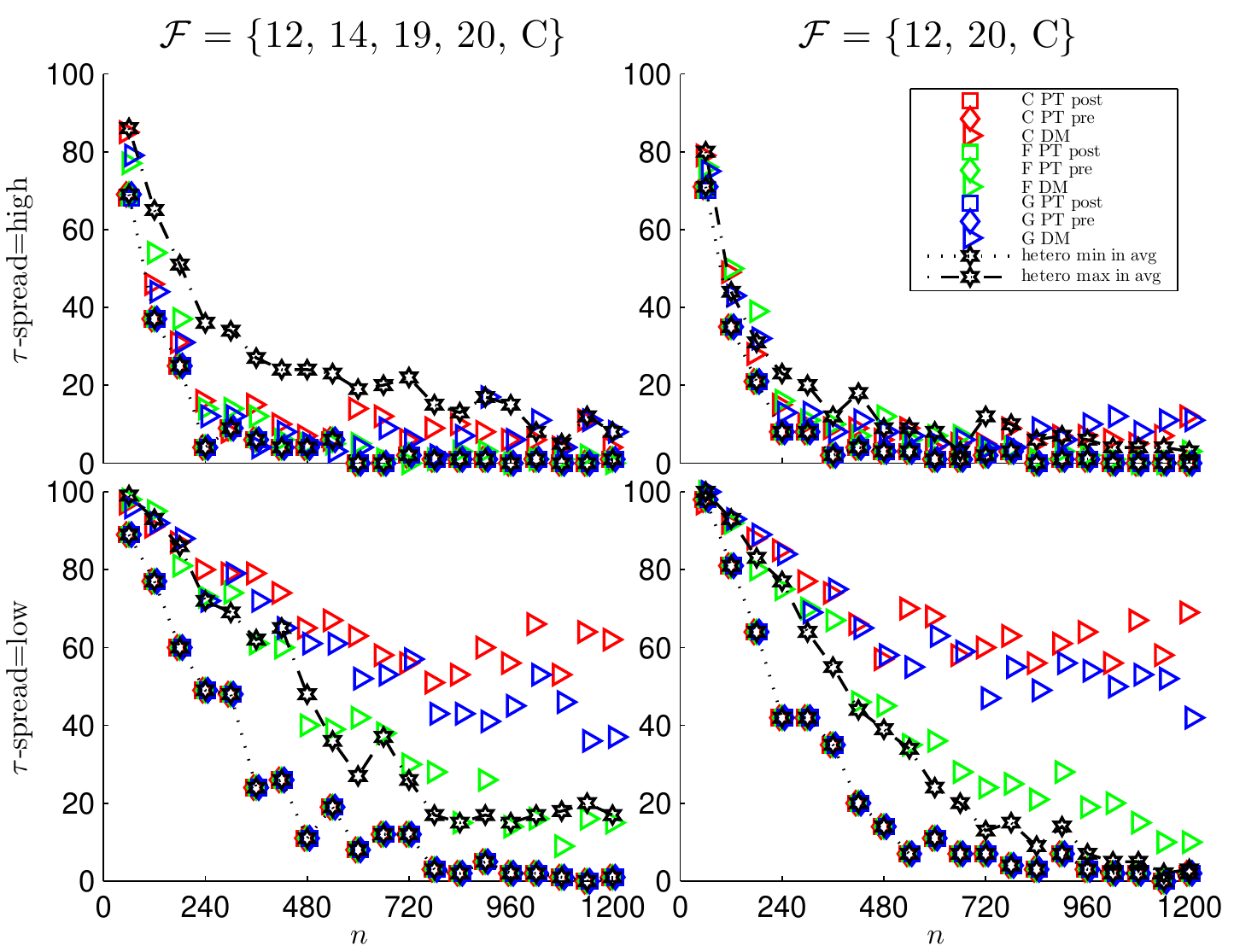}
\includegraphics[width=0.49\textwidth]{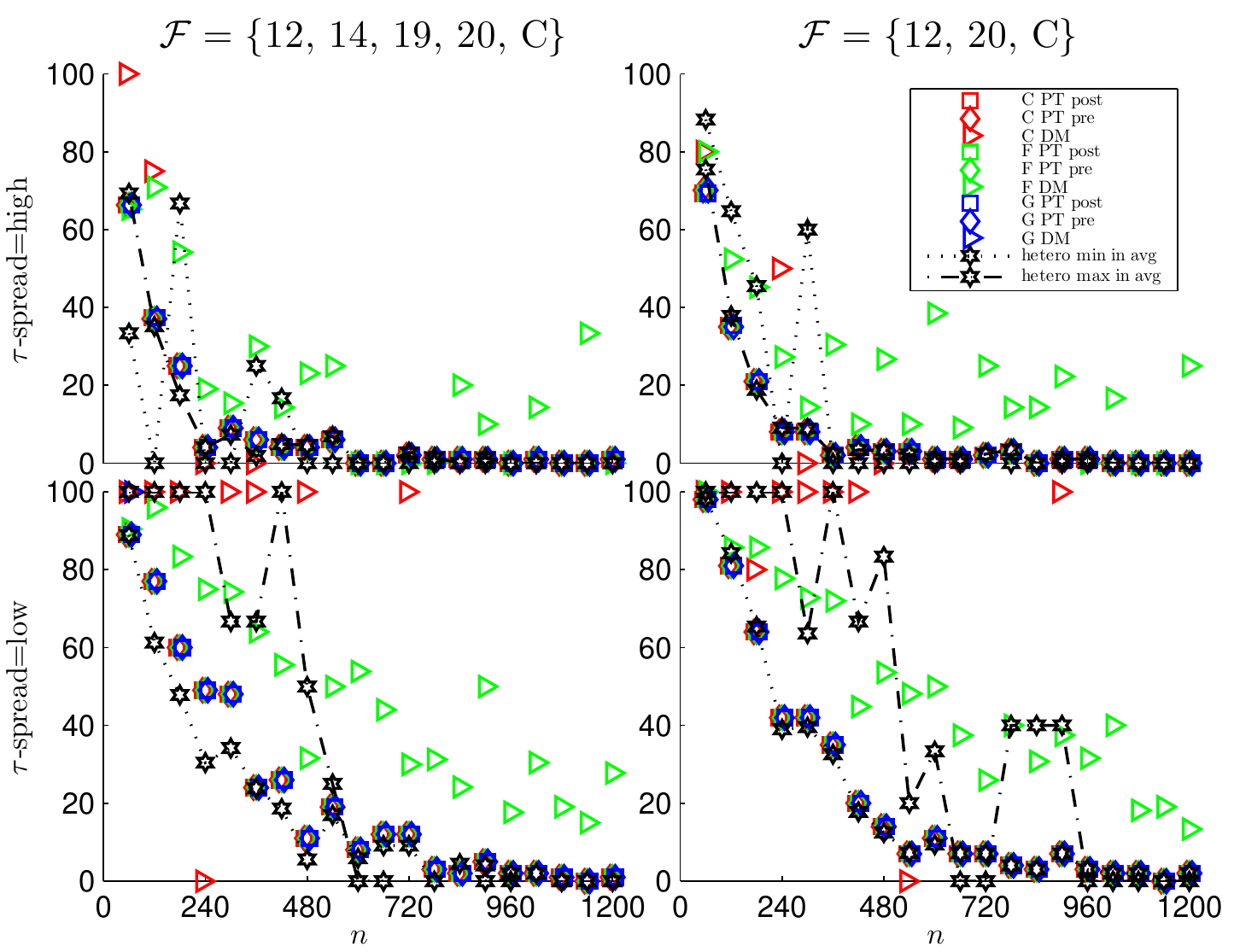}
\caption{The false structure ratio for the Re-est=KTauAvg and \#Forks=unknown homogenous estimators  for the four 15-HAC models. The optimistic ones are at the left-hand and the pessimistic ones at the right-hand.}
\label{fig:false_structure_ratio_homo}
\centering
\end{figure}

\begin{figure}[t]
\centering
\includegraphics[width=0.49\textwidth]{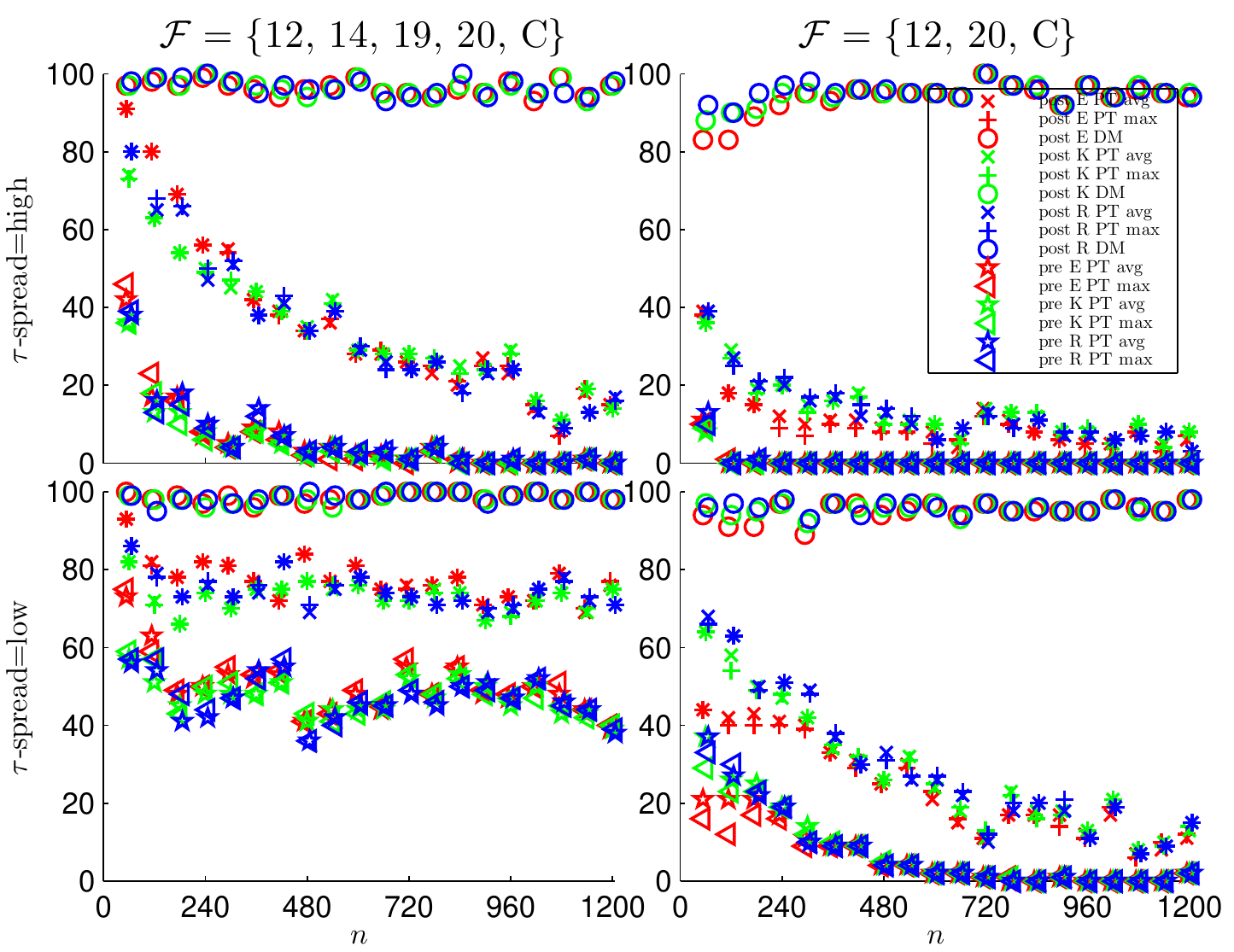}
\includegraphics[width=0.49\textwidth]{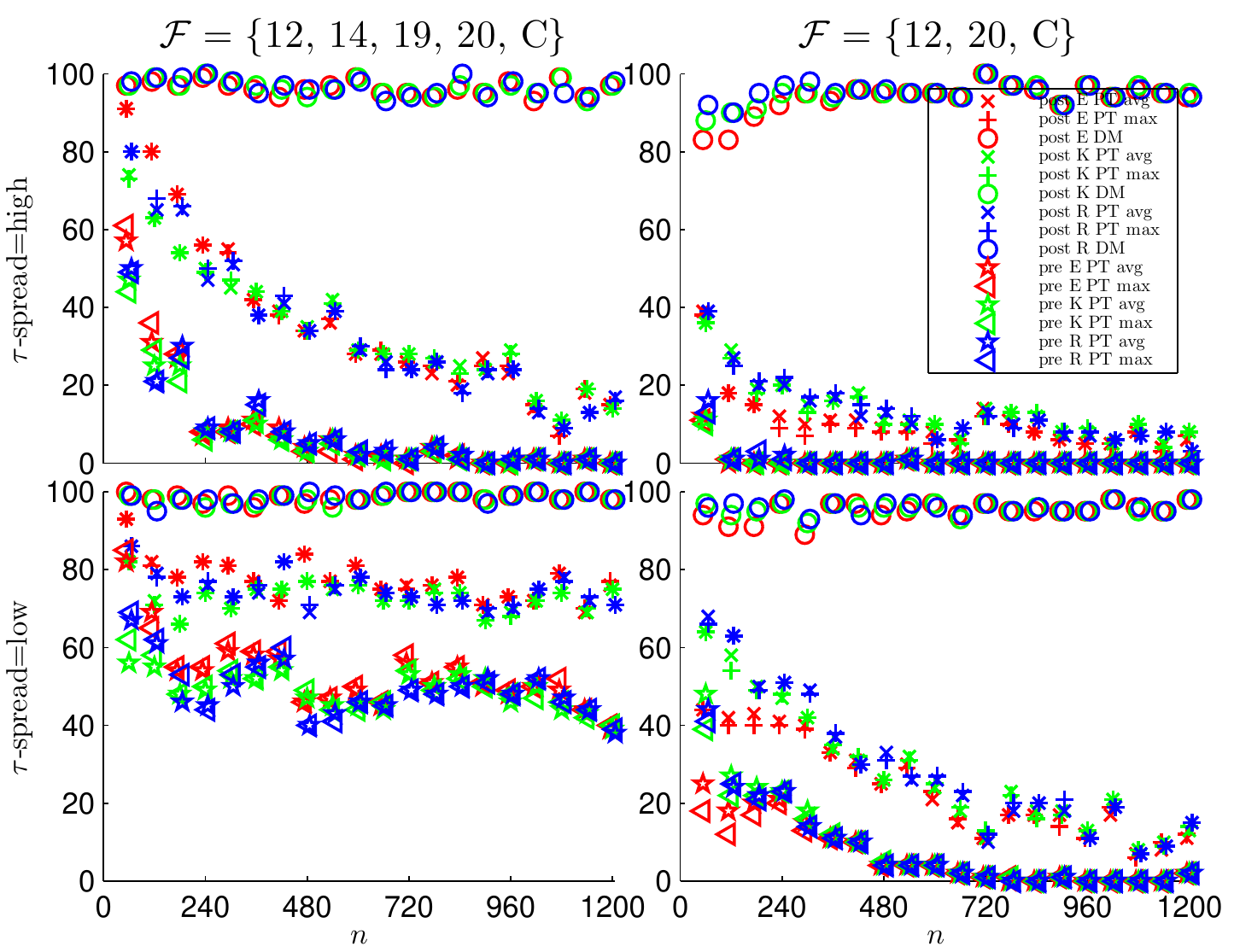}
\includegraphics[width=0.49\textwidth]{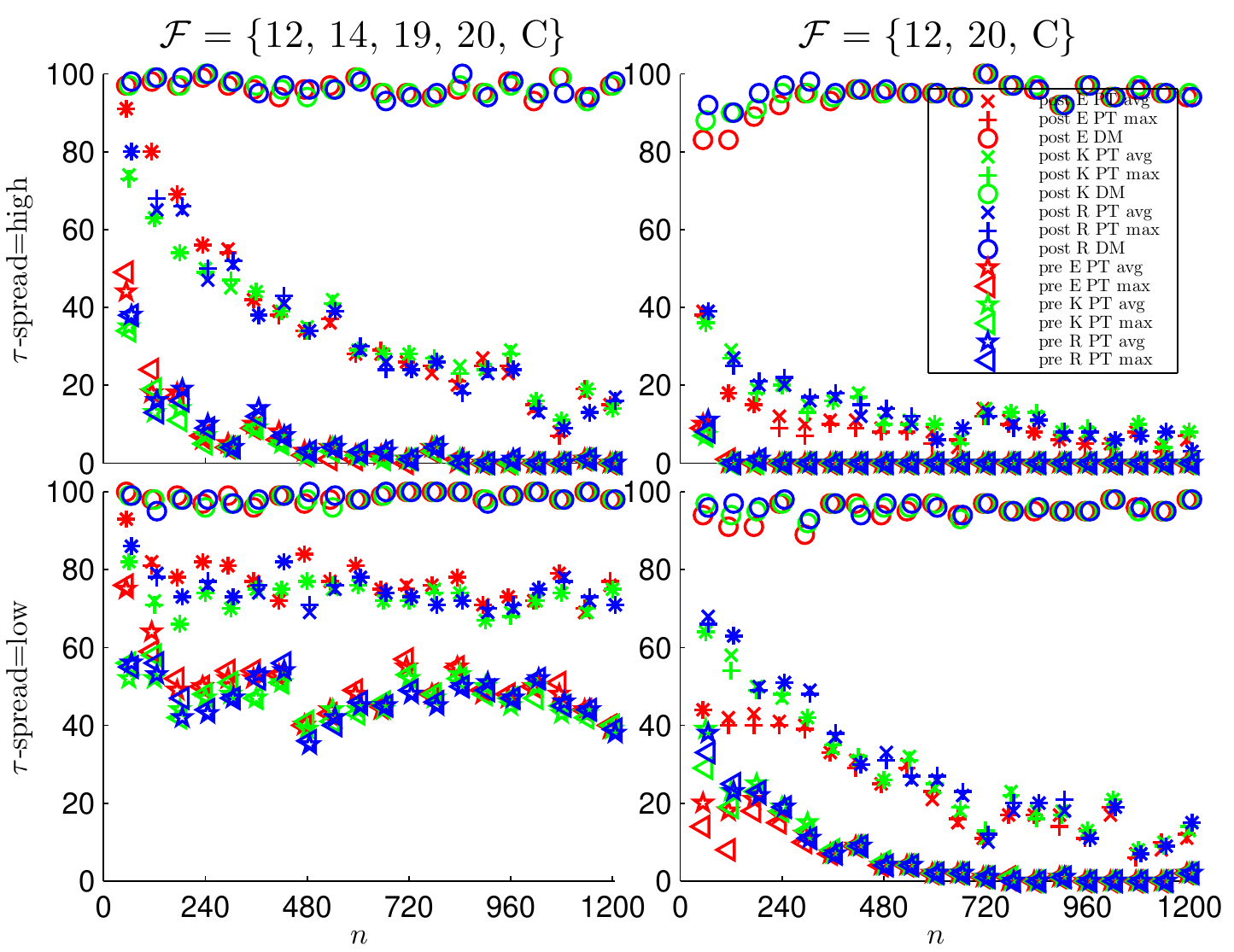}
\includegraphics[width=0.49\textwidth]{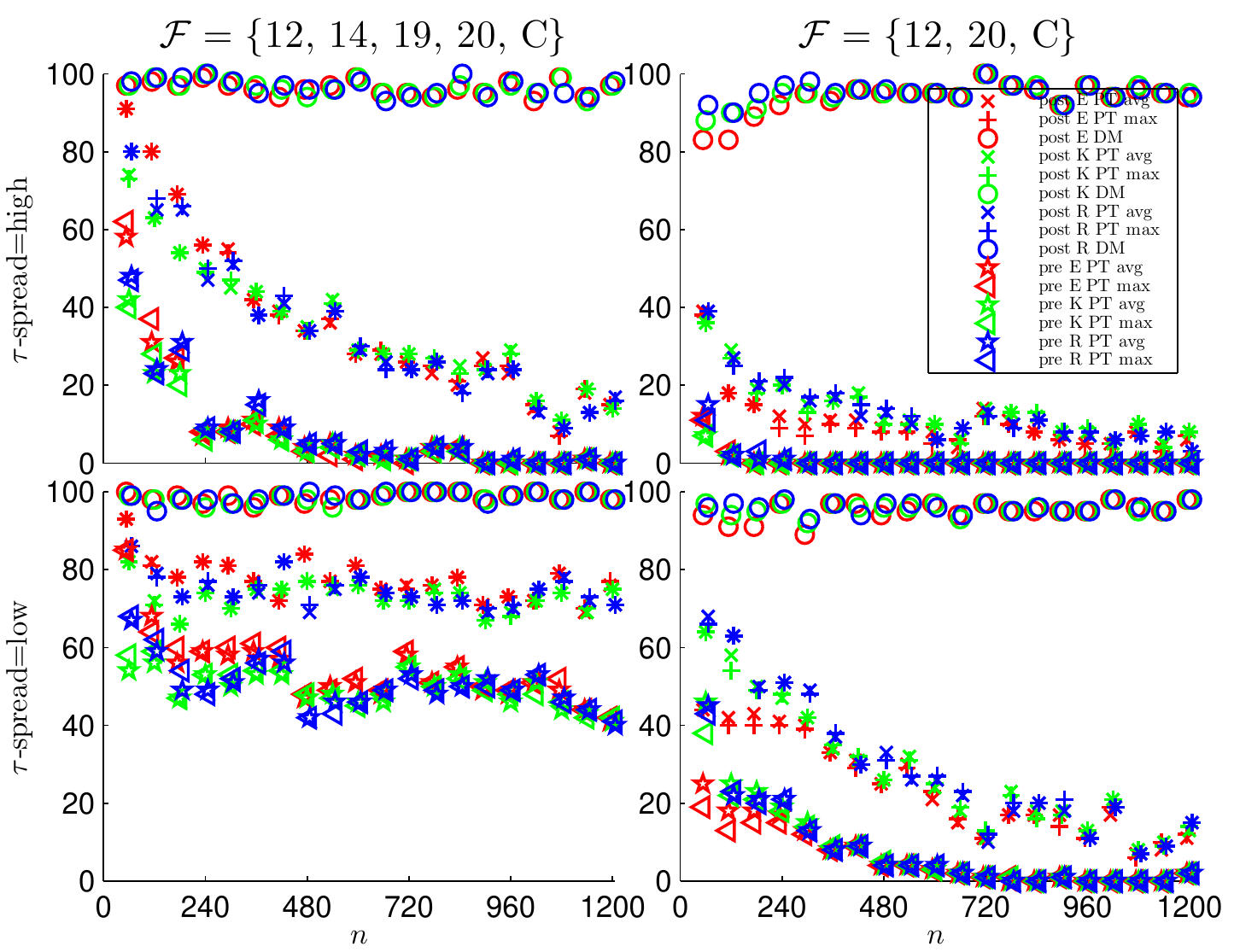}
\caption{Rejection rate for the four 15-HAC models and the pessimistic estimators. The left-hand and right-hand sub-figures correspond to \#Forks=known and to \#Forks=unknown, respectively, and the top and bottom sub-figures correspond to KTauAvg and to TauMin, respectively.}
\label{fig:rejection_rate_pes}
\centering
\end{figure}

The following can be observed: 
\begin{itemize}
	\item For the optimistic estimators, the ratio decreases as $n$ increases. This decrease depends on $\#\FF$. In most of the cases, the lower $\#\FF$, the faster the ratio decreases. Also, this decrease depends on the $\tau$-spread. The ratio for $\tau$-spread=low is mostly larger than for $\tau$-spread=high. Less formally, the estimators are more successful to find the true structure if there are less families and the parameters are far from each other. 
A next observation	is that the Coll=pre estimators clearly outperform the  Coll=post estimators.
Also, the ratio is the same for all Coll=pre estimators. This is due to the fact that all these estimators estimate the structure independently on $S_n$ and $g$, and always 100 out of $N =100$ estimates are obtained (note that the latter statement does not hold for the pessimistic estimators and thus the ratio for the Coll=pre pessimistic estimators can differ).
Then, as expected, the \#Forks=unknown estimators are less successful to estimate the true structure than the \#Forks=known estimators. However, this difference vanishes as $n$ increases, which justifies viability of our collapsing approach introduced in Section \ref{sec:collapsing}. To support this claim, we provide Figure \ref{fig:nForks_false_structure_ratio}. There, all information from Figure \ref{fig:false_struc_ratio_opt} is aggregated along each plot, and each plot shows the average of the 15 corresponding evaluations, e.g., the graph denoted \#Forks=known \& Re-est=KTauAvg corresponds to the average (for each $n$) of all 15 evaluation depicted in the corresponding plot in Figure \ref{fig:false_struc_ratio_opt}. The underlying model can be identified in the same way as in Figure \ref{fig:false_struc_ratio_opt}.	
Also, observe that for  \#Forks=unknown, the KTauAvg estimators mostly show lower average ratio than the TauMin estimators.
Finally, observe that the benchmark homogeneous evaluation coincide with the Coll=pre evaluations. 
	\item For the pessimistic estimators, most of the claims for the optimistic estimators can be adopted. What differs is that the evaluations for these estimators are more similar than for the optimistic estimators, e.g., observe that the Coll=post estimators are much closer (lower in the ratio) to the Coll=pre estimators than their optimistic counterparts for a given $n$. This observation is also visible in Figure \ref{fig:nForks_false_structure_ratio}, e.g, compare the ratio for \#Forks=known and a low $n$. One can also observe that, frequently, DM-based estimators are substantially higher in the ratio than the PT-based estimators. This is mostly given by the fact that for this attitude, very few (out of $N=100$) estimates are returned, which is illustrated in Figure \ref{fig:rejection_rate_pes}. If some of these few ones do not have the true structure, this can substantially affect the ratio. Another observation following from Figure \ref{fig:rejection_rate_pes} is that there are three clearly distinct groups of estimators, namely 1) the Coll=pre group, 2) the Coll=post \& Alg=PT group and 3) the Coll=post \& Alg=DM group, where the first one led the the least rejections, the second led to more rejections, and the third one even does not decrease in the rejection rate below 80\% no matter how high $n$ is. This observation is very similar to what can be observed in the small experiment described in Section \ref{sec:homo_HAC_estim}. 
\end{itemize}

% obtained by
% plotevaluation_nforks(fileName, 'trueStrucRatio', true, 19:24, true, 'none', false);
% plotevaluation_nforks(fileName, 'trueStrucRatio', false, 19:24, true, 'none', false);

Considering again the plots in Figures \ref{fig:false_struc_ratio_opt} and \ref{fig:false_struc_ratio_pes}, one can see that some groups of the heterogeneous estimators perform better more frequently than other groups. This is evident particularly for the optimistic estimators.
To quantify this observation, we provide a \emph{regression tree} \cite{CART1984}, which shows the features of an estimator that substantially affect the false structure ratio and how they affect this ratio.
The regression tree is trained on a data table that contains 6 columns, where the first 5 columns are the \emph{inputs} \cite{CART1984} of the tree. They describe an estimator by its features, which are Coll $\in$ \{pre, post\} determining if pre- or post-collapsing is used, respectively, Re-est $\in$ \{KTauAvg, TauMin\}, Att(itude) $\in$ \{opt(imistic), pes(simistic)\}, Sn $\in$ \{E, K, R\} determining which one of the statistics $S_n^{(E)}$, $S_n^{(K)}$ and $S_n^{(R)}$ is used as Input 6, and Alg(orithm) $\in$ \{PT(avg), PT(max), DM\} determining the algorithm and $g$ (Input 5) if the algorithm is PT. We omit the feature \#Forks as it is unrealistic to choose an estimator such that \#Forks=known. Also, given a data sample, it is unrealistic to arbitrarily choose $d$, $n$, $\tau$-spread or $\#\FF$. Hence the features used for building the tree are those that one can arbitrarily choose when looking for an appropriate estimator out of the considered ones.
The last column is the \emph{output} of the tree and contains, given an estimator, its rank among the 60 heterogeneous \#Forks=unknown estimators according to a selected evaluation criterion for each HAC model and $i \in \{1, ..., 20\}$, i.e., the table contains $60 \times 12 \times 20 = 14~400$ rows, which are thus based on the realistic half (\#Forks=unknown) of all ($d \in \{5, 10, 15\}$) generated heterogeneous estimates.
Also, in other words, given an estimator, such a tree predicts the rank of the estimator among the 60 heterogeneous estimators according to a given evaluation criterion. Finally, note that all regression trees depicted below are generated using the \texttt{fitrtree} function implemented in MATLAB and have been pruned to the the best level \cite{CART1984} using the 10-fold cross-validation. 

The regression tree obtained for the false structure ratio is depicted in Figure \ref{fig:false_struc_ratio_tree}. 
Observe that it confirms the observations mentioned above, i.e., the best (lowest in rank) estimators are the ones with Coll=pre and we can moreover infer from that the feature Coll is for the false structure ratio the most affecting feature out of the considered ones.
If Coll=post, the tree predicts for the Att=pes estimators a rank better than for the Att=opt estimators.
Finally, if Coll=post \& Att=opt, the tree predicts a better rank for the Alg=PT estimators, no matter if $g$=avg or $g$=max.
Also, as a rank predicted by the tree does neither depend on the features Re-est, Sn nor $g$, we can infer that these features do not affect the rank as much as the remaining ones. 
%\begin{figure}[hb]
%\centering
%\includegraphics[width=1\textwidth]{tree-trueStrucRatio-eps-converted-to.pdf}
%\caption{The regression tree for the false structure ratio ranks based on the five features of the heterogeneous estimators.}
%\label{fig:false_struc_ratio_tree}
%\centering
%\end{figure}

Finally, to address the robustness of the estimation approaches against misspecification of the underlying families, we show the ratio for the homogeneous estimators, but, for simplicity, only for the Re-est=KTauAvg and \#Forks=unknown estimators, see Figure \ref{fig:false_structure_ratio_homo}. Note that also the best and the worst heterogeneous evaluation (chosen analogously to homo min in avg) is shown, denoted by \textbf{hetero min in avg} and \textbf{hetero max in avg}, respectively.
%\begin{figure}[htb]
%\centering
%\includegraphics[width=0.49\textwidth]{d15-trueStrucRatio-nForksUnknownKTauAvg-isOpt1-model9to12-homo-eps-converted-to.pdf}
%\includegraphics[width=0.49\textwidth]{d15-trueStrucRatio-nForksUnknownKTauAvg-isOpt0-model9to12-homo-eps-converted-to.pdf}
%\caption{The false structure ratio for the Re-est=KTauAvg and \#Forks=unknown homogenous estimators  for the four 15-HAC models. The optimistic ones are at the left-hand and the pessimistic ones at the right-hand.}
%\label{fig:false_structure_ratio_homo}
%\centering
%\end{figure}
The observations are quite clear. The Alg=PT estimators are robust in the structure estimation against misspecification of the underlying family(ies), whereas the Alg=ML estimators are not. E.g., 
observe the similarity of the evaluations for the homogeneous Alg=PT estimators and hetero min in average, and  also that the Alg=ML estimators show worse results then the worst heterogeneous estimator.  
Considering the Alg=ML pessimistic estimators, the estimators assuming the Gumbel or the Clayton family are very prone to rejection, and for higher $n$, no structure was returned by such an estimator -- note that if none of the $N = 100$ estimates for a given $n$ and estimator was returned, no evaluation mark is shown. Finally, observe that the evaluation of the Coll=pre and Coll=post estimators look identical, however note that generally they are not the same.

%\clearpage

\subsubsection{Families}
\label{sec:exps_fams}
\begin{figure}[t]
\centering
\includegraphics[width=0.49\textwidth]{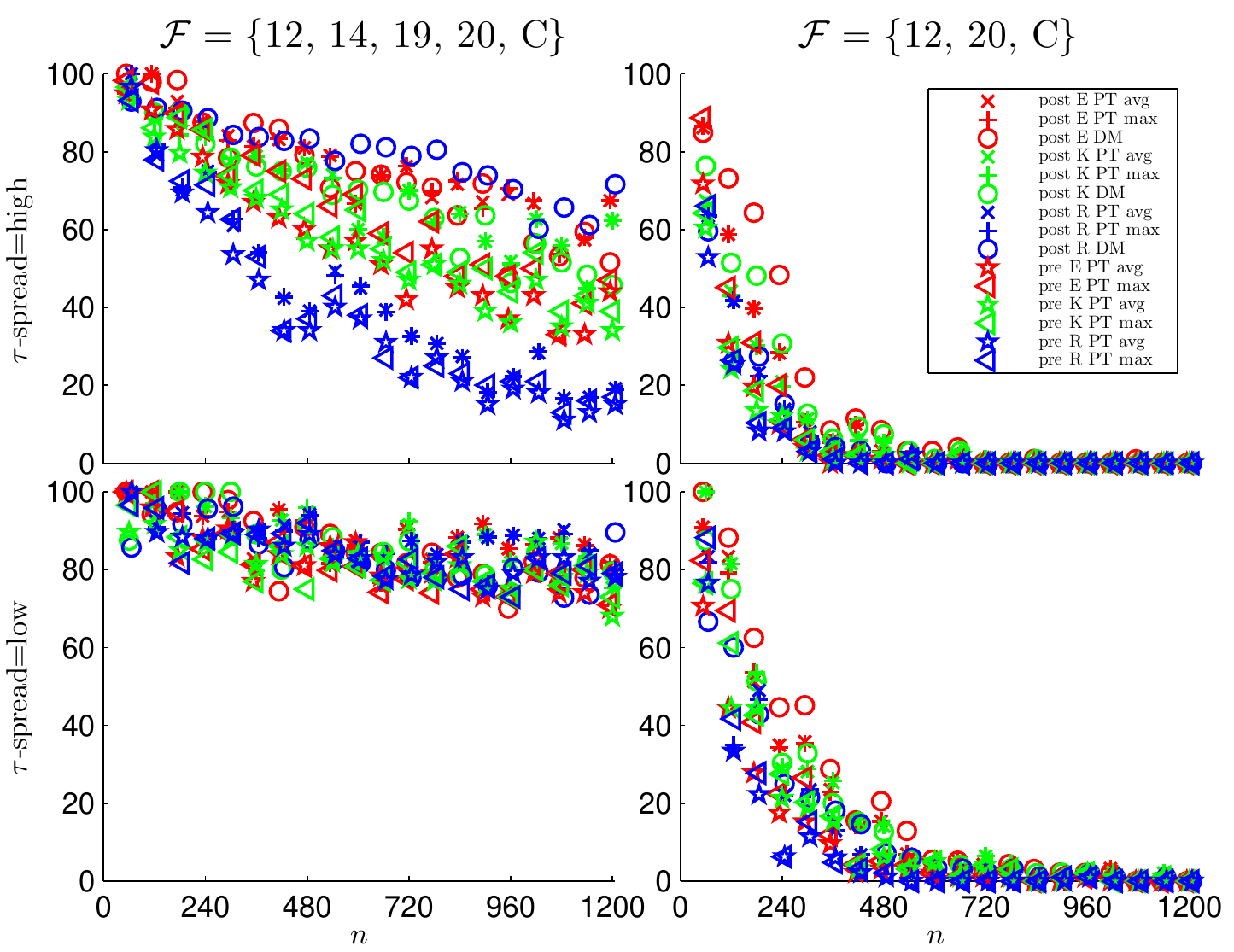}
\includegraphics[width=0.49\textwidth]{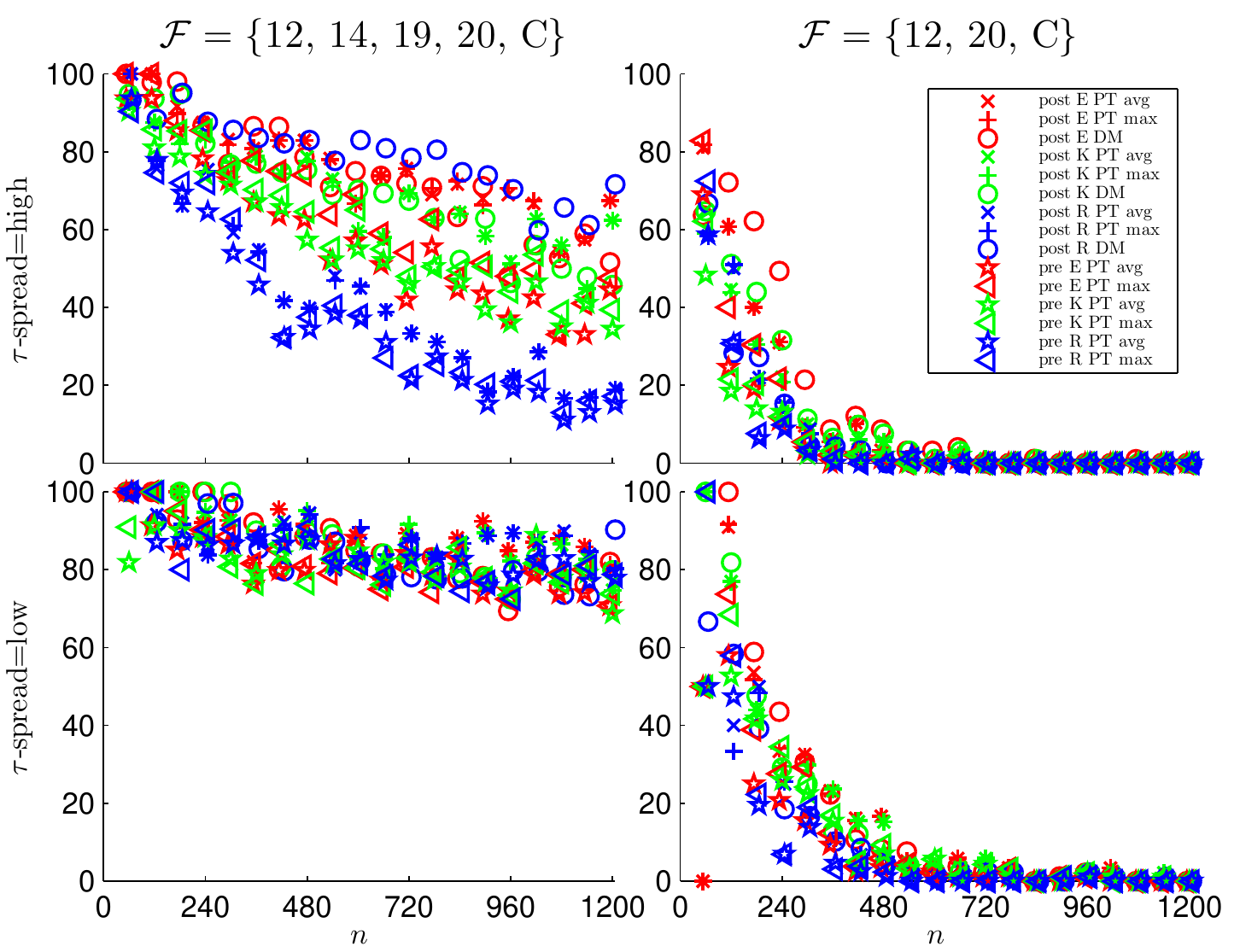}
\includegraphics[width=0.49\textwidth]{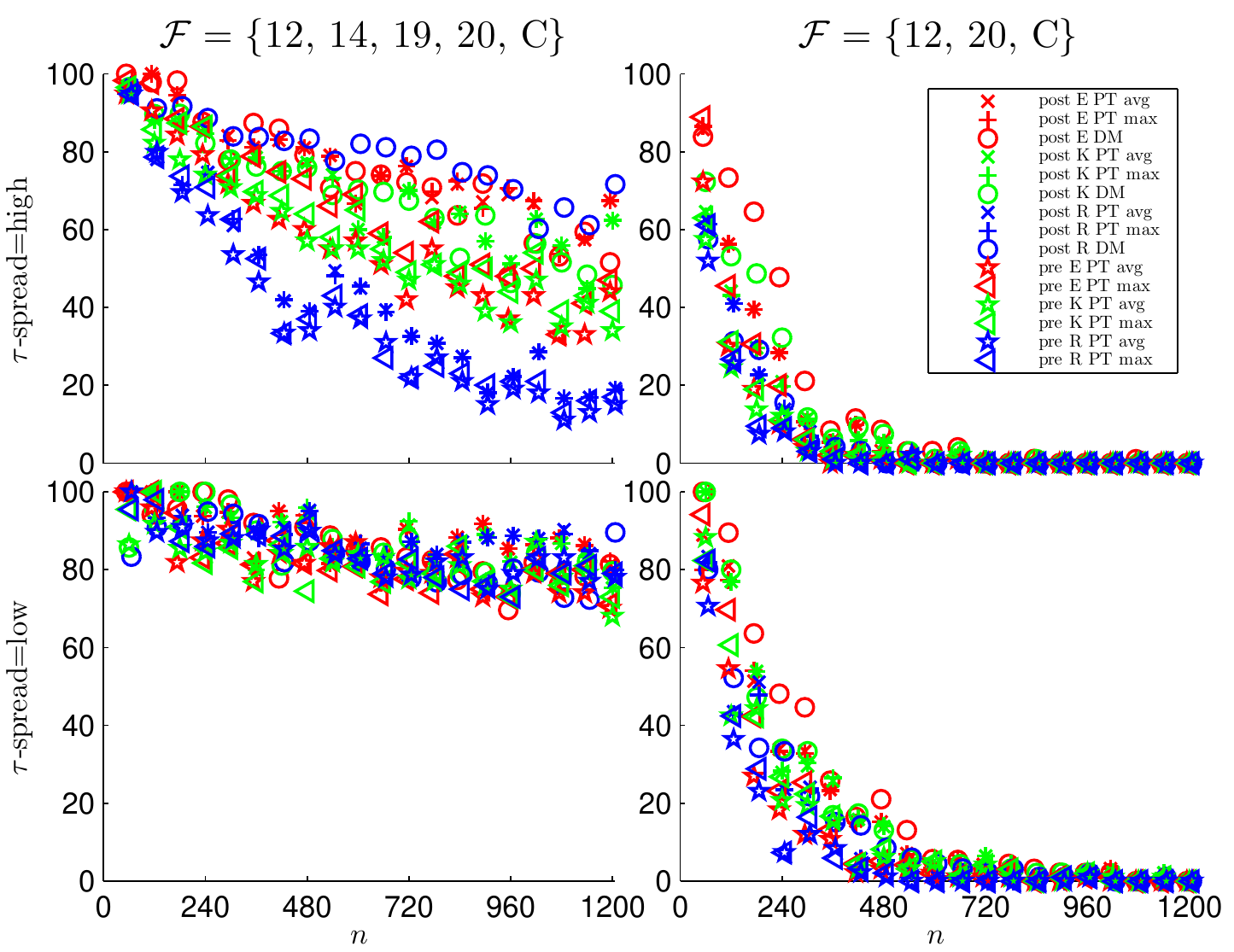}
\includegraphics[width=0.49\textwidth]{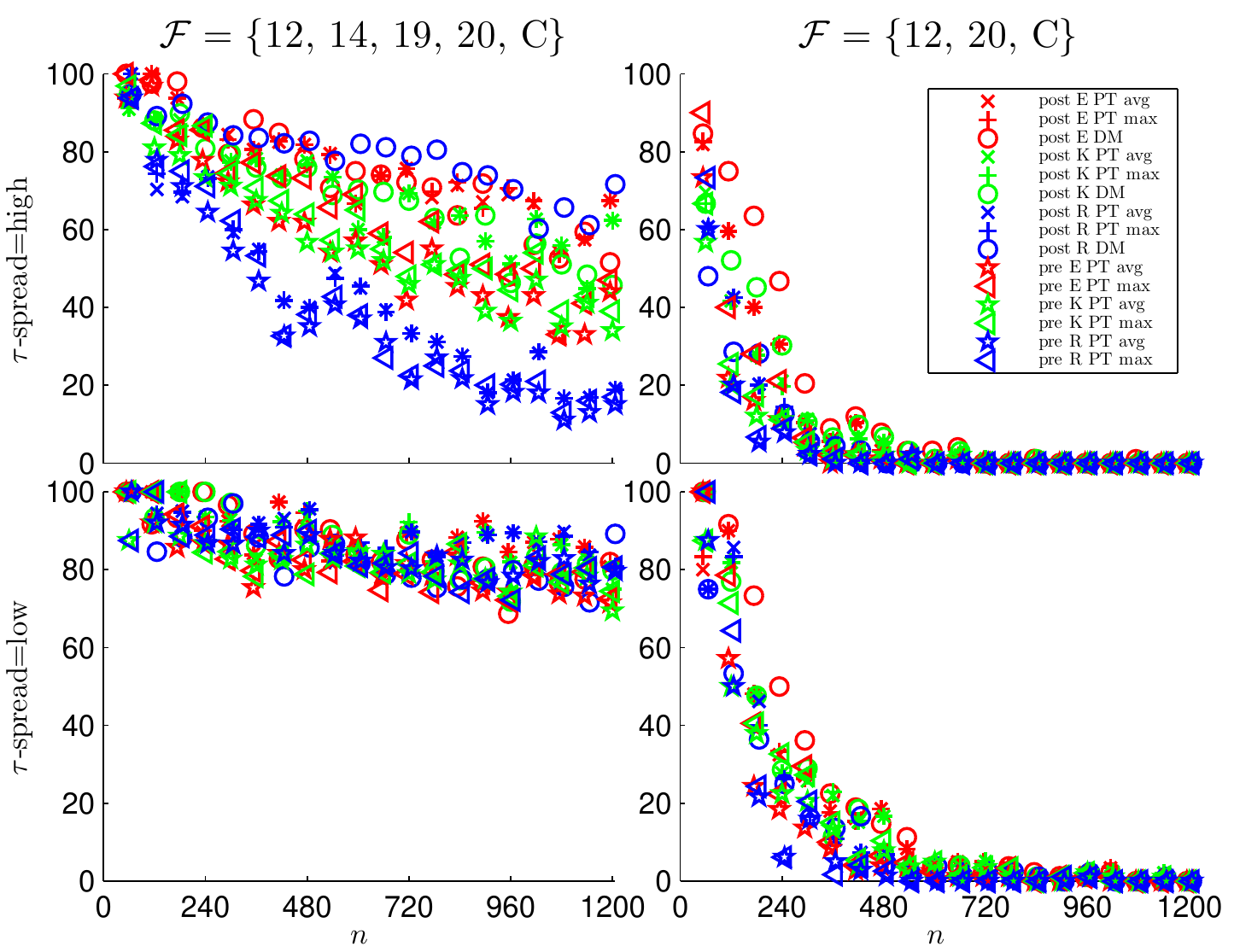}
\caption{The false families ratio (in \%) for the four 15-HAC models and the optimistic estimators. The left-hand and right-hand sub-figures correspond to \#Forks=known and to \#Forks=unknown, respectively, and the top and bottom sub-figures correspond to KTauAvg and to TauMin, respectively.}
\label{fig:false_fam_ratio_opt}
\centering
\end{figure}

\begin{figure}[htb]
\centering
\includegraphics[width=0.49\textwidth]{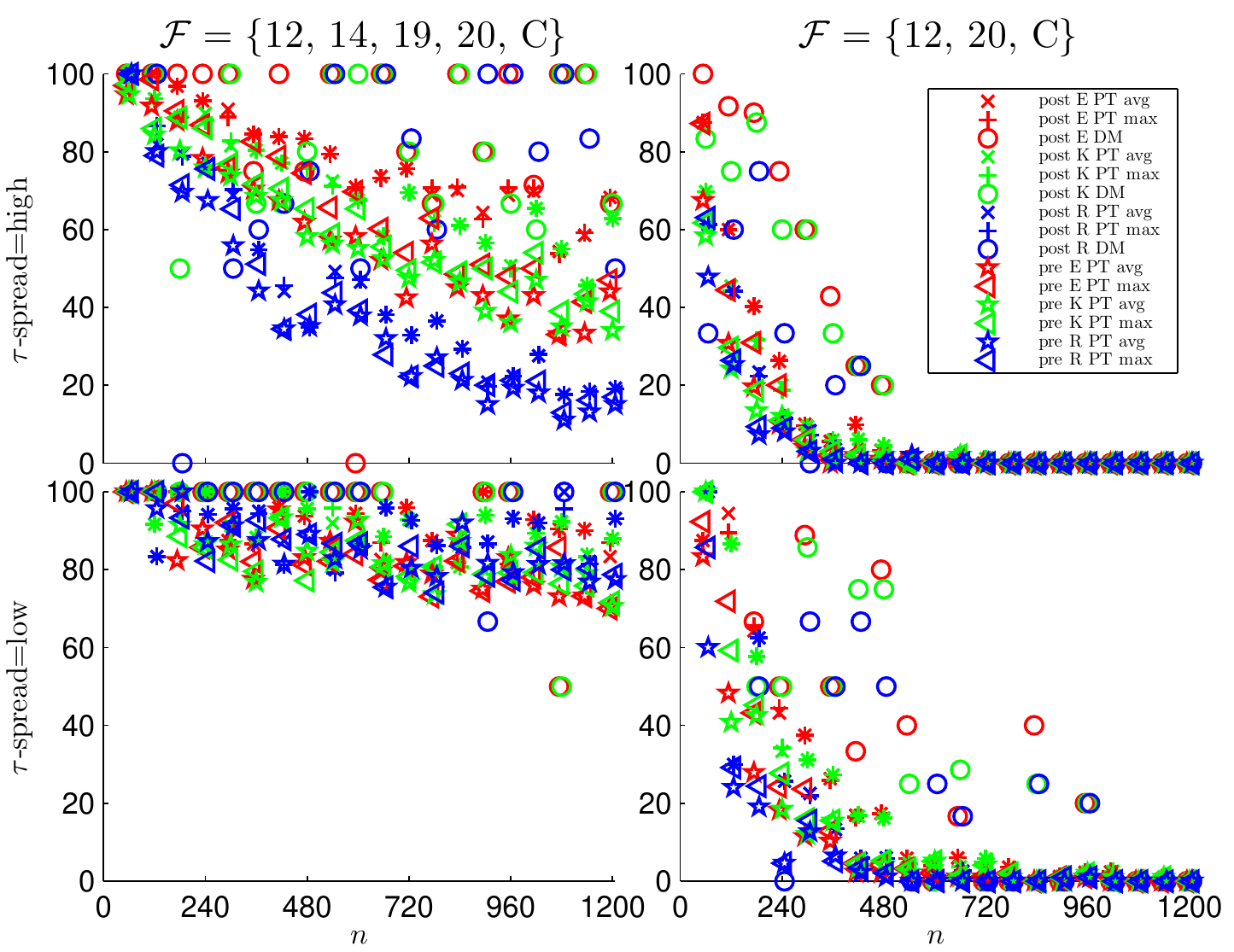}
\includegraphics[width=0.49\textwidth]{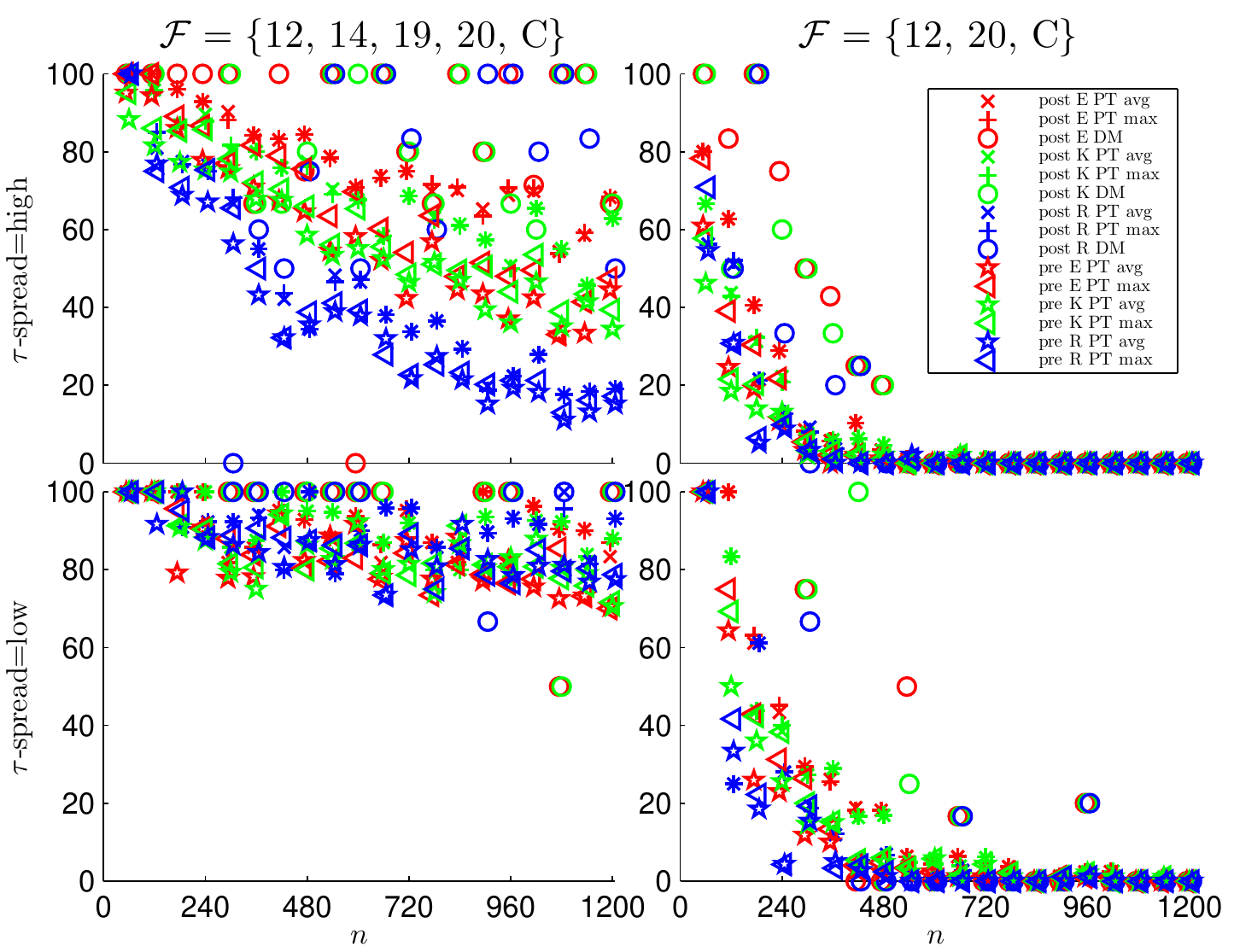}
\includegraphics[width=0.49\textwidth]{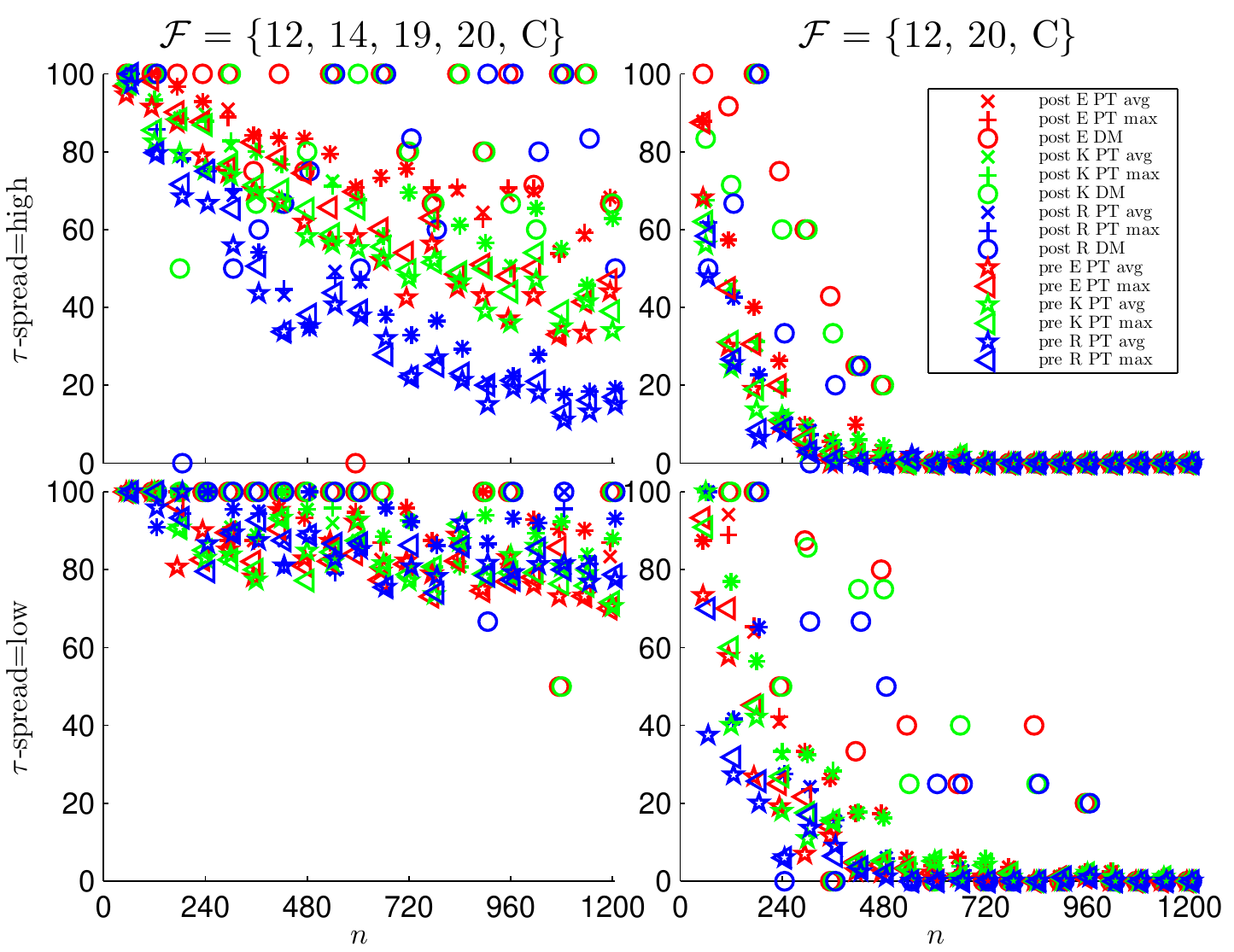}
\includegraphics[width=0.49\textwidth]{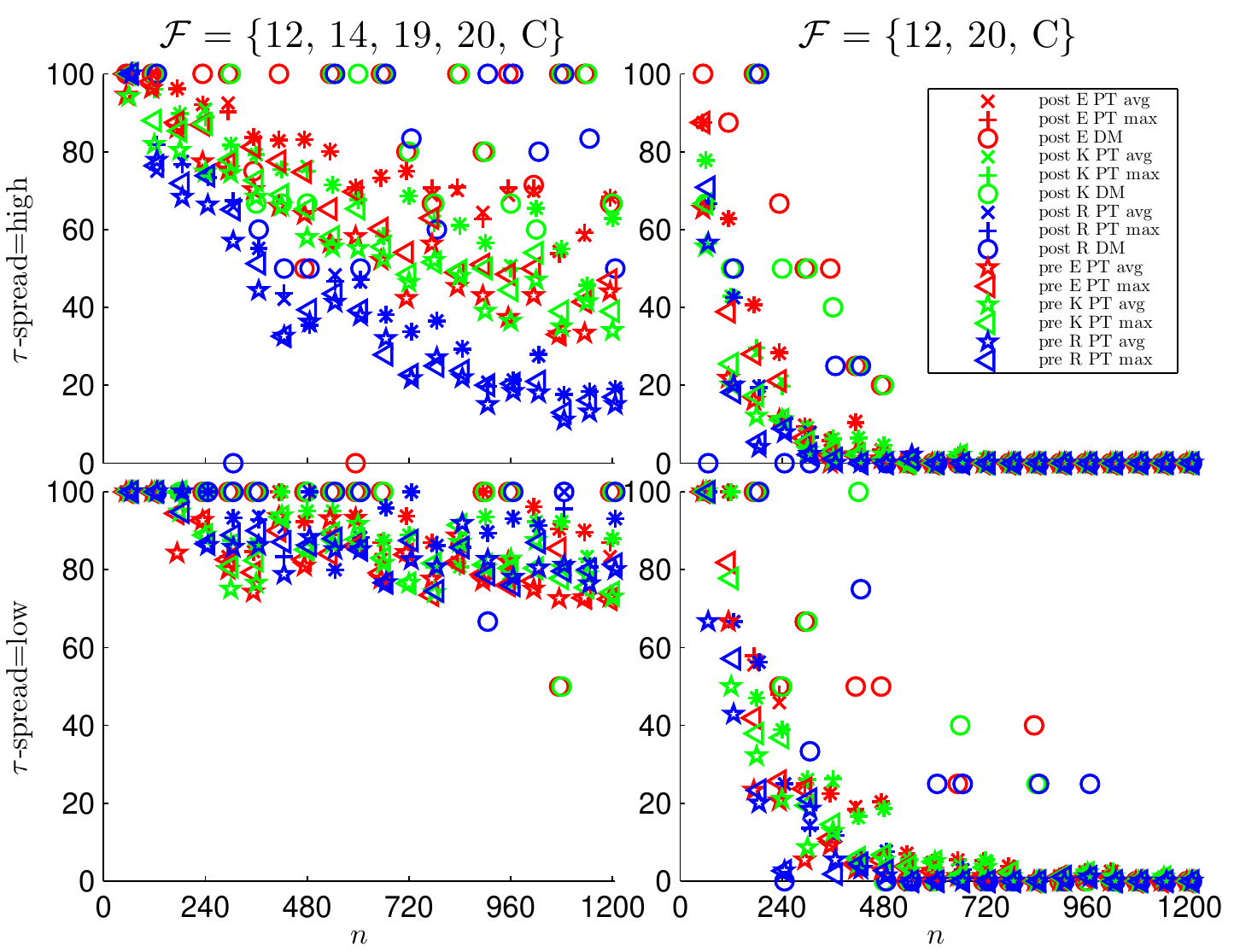}
\caption{The false families ratio (in \%) for the four 15-HAC models and the pessimistic estimators. The left-hand and right-hand sub-figures correspond to \#Forks=known and to \#Forks=unknown, respectively, and the top and bottom sub-figures correspond to KTauAvg and to TauMin, respectively.}
\label{fig:false_fam_ratio_pes}
\centering
\end{figure}

The performance of the estimators for estimating the true families is evaluated by the false families ratio. This evaluation for $d = 15$ is depicted in Figures \ref{fig:false_fam_ratio_opt} and \ref{fig:false_fam_ratio_pes}. 

\begin{figure}[t]
\centering
\includegraphics[width=0.49\textwidth]{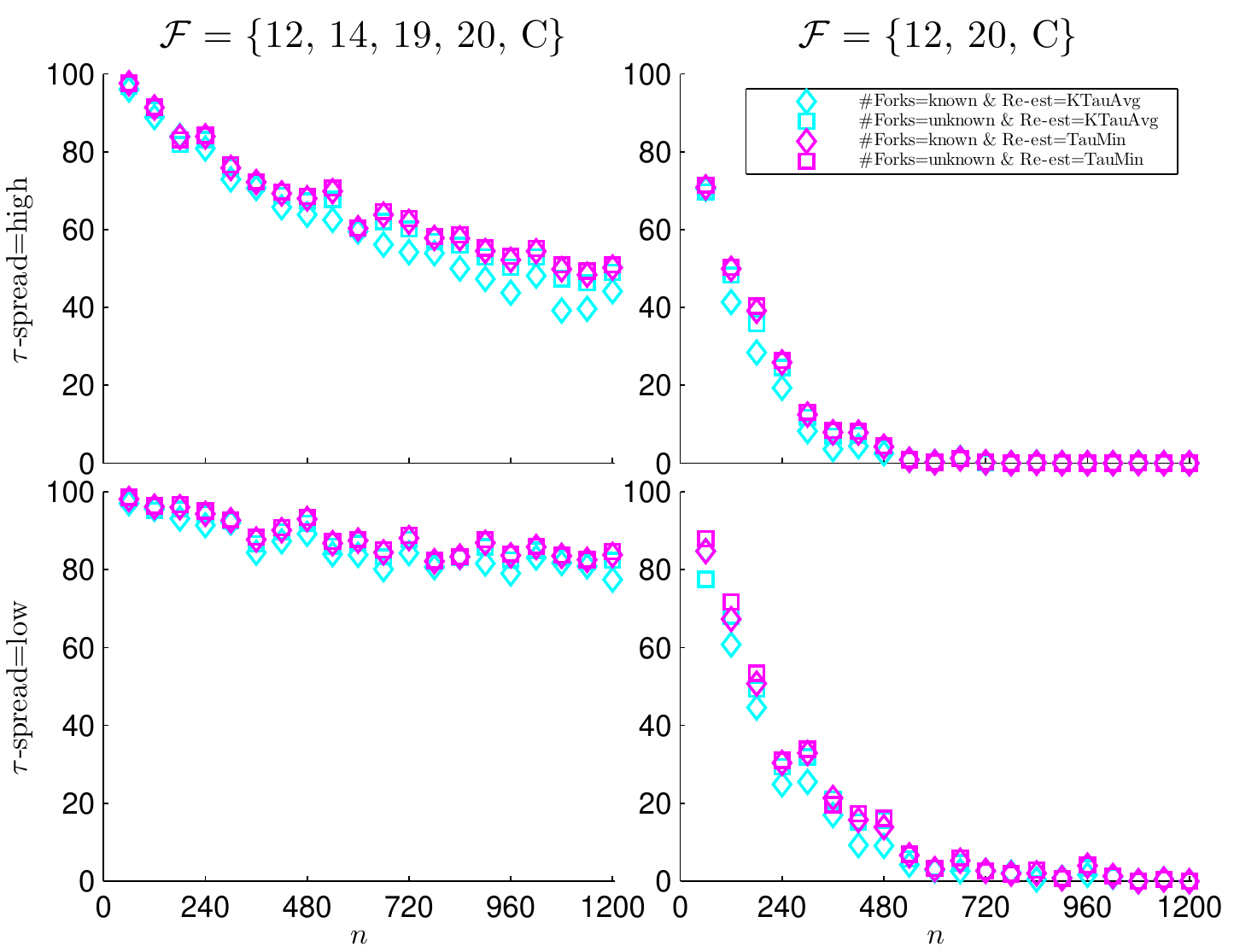}
\includegraphics[width=0.49\textwidth]{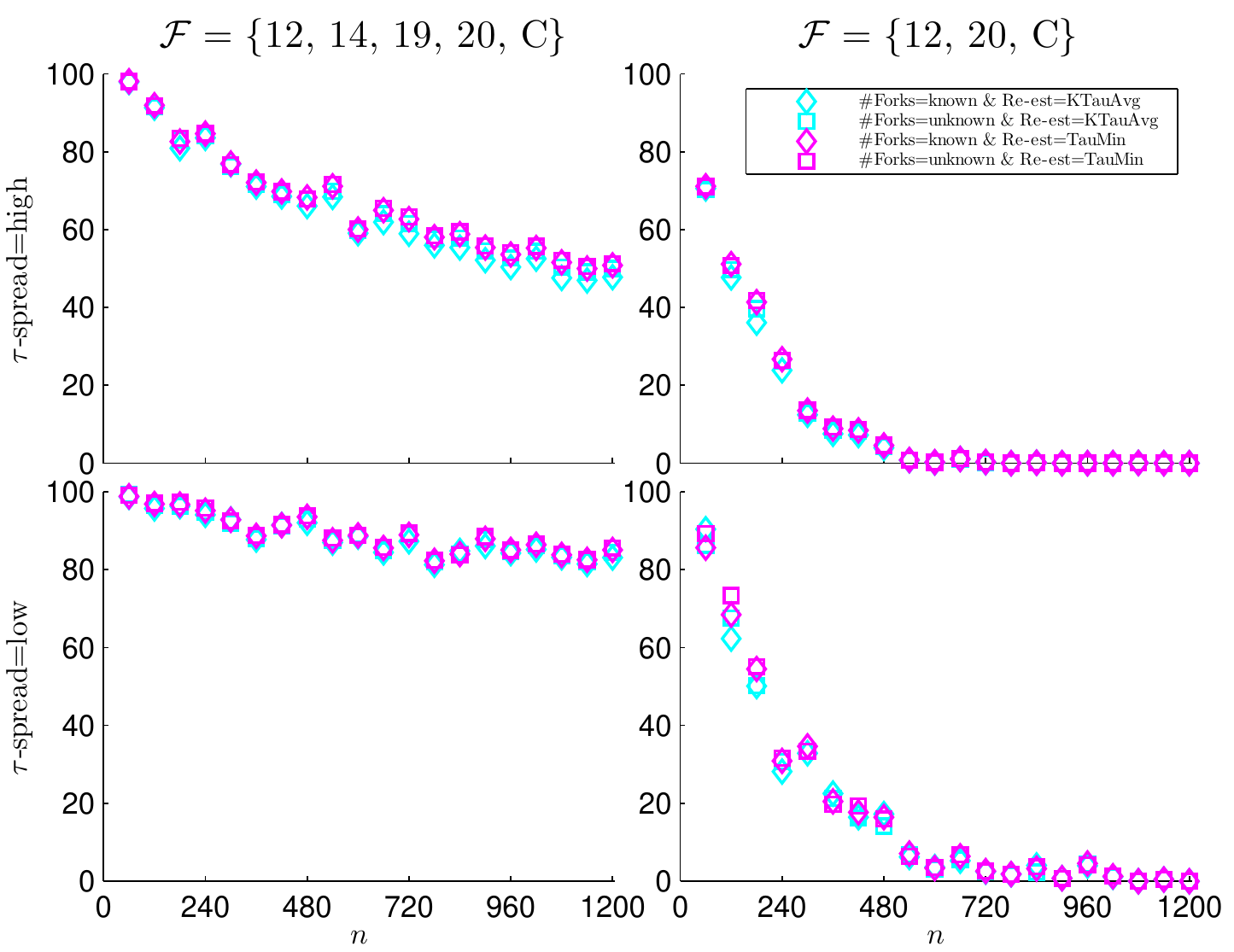}
\caption{Aggregated (by the average) false families ratio computed for the evaluations depicted in Figure \ref{fig:false_fam_ratio_opt} - the optimistic attitude, here depicted at the left-hand - and Figure \ref{fig:false_fam_ratio_pes} - the pessimistic attitude,  here depicted at the right hand.}
\label{fig:nForks_false_families_ratio}
\includegraphics[width=1\textwidth]{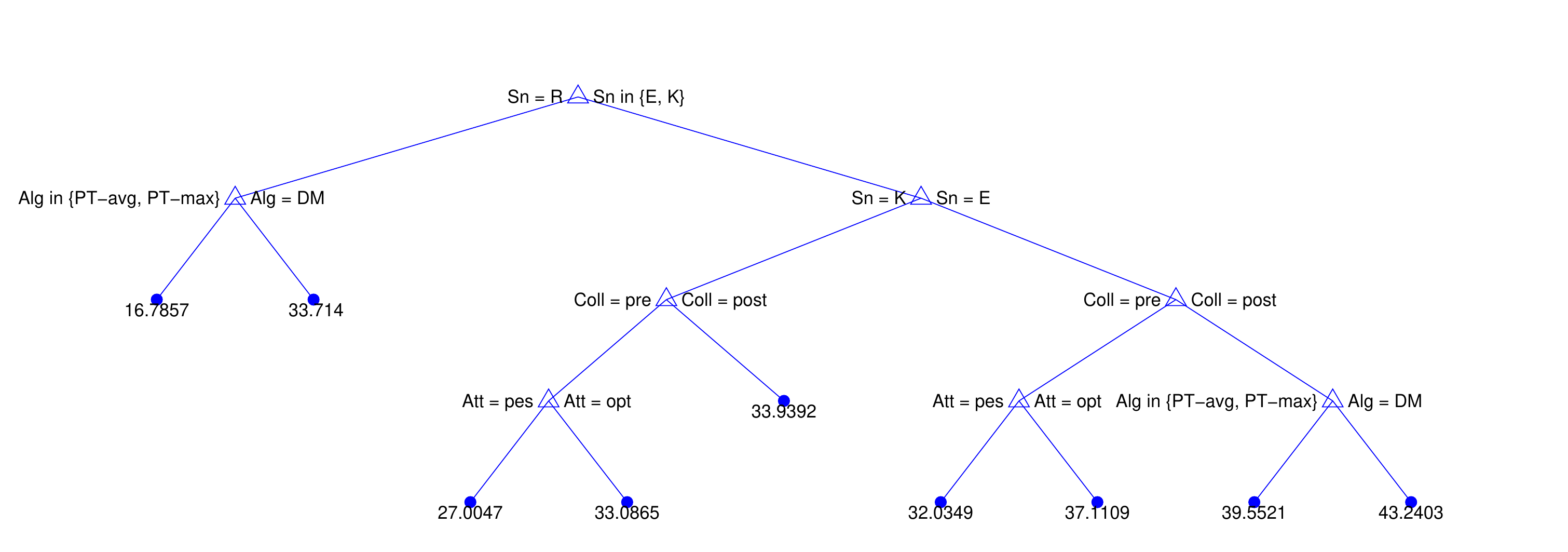}
\caption{The regression tree for the false families ratio ranks based on the five features of the heterogeneous estimators.}
\label{fig:false_fam_ratio_tree}
\centering
\end{figure}

The following can be observed: 
\begin{itemize}
	\item For the optimistic estimators, the ratio decreases as $n$ increases. This decrease strongly depends on $\#\FF$, e.g., compare each two plots corresponding to $\tau$-spread=low, i.e., removing two families out of $\FF$ substantially influences the ability of the estimators to estimate the true families. Dependence of the ratio on $\tau$-spread is also evident,  e.g., compare each two plots corresponding to $\#\FF = 5$.
Considering the Re-est and \#Forks features, differences are better visible from Figure \ref{fig:nForks_false_families_ratio}. 
This figure shows for the \#Forks=unknown \& Re-est=KTauAvg (realistic) estimators better results than for the \#Forks=known \& Re-est=TauMin (unrealistic) estimators, which is an interesting observation that suggests dominance of the KTauAvg approach over the TauMin approach under the optimistic attitude. 
Also, to get the ratio close to 0, the estimators need much more data ($n$) for $\#\FF=5$ then for $\#\FF=3$, however, this also depends on Sn, see again Figure \ref{fig:false_fam_ratio_opt}. There, for Alg=PT, we observe that the Sn=R estimators frequently perform better than the Sn=K estimators, and the latter ones perform better than the Sn=E estimators. We also observe that the Alg=PT estimators perform better than the Alg=DM estimators.
	\item For the pessimistic estimators, a majority of the observations for the optimistic estimators can be adopted.
	The most visible difference is that the Alg=DM estimators perform worse then in the optimistic case, which can be explained by looking at Figure \ref{fig:rejection_rate_pes}, i.e., as these estimators show the rejection rate close to 100, it frequently happens that none of the few ones that have not been rejected has the true families and thus the ratio is equal to 100\%. In Figure \ref{fig:nForks_false_families_ratio}, we observe more similar results among the considered averages than for the optimistic attitude, e.g, compare the plots for $\#\FF=3$.
\end{itemize}

% obtained by
% plotevaluation_nforks(fileName, 'trueFamRatio', true, 9:12, true, 'none', false);
% plotevaluation_nforks(fileName, 'trueFamRatio', false, 9:12, true, 'none', false);

We now provide a regression tree created analogously to the one provided in Section \ref{sec:exps_struc}, with the only change that the false families ratio is used instead of the false structure ratio. The tree is depicted in Figure \ref{fig:false_fam_ratio_tree}.
%\begin{figure}[htb]
%\centering
%\includegraphics[width=1\textwidth]{tree-trueFamRatio-eps-converted-to.pdf}
%\caption{The regression tree for the false families ratio ranks based on the five features of the heterogeneous estimators.}
%\label{fig:false_fam_ratio_tree}
%\centering
%\end{figure}
From the tree, we can infer that the most influencing feature on the ratio is Sn, where R shows results better than for E and K, and also K show better results than E. Note that this is seemingly a discrepancy between our results and the results presented in \cite{Gen09}, where $S_n^{(E)}$ performed in general better than $S_n^{(K)}$ and $S_n^{(R)}$. A closer look, however, reveals that for their results concerning only Archimedean copulas, $S_n^{(K)}$ and $S_n^{(R)}$ performed well compared with $S_n^{(E)}$.
Another observation based on the tree is that if Sn=R, the tree predicts the best rank for the Alg=PT estimators. 
Also note that as the tree does not contain Re-est nor dependency on $g$, we can infer that these features do not affect the rank as much as the remaining ones.

%\clearpage

\subsubsection{Parameters}
\label{sec:exps_params}
\begin{figure}[htb]
\centering
\includegraphics[width=0.49\textwidth]{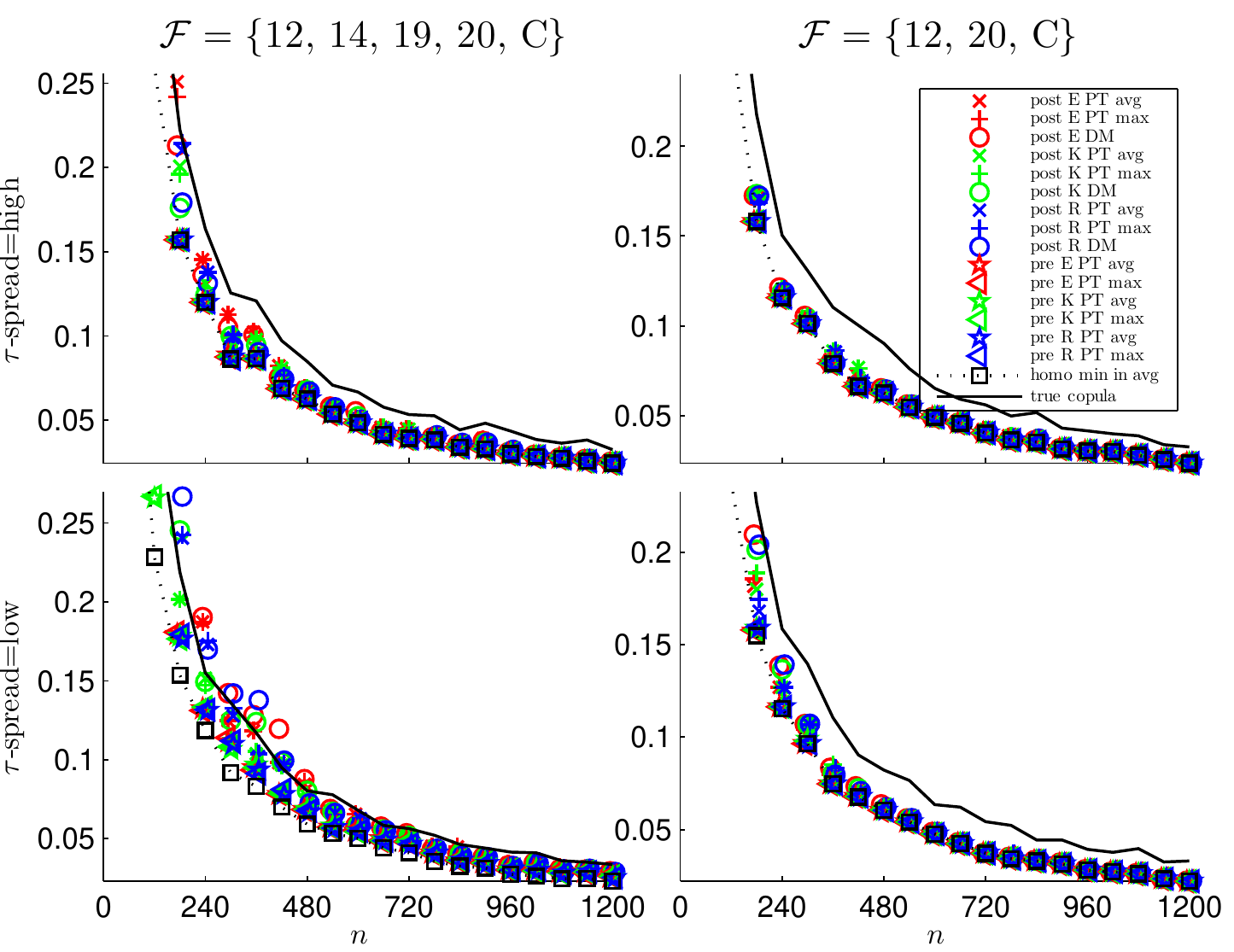}
\includegraphics[width=0.49\textwidth]{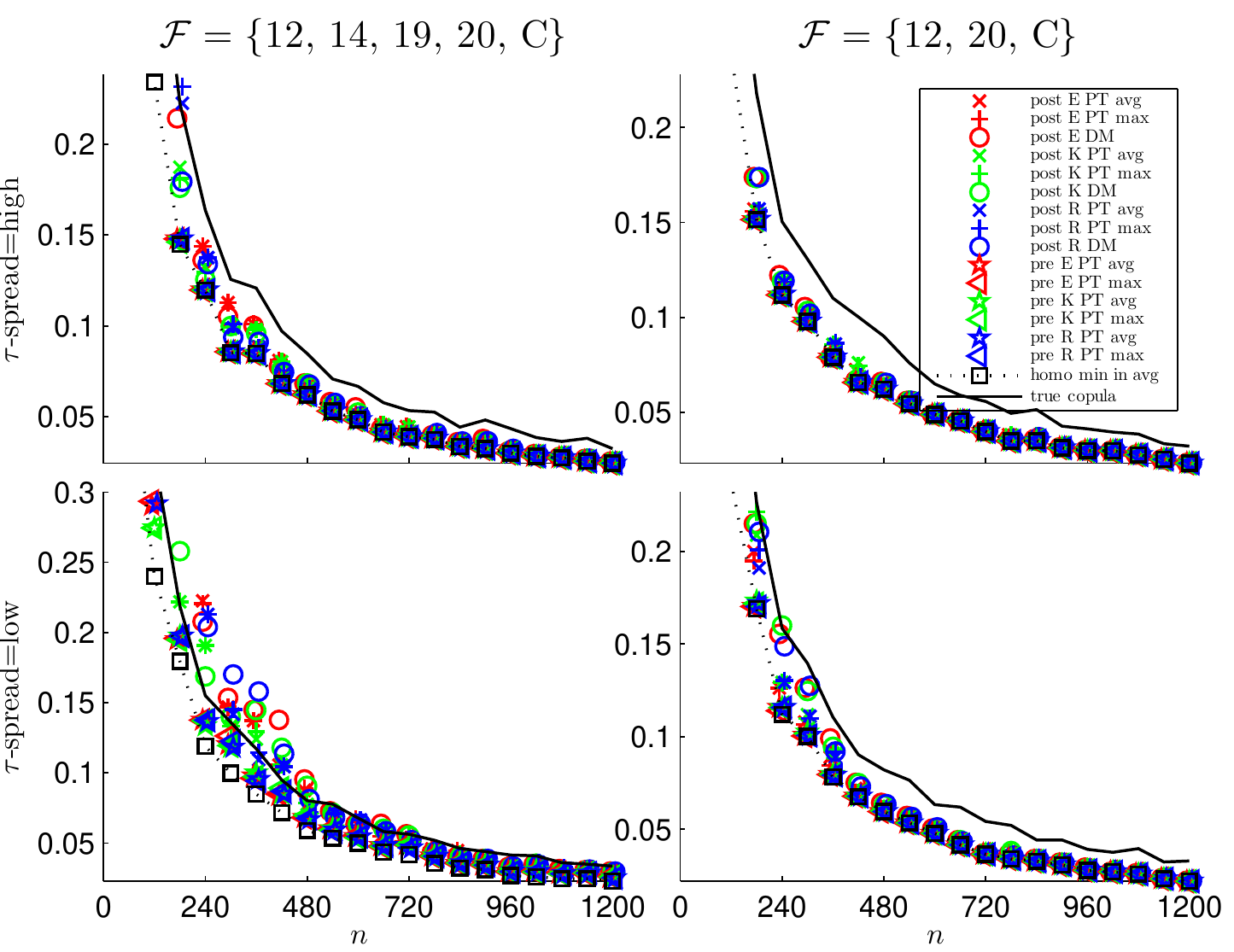}
\includegraphics[width=0.49\textwidth]{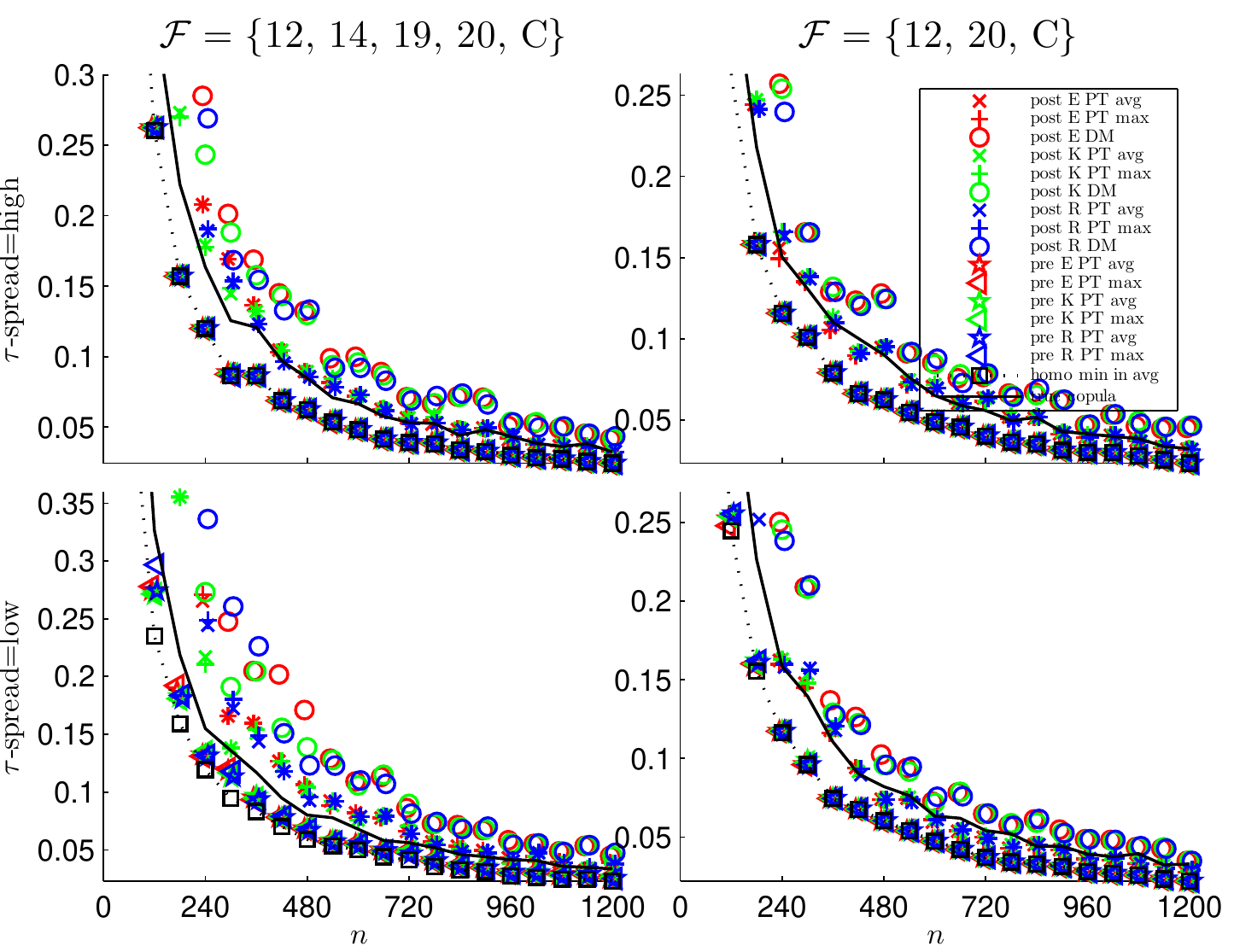}
\includegraphics[width=0.49\textwidth]{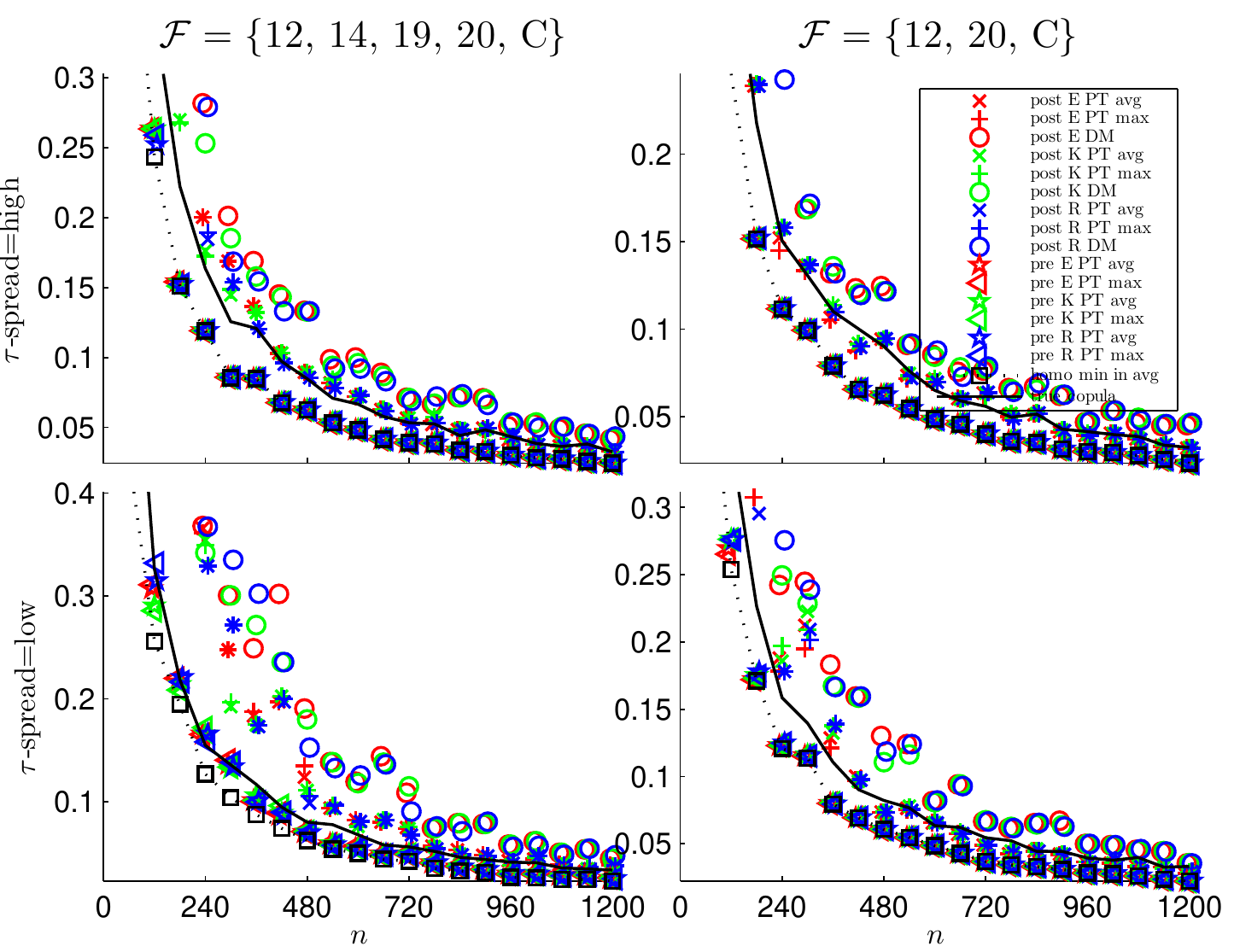}
\caption{The tau distance median for the four 15-HAC models and the optimistic estimators. The left-hand and right-hand sub-figures correspond to \#Forks=known and to \#Forks=unknown, respectively, and the top and bottom sub-figures correspond to KTauAvg and to TauMin, respectively.}
\label{fig:tau_distance_median_opt}
\centering
\end{figure}

\begin{figure}[hbt]
\centering
\includegraphics[width=0.49\textwidth]{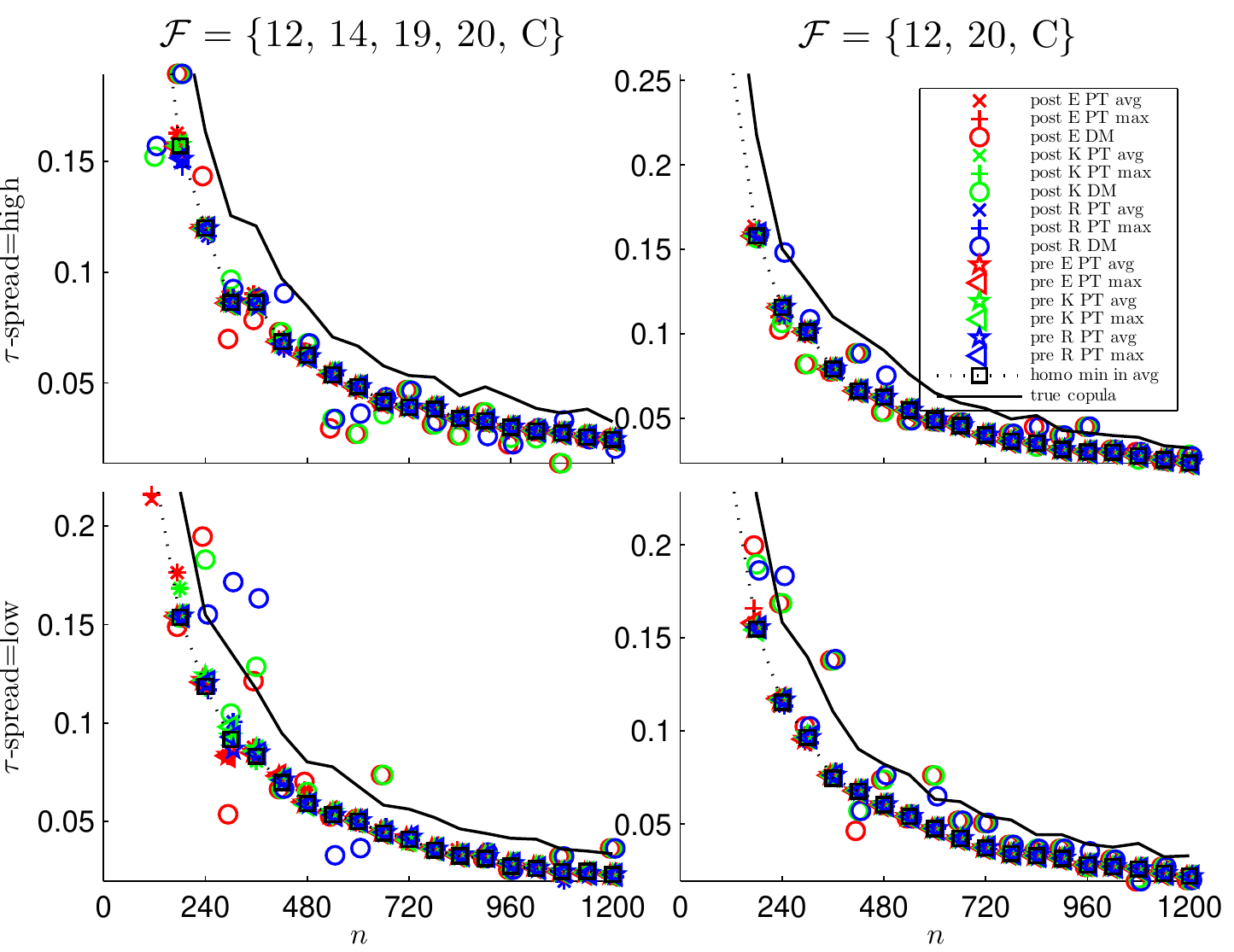}
\includegraphics[width=0.49\textwidth]{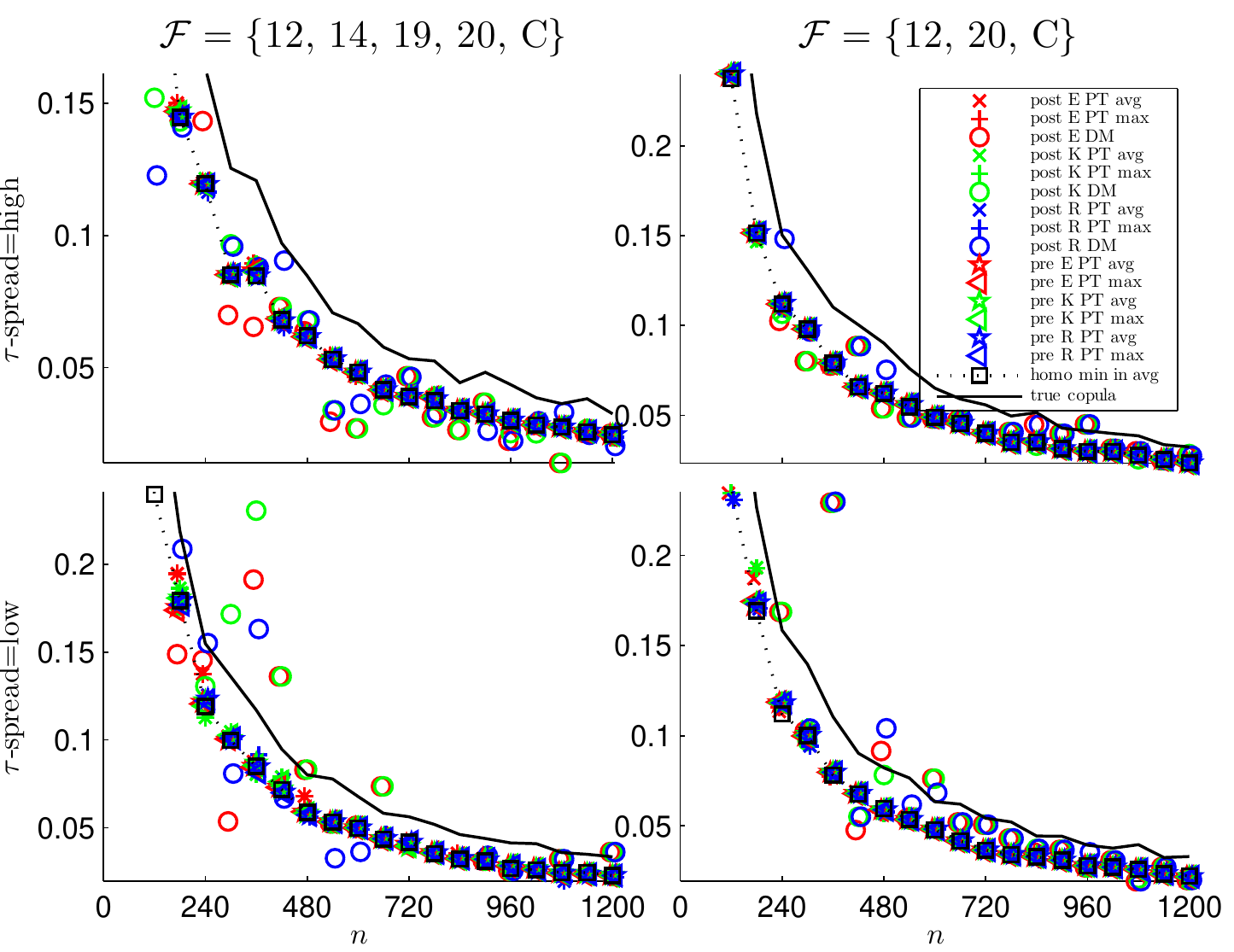}
\includegraphics[width=0.49\textwidth]{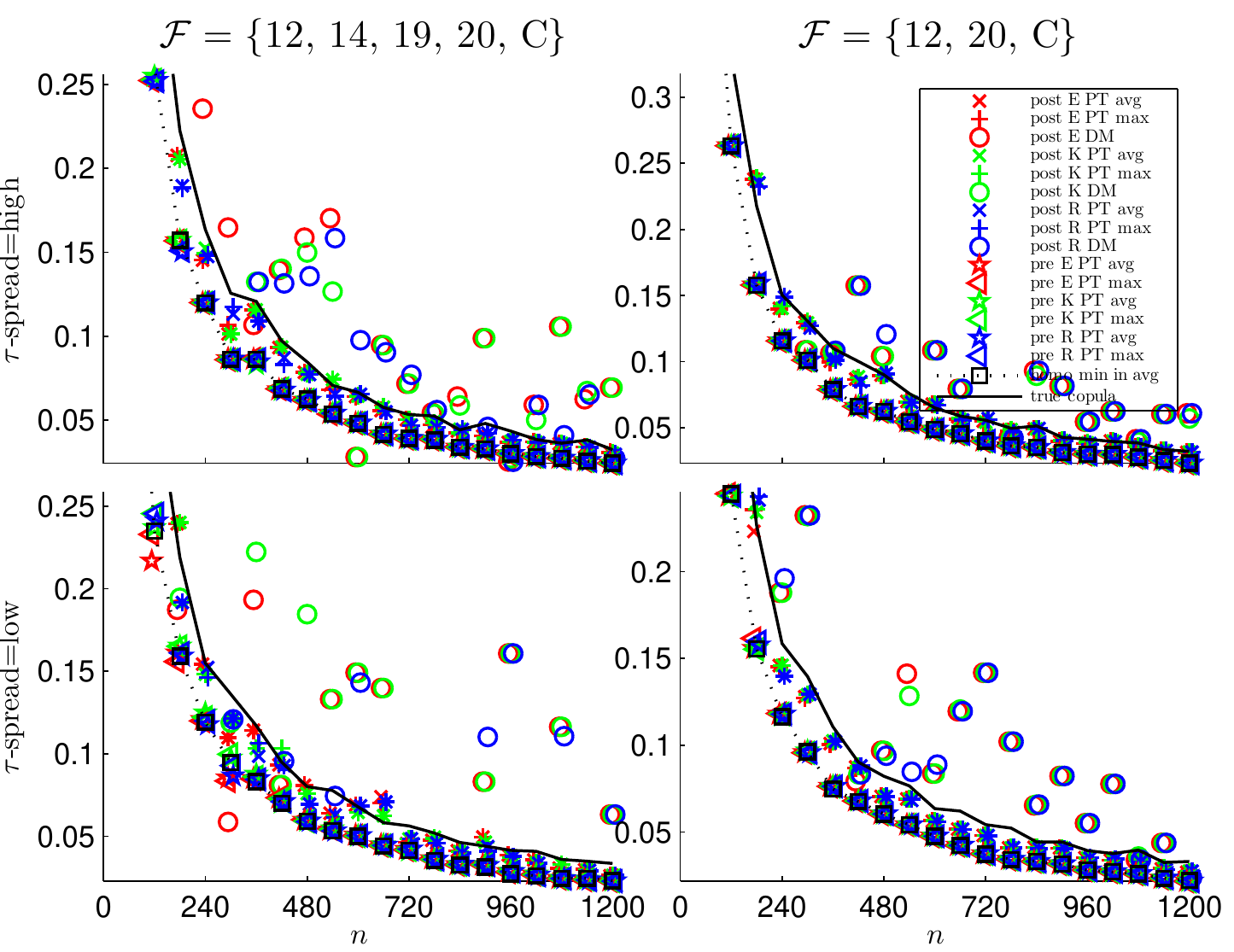}
\includegraphics[width=0.49\textwidth]{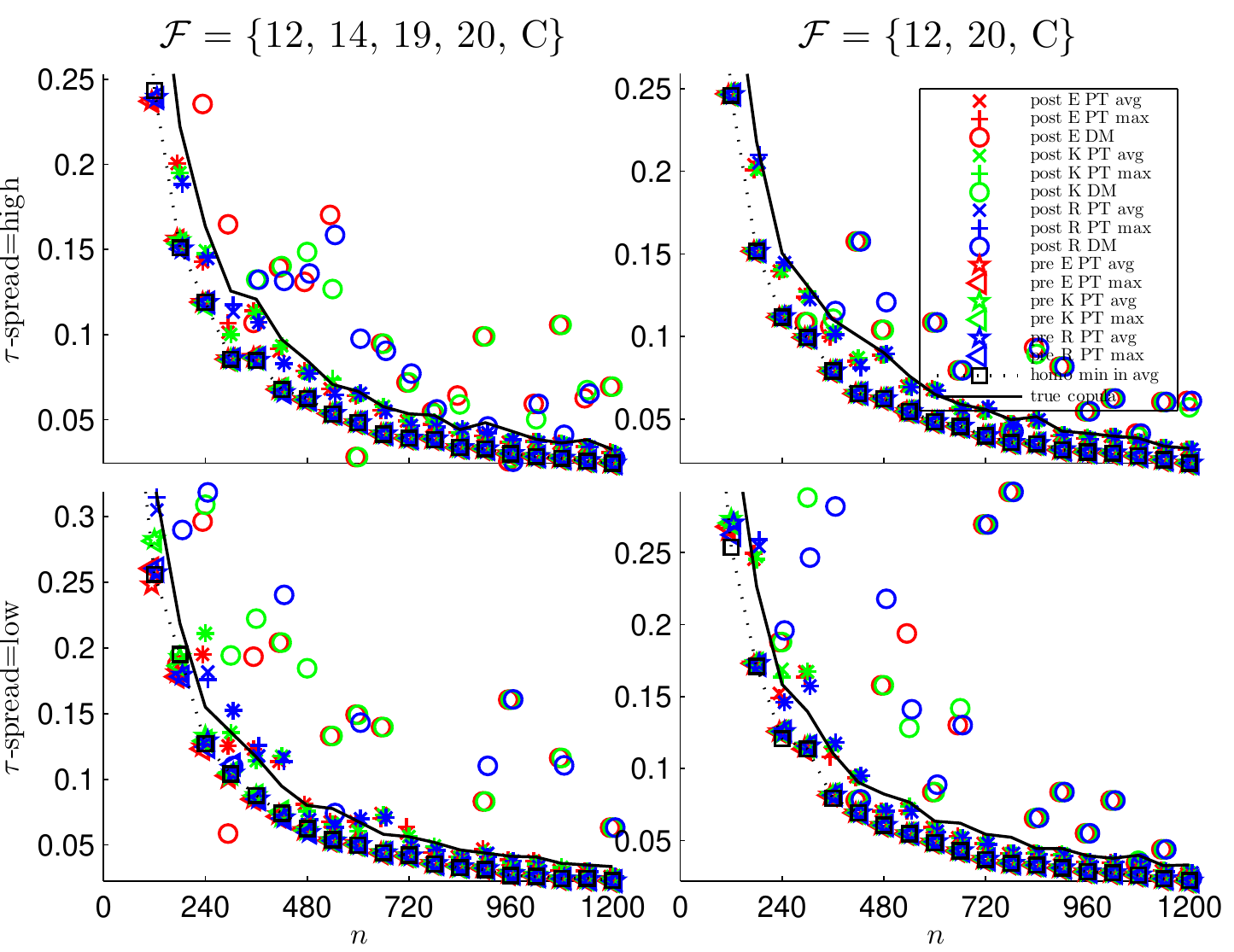}
\caption{The tau distance median for the four 15-HAC models and the pessimistic estimators. The left-hand and right-hand sub-figures correspond to \#Forks=known and to \#Forks=unknown, respectively, and the top and bottom sub-figures correspond to KTauAvg and to TauMin, respectively.}
\label{fig:tau_distance_median_pes}
\centering
\end{figure}

Precision of the estimators in the estimation of the parameters is evaluated by the tau distance median. This evaluation for $d = 15$ is depicted in Figures \ref{fig:tau_distance_median_opt} and \ref{fig:tau_distance_median_pes}. As this evaluation is not restricted to some limited interval, we do not show 10\% of the largest evaluations, to allow one to better distinguish among the other results.

\begin{figure}
\centering
\includegraphics[width=0.49\textwidth]{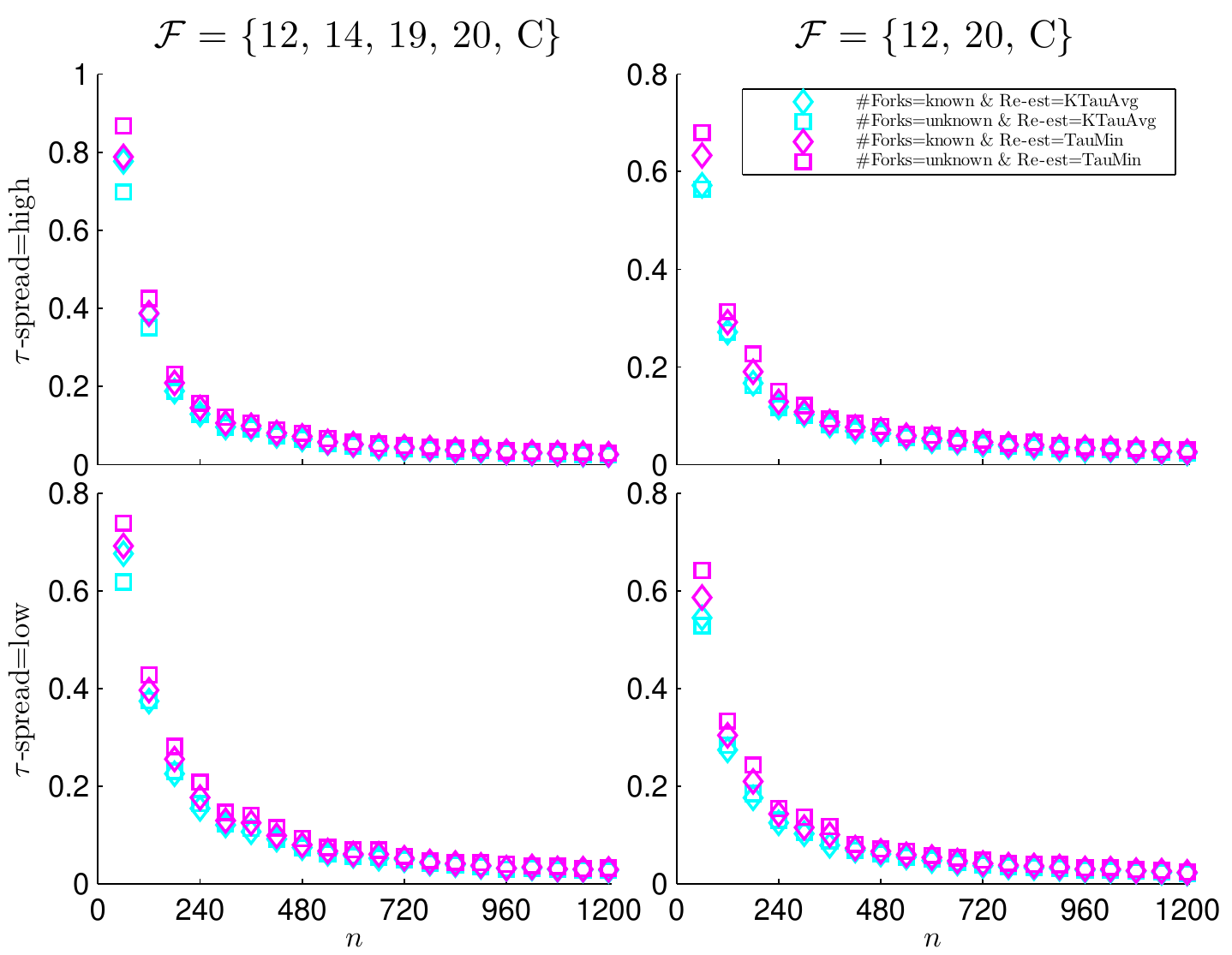}
\includegraphics[width=0.49\textwidth]{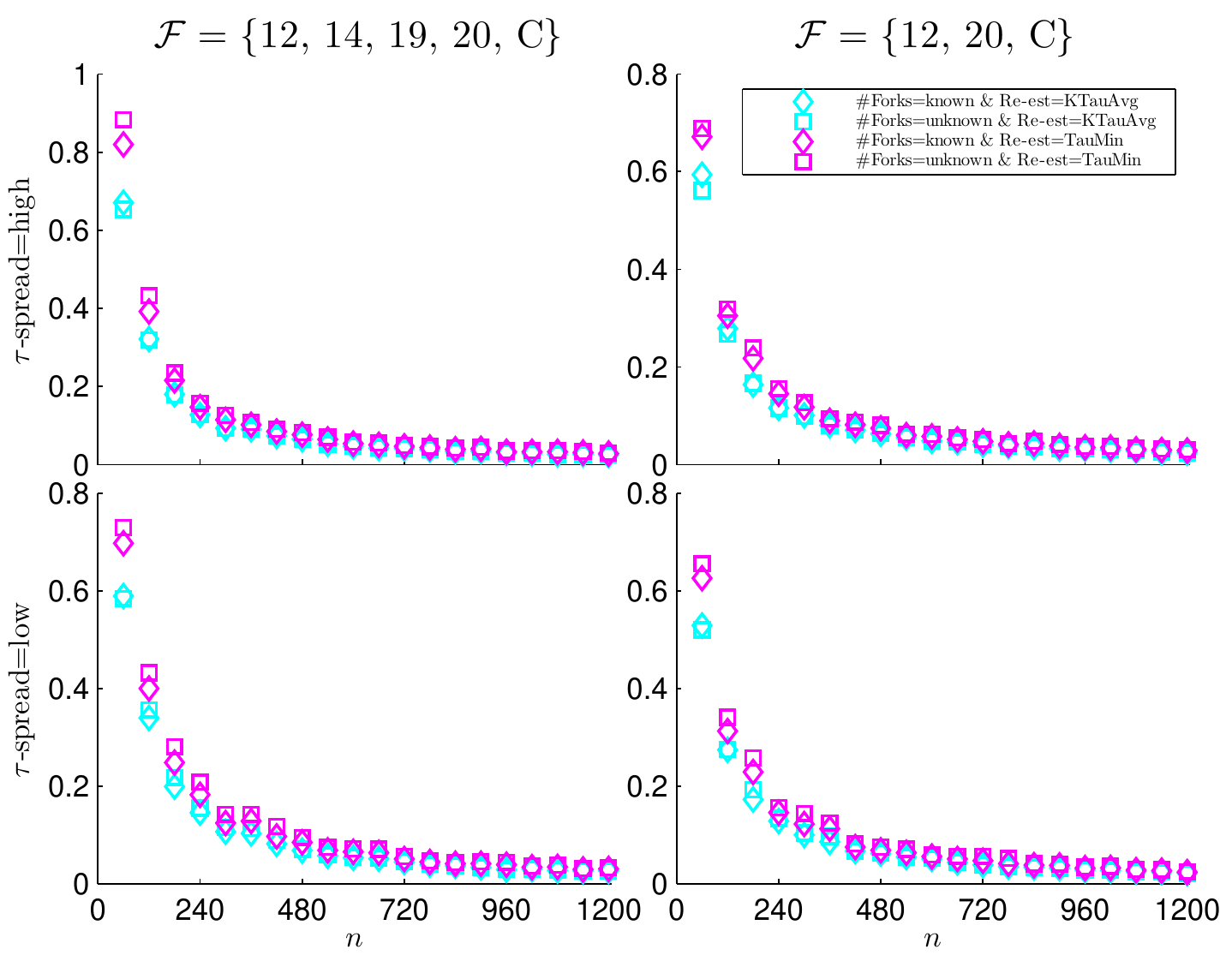}
\caption{Aggregated (by the average) tau distance median computed for the evaluations depicted in Figure \ref{fig:tau_distance_median_opt} - the optimistic attitude, here depicted at the left-hand - and Figure \ref{fig:tau_distance_median_pes} - the pessimistic attitude,  here depicted at the right hand.}
\label{fig:nForks_tau_dist_median}
\includegraphics[width=1\textwidth]{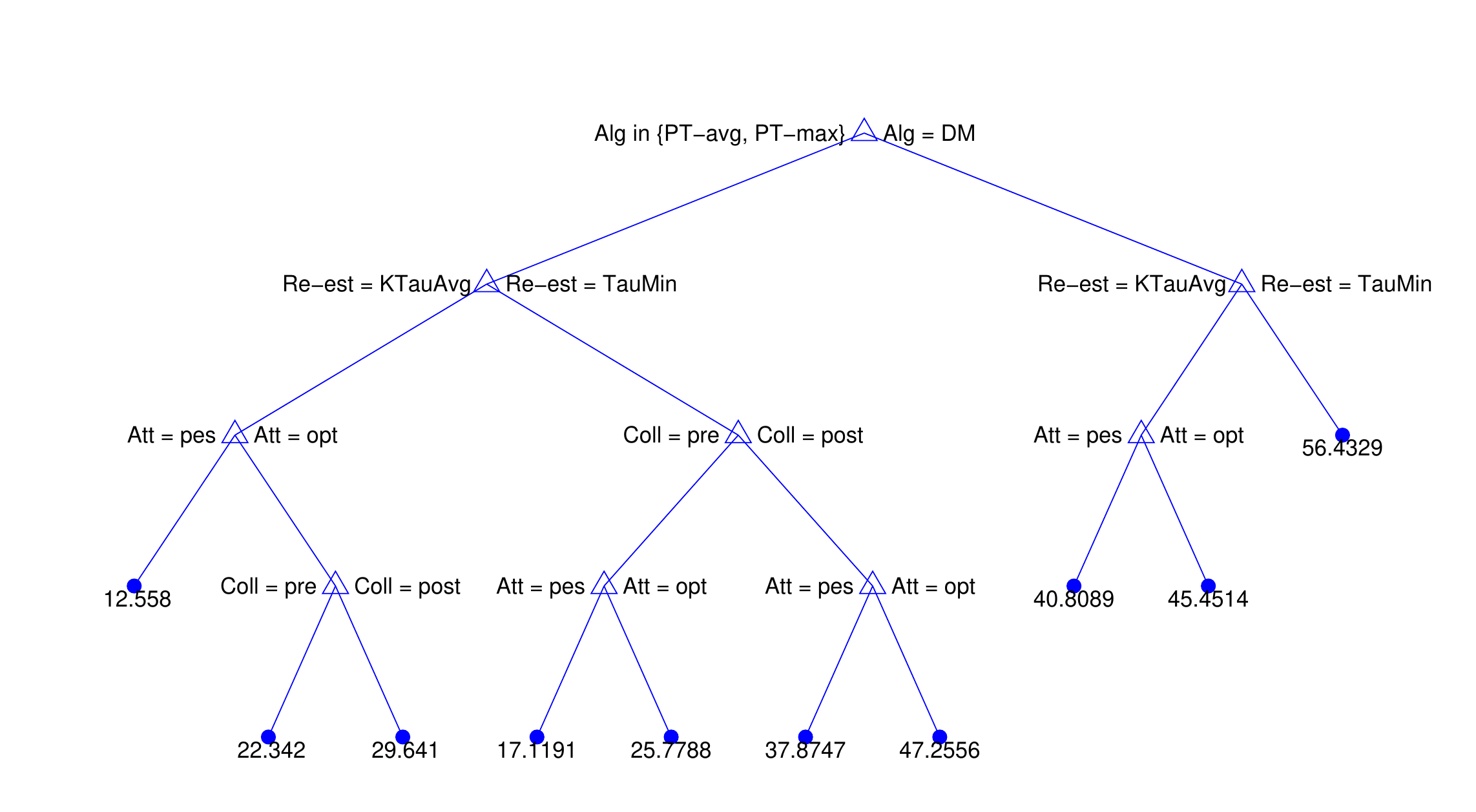}
\caption{The regression tree for the tau distance median ranks based on the five features of the heterogeneous estimators.}
\label{fig:tau_distance_median_tree}
\includegraphics[width=0.49\textwidth]{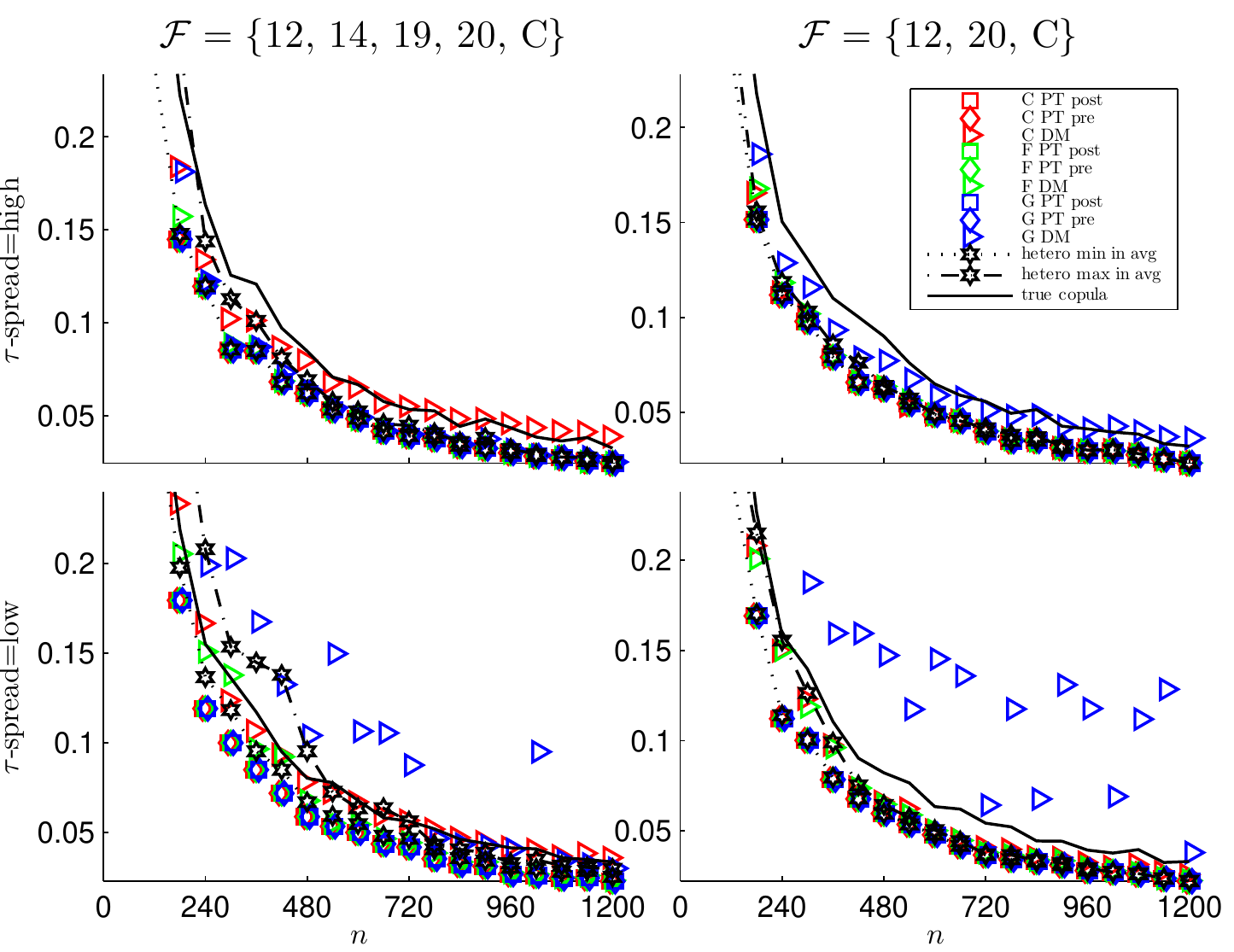}
\includegraphics[width=0.49\textwidth]{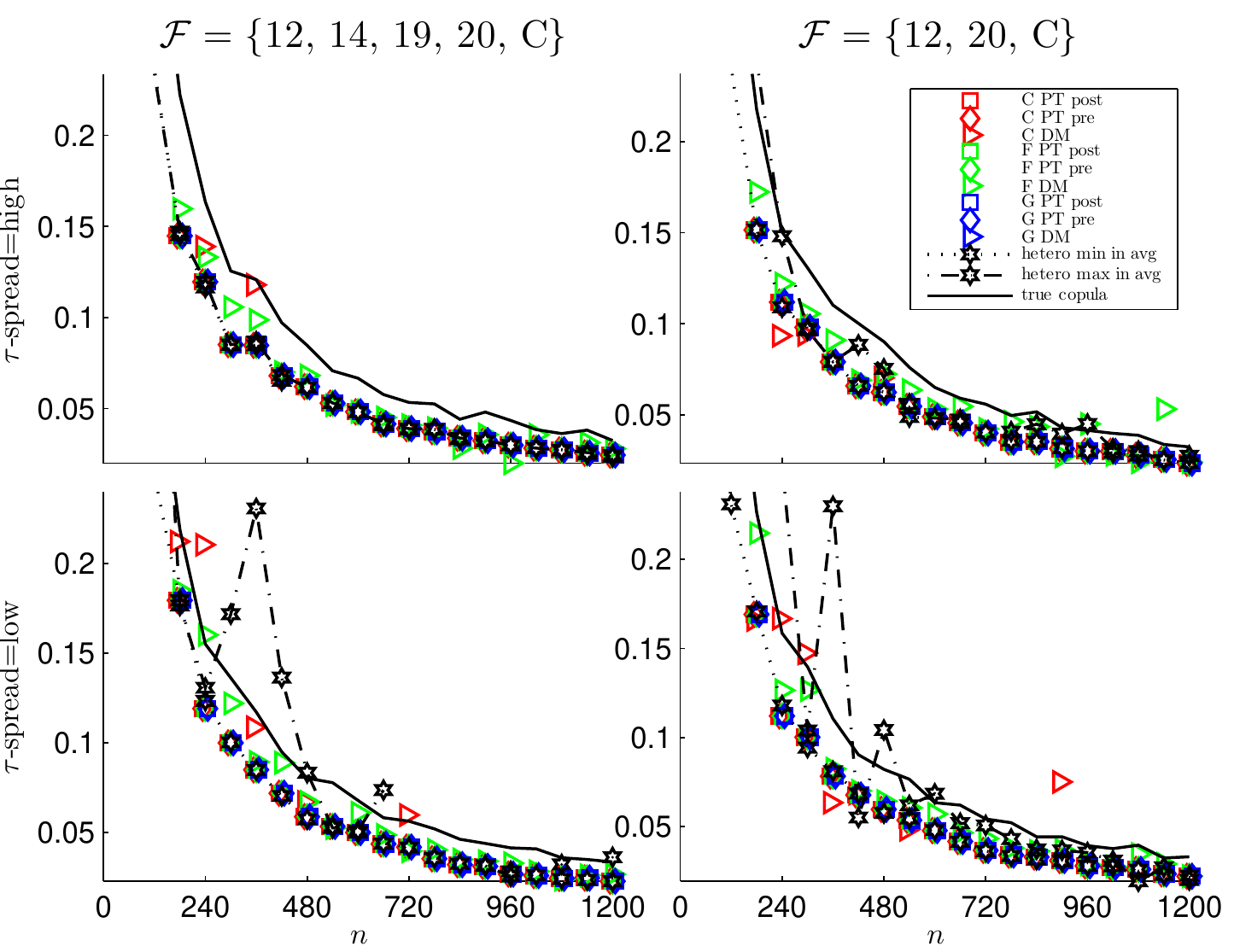}
\caption{The tau distance median for the Re-est=KTauAvg and \#Forks=unknown homogenous estimators  for the four 15-HAC models. The optimistic ones are at the left-hand and the pessimistic ones at the right-hand.}
\label{fig:tau_distance_median_homo}
\centering
\end{figure}

The following can be observed: 
\begin{itemize}
	\item For the optimistic estimators, the ratio converges to 0 as $n$ increases. 
	For Re-est=KTauAvg, we observe very similar results for all corresponding estimators. Also, the results are mostly better than for the true copula. Here, consider that we work on the pseudo-observations given by \eqref{eq:pseudo_observations}, see also Input 1 in Algorithm \ref{alg:hetero_HAC_estim}, and not directly on the observations from the model. For more details on this issue, see also \cite{Hofert13}. This realistic approach however substantially affects the data, particularly for low $n$, which explains our observation.
On the other hand, this observation justifies viability of these estimators for parameter estimation. We also observe that the benchmark homogeneous evaluation sometimes shows better results than for the best heterogeneous estimator. As we assume the optimistic attitude, the parameters in the heterogeneous estimates might be biased more by the trim function than in the homogeneous estimates as the s.n.c.~is more restrictive in the heterogeneous case; compare Tables \ref{tab:geners} and \ref{tab:geners_comb}. We thus get slightly worse precision for the parameters. However, as will be seen in Section \ref{sec:exps_gof}, this is a trade-off for better fit than of the homogeneous estimators.
	For Re-est=TauMin, we observe a difference between the Coll=pre estimators and the remaining ones. 
	The results are relatively similar for $\#\FF = 5$ and $\#\FF = 3$. For different $\tau$-spreads, a better performance can be observed for $\tau$-spread=high.  
In average, KTauAvg clearly outperforms TauMin, as is additionally illustrated by Figure \ref{fig:nForks_tau_dist_median};
			\item For the pessimistic estimators, a majority of the claims for the optimistic estimators can be adopted. 
			The most visible difference is that the Alg=DM estimators perform worse then in the optimistic case, which can be explained in the same way in Section \ref{sec:exps_fams}, and also by looking at Figure \ref{fig:rejection_rate_pes}.
			Considering the aggregated results depicted in Figure \ref{fig:nForks_tau_dist_median}, we observe slightly greater difference between KTauAvg and TauMin.
\end{itemize}
In Figure \ref{fig:nForks_tau_dist_median}, also observe how the difference between \#Forks=known and \#Forks=unknown vanishes for both attitudes with increasing $n$.

% obtained by
% plotevaluation_nforks(fileName, 'tauDistEvalAllMedian', true, 19:24, true, 'none', false);
% plotevaluation_nforks(fileName, 'tauDistEvalAllMedian', false, 19:24, true, 'none', false);

We again provide a regression tree created analogously to the one provided in Section \ref{sec:exps_struc}, with the only change that the tau distance median is used instead of the false structure ratio. The tree is depicted in Figure \ref{fig:tau_distance_median_tree}.
The tree again confirms our observations for $d=15$. Moreover, we observe that the best estimators are those with Alg in \{PT-avg, PT-max\} (or simply Alg=PT) \& Re-est=KTauAvg \& Att=pes. Considering the best optimistic estimators, the Alg=PT \& Re-est=KTauAvg ones with Coll=pre show the better rank.
Also note that a prediction according to the tree is neither dependent on Sn nor on $g$.
%\begin{figure}[htb]
%\centering
%\includegraphics[width=1\textwidth]{tree-tauDistEvalAllMedian-eps-converted-to.pdf}
%\caption{The regression tree for the tau distance median ranks based on the five features of the heterogeneous estimators.}
%\label{fig:tau_distance_median_tree}
%\centering
%\end{figure}

%\begin{figure}[htb]
%\centering
%\includegraphics[width=0.49\textwidth]{d15-tauDistEvalAllMedian-nForksUnknownKTauAvg-isOpt1-model9to12-homo-eps-converted-to.pdf}
%\includegraphics[width=0.49\textwidth]{d15-tauDistEvalAllMedian-nForksUnknownKTauAvg-isOpt0-model9to12-homo-eps-converted-to.pdf}
%\caption{The tau distance median for the Re-est=KTauAvg and \#Forks=unknown homogenous estimators  for the four 15-HAC models. The optimistic ones are at the left-hand and the pessimistic ones at the right-hand.}
%\label{fig:tau_distance_median_homo}
%\centering
%\end{figure}

To address the robustness of the estimation of the parameters against misspecification of the underlying families, an analogue of Figure \ref{fig:false_structure_ratio_homo} for the tau distance median is depicted in Figure \ref{fig:tau_distance_median_homo}.
The observations are again quite clear. The Alg=PT estimators are robust in the parameter estimation against misspecification of the underlying family(ies), whereas the Alg=ML estimators not. Also, considering the Alg=ML pessimistic estimators, the estimators assuming the Gumbel or the Clayton family are very prone to rejection and thus most of estimates have been rejected for for larger $n$ and thus no evaluation mark is presented in these cases. The evaluations for Coll=pre and Coll=post again look identical.

%\clearpage
%-------------------------------------------------------------------------------------------------------

\subsubsection{Goodness of fit}
\label{sec:exps_gof}
\begin{figure}[tb]
\centering
\includegraphics[width=0.49\textwidth]{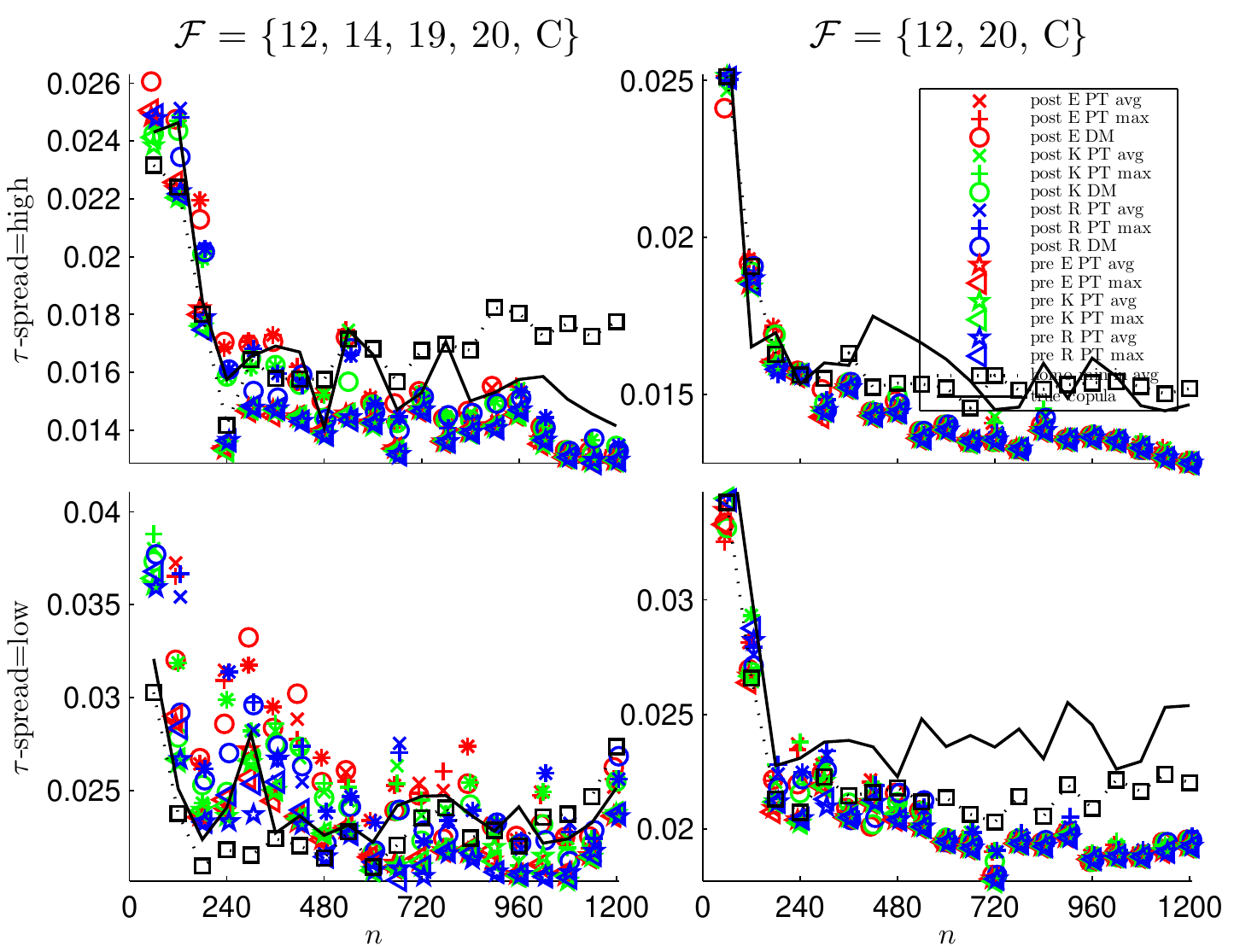}
\includegraphics[width=0.49\textwidth]{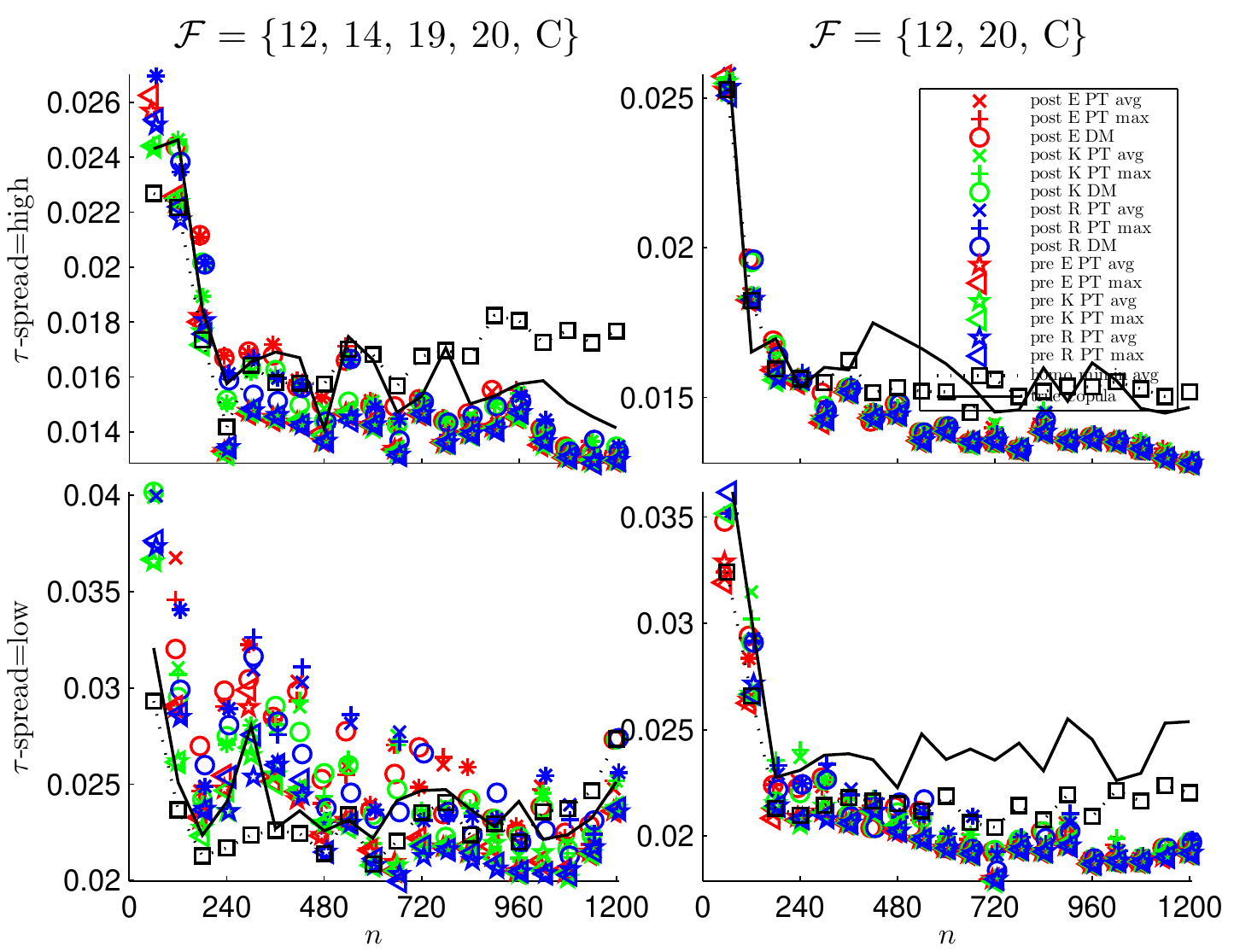}
\includegraphics[width=0.49\textwidth]{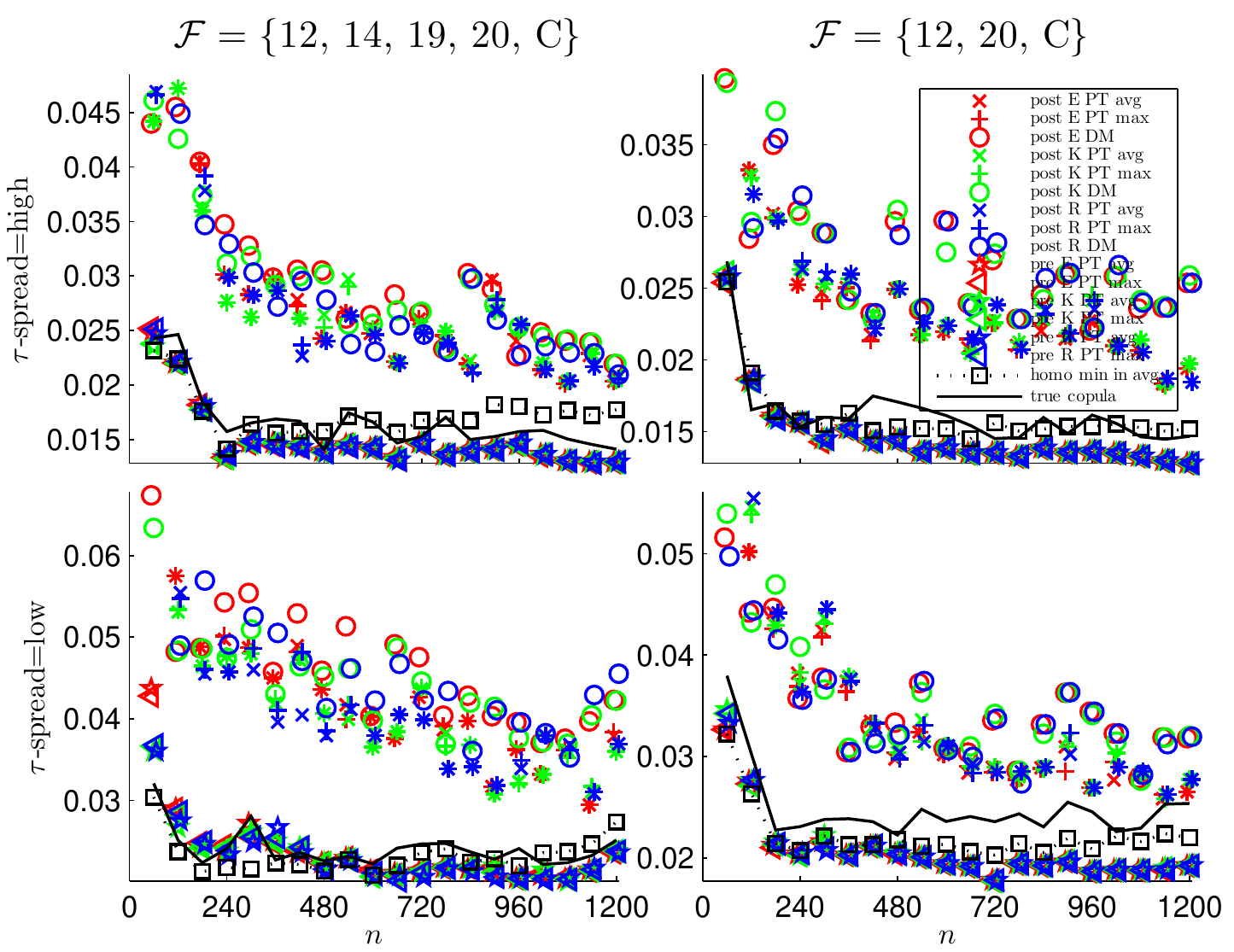}
\includegraphics[width=0.49\textwidth]{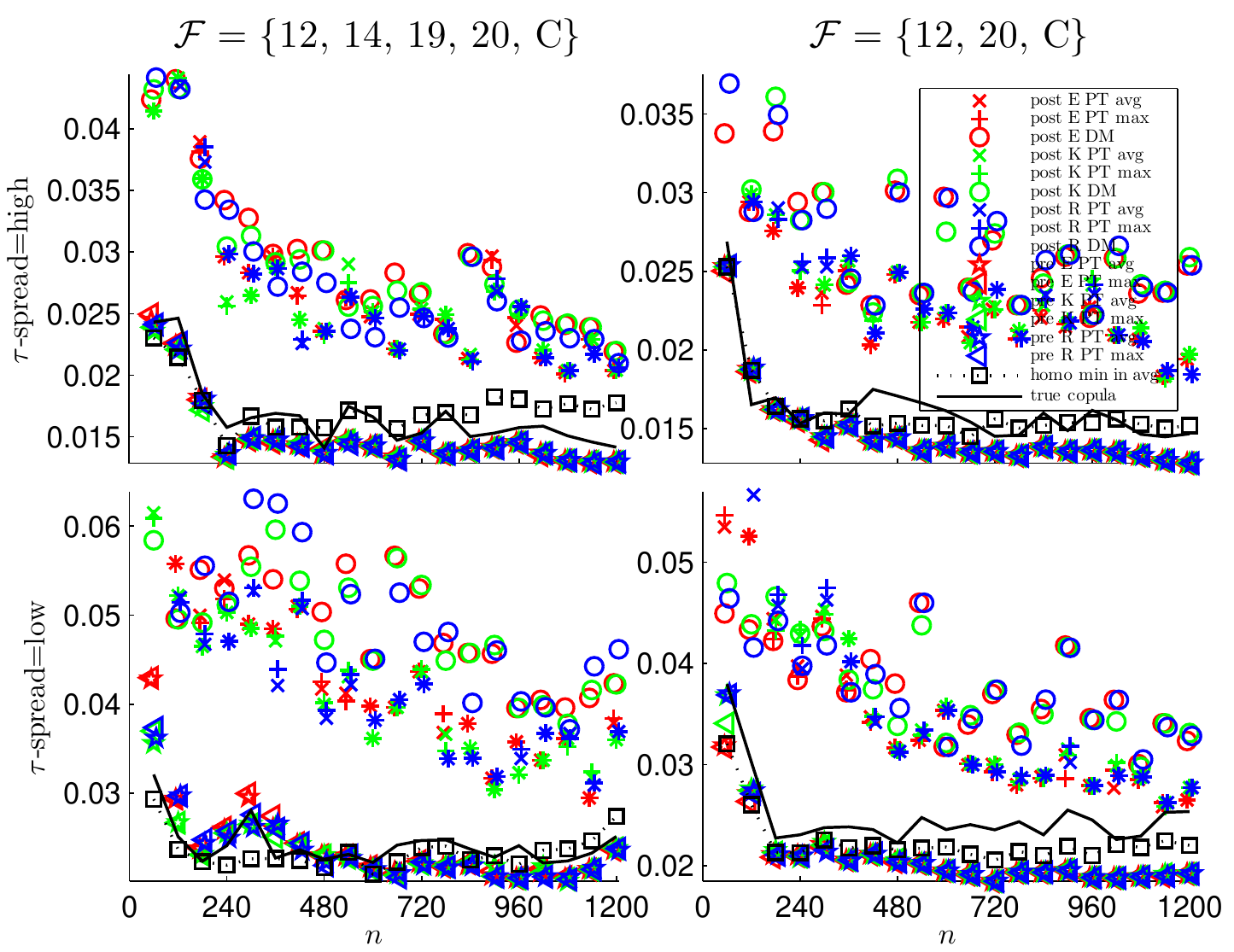}
\caption{GoF median for the four 15-HAC models and the optimistic estimators. The left-hand and right-hand sub-figures correspond to \#Forks=known and to \#Forks=unknown, respectively, and the top and bottom sub-figures correspond to KTauAvg and to TauMin, respectively.}
\label{fig:gof_median_opt}
\centering
\end{figure}

\begin{figure}[hbt]
\centering
\includegraphics[width=0.49\textwidth]{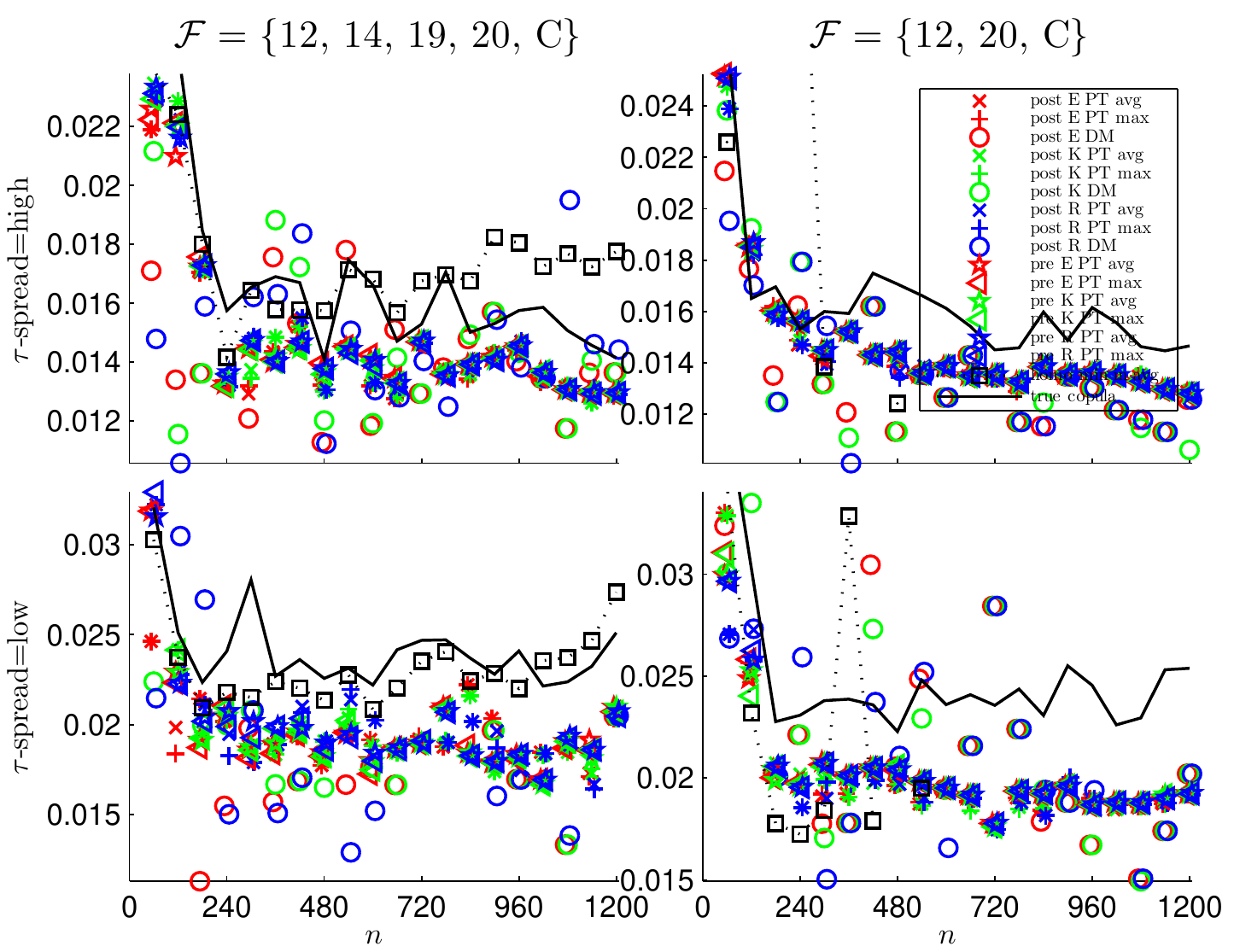}
\includegraphics[width=0.49\textwidth]{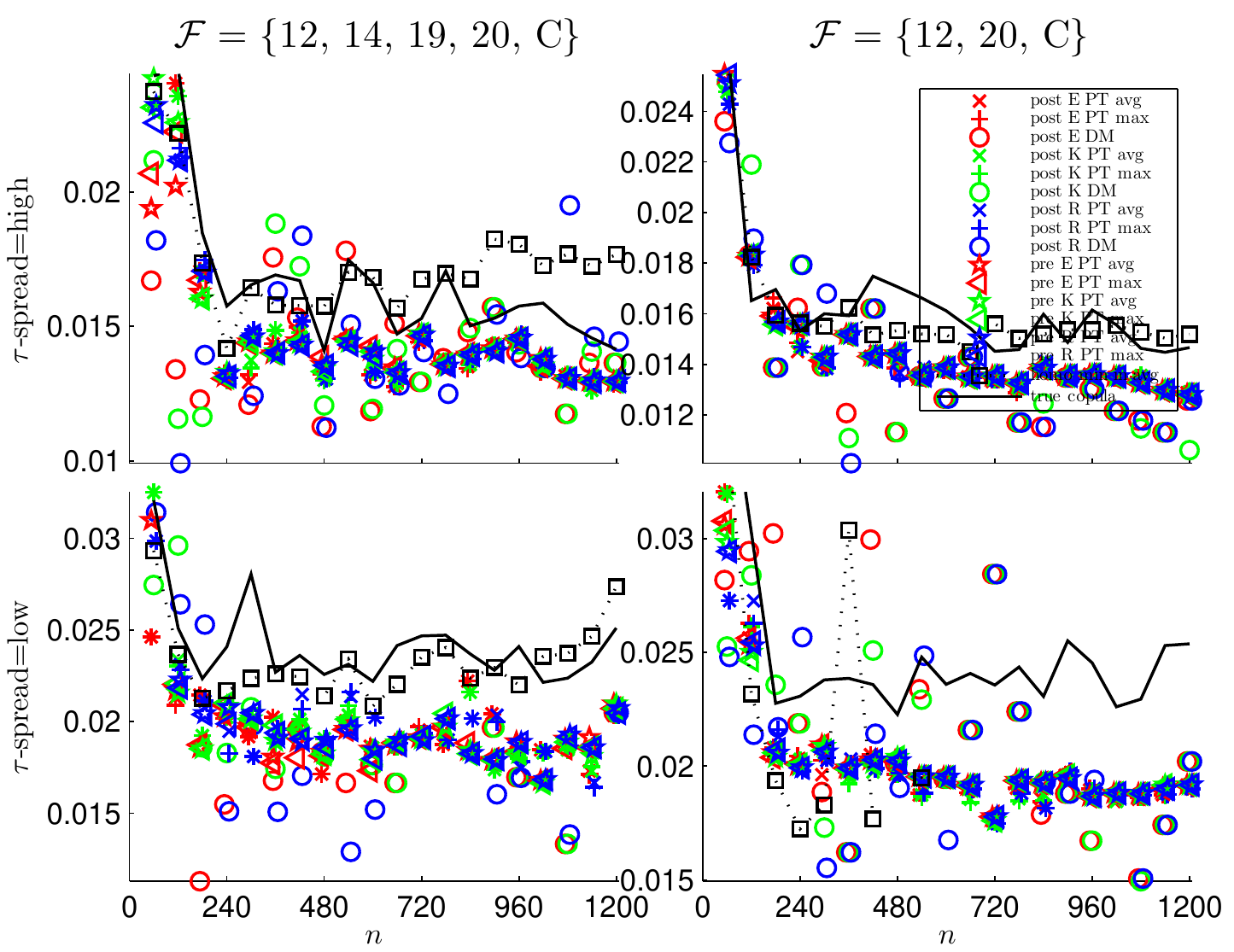}
\includegraphics[width=0.49\textwidth]{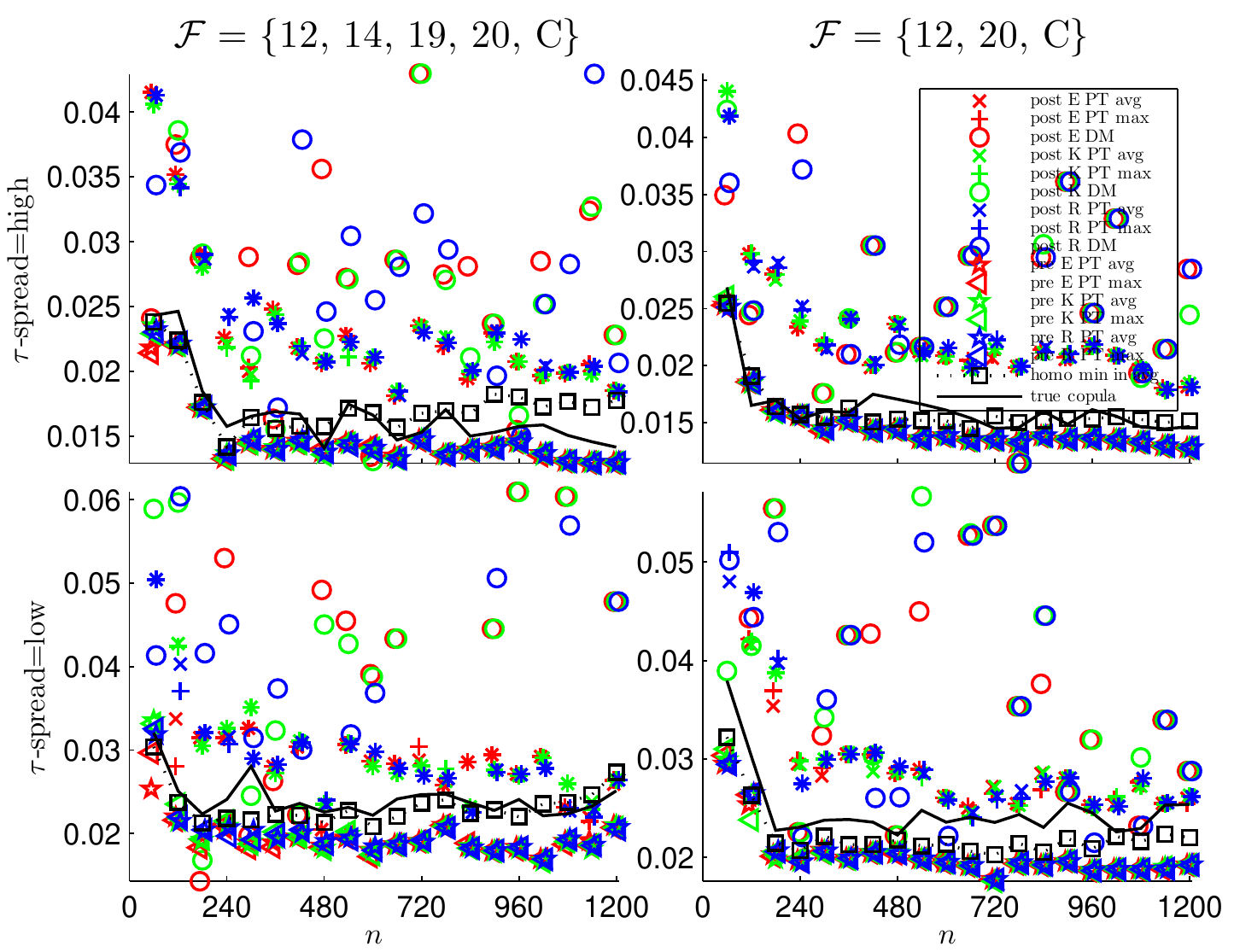}
\includegraphics[width=0.49\textwidth]{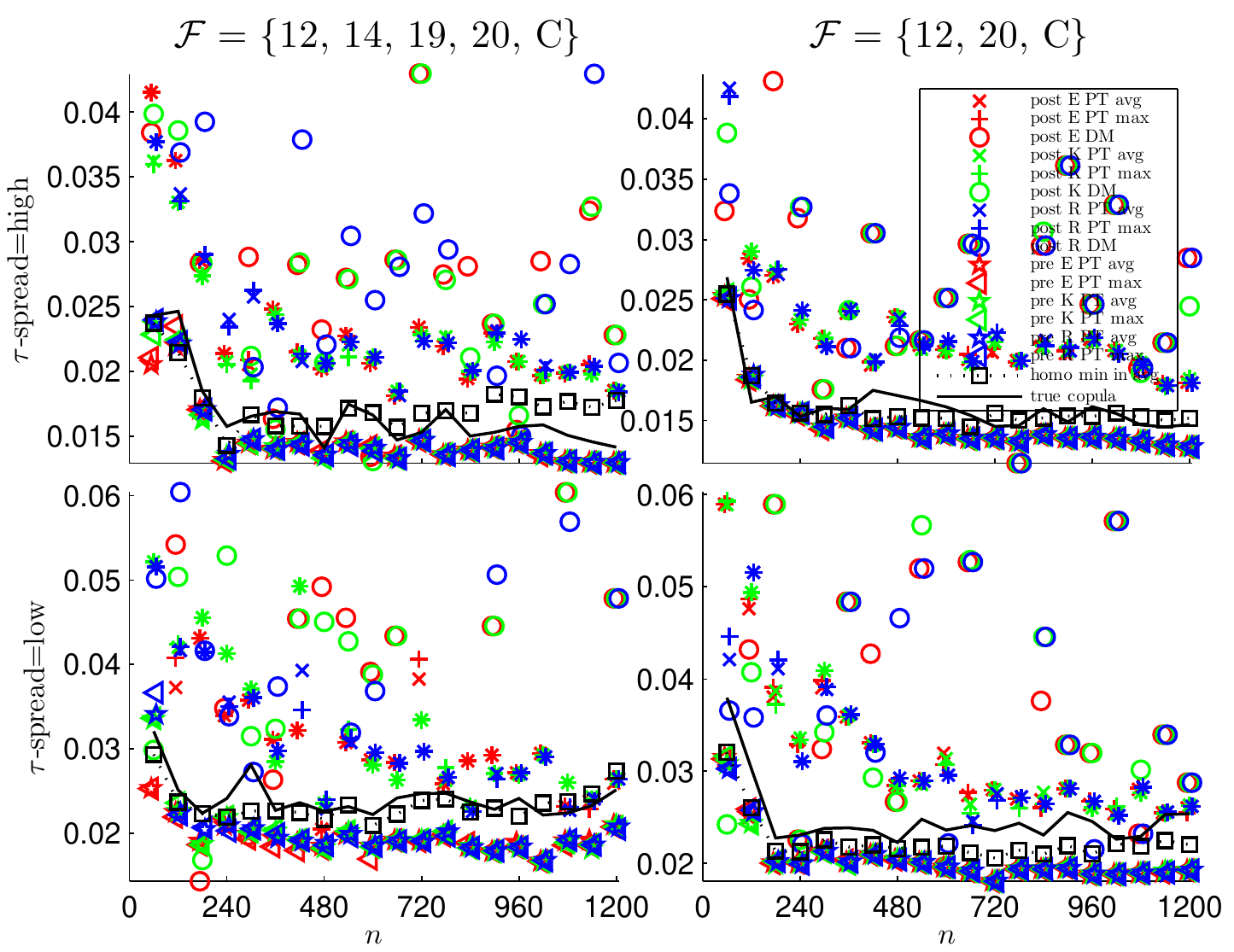}
\caption{GoF median for the four 15-HAC models and the pessimistic estimators. The left-hand and right-hand sub-figures correspond to \#Forks=known and to \#Forks=unknown, respectively, and the top and bottom sub-figures correspond to KTauAvg and to TauMin, respectively.}
\label{fig:gof_median_pes}
\centering
\end{figure}

The overall fit generated by the estimators is evaluated by the GoF median. This evaluation for $d = 15$ is depicted in Figures \ref{fig:gof_median_opt} and \ref{fig:gof_median_pes}. As this evaluation is not restricted to some limited interval, we do not show 2\% of the largest evaluations to allow for better distinguishing among the other results.

\begin{figure}
\centering
\includegraphics[width=0.49\textwidth]{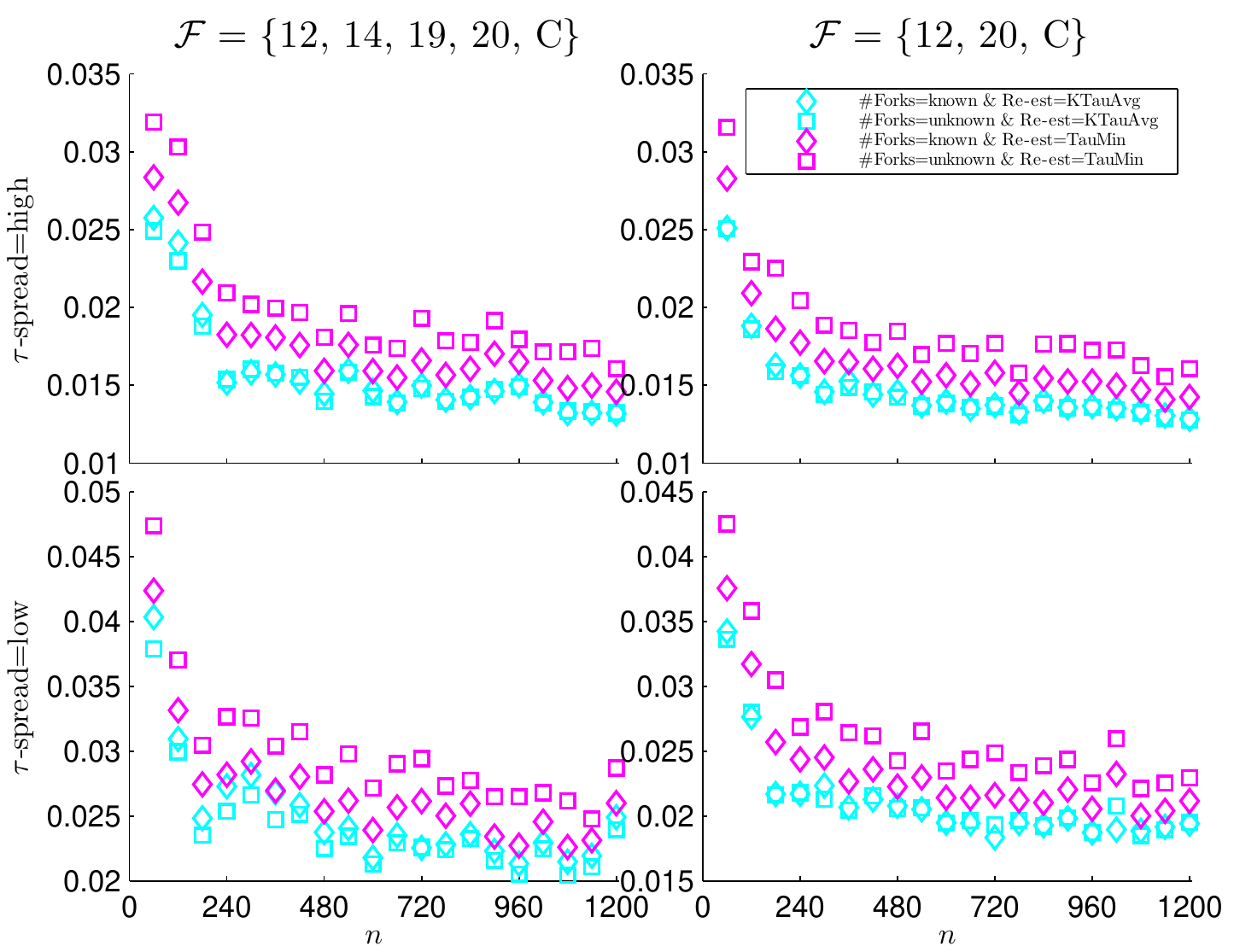}
\includegraphics[width=0.49\textwidth]{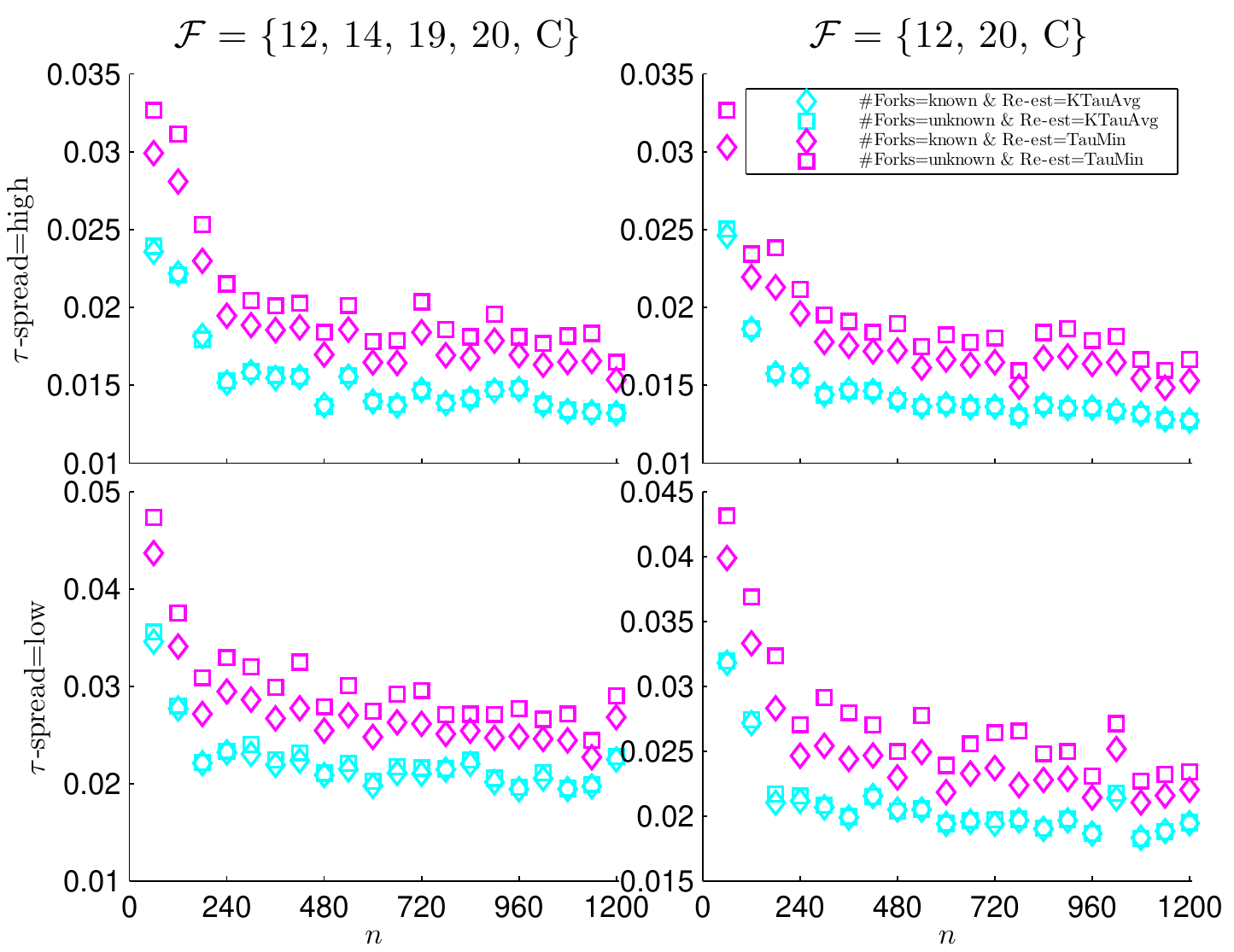}
\caption{Aggregated (by the average) GoF median computed for the evaluations depicted in Figure \ref{fig:gof_median_opt} - the optimistic attitude, here depicted at the left-hand - and Figure \ref{fig:gof_median_pes} - the pessimistic attitude,  here depicted at the right hand.}
\label{fig:nForks_gof_median}
\includegraphics[width=1\textwidth]{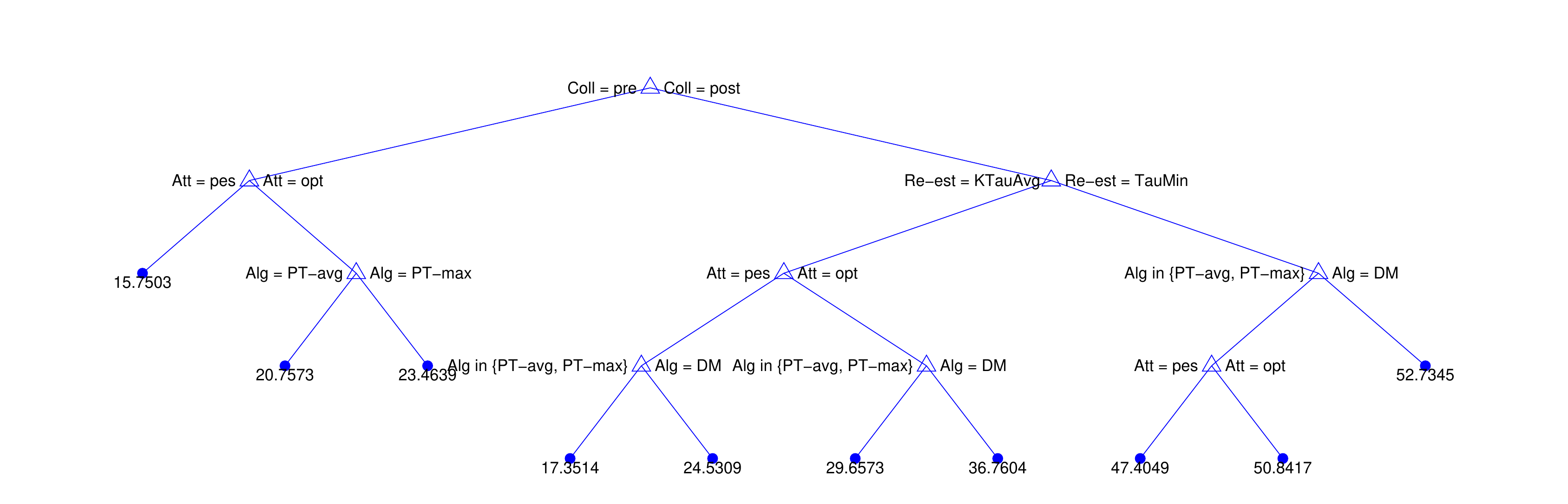}
\caption{The regression tree for the GoF median ranks based on the five features of the heterogeneous estimators.}
\label{fig:gof_median_tree}
\includegraphics[width=0.49\textwidth]{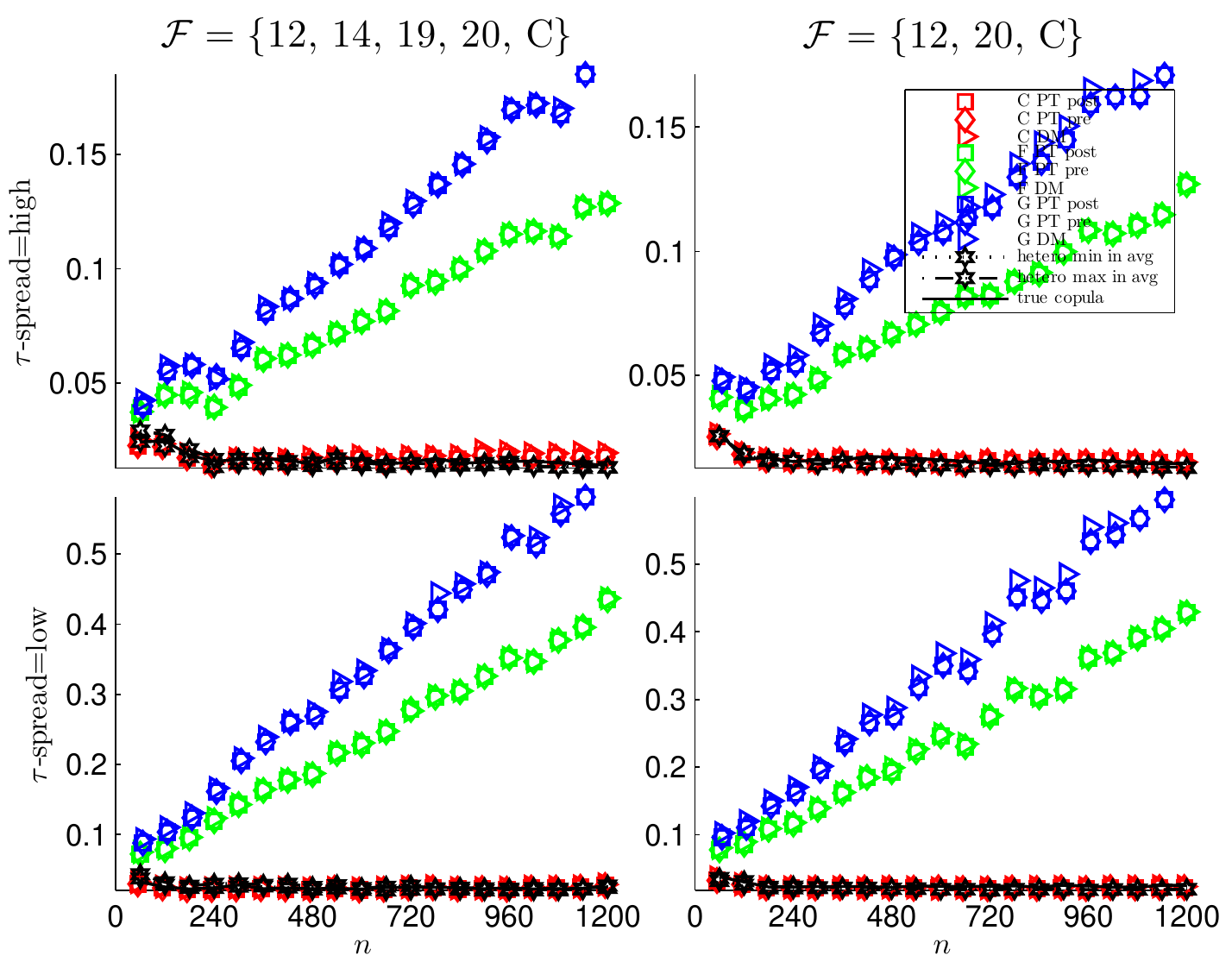}
\includegraphics[width=0.49\textwidth]{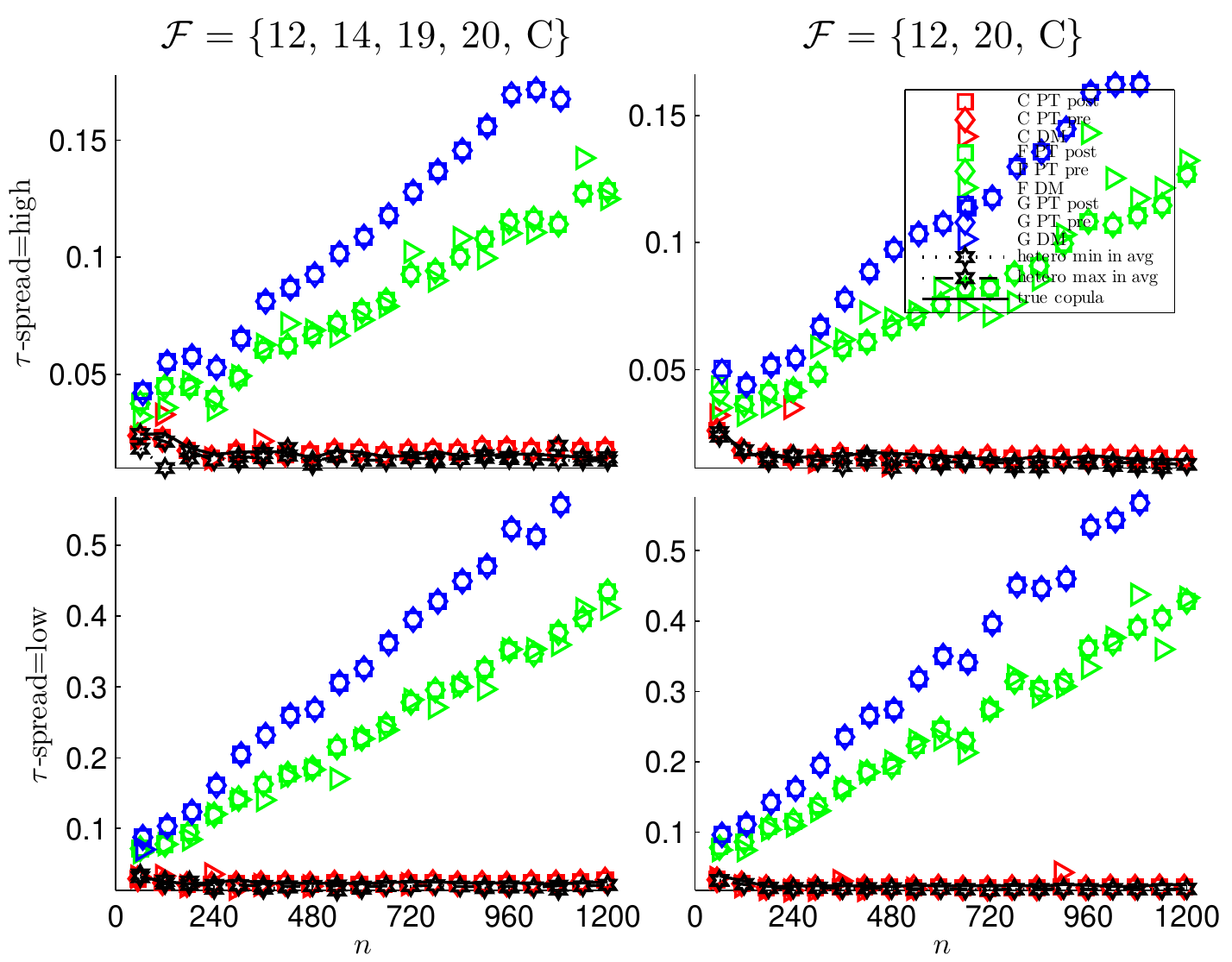}
\caption{The GoF median for the Re-est=KTauAvg and \#Forks=unknown homogenous estimators  for the four 15-HAC models. The optimistic ones are at the left-hand and the pessimistic ones at the right-hand.}
\label{fig:gof_median_homo}
\centering
\end{figure}

The following can be observed: 
\begin{itemize}
	\item For the optimistic estimators, the evaluation mostly decreases as $n$ increases. For Re-est=KTauAvg, if $\#\FF=3$, the evaluations are substantially lower than for the true copula evaluation. If $\#\FF=5$, such an observation can be done for the Coll=pre estimators. 
For $\tau$-spread=low and $\#\FF=5$, we observe that the estimators need much more data than for the other variants of $\tau$-spread and $\#\FF$ to be get smaller than the true copula (in the evaluation).
The homo min in avg evaluation mostly increases as $n$ increases, i.e., as the amount of the data increases, 
misspecification of	the underlying families has increasing impact on GoF. Also, recall that in Section \ref{sec:exps_params}, we frequently observed for homo min in avg better results in the tau distance median than for the best heterogeneous estimators. Here, we observe the other side of the mentioned trade-off, i.e., even if the parameters are for the best homogeneous estimator closer to the true parameters than for the best heterogeneous estimators, the best heterogeneous estimators still show better results in GoF than the best heterogeneous estimator.
This observation confirms that correct estimation of the underlying families is crucial for GoF.
For Re-est=TauMin, we observe substantially worse performance for the estimators that are not Coll=pre. This confirms the dominance of the Coll=pre estimators in GoF over the remaining ones.
The difference between KTauAvg and TauMin is also visible from the aggregated evaluations depicted in Figure \ref{fig:nForks_gof_median}, where also the supremacy of KTauAvg over TauMin is obvious. This supremacy of KTauAvg is also supported by the observation that the difference between the evaluations for \#Forks=known and \#Forks=unknown is much smaller than for TauMin.
		\item For the pessimistic estimators, almost all claims done for the optimistic estimators can be adopted. The main difference can be observed for the Alg=DM estimators, which now differ more from the remaining ones. In Figure \ref{fig:nForks_gof_median}, we do not observe any substantial deviations from the optimistic version.
\end{itemize}

% obtained by
% plotevaluation_nforks(fileName, 'gofEvalAllMedian', true, 19:24, true, 'none', false);
% plotevaluation_nforks(fileName, 'gofEvalAllMedian', false, 19:24, true, 'none', false);

We again provide a regression tree created analogously to the one provided in Section \ref{sec:exps_struc}, with the only change that the GoF median is used instead of the false structure ratio. The tree is depicted in Figure \ref{fig:gof_median_tree}.
The tree is again in accordance with our observations done for $d=15$. Moreover, it shows the best predicted rank for the Coll=pre \& Att=pes estimators. If one wants to consider an optimistic estimator, the best rank is predicted for the  Coll=pre \& Alg=PT-avg estimators. Note that this is the only time we observe a substantial influence of $g$ on a predicted rank, and the better rank is predicted for $g$=avg.

%\begin{figure}[htb]
%\centering
%\includegraphics[width=1\textwidth]{tree-gofEvalAllMedian-eps-converted-to.pdf}
%\caption{The regression tree for the GoF median ranks based on the five features of the heterogeneous estimators.}
%\label{fig:gof_median_tree}
%\centering
%\end{figure}

To address robustness of the estimators against misspecification of the underlying families, an analogue of Figure \ref{fig:false_structure_ratio_homo} is depicted in Figure \ref{fig:tau_distance_median_homo} for the GoF median.
The following can be observed. The more the family assumed by an estimator is distinct from the underlying families, the worse fit  for these estimates is obtained. Again, considering the Alg=ML pessimistic estimators, the estimators assuming the Gumbel or the Clayton family are very prone to rejection and thus most of the estimates have been rejected for larger $n$ (and thus no evaluation mark is presented in these cases). We again do not observe any difference between the Coll=pre and Coll=post estimators.

%\begin{figure}[htb]
%\centering
%\includegraphics[width=0.49\textwidth]{d15-gofEvalAllMedian-nForksUnknownKTauAvg-isOpt1-model9to12-homo-eps-converted-to.pdf}
%\includegraphics[width=0.49\textwidth]{d15-gofEvalAllMedian-nForksUnknownKTauAvg-isOpt0-model9to12-homo-eps-converted-to.pdf}
%\caption{The GoF median for the Re-est=KTauAvg and \#Forks=unknown homogenous estimators  for the four 15-HAC models. The optimistic ones are at the left-hand and the pessimistic ones at the right-hand.}
%\label{fig:gof_median_homo}
%\centering
%\end{figure}

%-------------------------------------------------------------------------------------------------------

%\clearpage

\subsubsection{Summary}
\begin{itemize}
\item If the underlying families \emph{are} specified correctly, as $n$ increases, all considered estimators improve (in $n$) in finding the true structure and families and in the precision of the estimated parameters. Considering GoF, the Coll=pre estimators show mostly better results than the remaining estimators including the homogeneous estimators.
\item If the underlying families \emph{are not} specified correctly, the homogeneous Alg=PT estimators improve (in $n$) in finding the true structure and in the precision of the estimated parameters. Such a claim cannot be done for the homogeneous Alg=DM estimators. Considering GoF, the more the underlying families are distinct from the family assumed by an estimator, the worse fit of the estimates is obtained.
\item The improvement based on increasing $n$ substantially depends on how many different underlying families are in the underlying HAC and how close the parameters in the underling HAC are.
\item As $n$ increases, we mostly observe vanishing differences between the results obtained for \#Forks=known and \#Forks=unknown. This suggests viability of our collapsing approach proposed in Section \ref{sec:collapsing}.
\item In most of the evaluations, we observe dominance of Coll=pre over Coll=post, most significantly for the GoF median evaluation criterion. For the homogeneous estimators, influence of the feature Coll has not been observed.
\item In most of the evaluations, we observed dominance of Re-est=KTauAvg over Re-est=TauMin. This illustrates how important  it is to re-estimate the parameter after collapsing two nodes into one.
\item In most of the evaluations, we observed dominance of Alg=PT over Alg=DM. For the homogeneous Alg=PT estimators, we also observed the robustness against misspecification of the underlying families in estimating the true structure and parameters.
\item Considering the Alg=PT estimators and $g$, a substantial difference between the performance of the Alg=PT-avg and Alg=PT-max estimators has been observed only once for the Coll=Pre \& Att=opt estimators in the tree obtained for the GoF median evaluation criterion, where $g$=avg has shown better performance than $g$=max.
\item Considering the attitude, on the one hand, the construction of the optimistic estimators assures that an estimate is always returned. On the other hand, the pessimistic estimators show better results more frequently than the optimistic ones. This can be expected as the pessimistic estimators never trim parameters to satisfy the s.n.c. However, as there is no guarantee that a pessimistic estimator will be better than its optimistic version, see Section \ref{sec:hetero_example} for two opposing examples, 
we suggest to use both (optimistic and pessimistic) versions of an estimator, and if the pessimistic one returns an estimate, choose the better one according to some selected evaluation, e.g., according to GoF.
We also observed that the rejection rate of the pessimistic estimators strongly depends on the features Coll and Alg, see the discussion related to Figure \ref{fig:rejection_rate_pes} in Section \ref{sec:exps_struc}.
\item Considering Sn, we observed better results in the false families ratio for the Sn=R estimators  than for the Sn=K estimators, and also better results in the same ratio for the Sn=K estimators than for the Sn=E estimators. In the regression trees generated for the remaining evaluation criteria, we have not observed any dependency on Sn.
\item Considering $d$, while analyzing the results for $d \in \{5, 10\}$, we have not observed any substantial discrepancies from the observations made for $d=15$.
\end{itemize}

To summarize, we suggest the Coll=pre \& Re-est=KTauAvg \& Alg=PT-avg \& Sn=R estimators 
as they have shown good results throughout all of the considered evaluation criteria and due to the robustness of the homogeneous Alg=PT estimators against misspecification of the underlying families. 
If one is interested in a particular evaluation criterion, the regression trees provided above can serve as a guide for choosing an appropriate heterogeneous HAC estimator.

It is also important to note that, even if we observed substantially better results for the estimators based on the inversion of Kendall's tau (Alg=PT) than for the ML estimators, one cannot conclude that one should rather avoid using ML estimation in HAC estimation. We have not given any supporting evidence for such a claim. Consider that the two proposed algorithms involve many other aspects that substantially affect a resulting estimate, e.g., the diagonal transformation, hence one cannot draw a conclusion about a particular aspect of an algorithm (but only about the whole algorithm).

However, the results can be considered as an evidence supporting the claim that it is an advantage to estimate as many components of a HAC (e.g., the structure and the parameters, more precisely, the $\tau$-version of the parameters, including collapsing and re-estimation, i.e., using the Coll=pre \& Alg=PT \& Re-est=KTauAvg estimators) as possible without making an assumption about the underlying families. Such a claim is in accordance with the results for both $\FF$-known and $\FF$-unknown scenarios.

\section{Conclusion} \label{sec:conclusion}
We presented a new approach for estimating HACs, with focus on heterogeneous HAC estimation. It generalizes an existing approach to homogeneous HACs estimation presented in \cite{goreckihofertholena2016approachjiis}, but can equally well be applied to other homogeneous HAC estimators. The latter was done in Appendix \ref{app:diagonal}, 
where the proposed framework was applied to another existing ML-based homogeneous HAC estimator \cite{Okh13,goreckihofertholena2014a,okhrin2015package}. The new approach was summarized by an algorithm in  pseudo-code and its theoretical justification was given by Theorem \ref{thm:alg_returns_HAC}, which shows that under weak conditions the algorithm returns a function satisfying the sufficient nesting condition in arbitrary dimensions. Two scenarios of Theorem \ref{thm:alg_returns_HAC} allowing for estimation of HACs involving up to five and up to four different Archimedean families are presented in Theorems \ref{thm:L12} and  \ref{thm:L123}, respectively. 

Moreover, a contribution concerning collapsing of HAC structures was proposed. The proposed collapsing strategy has an advantage that, by contrast to the strategies proposed in \cite{Okh13,uyttendaele2016estimation}, no pre-defined threshold is needed before a collapsing process has begun. Also, a complementary contribution concerning re-estimation of collapsed HACs based on the Kendall correlation matrix was proposed, together with a new theory describing a close relationship between clustering-based estimation of HAC structures and the sufficient nesting condition. The latter was summarized in Theorem \ref{thm:alg_homo_returns_HAC} and led to a simplification of an existing method for homogeneous HAC estimation proposed in \cite{goreckihofertholena2016approachjiis}.

Complementary to these theoretical contributions, precision and robustness of the proposed approaches were illustrated by experiments on simulated data, based on 12 heterogeneous HAC models with up to five different parametric families of Archimedean generators involved in a single HAC model, 120 heterogeneous and 72 homogeneous HAC estimators. 
The results of these experiments confirmed that the proposed approach is able, apart from properly determining the structure and estimating the parameters of a HAC, to properly estimate different parametric families of Archimedean generators of a heterogeneous HAC. 
The results also confirmed viability of the proposed approach to collapsing, and supremacy of the proposed approach to re-estimation over the approach proposed in \cite{uyttendaele2016estimation}.
Moreover, the considered estimators were compared among each other according to five evaluations criteria, and
suggestions about which estimator to choose were provided, also in a form of regression trees.

This work, of course, opens many new questions, e.g.:
\begin{itemize}
	\item What if one uses in Step 5 of Algorithm \ref{alg:hetero_HAC_estim} the ML estimator instead of the inversion of Kendall's tau? Or, what if one does exactly the opposite in Step 5 of Algorithm \ref{alg:diag_hetero_HAC_estim}?
	\item What if one substitutes the approach to GoF testing used in Algorithm \ref{alg:diag_hetero_HAC_estim} by the approach to GoF testing used in Algorithm \ref{alg:hetero_HAC_estim}? In other words, what if one substitutes Step 7 of Algorithm \ref{alg:diag_hetero_HAC_estim} by Step 7 of Algorithm \ref{alg:hetero_HAC_estim}?
	\item What if one uses the pre-collapsed structure obtained using Algorithm \ref{alg:structure_estim} and KTauAvg in Algorithm \ref{alg:diag_hetero_HAC_estim} instead of the structure estimated there based on the diagonal transformation?
	\item Can these variants be somehow merged (e.g., using some weight for each variant) together in order the get even better fits?
\item In \cite{Hof11,Hof10book}, an approach allowing to construct HACs in even more generality (based on \emph{general nesting transformations of generators} \cite[p. 110]{Hof10book}) is proposed. To which extent can the results proposed in this work be applied to these HACs?
\item Or how will the proposed estimators behave on real-world data?
\end{itemize}
Answering these questions should be considered in further research. %Finally note that a MATLAB toolbox implementing all considered estimators can be obtained from the authors.

%\section*{Acknowledgment}
%Marius Hofert
%acknowledges support from the Natural Sciences and
%Engineering Research Council of Canada (NSERC Discovery Grant
%numbers 5010 and 435844, respectively).

%As future research, we consider to use wide range of GoF statistics and also investigate other possibilities of HAC structure determination.

%Copulas are a feasible tool for the modeling of complex patters. A popular alternative to Gaussian copulas, the hierarchical Archimedean copulas, are convenient copula models even in high dimensions due to their flexibility and rather limited number of parameters. Despite their popularity, a general approach for their estimation has been addressed only in one recently published paper\cite{Okh13}, which proposes several methods for the estimation task.

%We propose an alternative approach to structure determination and estimation of a hierarchical Archimedean copula, which combines the advantages and avoids the disadvantages of the previously mentioned methods in the terms of the correctly determined structures ratio, the goodness-of-fit of the estimates, and computation time. This is confirmed in the experiments on simulated data performed for different dimensions and copula models. 

%\input{diagram}

 \bibliographystyle{tfnlm}

\begin{thebibliography}{10}
\providecommand{\url}[1]{\normalfont{#1}}
\providecommand{\urlprefix}{Available from: }

\bibitem{Skl59}
Sklar~A. Fonctions de r\'{e}partition a n dimensions et leurs marges. Publ Inst
  Stat Univ Paris. 1959;\hspace{0pt}8:229--231.

\bibitem{Nel06}
Nelsen~R. An introduction to copulas. 2nd ed. Springer; 2006.

\bibitem{Joe97}
Joe~H. Multivariate models and dependence concepts. London: Chapman \& Hall;
  1997.

\bibitem{Hofert13}
Hofert~M, M{\"a}chler~M, McNeil~AJ. Archimedean copulas in high dimensions:
  Estimators and numerical challenges motivated by financial applications.
  Journal de la Soci{\'e}t{\'e} Fran{\c{c}}aise de Statistique.
  2013;\hspace{0pt}154(1):25--63.

\bibitem{Hof11}
Hofert~M. Efficiently sampling nested {A}rchimedean copulas. Computational
  Statistics and Data Analysis. 2011;\hspace{0pt}55(1):57--70.

\bibitem{Hof2012stoch}
Hofert~M. A stochastic representation and sampling algorithm for nested
  {A}rchimedean copulas. Journal of Statistical Computation and Simulation.
  2012;\hspace{0pt}82(9):1239--1255.

\bibitem{Okh13}
Okhrin~O, Okhrin~Y, Schmid~W. On the structure and estimation of hierarchical
  {A}rchimedean copulas. Journal of Econometrics.
  2013;\hspace{0pt}173(2):189--204.
  \urlprefix\url{http://www.sciencedirect.com/science/article/pii/S0304407612002667}.

\bibitem{Gor14structure}
G\'{o}recki~J, Hole\v{n}a~M. Structure determination and estimation of
  hierarchical {A}rchimedean copulas based on {K}endall correlation matrix. In:
  Appice~A, Ceci~M, Loglisci~C, et~al., editors. New frontiers in mining
  complex patterns. Springer International Publishing; 2014. Lecture Notes in
  Computer Science; p. 132--147.

\bibitem{goreckihofertholena2014a}
G{\'o}recki~J, Hofert~M, Hole{\v{n}}a~M. On the consistency of an estimator for
  hierarchical {A}rchimedean copulas. In: Tala{\v{s}}ov{\'a}~J, Stoklasa~J,
  Tal{\'a}{\v{s}}ek~T, editors. 32nd international conference on mathematical
  methods in economics. Palack{\'y} University, Olomouc; 2014. p. 239--244.

\bibitem{goreckihofertholena2016approachjiis}
G{\'o}recki~J, Hofert~M, Hole{\v{n}}a~M. An approach to structure determination
  and estimation of hierarchical {A}rchimedean copulas and its application to
  bayesian classification. Journal of Intelligent Information Systems.
  2016;\hspace{0pt}46(1):21--59.

\bibitem{embrechts2016}
Embrechts~P, Hofert~M, Wang~R. Bernoulli and tail-dependence compatibility. Ann
  Appl Probab. 2016 06;\hspace{0pt}26(3):1636--1658.
  \urlprefix\url{http://dx.doi.org/10.1214/15-AAP1128}.

\bibitem{uyttendaele2016estimation}
Uyttendaele~N, et~al. On the estimation of nested {A}rchimedean copulas: A
  theoretical and an experimental comparison. UCL; 2016. Report no.:.

\bibitem{McNei09}
McNeil~AJ, Ne\v{s}lehov\'{a}~J. Multivariate {A}rchimedean copulas,
  \emph{d}-monotone functions and $l_1$-norm symmetric distributions. The
  Annals of Statistics. 2009;\hspace{0pt}37:3059--3097.

\bibitem{widder1946}
Widder~DV. The laplace transform. Princeton University Press; 1946.

\bibitem{Hof10book}
Hofert~M. Sampling nested {A}rchimedean copulas with applications to cdo
  pricing. Suedwestdeutscher Verlag fuer Hochschulschriften; 2010.

\bibitem{Hof11CDO}
Hofert~M, Scherer~M. {CDO} pricing with nested {A}rchimedean copulas.
  Quantitative Finance. 2011;\hspace{0pt}11(5):775--787.

\bibitem{Holena13}
Hole\v{n}a~M, \v{S}\v{c}avnick\'{y}~M. Application of copulas to data mining
  based on observational logic. In: ITAT 2013: Information Technologies -
  Applications and Theory Workshops, Posters, and Tutorials.; Donovaly,
  Slovakia. North Charleston : CreateSpace Independent Publishing Platform;
  2013. p. 77--85.

\bibitem{McNeil08}
McNeil~AJ. Sampling nested {A}rchimedean copulas. Journal of Statistical
  Computation and Simulation. 2008;\hspace{0pt}78(6):567--581.

\bibitem{holenascavnickybajer2015usingcopulas}
Hole\v{n}a~M, Bajer~L, \v{S}\v{c}avnick\'{y}~M. Using copulas in data mining
  based on the observational calculus. IEEE Transactions on Knowledge and Data
  Engineering. 2015;\hspace{0pt}Accepted for publication.

\bibitem{rezapour2015construction}
Rezapour~M. On the construction of nested {A}rchimedean copulas for
  $d$-monotone generators. Statistics \& Probability Letters.
  2015;\hspace{0pt}101:21--32.

\bibitem{Hof08}
Hofert~M. Sampling {A}rchimedean copulas. Computational Statistics \& Data
  Analysis. 2008;\hspace{0pt}52(12):5163 -- 5174.

\bibitem{Hering2010}
Hering~C, Hofert~M, Mai~JF, et~al. Constructing hierarchical {A}rchimedean
  copulas with {L}\`{e}vy subordinators. Journal of Multivariate Analysis.
  2010;\hspace{0pt}101(6):1428 -- 1433.
  \urlprefix\url{http://www.sciencedirect.com/science/article/pii/S0047259X09001961}.

\bibitem{Gen93}
Genest~C, Rivest~LP. Statistical inference procedures for bivariate archimedean
  copulas. Journal of the American statistical Association.
  1993;\hspace{0pt}88(423):1034--1043.

\bibitem{cramer1928composition}
Cram{\'e}r~H. On the composition of elementary errors: First paper:
  Mathematical deductions. Scandinavian Actuarial Journal.
  1928;\hspace{0pt}1928(1):13--74.

\bibitem{Gen09}
Genest~C, R\'{e}millard~B, Beaudoin~D. Goodness-of-fit tests for copulas: A
  review and a power study. Insurance: Mathematics and Economics.
  2009;\hspace{0pt}44(2):199--213.

\bibitem{CART1984}
Breiman~L, Freidman~J, Olshen~R, et~al. Classification and regression trees.
  Wadsworth; 1984.

\bibitem{Gor13MME}
G\'{o}recki~J, Hole\v{n}a~M. An alternative approach to the structure
  determination of hierarchical {A}rchimedean copulas. Jihlava; 2013. p. 201 --
  206; Proceedings of the 31st International Conference on Mathematical Methods
  in Economics (MME 2013).

\bibitem{Clar09}
Clarke~B, Fokoue~E, Zhang~HH. Principles and theory for data mining and machine
  learning. Springer; 2009.

\bibitem{batagelj1981ahc}
Batagelj~V. Note on ultrametric hierarchical clustering algorithms.
  Psychometrika. 1981;\hspace{0pt}46(3):351--352.
  \urlprefix\url{http://dx.doi.org/10.1007/BF02293743}.

\bibitem{Koj10a}
Kojadinovic~I, Yan~J. Modeling multivariate distributions with continuous
  margins using the copula r package. Journal of Statistical Software.
  2010;\hspace{0pt}34(9):1--20.

\bibitem{OkhrinRistig2014HACinR}
Okhrin~O, Ristig~A. Hierarchical {A}rchimedean copulae: The {HAC} package.
  Journal of Statistical Software. 2014 6;\hspace{0pt}58(4).
  \urlprefix\url{http://www.jstatsoft.org/v58/i04}.

\bibitem{Okhrin2015Conditional}
Okhrin~O, Ristig~A, Sheen~JR, et~al. Conditional systemic risk with penalized
  copula. Berlin; 2015. SFB 649 Discussion Paper 2015-038.
  \urlprefix\url{http://hdl.handle.net/10419/121999}.

\bibitem{Segers2014nonparametric}
Segers~J, Uyttendaele~N. Nonparametric estimation of the tree structure of a
  nested {A}rchimedean copula. Computational Statistics \& Data Analysis.
  2014;\hspace{0pt}72:190--204.

\bibitem{okhrin2015package}
Okhrin~O, Ristig~A. Package '{HAC}'.
  2015;\hspace{0pt}\urlprefix\url{ftp://journal.r-project.org/pub/R/web/packages/HAC/HAC.pdf}.

\end{thebibliography}

\begin{appendices}

\section{Proofs (Lemmas  \ref{lem:nnc_equals_snc}, \ref{lem:D_rs_is_monotonic}, \ref{lem:n2f12}, \ref{lem:third_class}, \ref{lem:n2f123}  and Theorems \ref{thm:alg_homo_returns_HAC}, \ref{thm:alg_returns_HAC}, \ref{thm:L12}, \ref{thm:L123})}
\label{app:proofs}

\begin{pf} (\textsc{of Lemma \ref{lem:nnc_equals_snc}})
Let \eqref{eq:in_Theta_a} holds and $\HAC$ be an $a$-homogeneous function given by \eqref{eq:function_hac} with the labeling $\lambda$ defined $\lambda(i) = \psi^{(a, \tau_{(a)}^{-1}(\tau_i))}$ for all $i \in \{d+1,..., d +k\}$. 

Let $v$ and $\tilde{v}$ be two forks from $\VV$ such that $\tilde{v} \in \wedge(v)$. 
Assume that $\tau(\lambda(v)) \leq \tau(\lambda(\tilde{v}))$, i.e., assume \eqref{eq:nnc} for particular $v$ and $\tilde{v}$. 
This can be rewritten to $\tau(\psi^{(a, \tau_{(a)}^{-1}(\tau_v))}) \leq \tau(\psi^{(a, \tau_{(a)}^{-1}(\tau_{\tilde{v}}))})$. 
Using 1), it follows that $\tau_{(a)}(\tau_{(a)}^{-1}(\tau_v))) \leq \tau_{(a)}( \tau_{(a)}^{-1}(\tau_{\tilde{v}}))$.
Using 2), it follows that $\tau_{(a)}^{-1}(\tau_v) \leq  \tau_{(a)}^{-1}(\tau_{\tilde{v}})$.
Using 3), $\tau_{(a)}^{-1}(\tau_v) \leq  \tau_{(a)}^{-1}(\tau_{\tilde{v}})$ is equivalent to $( \psi^{(a,\tau_{(a)}^{-1}(\tau_v))}, \psi^{(a,\tau_{(a)}^{-1}(\tau_{\tilde{v}}))}) \in \PSI$, which establishes the proof.
\end{pf}

\centerline{$\square$}

\begin{pf} (\textsc{of Lemma \ref{lem:D_rs_is_monotonic}})
Clearly, $p$ represents the combined cluster $(s,t)$ (through $\downarrow(p)$), i.e., $D_{r(s,t)} =  D_{rp}$. 
As $-D_{rp} = \avg((\tau^n_{\tilde{i}\tilde{j}})_{(\tilde{i},\tilde{j}) \in \downarrow(r) \times \downarrow(p)})$, and as $\downarrow(p) = \downarrow(s) \cup  \downarrow(t)$, $-D_{rp}$ turns to $\avg((\tau^n_{\tilde{i}\tilde{j}})_{(\tilde{i},\tilde{j}) \in \downarrow(r) \times (\downarrow(s) \cup  \downarrow(t))}) = \avg((\tau^n_{\tilde{i}\tilde{j}})_{(\tilde{i},\tilde{j}) \in (\downarrow(r) \times \downarrow(s)) \cup  (\downarrow(r) \times \downarrow(t)}))) = \avg((\tau^n_{\tilde{i}\tilde{j}})_{(\tilde{i},\tilde{j}) \in \downarrow(r) \times \downarrow(s)},$ $(\tau^n_{\tilde{i}\tilde{j}})_{(\tilde{i},\tilde{j}) \in \downarrow(r) \times \downarrow(t)})$. 
%Obviously, the vector $((\tau^n_{\tilde{i}\tilde{j}})_{(\tilde{i},\tilde{j}) \in \downarrow(r) \times \downarrow(s)})$ contains $\#\downarrow(r) \cdot \#\downarrow(s)$ elements and the vector $(\tau^n_{\tilde{i}\tilde{j}})_{(\tilde{i},\tilde{j}) \in \downarrow(r) \times \downarrow(t)}$ contains $\#\downarrow(r) \cdot\#\downarrow(t)$ elements. 
Let
\begin{flalign}
m = \#\downarrow(r) \cdot\#\downarrow(s) +\#\downarrow(r) \cdot\#\downarrow(t).
\label{eq:D_rs_m}
\end{flalign} 
Consider that $\avg((\tau^n_{\tilde{i}\tilde{j}})_{(\tilde{i},\tilde{j}) \in \downarrow(r) \times \downarrow(s)}) = \frac{1}{\#\downarrow(r) \#\downarrow(s)}\sum((\tau^n_{\tilde{i}\tilde{j}})_{(\tilde{i},\tilde{j}) \in \downarrow(r) \times \downarrow(s)})$ and $\avg((\tau^n_{\tilde{i}\tilde{j}})_{(\tilde{i},\tilde{j}) \in \downarrow(r) \times \downarrow(t)})$ $=$ $\frac{1}{\#\downarrow(r) \#\downarrow(t)}$ $\sum ((\tau^n_{\tilde{i}\tilde{j}})_{(\tilde{i},\tilde{j}) \in \downarrow(r) \times \downarrow(t)})$, where $\sum(a_1, ..., a_w)$ denotes $\sum_{i=1}^{w}a_i$.
Hence, we continue with $\avg((\tau^n_{\tilde{i}\tilde{j}})_{(\tilde{i},\tilde{j}) \in \downarrow(r) \times \downarrow(s)},$ $(\tau^n_{\tilde{i}\tilde{j}})_{(\tilde{i},\tilde{j}) \in \downarrow(r) \times \downarrow(t)})$ $= \frac{1}{m}\bigl(\sum((\tau^n_{\tilde{i}\tilde{j}})_{(\tilde{i},\tilde{j}) \in \downarrow(r) \times \downarrow(s)}) + \sum((\tau^n_{\tilde{i}\tilde{j}})_{(\tilde{i},\tilde{j}) \in \downarrow(r) \times \downarrow(t)})\bigr) = $ 
%$\sum((\frac{1}{m}\tau^n_{\tilde{i}\tilde{j}})_{(\tilde{i},\tilde{j}) \in \downarrow(r) \times \downarrow(s)},$ $(\frac{1}{m}\tau^n_{\tilde{i}\tilde{j}})$ $_{(\tilde{i},\tilde{j}) \in \downarrow(r) \times \downarrow(t)}) = $
$\sum((\frac{1}{m}\tau^n_{\tilde{i}\tilde{j}})_{(\tilde{i},\tilde{j}) \in \downarrow(r) \times \downarrow(s)})+$ $\sum((\frac{1}{m}\tau^n_{\tilde{i}\tilde{j}})_{(\tilde{i},\tilde{j}) \in \downarrow(r) \times \downarrow(t)})$ $ = $
$\sum((\frac{\#\downarrow(r) \#\downarrow(s)}{m\#\downarrow(r) \#\downarrow(s)}\tau^n_{\tilde{i}\tilde{j}})$ $_{(\tilde{i},\tilde{j}) \in \downarrow(r) \times \downarrow(s)}) + $ $\sum((\frac{\#\downarrow(r) \#\downarrow(t)}{m\#\downarrow(r) \#\downarrow(t)}\tau^n_{\tilde{i}\tilde{j}})_{(\tilde{i},\tilde{j}) \in \downarrow(r) \times \downarrow(t)})$ = 
$\frac{\#\downarrow(r) \#\downarrow(s)}{m}$ $\avg((\tau^n_{\tilde{i}\tilde{j}})_{(\tilde{i},\tilde{j}) \in \downarrow(r) \times \downarrow(s)})+\frac{\#\downarrow(r) \#\downarrow(t)}{m}\avg((\tau^n_{\tilde{i}\tilde{j}})_{(\tilde{i},\tilde{j}) \in \downarrow(r) \times \downarrow(t)})$
which finally turns to $-\frac{\#\downarrow(r) \#\downarrow(s)}{m}$ $D_{rs} - \frac{\#\downarrow(r) \#\downarrow(t)}{m}D_{rt}$.
Thus $D_{r(s,t)} = \frac{\#\downarrow(r) \#\downarrow(s)}{m}D_{rs} + \frac{\#\downarrow(r) \#\downarrow(t)}{m}D_{rt}$. Considering \eqref{eq:D_rs_m} establishes the proof. %$\hfill~ \hfill\qed$
\end{pf}

\centerline{$\square$}

%begin from Theorem 3
%\setcounter{thm2}{2}

\begin{pf} (\textsc{of  Theorem \ref{thm:alg_homo_returns_HAC}})
Corollary \ref{cor:alg_satisfies_nnc} implies that for any $(\tau^n_{ij})$ the triplet $(\hat{\VV}, \hat{\EE}, (\hat{\tau}_k)_{d+1}^{2d-1})$ returned by Algorithm \ref{alg:structure_estim} satisfies \eqref{eq:nnc}  provided $\tau_k = \hat{\tau}_k$ for all $k \in \{d+1, ..., 2d-1\}$.
Under \eqref{eq:hat_in_Theta_a} and the assumptions on $a$, it follows from Lemma \ref{lem:nnc_equals_snc} that $C_{(\hat{\VV}, \hat{\EE}, \hat{\lambda})}$ with the labeling $\hat{\lambda}$ defined $\hat{\lambda}(k) = \psi^{(a, \tau_{(a)}^{-1}(\hat{\tau}_k))}$ for all $k \in \{d+1,..., 2d -1\}$ satisfies \eqref{eq:holena_nesting_cond_simple}. %$\hfill\qed$
\end{pf}

%\begin{thm2} 
%Let $\mathcal{F} \subseteq \FALL$,  $\mathcal{N}_0 \in \FR$ be Archimedean, $\mathcal{N}^2_{\mathcal{F}}$ be given by Definition \ref{def:n2} and the condition
%\begin{flalign}
%\exists (a_1, \theta_1), a_1 \in \mathcal{F}, \theta_1 \in [0, +\infty) ~\textrm{such that}~
%(a_1, \theta_1) ~ \tilde{\in} ~ \mathcal{N}_0 \tilde{\cap} \mathcal{N}^2_{\mathcal{F}}(a_2, \theta_2)  \tilde{\cap}   \mathcal{N}^2_{\mathcal{F}}(a_3, \theta_3)
%\end{flalign}
%holds for all $(a_2, \theta_2) \tilde{\in} \mathcal{N}_0$ and $(a_3, \theta_3) \tilde{\in} \mathcal{N}_0$. Then, given any inputs 1) and 5)-8), Algorithm \ref{alg:hetero_HAC_estim} returns a triplet $(\hat{\VV}, \hat{\EE}, \hat{\lambda})$ satisfying \eqref{eq:holena_nesting_cond}, i.e., $C_{(\hat{\VV}, \hat{\EE}, \hat{\lambda})}$ is a copula.
%\end{thm2}

\centerline{$\square$}

\begin{pf}(\textsc{of Theorem \ref{thm:alg_returns_HAC}})
%how algorithm works
%We use elements of $\FR$ to represent the parameter constraints for the parent of $i \in \hat{\VV}$ in the following way. In Input 3), the algorithm asks for $\mathcal{N}_0 \in \FR$. In Step 2, we partially define the mapping $\mathcal{N}^1 ~:~ \hat{\VV} \rightarrow \FR$ by $\mathcal{N}^1(i) = \mathcal{N}_0, i = 1, ..., d$, i.e., the parameter constraints for the parent of the leaf $i$ are stored in $\mathcal{N}^1(i)$. For $i \in \{d+1, ..., 2d-1\}$, $\mathcal{N}^1$ is defined during the algorithm.

We will show that the output $(\hat{\VV}, \hat{\EE}, \hat{\lambda})$ satisfies \eqref{eq:holena_nesting_cond}  for any inputs 1), 5) and 6). Given $k \in \{1, ..., d-1\}$, the children of the fork $d+k$ is $\{i, j\}$ (Step 3). 
Without a loss of generality, we can assume that $i < j$.
Hence showing that $(\hat{\VV}, \hat{\EE}, \hat{\lambda})$ satisfies \eqref{eq:holena_nesting_cond} transforms to showing either, if $i \leq d$ and $j >d$, then
\begin{flalign}
\label{eq:holena_nesting_cond_transformed1}
 (\hat{\lambda}(d+k), \hat{\lambda}(j)) \in \tilde{\Psi}^2_{+\infty},
\end{flalign}
or, if $i, j > d$, then
\begin{flalign}
\label{eq:holena_nesting_cond_transformed2}
(\hat{\lambda}(d+k), \hat{\lambda}(i)) \in \tilde{\Psi}^2_{+\infty} ~\&~ (\hat{\lambda}(d+k), \hat{\lambda}(j)) \in \tilde{\Psi}^2_{+\infty},
\end{flalign}
for all $k \in \{1, ..., d-1\}$.

The proof is performed by induction according to $k \in \{1, ..., d-1\} $. Assume the first loop ($k = 1$). In Step 4, as $i, j \in \{1, ..., d\}$ (see Step 3), it follows that $\mathcal{N} = \mathcal{N}^1(i) \tilde{\cap} \mathcal{N}^1(j) = \mathcal{N}_0 \tilde{\cap} \mathcal{N}_0 = \mathcal{N}_0$, i.e, $\NNN$ is Archimedean. 
Due to trimming in Step 6, it holds that $(a_l, \hat{\theta}_l) \tilde{\in} \mathcal{N}, l = 1, ..., \#\mathcal{N}$, what assures that $\hat{\lambda}(d+1)$ (Step 8) is a c.m. generator, see Definition \ref{def:set_of_generators}. In Step 9, define $\mathcal{N}^1(d+1) := \mathcal{N} \tilde{\cap} \mathcal{N}^2_{\FF}(a_{l^*}, \hat{\theta}_{l^*})$. 
As $\mathcal{N} \tilde{\cap} \mathcal{N}^2_{\FF}(a_{l^*}, \hat{\theta}_{l^*}) = \mathcal{N} \tilde{\cap} \mathcal{N}^2_{\FF}(a_{l^*}, \hat{\theta}_{l^*}) \tilde{\cap} \mathcal{N}^2_{\FF}(a_{l^*}, \hat{\theta}_{l^*})$, see Remark \ref{rmk:semigroup}, and $(a_{l^*}, \hat{\theta}_{l^*})\in \mathcal{N}_0$, it follows from \eqref{eq:non_empty_cond} that $\mathcal{N}^1(d+1) \neq \emptyset$. Hence, for any $(a, \theta) \tilde{\in} \mathcal{N}^1(d+1)$ holds that  $(\psi^{(a, \theta)}, \psi^{(a_{l^*}, \hat{\theta}_{l^*})}) \in \tilde{\Psi}^2_{+\infty}$, see Definition \ref{def:n2}.

The induction step. In a $k$-th loop, $k \geq 2$, assume that for all $(a,\theta)\tilde{\in}\mathcal{N}^1(d+q)$, the pair $(\psi^{(a,\theta)}, \hat{\lambda}(d+q)) \in \tilde{\Psi}^2_{+\infty}$ for all $q \in \{1, .., k - 1\}$. After Step 3 (without loss of generality, assume $i < j$), distinguish 3 cases:

\begin{enumerate}
\item $i, j \leq d$. In this case, the computation of $\hat{\lambda}(d+k)$ is analogous to the first ($k = 1$) loop, and thus $\hat{\lambda}(d+k) \in \Psi_{+\infty}$. Step 9 assures that for any $(a, \theta) \tilde{\in} \mathcal{N}^1(d+k)$ holds that $(\psi^{(a, \theta)},\hat{\lambda}(d+k)) \in \tilde{\Psi}^2_{+\infty}$. Also, it follows from \eqref{eq:non_empty_cond} that $\mathcal{N}^1(d+k) \neq \emptyset$.

\item $i \leq d$ and $j > d$. Thus $\mathcal{N}^1(i) = \mathcal{N}_0$. Use the induction, which assures that $\forall (a, \theta) \tilde{\in} \mathcal{N}^1(j)$, $(\psi^{(a, \theta)}, \hat{\lambda}(j)) \in \tilde{\Psi}^2_{+\infty}$.
Then in Step 4, $\mathcal{N} := \mathcal{N}_0 \tilde{\cap} \mathcal{N}^1(j)$. 
Due to trimming in Step 6, it holds that $(a_l, \hat{\theta}_l) \tilde{\in} \mathcal{N}, l = 1, ..., \#\mathcal{N}$. Hence, $\hat{\lambda}(d+k)$ (obtained in Step 8) is a c.m. generator (it follows from the fact that  $(a_l, \hat{\theta}_l) \tilde{\in} \mathcal{N}_0, ~ l = 1, ..., \# \mathcal{N}$), and that ($\hat{\lambda}(d+k), \hat{\lambda}(j))  \in \tilde{\Psi}^2_{+\infty}$ (this follows from the fact that  $(a_l, \hat{\theta}_l) \tilde{\in} \mathcal{N}^1(j), ~ l = 1, ..., \# \mathcal{N}$). Step 9 assures that for any $(a, \theta) \tilde{\in} \mathcal{N}^1(d+k)$ holds that 
$(\psi^{(a, \theta)},\hat{\lambda}(d+k)) \in \tilde{\Psi}^2_{+\infty}$.
The condition \eqref{eq:non_empty_cond} assures that $\mathcal{N}^1(d+k) \neq \emptyset$. 
%The case where $j \leq d, i > d$ is not possible due to the condition for the domain of maximization in Step 5.

\item
$i, j > d$. Use the induction, which assures that $\forall (a, \theta) \tilde{\in} \mathcal{N}^1(i)$, $(\psi^{(a, \theta)}, \hat{\lambda}(i)) \in \tilde{\Psi}^2_{+\infty}$ and $\forall (a, \theta) \tilde{\in} \mathcal{N}^1(j)$, $(\psi^{(a, \theta)}, \hat{\lambda}(j)) \in \tilde{\Psi}^2_{+\infty}$.
In Step 4, $\mathcal{N} := \mathcal{N}^1(i) \tilde{\cap} \mathcal{N}^1(j)$. 
Due to trimming in Step 6, it holds that $(a_l, \hat{\theta}_l) \tilde{\in} \mathcal{N}, l = 1, ..., \#\mathcal{N}$. 
Consequently, $\hat{\lambda}(d+k)$ (obtained in Step 8) is a c.m. generator and holds that
($\hat{\lambda}(d+k), \hat{\lambda}(i)) \in \tilde{\Psi}^2_{+\infty}$ and  $(\hat{\lambda}(d+k), \hat{\lambda}(j)) \in \tilde{\Psi}^2_{+\infty}$. 
Step 9 assures that for any $(a, \theta) \tilde{\in} \mathcal{N}^1(d+k)$ holds that 
$(\psi^{(a, \theta)},\hat{\lambda}(d+k)) \in \tilde{\Psi}^2_{+\infty}$ and
and the condition \eqref{eq:non_empty_cond} assures that $\mathcal{N}^1(d+k) \neq \emptyset$. 
\end{enumerate}
Finally, let $k \in \{1, ..., d-1\}$ and $\wedge(d+k) = \{i, j\}, ~ i < j$. Also, let, analogously to the convention stated in Remark \ref{rem:convention}, $\hat{\lambda}(d+k) = \psi^{(a_{d+k}, \theta_{d+k})}$. Then, it holds that $(a_{d+k}, \theta_{d+k}) \tilde{\in} \NNN^1(j)$ provided $i \leq d$ and $j > d$, and thus $(\hat{\lambda}(d+k), \hat{\lambda}(j)) \in \PSI$. Also,  it holds that $(a_{d+k}, \theta_{d+k}) \tilde{\in} \NNN^1(i) \tilde{\cap} \NNN^1(j)$ provided $i, j> d$ and thus $(\hat{\lambda}(d+k), \hat{\lambda}(i)) \in \PSI$ and $(\hat{\lambda}(d+k), \hat{\lambda}(j)) \in \PSI$.
%$\hfill\qed$
\end{pf}

\centerline{$\square$}

%\begin{lemma2} 
%Let $\mathcal{N}_{\mathcal{F}_{24}} \in \FR$ be $\mathcal{F}_{24}$-Archimedean. Then the mapping defined by
%\begin{equation}
%\mathcal{N}^2_{\mathcal{F}_{24}}(a, \theta) = \left\{ \begin{array}{ll}
%\{ (\textrm{C}, (0, \theta] ) \} & 		\textrm{if} ~ a = \textrm{C}\\
%\{ (\textrm{C}, (0, 1] ), (12, [1, \theta]\} &  \textrm{if} ~ a = 12\\
%\{ (\textrm{C}, (0, \frac{1}{\theta}] )\} & \textrm{if} ~ a = 14\\
%\{ (\textrm{C}, (0, 1] ), (19, [0, \theta]\} & \textrm{if} ~ a = 19\\
 %\{ (\textrm{C}, (0, \theta] ), (20, [0, \theta])\} & \textrm{if}  ~ a = 20\\
%\end{array} \right.
%\end{equation}
%for all $(a, \theta) \tilde{\in} \mathcal{N}_{\mathcal{F}_{24}}$ is the mapping given by Definition \ref{def:n2} for $\mathcal{F} = \mathcal{F}_{24}$.
%\end{lemma2}

\begin{pf} (\textsc{of Lemma \ref{lem:n2f12}})
Start with the $a$ = C case. We show for all (C, $\theta_1$), (C, $\theta_2) \tilde{\in}  \NNN_{\mathcal{F}_{24}}$ that 
(C$,\theta_1)\tilde{\in}\mathcal{N}^2_{\mathcal{F}_{24}}(\textrm{C},\theta_2)$, if and only if $(\psi^{(\textrm{C}, \theta_1)}, \psi^{(\textrm{C}, \theta_2)}) \in \PSI$. Note that it follows from Definition \ref{def:n2} that $(a_1, \theta_1) \tilde{\in} \mathcal{N}^2_{\mathcal{F}_{24}}(\textrm{C}, \theta_2)$ does not hold for any $a_1 \in \{12, 14, 19, 20\}, \theta_1 \in [0, +\infty), \theta_2 \in (0, +\infty)$.
\\$(\Rightarrow)$ 
Assume that (C$, \theta_1) \tilde{\in} \mathcal{N}^2_{\mathcal{F}_{24}}(\textrm{C},\theta_2), \theta_2 \in (0, +\infty)$. As
$\mathcal{N}^2_{\mathcal{F}_{24}}(\textrm{C}, \theta_2) = $ $\{ (\textrm{C}, (0,\theta_2] ) \}$, according to Definition \ref{def:nesting_semigroup}, there exists $r \in \RR$ such that (C, $r) \in \{ (\textrm{C}, (0, \theta_2] ) \}$ and $\theta_1 \in r$. It implies that $r = (0, \theta_2]$ and thus $\theta_1 \in (0, \theta_2]$ and hence, $(\psi^{(\textrm{C}, \theta_1)}, \psi^{(\textrm{C}, \theta_2)}) \in \PSI$, cf. Table \ref{tab:geners}. 
\\$(\Leftarrow)$
For any $\theta_1, \theta_2 \in (0, +\infty)$, assume that $(\psi^{(\textrm{C}, \theta_1)}, \psi^{(\textrm{C}, \theta_2)}) \in \PSI$. Then $\theta_1 \leq \theta_2$, which assures that (C$, \theta_1) \tilde{\in} \mathcal{N}^2_{\mathcal{F}_{24}}(\textrm{C}, \theta_2)$.\\

\noindent Now the $a$ = 12 case. Firstly, we show, similarly to the $a = $ C case, for all (12, $\theta_1$), (12, $\theta_2) \tilde{\in}  \NNN_{\mathcal{F}_{24}}$ that 
(12$,\theta_1)\tilde{\in}\mathcal{N}^2_{\mathcal{F}_{24}}(12,\theta_2)$, if and only if $(\psi^{(\textrm{12}, \theta_1)}, \psi^{(\textrm{12}, \theta_2)}) \in \PSI$. 
\\$(\Rightarrow)$ 
Assume that (12$, \theta_1) \tilde{\in} \mathcal{N}^2_{\mathcal{F}_{24}}($12$,\theta_2), \theta_2 \in [1, +\infty)$. As
$\mathcal{N}^2_{\mathcal{F}_{24}}($12$, \theta_2) = \{ (\textrm{C}, (0, 1]),$ $(12, [1, \theta_2] ) \}$, according to Definition \ref{def:nesting_semigroup}, there exists (12, $r) \in \{ (\textrm{C}, (0, 1]),$ $(12, [1, \theta_2] ) \}$ such that $\theta_1 \in r$. It implies that $\theta_1 \in [1, \theta_2]$ and hence, $(\psi^{(\textrm{12}, \theta_1)}, \psi^{(\textrm{12}, \theta_2)}) \in \PSI$, cf. Table \ref{tab:geners}. 
\\$(\Leftarrow)$
For any $\theta_1, \theta_2 \in [1, +\infty)$, assume that $(\psi^{(\textrm{12}, \theta_1)}, \psi^{(\textrm{12}, \theta_2)}) \in \PSI$. Then $\theta_1 \leq \theta_2$, which assures that (12$, \theta_1) \tilde{\in} \mathcal{N}^2_{\mathcal{F}_{24}}($12$, \theta_2)$.\\

Now we show for all (C, $\theta_1$), (12, $\theta_2) \tilde{\in}  \mathcal{F}_{24}$ that 
(C$,\theta_1)\tilde{\in}\mathcal{N}^2_{\mathcal{F}_{24}}(12,\theta_2)$, if and only if $(\psi^{(\textrm{C}, \theta_1)}, \psi^{(\textrm{12}, \theta_2)}) \in \PSI$.
Note that it follows from Definition \ref{def:n2} that $(a_1, \theta_1) \tilde{\in} \mathcal{N}_{\mathcal{F}_{24}}(12, \theta_2)$ does not hold for any $a_1 \in \{14, 19, 20\}, \theta_1 \in [0, +\infty), \theta_2 \in [1, +\infty)$.
\\$(\Rightarrow)$ 
Assume that (C$, \theta_1) \tilde{\in} \mathcal{N}^2_{\mathcal{F}_{24}}($12$,\theta_2), ~\theta_2 \in [1, +\infty)$. As
$\mathcal{N}^2_{\mathcal{F}_{24}}($12$, \theta_2) = \{ (\textrm{C}, (0, 1]),$ $(12, [1, \theta_2] ) \}$, according to Definition \ref{def:nesting_semigroup}, there exists (C, $r) \in \{ (\textrm{C}, (0, 1]), (12, [1, \theta_2] ) \}$ such that $\theta_1 \in r$. It implies that $\theta_1 \in (0, 1]$ and hence, $(\psi^{(\textrm{C}, \theta_1)}, \psi^{(\textrm{12}, \theta_2)}) \in \PSI$, cf. Table \ref{tab:geners_comb}. 
\\$(\Leftarrow)$
For any $\theta_1 \in (0, +\infty), \theta_2 \in [1, +\infty)$, assume that $(\psi^{(\textrm{C}, \theta_1)}, \psi^{(\textrm{12}, \theta_2)}) \in \PSI$. Then $\theta_1 \in (0, 1]$, cf. Table \ref{tab:geners_comb}, which assures that (C$, \theta_1) \tilde{\in} \mathcal{N}^2_{\mathcal{F}_{24}}($12$, \theta_2)$.\\	

\noindent The proof for the rest of the cases ($a$ = 14, 19, 20) is analogous. %$\hfill\qed$\\
\end{pf}

\centerline{$\square$}

%\begin{thm2}
%Given the inputs $\mathcal{F} = \mathcal{F}_{24}, ~\mathcal{N}^2_{\mathcal{F}} = \mathcal{N}^2_{\mathcal{F}_{24}}, ~\mathcal{N}_0 = \mathcal{N}_{\mathcal{F}_{24}}$ and any inputs 1) and 5)-7), Algorithm \ref{alg:hetero_HAC_estim} returns the triplet $(\hat{\VV}, \hat{\EE}, \hat{\lambda})$ satisfying \eqref{eq:holena_nesting_cond}.
%\end{thm2}

\begin{pf} (\textsc{of Theorem \ref{thm:L12}})  Firstly, we will show that \eqref{eq:non_empty_cond} holds for all $(a_2, \theta_2) \tilde{\in} {\NNN_{\mathcal{F}_{24}}}$ and $(a_3, \theta_3) \tilde{\in}$ $\mathcal{N}_{\mathcal{F}_{24}}$. We know that (C$, (0, +\infty)) \in \NNN_{\mathcal{F}_{24}}$. Also, it follows from the explicit representation of $\mathcal{N}^2_{\mathcal{F}_{24}}$ given in Lemma \ref{lem:n2f12} that there exist $\theta_1, \theta_2 \in (0, +\infty)$  such that      (C, $ \theta_1) \tilde{\in} \mathcal{N}^2_{\mathcal{F}_{24}}(a_2, \theta_2)$ and (C, $\theta_2) \tilde{\in} \mathcal{N}^2_{\mathcal{F}_{24}}(a_3, \theta_3)$. Hence, for any $0 < \theta \leq \min(\theta_1, \theta_2)$, it holds that
(C$, \theta) ~ \tilde{\in} ~ \mathcal{N}_{\mathcal{F}_{24}} \tilde{\cap} \mathcal{N}^2_{\mathcal{F}_{24}}(a_2, \theta_2)  \tilde{\cap}   \mathcal{N}^2_{\mathcal{F}_{24}}(a_3,$ $ \theta_3)$, see also Lemma \ref{lem:n2f12}. Now, applying Theorem \ref{thm:alg_returns_HAC} for $\mathcal{F} = \mathcal{F}_{24}, ~\mathcal{N}^2_{\mathcal{F}} = \mathcal{N}^2_{\mathcal{F}_{24}}$ and $\mathcal{N}_0 = \mathcal{N}_{\mathcal{F}_{24}}$, the statement is proved. %$\hfill\qed$
\end{pf}

\centerline{$\square$}

%\begin{lemma2} 
%Let  $\mathcal{F}\subseteq \FALL$ be such that C $\in \FF$ or 20 $\in \FF$, $\HAC$ be a $\FF$-heterogeneous $d$-HAC satisfying \eqref{eq:holena_nesting_cond}. By convention, $\lambda(i) = \psi^{(a_i, \theta_i)}$ for all $i \in \{d+1, ..., 2d-1\}$. Also, let for all $i, j \in \{d+1, ..., 2d-1\}$ such that $i = \uparrow(j)$ be satisfied $(a_i, a_j) \in \FNEST$ (see \eqref{eq:known_nestables}).
%Then it holds that, if $a_i$ = C or $a_i$ = 20 for some $i \in \{d+1, ..., 2d-1\}$ and $\theta_i < 1$, then $a_j \neq $ A for all $j \in \{d+1, ..., 2d-1\}$.
%\end{lemma2}

\begin{pf}(\textsc{of Lemma \ref{lem:third_class}})
Consider $i, j \in \{d+1, ..., 2d-1\}$ such that $i = \uparrow(j)$.
If $a_j =$ A, then $a_i = $ A, cf. Tables \ref{tab:geners}, \ref{tab:geners_comb}. It in turn implies that for all $m \in \Uparrow(q)$ holds that $a_m =$ A provided $q \in \{d+1, ..., 2d-1\}$ such that $a_q =$ A.
Similarly, if $a_j =$ C such that $\theta_j  < 1$, then $a_i =$ C and $\theta_i \leq \theta_j$.
Note that it is not possible that $a_j = $ A as the assumed value of $\theta_j$ violates the s.n.c. for the combination (A, C), cf. Table \ref{tab:geners_comb}. It in turn implies that for all $m \in \Uparrow(q)$ holds that $a_m =$ C and $\theta_m < 1$ provided $q \in \{d+1, ..., 2d-1\}$ such that $a_q =$ C and $\theta_q < 1$. The same consideration can be made for $a_j$ = 20, i.e., for all $m \in \Uparrow(q)$ holds that $a_m =$ 20 and $\theta_m < 1$ provided $q \in \{d+1, ..., 2d-1\}$ such that $a_q =$ 20 and $\theta_q < 1$.

Now assume that \{A, C\} $\subseteq \mathcal{F}$ and $a_i =$ A, $a_j =$ C, $\theta_j < 1$,  and $i, j \in \{d+1, ..., 2d-2\}, ~i \neq j$.
As mentioned above, it holds that for all $m \in \Uparrow(i)$ that $a_m =$ A and for all  $q \in \Uparrow(j)$ that  $a_q =$ C. However, this contradicts for  $a_{2d-1}$  as $2d-1 \in \Uparrow(i) \cap \in \Uparrow(j)$. A proof for the case in which $a_{2d-1} =$ A or for the case when $a_{2d-1} =$ C and $\theta_{2d-1} < 1$ is clear.
A proof for the case \{A, 20\} $\subseteq \mathcal{F}$ is analogous. %$\hfill\qed$\\
\end{pf}

\centerline{$\square$}
%\begin{lemma2}
%Let $\tilde{\mathcal{N}}_{\mathcal{F}_{1234}} \in \FR$ be $\mathcal{F}_{1234}$-Archimedean. Then the mapping defined by
%\begin{equation}
%\mathcal{N}^2_{\mathcal{F}_{1234}}(a, \theta) = \left\{ \begin{array}{ll}
%\{ (\textrm{A}, [0, \theta] ) \} & 		\textrm{if} ~ a = \textrm{A}\\
%\{ (\textrm{C}, (0, \theta] ) \} & 		\textrm{if} ~ a = \textrm{C}, ~ \theta < 1 \\
%\{ (\textrm{A}, [0, 1) ), (\textrm{C}, (0, \theta] ) \} & 		\textrm{if} ~ a = \textrm{C}, ~ \theta \geq 1 \\
%\{ (\textrm{A}, [0, 1) ), (\textrm{C}, (0, 1] ), (19, (0, \theta]\} & \textrm{if} ~ a = 19\\
 %\{ (\textrm{C}, (0, \theta] ), (20, (0, \theta])\} & \textrm{if}  ~ a = 20,~\theta < 1\\
 %\{ (\textrm{A}, [0, 1) ), (\textrm{C}, (0, \theta] ), (20, (0, \theta])\} & \textrm{if}  ~ a = 20, ~ \theta \geq 1 \\
%\end{array} \right.
%\end{equation}
%for all $(a, \theta) \tilde{\in} \tilde{\mathcal{N}}_{\mathcal{F}_{1234}}$ is the mapping from Definition \ref{def:n2} for $\mathcal{F} = \mathcal{F}_{1234}$.
%\end{lemma2}

\begin{pf}(\textsc{of Lemma \ref{lem:n2f123}})
Using the consequences of Lemma \ref{lem:third_class}, the proof is analogous to the proof of Lemma \ref{lem:n2f12}.%$\hfill\qed$
\end{pf}

\centerline{$\square$}

%\begin{thm2}
%Given the inputs $\mathcal{F} = \mathcal{F}_{1234},~ \mathcal{N}^2_{\mathcal{F}} = \mathcal{N}^2_{\mathcal{F}_{1234}}, ~\mathcal{N}_0 = \mathcal{N}_{\mathcal{F}_{1234}}$ and any inputs 1) and 5)-7), Algorithm \ref{alg:hetero_HAC_estim} returns the triplet $(\hat{\VV}, \hat{\EE}, \hat{\lambda})$ satisfying \eqref{eq:holena_nesting_cond}.
%\end{thm2}

\begin{pf}(\textsc{of Lemma \ref{thm:L123}}) Firstly, we will show that \eqref{eq:non_empty_cond} holds for all $(a_2, \theta_2) \tilde{\in} {\NNN_{\mathcal{F}_{1234}}}$ and $(a_3, \theta_3)$ $\tilde{\in} \mathcal{N}_{\mathcal{F}_{1234}}$. We know that (A$, [0, 1)) \in \NNN_{\mathcal{F}_{1234}}$. Also, it follows from the explicit representation of $\mathcal{N}^2_{\mathcal{F}_{1234}}$ given in Lemma \ref{lem:n2f123} (and regarding the consequences of Lemma \ref{lem:third_class}) that there exist $\theta_1, \theta_2 \in [0, 1)$  such that      (A, $ \theta_1) \tilde{\in} \mathcal{N}^2_{\mathcal{F}_{1234}}(a_2, \theta_2)$ and (A, $\theta_2) \tilde{\in} \mathcal{N}^2_{\mathcal{F}_{1234}}(a_3, \theta_3)$. Hence, for any $0 \leq \theta \leq \min(\theta_1, \theta_2)$, it holds that
(A$, \theta) ~ \tilde{\in} ~ \mathcal{N}_{\mathcal{F}_{1234}}$ $\tilde{\cap} \mathcal{N}^2_{\mathcal{F}_{1234}}(a_2, \theta_2)  \tilde{\cap}   \mathcal{N}^2_{\mathcal{F}_{1234}}(a_3,$ $ \theta_3)$, see also Lemma \ref{lem:n2f123}. Now, applying Theorem \ref{thm:alg_returns_HAC} for $\mathcal{F} = \mathcal{F}_{1234}, ~\mathcal{N}^2_{\mathcal{F}} = \mathcal{N}^2_{\mathcal{F}_{1234}}$ and $\mathcal{N}_0 = \mathcal{N}_{\mathcal{F}_{1234}}$, the statement is proved. %$\hfill\qed$
\end{pf}

\section{The diagonal heterogeneous  HAC estimator}
\label{app:diagonal}

\begin{algorithm}[t]
\floatname{algorithm}{Algorithm}
\caption{The diagonal heterogeneous HAC estimation}
\label{alg:diag_hetero_HAC_estim}

\begin{algorithmic}
\renewcommand{\algorithmicrequire}{\textbf{Input:}}
\renewcommand{\algorithmicensure}{\textbf{Output:}}

\REQUIRE 
\STATE the same inputs as for Algorithm \ref{alg:hetero_HAC_estim} except 5), which is not necessary here

\STATE ~
\renewcommand{\algorithmicensure}{\textbf{The estimation:}}
\ENSURE

\STATE 1. compute $(\tau^n_{ij})$ for  $(U_{i1}, ..., U_{id}),~ i = 1, ..., n$
\STATE 2.  $\hat{\VV} = \{1, ..., 2d-1\}, ~\hat{\EE} = \emptyset,~ \III = \{1, ..., d\}$
\STATE 3. $\mathcal{N}^1(k) := \mathcal{N}_0, ~k = 1, ...,d$

\FOR{$k = 1, ..., d - 1$}

\STATE 3. find two nodes corresponding to the maximum to join, i.e., 
\STATE ~~~ $(i, j) :=\argmax\limits_{\tilde{i} < \tilde{j},~ \tilde{i} \in \mathcal{I}, ~\tilde{j} \in \mathcal{I}} (\tau^n_{\tilde{i}\tilde{j}})$
\STATE 4. (*) $\{(a_1, r_1), ..., (a_{\#\mathcal{N}}, r_{\#\mathcal{N}})\} = \mathcal{N} := \mathcal{N}^1(i) \tilde{\cap} \mathcal{N}^1(j)$ %... $\tilde{\cap}$ is defined in Definition \ref{def:nesting_semigroup}
\FOR{$l = 1, ..., \# \mathcal{N}$}
\STATE 5. estimate the parameter using ML assuming the family $a_l$, i.e., 
\STATE ~~~ $\hat{\theta}_l :=  \textrm{MLE}( (U_{\bullet i}, U_{\bullet j}), \psi^{(a_l, \theta)})$ 
\STATE 6. $\hat{\theta}_l := \textrm{trim} (\hat{\theta}_l,r_l)$
\ENDFOR	
	\STATE 7. $l^* := \argmin\limits_{l \in \{1, ..., \#\mathcal{N}\}} S_n( (U_{\bullet i}, U_{\bullet j}), \psi^{(a_l, \theta_l)})$ %... $S_n^{g_2}$ is defined in Definition \ref{def:agg_s_n}
\STATE 8. $\hat{\lambda}(d+k) := \psi^{(a_{l^*}, \hat{\theta}_{l^*})}$ 
\STATE 9. (**) $\mathcal{N}^1(d+k) := \mathcal{N} \tilde{\cap} \mathcal{N}^2_{\mathcal{F}}(a_{l^*}, \hat{\theta}_{l^*})$ 
\STATE 10. compute new observations corresponding to the joint of $i$ and $j$, i.e., 
\STATE ~~~~ $U_{m, d+k} := \hat{\lambda}(d+k)(2\hat{\lambda}(d+k)^{-1}(\max\{U_{mi}, U_{mj}\})$ for all $m \in \{1, ..., n\}$
\STATE 11. compute $\tau^n_{d+k,s} (= \tau^n_{s,d+k})$ from $(U_{\bullet d+k}, U_{\bullet s})$ for all $s \in \{1, ..., d+k-1\}$
\STATE 12. $\mathcal{I} := \mathcal{I} \cup \{d + k\} \backslash \{i, j\}$
\STATE 13. $\hat{\EE} := \hat{\EE} \cup \{\{i, d+k\}, \{j, d+k\}\}$
\ENDFOR
\STATE ~

\renewcommand{\algorithmicensure}{\textbf{Output:}}
\ENSURE
\STATE $(\hat{\VV}, \hat{\EE}, \hat{\lambda})$
\STATE ~
\STATE (*) if $\mathcal{N}$ computed in this step is empty, stop 
\STATE (**) if $\mathcal{N}^1(d+k)$ computed in this step is empty, stop

\end{algorithmic}
\end{algorithm}

This section introduces another heterogeneous HAC estimator, which is based on the \emph{transformation using the diagonal of the AC} proposed in \cite{goreckihofertholena2014a} and is generalized to the heterogeneous case using the approach proposed here in Section \ref{sec:hetero_HAC_estim}.
The estimator is summarized by Algorithm \ref{alg:diag_hetero_HAC_estim} (note that the steps that are the same as in Algorithm \ref{alg:hetero_HAC_estim} are left without a comment).
There, the key step implementing the transformation is Step 10, where the diagonal transformation is used to compute a new vector of observations representing the copula corresponding to the fork $d+k$. 
Observe that the new vector is then used for the structure determination (Step 3), which is hence included directly in the algorithm (see Steps 2, 12 and 13).
Due to the transformation,  no aggregation functions, like $\avg$ or $g$ in Algorihtm \ref{alg:hetero_HAC_estim}, are needed as the parameter estimation (Step 5) and the GoF evaluation (Step 7) are always based only on one pair ($i$-th and $j$-th) of the data columns and not on (possibly) more than one pair (the pairs in $\downarrow\!(i) \times \downarrow\!(j)$ used in Algorithm \ref{alg:hetero_HAC_estim}). This, on the one hand, simplifies its implementation, but, on the other hand, does not assures the s.n.c.~even in the homogenous case (assuming $\#\FF = 1$), which thus has to be manually forced, as addressed in Section \ref{sec:homo_HAC_estim}.
In Step 5, we use the ML estimation, where $\textrm{MLE}( (U_{\bullet i}, U_{\bullet j}), \psi^{(a_l, \theta)})$ denotes the 2-AC ML parameter estimator based on the pair of data columns $(U_{\bullet i}, U_{\bullet j})$ (see Definition \ref{def:agg_s_n}) provided the 2-AC model is from the family $a_l$.
We choose this estimator to form an alternative to the $\tau$-based estimator we use in Algorithm \ref{alg:hetero_HAC_estim}.
Of course, other AC estimators can be used, however, note that these two (the ML and the $\tau$-based) estimators have shown the best efficiency in the experimental evaluation reported in \cite{Hofert13}.
%Finally, it is important to note that Theorems \ref{thm:alg_returns_HAC}, \ref{thm:L12} and \ref{thm:L123} also apply to this new algorithm.

\end{appendices}

\end{document}